\documentclass{icrc29}
\usepackage{graphicx,amssymb,amsmath,times,floatflt,wrapfig,epsfig}
\begin{document}

%
%

\begin{center}
{\Large \bf The IceCube Collaboration:  \\
        contributions to the\\ 
29$^{th}$ International Cosmic Ray Conference (ICRC 2005), \\
Pune, India, Aug. 2005}

\vskip 0.5 cm
{\large \bf The IceCube Collaboration}

{A.~Achterberg$^{t}$, 
M.~Ackermann$^{d}$, 
J.~Ahrens$^{k}$, 
D.W.~Atlee$^{h}$, 
J.N.~Bahcall*$^{u}$, 
X.~Bai$^{a}$, 
B.~Baret$^{s}$, 
M.~Bartelt$^{n}$, 
R.~Bay$^{i}$, 
S.W.~Barwick$^{j}$
% ...
\\
${(*)}$ Deceased\\
${(a)}$ Bartol Research Institute, University of Delaware, Newark, DE 19716 USA\\
${(b)}$ Department of Physics, University of Wuppertal, D-42119 Wuppertal, Germany\\
${(c)}$ Universit\'e Libre de Bruxelles, Science Faculty CP230, B-1050 Brussels, Belgium\\
${(d)}$ DESY, D-15735, Zeuthen, Germany\\
${(e)}$ Blackett Laboratory, Imperial College, London SW7 2BW, UK\\
${(f)}$ Dept. of Technology, Kalmar University, S-39182 Kalmar, Sweden\\
${(g)}$ Lawrence Berkeley National Laboratory, Berkeley, CA 94720, USA\\
${(h)}$ Dept. of Physics, Pennsylvania State University, University Park, PA 16802, USA\\
${(i)}$ Dept. of Physics, University of California, Berkeley, CA 94720, USA\\
${(j)}$ Dept. of Physics and Astronomy, University of California, Irvine, CA 92697, USA\\
${(k)}$ Institute of Physics, University of Mainz, Staudinger Weg 7, D-55099 Mainz, Germany\\
${(l)}$ Dept. of Physics, University of Maryland, College Park, MD 20742, USA\\
${(m)}$ University of Mons-Hainaut, 7000 Mons, Belgium\\
${(n)}$ Dept. of Physics, Universit\"at Dortmund, D-44221 Dortmund, Germany\\
${(o)}$ Dept. of Physics, University of Wisconsin, Madison, WI 53706, USA\\
${(p)}$ Dept. of Physics, University of Wisconsin, River Falls, WI 54022, USA\\
${(q)}$ Division of High Energy Physics, Uppsala University, S-75121 Uppsala, Sweden\\
${(r)}$ Dept. of Physics, Stockholm University, SE-10691 Stockholm, Sweden\\
${(s)}$ Vrije Universiteit Brussel, Dienst ELEM, B-1050 Brussels, Belgium\\
${(t)}$ Dept. of Physics and Astronomy, Utrecht University, NL-3584 CC Utrecht, NL\\
${(u)}$ Institute for Advanced Study, Princeton, NJ 08540, USA\\
${(v)}$ Dept. of Physics and Astronomy, University of Kansas, Lawrence, KS 66045, USA\\
${(w)}$ Dept. of Physics, Southern University, Baton Rouge, LA 70813, USA\\
${(x)}$ Dept. of Astronomy, University of Wisconsin, Madison, WI 53706, USA\\
${(y)}$ Dept. of Physics, Chiba University, Chiba 263-8522 Japan\\
${(z)}$ CTSPS, Clark-Atlanta University, Atlanta, GA 30314, USA\\
${(aa)}$ Dept. of Astronomy and Astrophysics, Pennsylvania State University, University Park, PA 16802, USA\\
${(ab)}$ Institut f\"ur Physik, Humboldt Universit\"at zu Berlin, D-12489 Berlin, Germany\\
${(ac)}$ Universit\`a di Bari, Dipartimento di Fisica, I-70126, Bari, Italy}
\end{center}
\vskip 2.cm
\section*{Acknowledgments}

{We acknowledge the support of the following agencies: National
Science Foundation--Office of Polar Programs, National Science
Foundation--Physics Division, University of Wisconsin Alumni Research
Foundation, Department of Energy, and National Energy Research
Scientific Computing Center (supported by the Office of Energy
Research of the Department of Energy), the NSF-supported TeraGrid 
systems at the San Diego Supercomputer
Center (SDSC), and the National Center for Supercomputing Applications
(NCSA); 
Swedish Research Council, 
Swedish Polar Research
Secretariat, and Knut and Alice Wallenberg Foundation, Sweden; German
Ministry for Education and Research, Deutsche Forschungsgemeinschaft
(DFG), Germany; Fund for Scientific Research (FNRS-FWO), Flanders
Institute to encourage scientific and technological research in
industry (IWT), and Belgian Federal Office for Scientific, Technical
and Cultural affairs (OSTC).}

\newpage

\begin{center}
{\large \bf Table of contents}
\end{center}
\begin{enumerate}
\item J.~Hodges for the IceCube Collaboration, {\it Search for Diffuse Flux of Extraterrestrial Muon Neutrinos using AMANDA-II Data from 2000 to 2003}
\item K.~M\"unich for the IceCube Collaboration, {\it Search for a diffuse flux of non-terrestrial muon neutrinos with the AMANDA detector}
\item L.~Gerhardt for the IceCube Collaboration, {\it Sensitivity of AMANDA-II to UHE Neutrinos}
\item M.~Ackermann and E.~Bernardini for the IceCube Collaboration, {\it An investigation of seasonal variations in the atmospheric neutrino rate with the AMANDA-II neutrino telescope}
\item M.~Ackermann, E.~Bernardini and T.~Hauschildt for the IceCube Collaboration, {\it Search for high energy neutrino point sources in the northern hemisphere with the AMANDA-II neutrino telescope}
\item M.~Ackermann, E.~Bernardini, T.~Hauschildt and E.~Resconi, {\it Multiwavelength comparison of selected neutrino point source candidates}
\item A.~Gro{\ss} and T.~Messarius for the IceCube Collaboration, {\it A source stacking analysis of AGN as neutrino point source candidates  with AMANDA}
\item J.~L.~ Kelley for the IceCube Collaboration, {\it A Search for High-energy Muon Neutrinos from the Galactic Plane with AMANDA-II}
\item K.~Kuehn for the IceCube Collaboration and the IPN Collaboration, {\it The Search for Neutrinos from Gamma-Ray Bursts with AMANDA}
\item M.~Stamatikos, J.~Kurtzweil and M.~J.~Clarke for the IceCube Collaboration, {\it 
Probing for Leptonic Signatures from GRB030329 with AMANDA-II}
\item B.~Hughey, I.~Taboada for the IceCube Collaboration, {\it Neutrino-Induced Cascades From GRBs With AMANDA-II}
\item D.~Hubert, A.~Davour, C.~de los Heros for the IceCube Collaboration, {\it Search for neutralino dark matter with the AMANDA neutrino detector}
\item  A.~Silvestri for the IceCube Collaboration, {\it Performance of AMANDA-II using Transient Waveform Recorders}
\item T.~Messarius for the IceCube Collaboration, {\it A software trigger for the AMANDA neutrino detector}
\item T.~K.~Gaisser for the IceCube Collaboration, {\it Air showers with IceCube: First Engineering Data}
\item D.~Chirkin for the IceCube Collaboration, {\it IceCube: Initial Performance}
\item H.~Miyamoto for the IceCube Collaboration, {\it Calibration and characterization of photomultiplier tubes
of the IceCube neutrino detector}
\item J.~A.~Vandenbroucke for the IceCube Collaboration, {\it Simulation of a Hybrid Optical/Radio/Acoustic Extension to IceCube for EeV Neutrino Detection}
\end{enumerate}

\newpage

\title[Search for Diffuse Flux of Extraterrestrial Muon Neutrinos using AMANDA-II Data from 2000 to 2003]{Search for Diffuse Flux of Extraterrestrial Muon Neutrinos using AMANDA-II Data from 2000 to 2003}
\author[J.~Hodges for the IceCube Collaboration] {J.~Hodges$^a$ for the IceCube Collaboration\\ 
       (a) Physics Dept. University of Wisconsin. Madison, WI 53706, USA \\
        }
\presenter{Presenter: J.~Hodges (hodges@icecube.wisc.edu), \
usa-hodges-J-abs1-og25-oral}

\maketitle

\begin{abstract}
The detection of extraterrestrial neutrinos would confirm predictions that
hadronic processes are occurring in high energy astrophysical sources such
as active galactic nuclei and gamma-ray bursters. Many models predict a
diffuse background flux of neutrinos that is within reach of the AMANDA-II
detector. Four years of experimental data (2000 to 2003) have been combined
to search for a diffuse flux of neutrinos assumed to follow an E$^{-2}$
energy dependence.  Event quality cuts and an energy cut were applied to
separate the signal hypothesis from the background of cosmic ray muons and
atmospheric neutrinos. The preliminary results of this four-year analysis
will be presented.
\end{abstract}

\sloppy

\section{Introduction and Motivation}

Currently most of the information on the universe comes from photons, however 
the detection of cosmic neutrinos would provide a new picture of distant 
regions of space. Neutrinos 
travel in straight lines, undeflected by magnetic fields. They interact rarely, 
making detection challenging. However, if detected, the direction of an 
extraterrestrial neutrino would point to the particle's origin. 
Many theoretical models predict that neutrinos originate in hadronic processes 
within high energy astrophysical sources such as active galactic nuclei and gamma-ray bursters.

This analysis searches for neutrinos from unresolved sources. This 
search for a diffuse flux of extraterrestrial neutrinos assumes 
an E$^{-2}$ energy spectrum. Although other models will be assumed in future work, 
this energy spectrum assumption is based on the theory of particle acceleration in 
strong shocks \cite{longair}. Protons experience first-order Fermi acceleration and interact 
with protons and photons. The resulting pions are thought to decay into neutrinos 
that keep the same energy spectrum as the primary \cite{hephalzen}. 

The AMANDA-II detector is a collection of 19 strings buried 
in the ice at the South Pole \cite{recopaper}. A total of 677 optical modules are attached to 
these strings between the depths of 1500-2000 m. Each optical module consists 
of a photomultiplier tube surrounded by a pressure-resistant glass sphere. 
AMANDA-II has been operating since 2000.

Downgoing muons created when cosmic rays interact in the atmosphere trigger the 
AMANDA-II detector at the rate of 80 Hz. This saturates any possible extraterrestrial signal 
that might be seen from the Southern Hemisphere. Hence, upgoing events travelling from the 
Northern Hemisphere to the detector are selected for this analysis. 
In this way, the Earth acts as a filter against cosmic ray 
muons \cite{nature}. Sky coverage is restricted to 2$\pi$ sr. However, above 
the PeV range, the field of view is reduced due to neutrino absorption in the Earth 
\cite{recopaper}.

Muons, created from interacting muon neutrinos, travel long distances in the ice while emitting 
Cherenkov light. The muon tracks are reconstructed from the detection times with 
a median space angle resolution of $2^{o}$ \cite{recopaper} when events from the 
highest cut selection are used. All neutrino flavors can cause hadronic 
or electromagnetic cascades which appear as a spherical point source of light, however 
they will not be considered here.

\section{Backgrounds}

The analysis involved the simulation of several different classes of background events.
Monte Carlo simulation of neutrinos in the ice was performed 
assuming an $E^{-1}$ spectrum. The spectrum was reweighted to model an $E^{-2}$ extraterrestrial 
neutrino signal with a $\nu_\mu$ test flux of $E^{2}dN/dE = 1\times$10$^{-6}$\;GeV~cm$^{-2}$~s$^{-1}$~sr$^{-1}$. 
Atmospheric neutrinos were simulated by reweighting the neutrino events to a 
steeper $E^{-3.7}$ spectrum. This spectral dependence is common to both atmospheric 
muons and neutrinos because they are both produced by cosmic ray interactions in the 
atmosphere. 

Sixty-three days of downgoing atmospheric muons were simulated 
with CORSIKA \cite{corsika}. Downgoing atmospheric muons can reach 
AMANDA depth, however they cannot 
penetrate the Earth from the other hemisphere. In contrast, atmospheric neutrinos from 
the Northern Hemisphere can reach the detector. Hence, atmospheric muons can be 
rejected with directional cuts, but atmospheric neutrinos cannot be distinguished from 
signal in this way. However, because the signal has a harder energy spectrum, both atmospheric muons 
and neutrinos can be separated from signal by energy-based cuts. 

It is also possible that two downgoing muons from independent cosmic ray interactions 
may occur in the detector within the same detector trigger window. If this occurs, 
the software will have to guess one incidence direction for the pattern of light 
produced by two different tracks. The resulting reconstructed direction may be 
upgoing. These events, known as coincident muons, were simulated for 826 days of livetime.

\section{Analysis Optimization and Sensitivity}

The 2000 to 2003 search encompassed 807 days of detector livetime.
As will be described below, the number of optical modules recording photons was used as an 
energy-related observable to distinguish extraterrestrial neutrinos from atmospheric 
backgrounds \cite{jodithesis}. Analysis cuts were optimized by studying the signal and 
background Monte Carlo. In order to satisfy blindness requirements, the Monte Carlo and 
data were checked for agreement with events triggering less than 80 optical modules 
(NChannel $<80$), hence leaving the high energy data unbiased.

The analysis began with 7.1$\times10^{9}$ data events which triggered the detector during 
this time. Most of these were downgoing muons. All data 
events underwent an initial track reconstruction in which the software picked a direction of 
the particle based on the pattern of light the detector recorded. All events with reconstructed 
zenith angles between $0^{o}$ (travelling straight down) and $80^{o}$ degrees (just above 
the horizon) were removed.

\begin{figure}[h]
\begin{center}
\mbox{
\includegraphics*[width=0.41\textwidth]{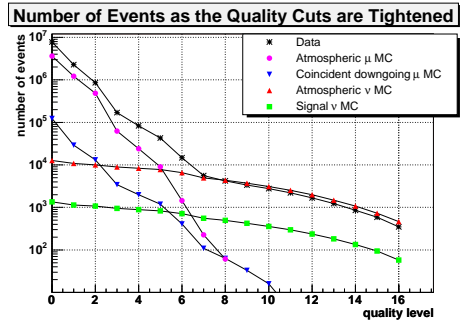}}
\caption{\label{fig:counts} Above: Number of events as the quality cuts are 
tightened for 807 days of detector livetime. 
Cuts become increasingly tighter to the right. The quality cuts were 
aimed at removing the downgoing atmospheric muons and coincident atmospheric 
muons from the data. Only events with less than 80 optical modules triggered are shown.}
\end{center}
\end{figure}

More computer-intensive track reconstructions were performed on the remaining events. 
Although zenith angle cuts required the events to be upgoing, many downgoing atmospheric 
muons were misreconstructed as upgoing and remained in the sample. To separate these events 
from the expected atmospheric and extraterrestrial neutrino signals, events were removed 
if they did not show the signature of a muon neutrino in the detector. Long muon tracks 
smoothly emit Cherenkov light. High quality events cause many hits that arrive close to 
the calculated time for unscattered photons. These hits must be spread evenly along the track and 
the log likelihood that the track is upgoing rather than downgoing must be high. As the 
requirements for these parameters were tightened, the number of data events began to look 
increasingly like the atmospheric and extraterrestrial signal Monte Carlo and less like 
the downgoing atmospheric muon simulation (see figure~\ref{fig:counts}). 
The quality cuts were optimized 
to reject the single and coincident muon backgrounds while preserving the expected signal. 
Quality cut level 11 was chosen to define the sample used for analysis. After a zenith angle 
cut at $80^{o}$ degrees, 7,769,850 data events remained in the low-energy (NChannel $<80$) 
sample, but the event quality cuts reduced this to 2207 data events.

\begin{figure}[h]
\begin{center}
\mbox{
\includegraphics*[width=0.42\textwidth]{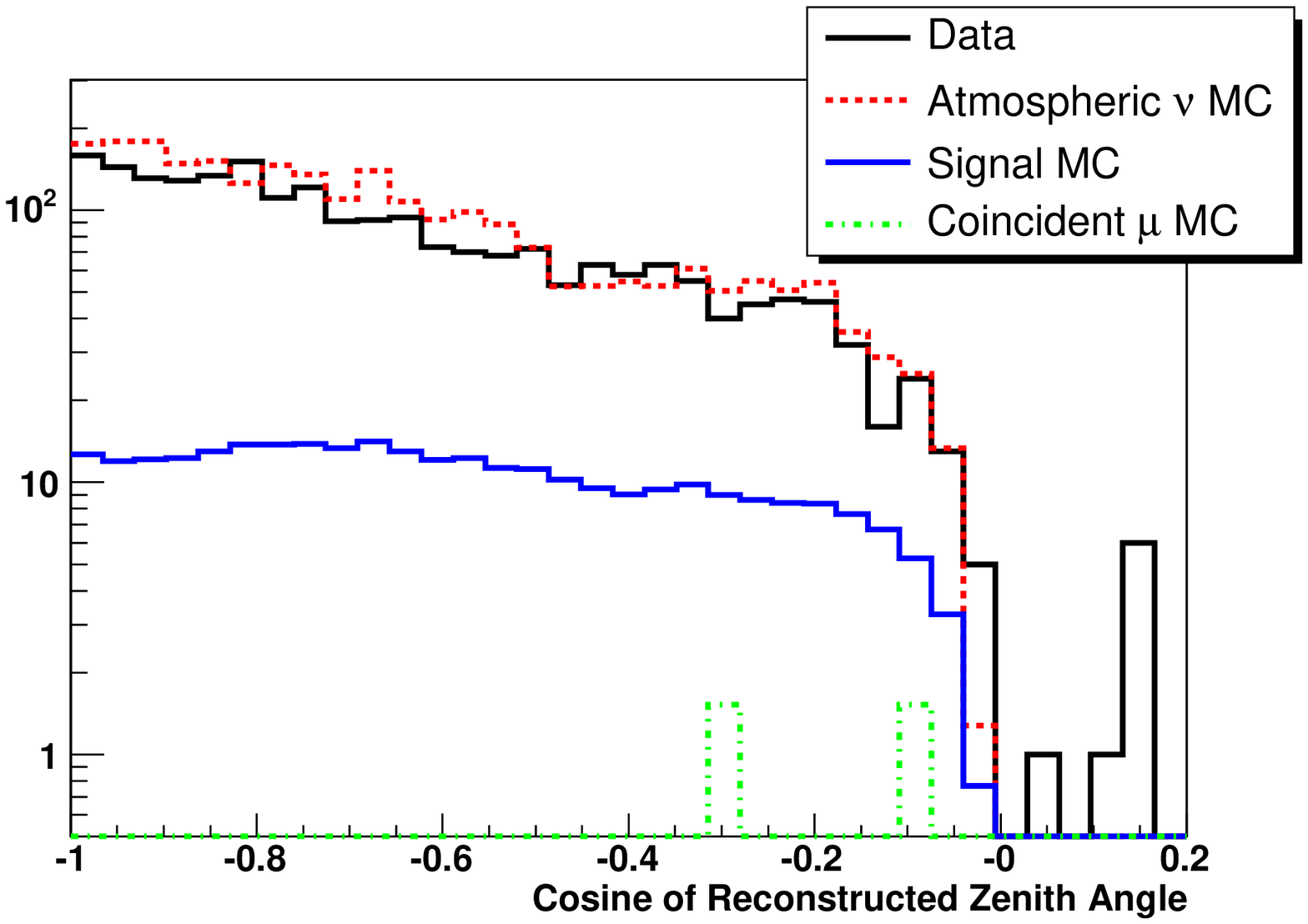}
\includegraphics*[width=0.42\textwidth]{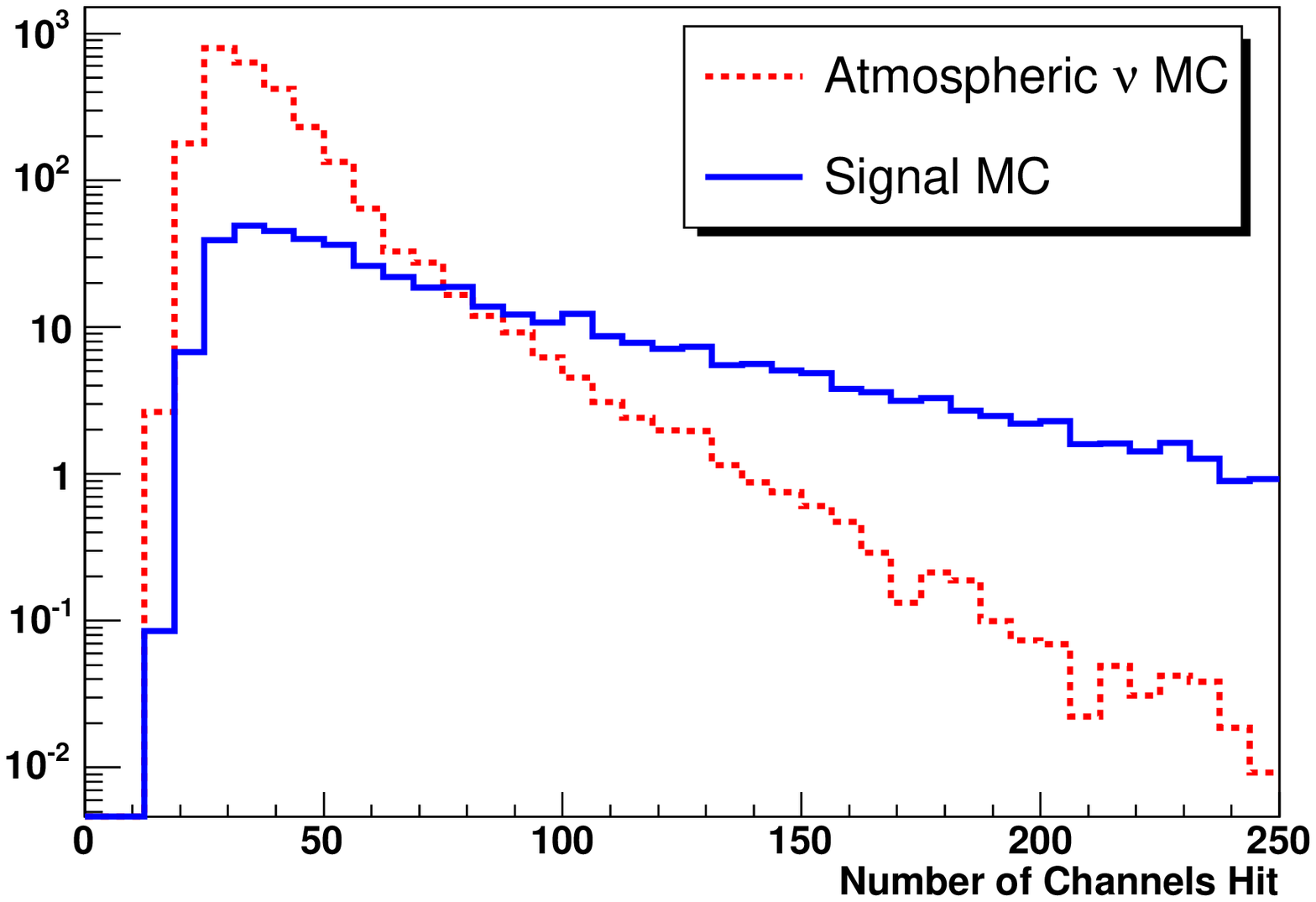}}
\caption{\label{fig:zenith} Left: Reconstructed zenith angle. The cosine of the 
zenith angle is shown for all low energy data and Monte Carlo. The number of events 
for 807 days of livetime is shown on the y-axis before the energy cut. After quality cuts, the 
data and predicted background show good agreement. At this 
high quality level, the downgoing atmospheric muons and coincident 
downgoing muons are nearly entirely removed. Right: Number of channels (optical modules) hit for
a signal test flux of $1\times$10$^{-6}$\;GeV~cm$^{-2}$~s$^{-1}$~sr$^{-1}$.
This energy related variable is used to separate signal neutrinos from the atmospheric 
$\nu_\mu$ background. }
\end{center}
\end{figure}

\begin{figure}[h]
\begin{center}
\mbox{
\includegraphics*[width=0.5\textwidth]{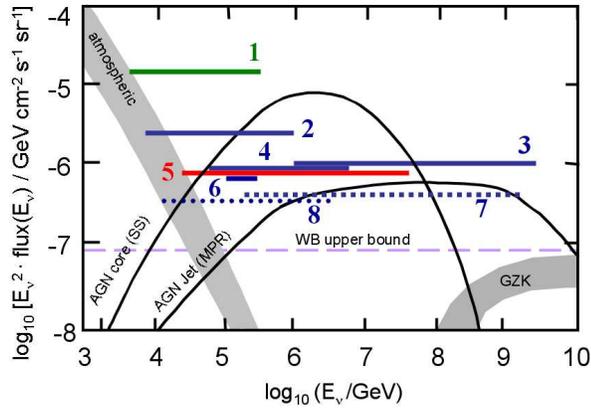}}
\caption{\label{fig:limits} All-flavor neutrino limits and sensitivity on 
an $E^{2}\frac{dN}{dE}$ plot. Neutrino oscillations are assumed. (1) The MACRO $\nu_\mu$ 
analysis for 5.8 years (limit multiplied by three for oscillations) \cite{macro}. 
(2) The AMANDA-B10 $\nu_\mu$ analysis from 1997 (multiplied by three for oscillations) \cite{97limit}. 
(3) AMANDA-B10 ultra-high energy neutrinos of all flavors 1997 \cite{uhe}. (4) AMANDA-II 
all-flavor cascade limit from 
2000 \cite{cascades}. (5) Baikal cascades 1998 - 2003 \cite{baikal}. (6) The preliminary results of 
the 2000 AMANDA-II $\nu_\mu$ analysis (multiplied by three for oscillations). The limit is derived 
after unfolding the atmospheric neutrino spectrum. (7) The sensitivity for 
AMANDA-II ultra-high energy neutrinos of all flavors 2000. 
(8) 2000 to 2003 AMANDA-II $\nu_\mu$ sensitivity 
(multiplied by three for oscillations).}
\end{center}
\end{figure}

The Feldman-Cousins method for calculating the 
average upper limit was applied \cite{feldcous}. The Model Rejection Factor is defined as 
the average upper limit divided by the number of predicted signal events \cite{mrp} for a 
$\nu_\mu$ signal test flux $E^{2}dN/dE = 1\times$10$^{-6}$\;GeV~cm$^{-2}$~s$^{-1}$~sr$^{-1}$. 
Using the simulation 
of signal and background, the Model Rejection Factor was calculated as a function of the 
number of optical modules hit in the detector. The minimum Model Rejection Factor, which 
indicates the best placement of the NChannel cut to separate signal from background, 
occurs for 
NChannel $\ge100$ (see figure~\ref{fig:zenith}). Above this cut, the 
expected 
background of 16.2 atmospheric neutrinos 
leads to an average upper limit of 8.19. Dividing this by the expected signal for the given 
test flux (86.3 events) leads to a final sensitivity on the $\nu_\mu$ flux of 
$1\times$10$^{-6}\times8.19/86.3$\;GeV~cm$^{-2}$~s$^{-1}$~sr$^{-1}$, or 
$9.5\times$10$^{-8}$\;GeV~cm$^{-2}$~s$^{-1}$~sr$^{-1}$. 
The sensitivity (multiplied by three for oscillations) is shown in 
figure~\ref{fig:limits} in relation to several 
other models and analyses. The signal Monte Carlo events that populate the final data set have true 
neutrino energies between 13 TeV and 3.2 PeV (90\% region).

\section{Conclusions}

With the cuts established and sensitivity determined, the high energy data can be 
studied. Limits obtained from the analysis of the complete data set will be presented at 
the conference. 

The sensitivity of this four-year analysis is improved by a factor of nine over the limit 
previously set for one year of experimental data (1997). Limits are closing in on the 
Waxman-Bahcall bound. In the future, several other models for extraterrestrial neutrino 
production will also be tested.

\newpage
\setcounter{section}{0}

%
\def\funit{GeV~cm$^{-2}$~s$^{-1}$~sr$^{-1}$}

\title[Search for a diffuse flux of non-terrestrial muon neutrinos with the AMANDA detector]{Search for a diffuse flux of non-terrestrial muon neutrinos with the AMANDA detector}

\author[K. M\"unich for the IceCube Collaboration] {K. M\"unich$^a$ for the IceCube collaboration\\
        (a) Institute for Physics, University of Dortmund, D-44221 Dortmund,
            Germany
        }
\maketitle
\section*{\Large Search for a diffuse flux of non-terrestrial muon neutrinos with the AMANDA detector}

\vskip 0.2cm
{\large K. M\"unich$^{\it a}$ for the IceCube Collaboration} \\
{\it (a) Institute for Physics, University of Dortmund, D-44221 Dortmund,
            Germany}
\vskip 0.2cm
{\large Presenter: K. M\"unich (muenich@physik.uni-dortmund.de), \  
ger-muenich-K-abs1-og25-oral}

\begin{abstract}
Over the past decade, many extragalactic source types have been suggested
as potential sources for the ultrahigh energy cosmic ray flux. Assuming
hadronic particle acceleration in these sources, a diffuse neutrino flux may
be produced along with the charged cosmic ray component. In the presence of a
high background of atmospheric neutrinos, no extragalactic neutrino signal has
been observed yet. In this paper, a new analysis to investigate with the
Antarctic Muon And Neutrino Detector Array (AMANDA-II) a possible extragalactic
component in addition to the atmospheric neutrino flux is presented. The
analysis is based on the year 2000 data. Using an unfolding method, it
is shown that the spectrum follows the atmospheric neutrino flux prediction
\cite{Volkova_80} up to energies above 100~GeV. A limit on the
extraterrestrial contribution is obtained from the application of a confidence
interval construction to the unfolding problem. 
\end{abstract}

\section{Introduction}
Neutrino-astrophysics has enlarged over the last years the knowledge 
of 
neutrinos and their properties. Current experiments are able
to measure the neutrino flux from the sun as well as the flux that is produced
by cosmic rays interacting with the atmosphere. The aim of high energy
neutrino experiments such as AMANDA \cite{Ahrens_03,Messarius_05,Amanda} is to observe an extraterrestrial
component of the neutrino spectrum. The AMANDA detector, located at the
geographical South Pole, uses the ice as the active volume.\\ 
Due to the high atmospheric neutrino flux at
energies $E_{\nu}>50$~GeV, a non-atmospheric component has not yet been observed.
The atmospheric flux decreases roughly with $E_{\nu}^{-3.7}$ as opposed
to the extragalactic contribution, which is expected to be around 1.7 powers
flatter, $E_{\nu}^{-2}$. Thus, it is predicted that an additional contribution
should become dominant at higher energies. The exact energy at which
the extraterrestrial flux exceeds the prediction of the atmospheric one is not
known due to the uncertainties in the source properties which would determine
the normalization of the neutrino flux. \\
The diffuse neutrino flux presented is measured with a combination of a neural
network and a regularized unfolding \cite{Blobel_84} as described in \cite{Geenen_03}. Since the
measured neutrino flux corresponds with the expectation of the atmospheric
neutrino flux up to an energy of 100~TeV, the question of additional
constituents and their exclusion has to be investigated. This paper describes
how an upper limit to the neutrino flux from extraterrestrial sources can be
obtained. It is shown how the unified approach of Feldman \&
Cousins can be applied to an unfolding problem to set a 90\% confidence
belt. Taking into account the statistical behavior of individual events, the probability
density functions are calculated using large MC statistics.
Finally, a limit on the diffuse muon neutrino and antineutrino flux from extragalactic sources is
presented. This limit gives the most restrictive estimate of
an upper bound of the neutrino flux among currently existing experiments.

\section{Method to obtain a 90\% confidence belt}

The neutrino energy spectrum is dominated by the
background of atmospheric neutrinos. By means of MC studies of
atmospheric neutrinos the number of events per energy interval can be estimated.
The lower energy threshold of examined events
for a potential neutrino signal can be optimized \cite{Hill_03}.
This leads to a limit on the non-atmospheric
neutrino flux using the number of measured events
above the optimized threshold.
The probability $P$ to measure $n$ events in a certain energy bin for a given mean
signal $\mu$ is calculated by using large MC statistics. $P$ is
also called probability density function, pdf. Its calculation is described
in the following paragraph.

\begin{figure}[h]
  \begin{center}
    \includegraphics[height=49mm]{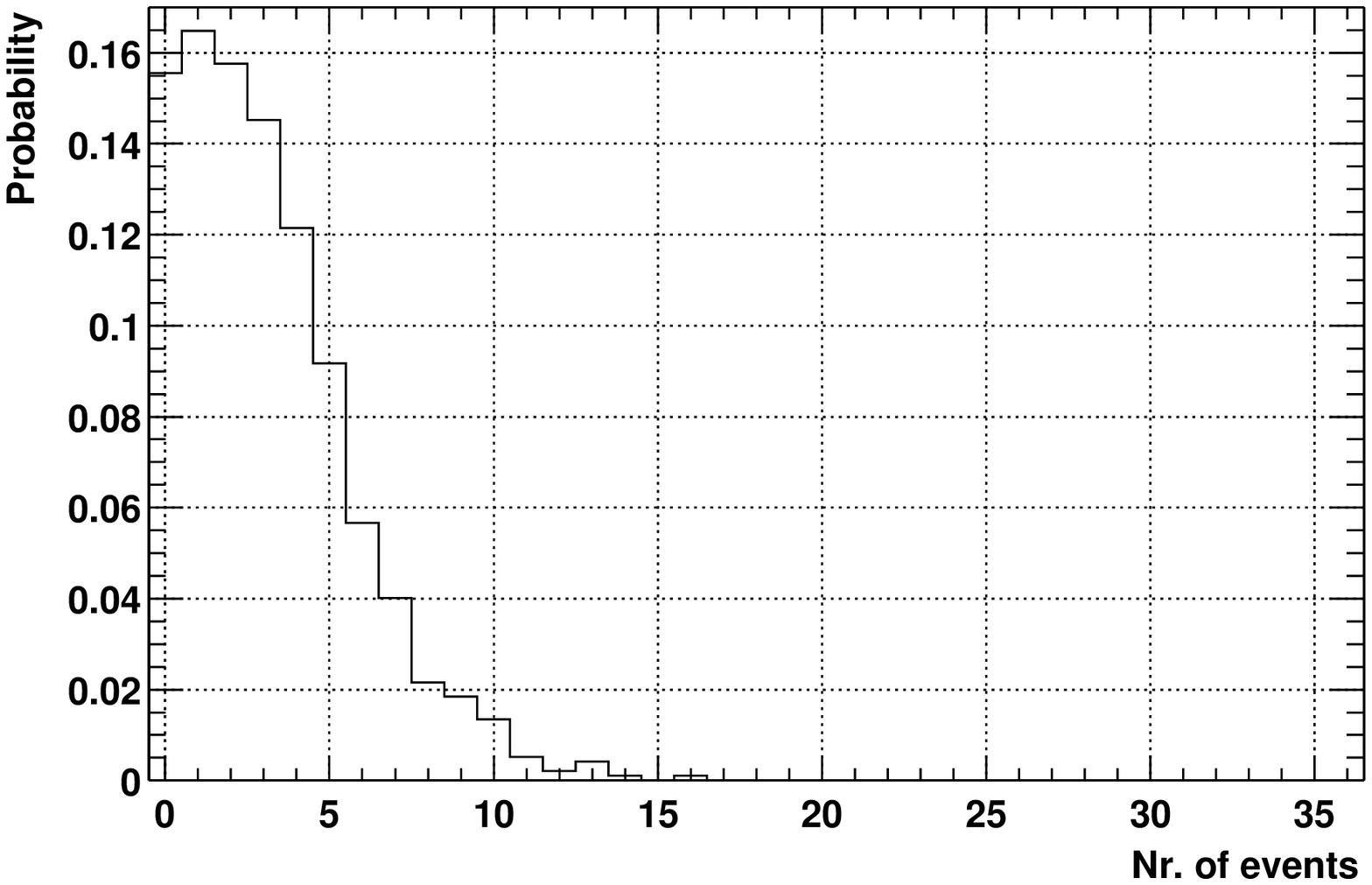}
    \includegraphics[height=49mm]{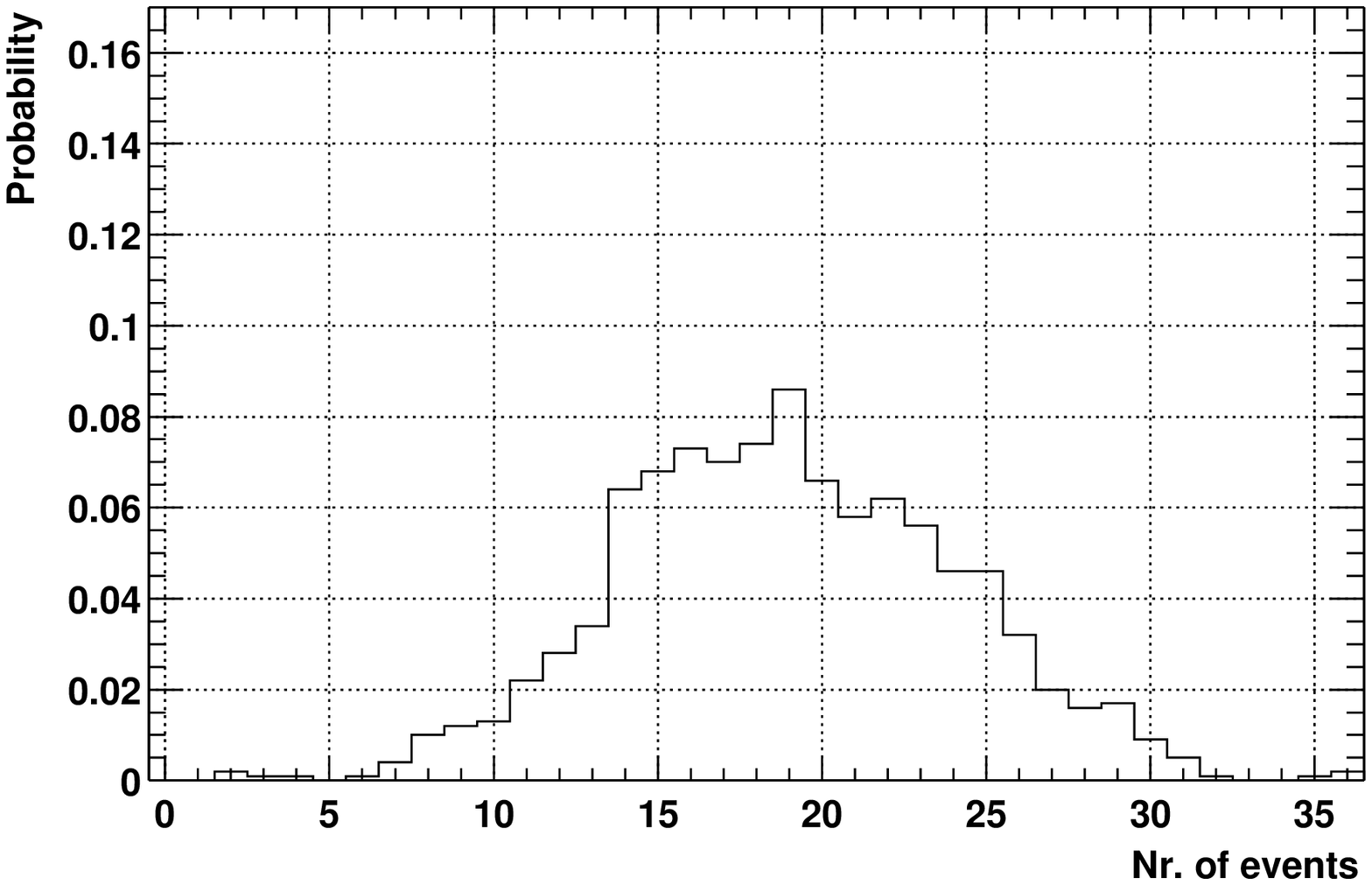}\\
   {\small (a)\hspace{7cm}(b)}
   \end{center}
   \vspace{-0.2pc}
   \caption{\label {pdf_distribution} Probability density function for two
    different signal contributions. $(a): 2.0 \cdot 10^{-7}$~\funit,\newline $(b): 10^{-6}$~\funit.}
\end{figure}

For the year 2000, 21 different signal contributions
ranging from $10^{-8}$~\funit\ to\newline $10^{-6}$~\funit\ are
used. The signal contribution $\mu$ is equal to the flux $\phi$ multiplied by $E^2$.
Each signal contribution $\mu$ is represented by 1000 one-year MC
experiments which are altogether equivalent to a data taking of 21000 years. The energy
is reconstructed using a combination of neural network and regularized
unfolding \cite{Geenen_03}. The resulting energy distribution is evaluated
for each of the 1000 MC experiments per fixed signal contribution,
resulting in 21000 energy distributions using all 21 signal
contributions. After applying an energy cut the event rate in the remaining
bins are summed up and histogramed. The normalized histograms give the
searched pdfs. In figure \ref{pdf_distribution} the pdf
for two different signal contributions is shown.

\begin{figure}[t]
  \begin{center}
    \includegraphics[height=50mm]{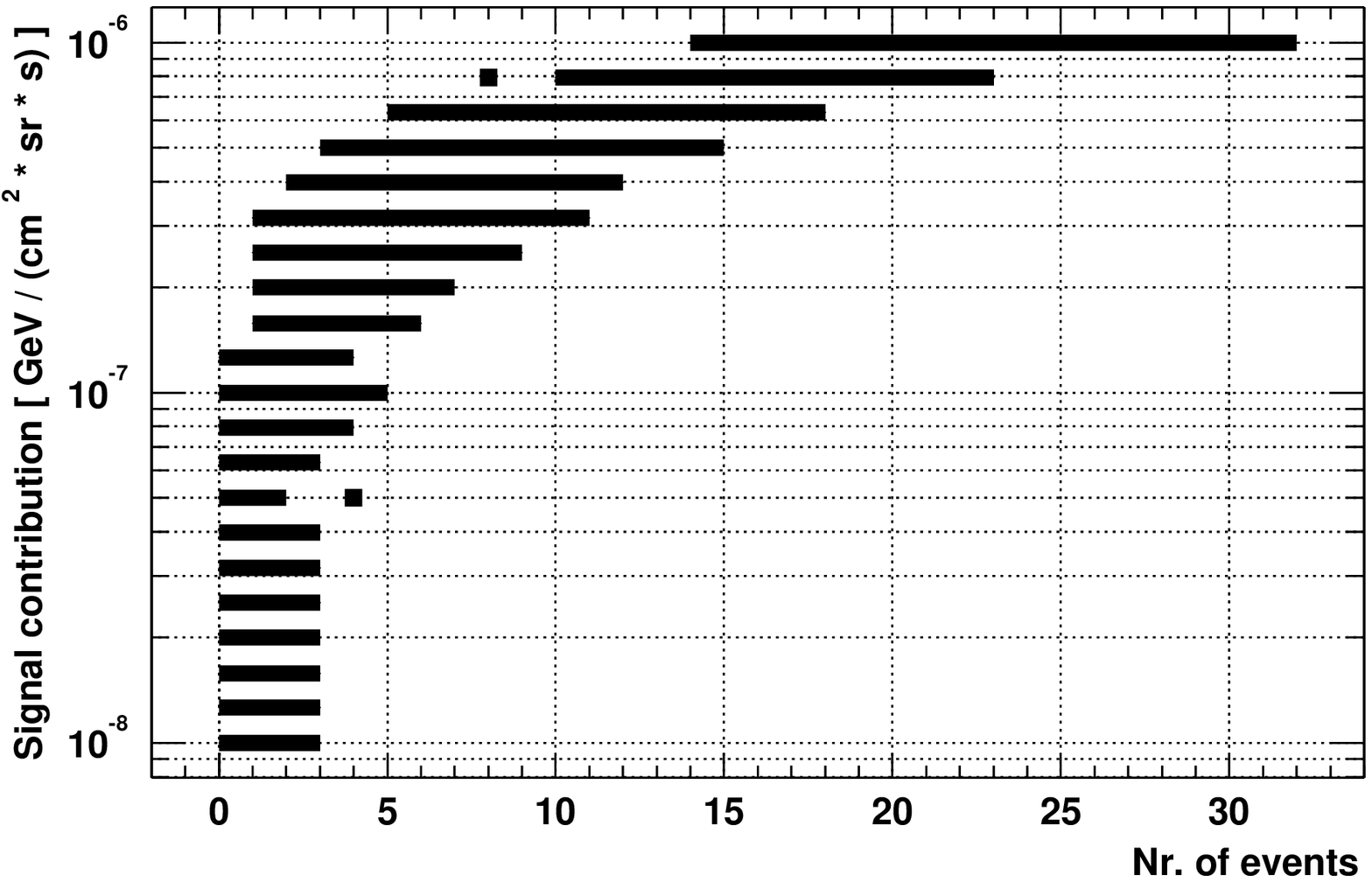}
    \includegraphics[height=50mm]{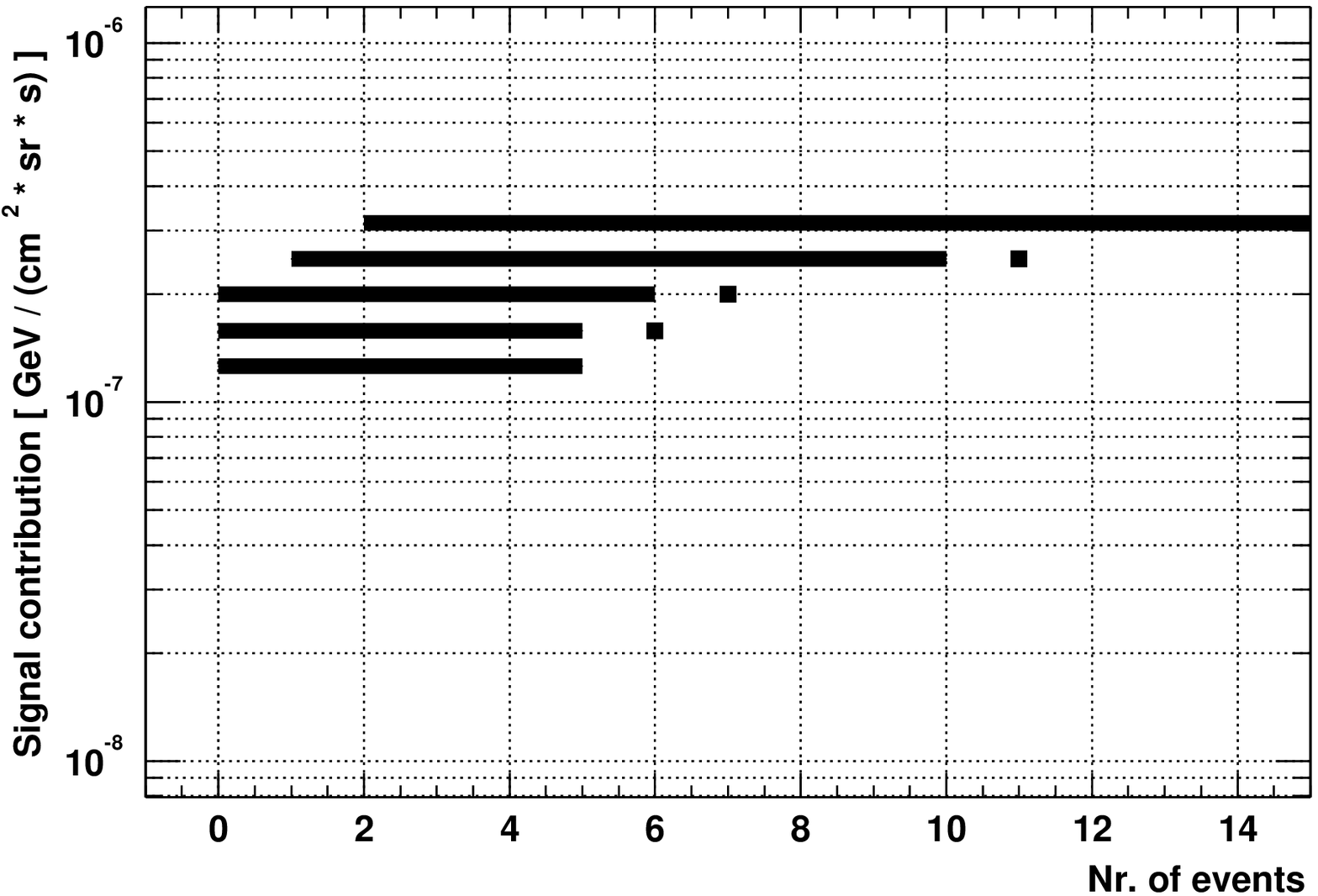}\\
   {\small (a)\hspace{7cm}(b)}
   \end{center}
   \vspace{-0.5pc}
   \caption{\label {conf_belt} Confidence belt in the signal
     contribution range of $(a):10^{-8}$\funit\ up to $10^{-6}$\funit\ and $(b):1.26 \cdot 10^{-7}$\funit\ up to $3.16
     \cdot 10^{-7}$\funit.}
\end{figure}

For generating a 90\% confidence belt the method described in the unified
approach by Feldman and Cousins~\cite{Feldman_98} is applied using the probability density functions described
above.
After reconstruction and unfolding of the energy, confidence belts for
different energy cuts are compared and the confidence belt for an energy cut
resulting in an energy range of
$100$~TeV $< E < 300$~TeV shows the best performance. The resulting
confidence belt for this cut is illustrated in figure \ref{conf_belt}(a). 

For data taken by AMANDA in the year 2000 optimized point source cuts \cite{Hauschild_03} and a zenith veto at 10 degrees below the horizon have been applied. The resulting sample consists of 570 neutrino events. With the method described above the energy is determined. 
Inspecting the energy distribution of the data leads to 0.36 events in the energy range of $100$~TeV $< E < 300$~TeV. Since the event numbers used for building the confidence belt displayed in figure \ref{conf_belt}(a) are integer, a limit for 0.36 events can only be derived using further interpolation methods.
To avoid this and to get a higher resolution the number of MC events 
have been enlarged by a factor of 10. This is done in the interesting signal contribution region from $1.26\cdot 10^{-7}$~\funit\ to  $3.16\cdot 10^{-7}$~\funit. With the resulting confidence belt presented in figure \ref{conf_belt}(b) a definite limit of $2.0\cdot10^{-7}$~\funit\ can be assigned to a event rate of 0.36.
 
\section{Discussion}
From the possible error contribution the systematic
uncertainties are dominating. The main contribution to the systematic
error is made by the uncertainty of the atmospheric neutrino flux 25\%, \cite{Wiebel_Sooth_98}.

Adding the smaller contributions as the uncertainty of the $\nu_{\mu}$
to $\mu$ cross section (ca. $10\%$) to this value
and including a maximal contamination by atmospheric neutrinos of the data set ($7\%$),
a total systematic error of 30\% has to be applied.\\
This leads to a limit for the AMANDA data of the year 2000 of 

$$\phi \cdot E^2 = 2.6 \cdot 10^{-7} GeV~cm^{-2}~s^{-1}~sr^{-1}.$$

\section{Conclusions}
\begin{figure}
  \begin{center}
    \includegraphics*[width=12cm]{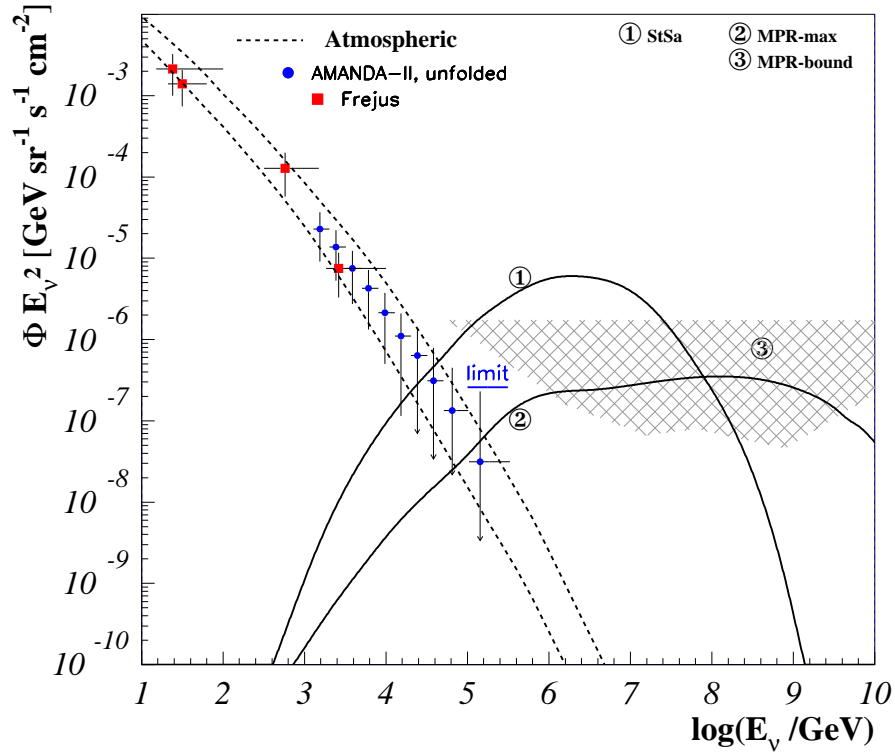}
   \end{center}
  \caption{Reconstructed neutrino spectra and resulting limit for the year
  2000 data compared with different flux models}
  \label{limit_plot}
\end{figure}

Fig.~\ref{limit_plot} shows the calculated spectrum and limit in the context
of different muon neutrino and anti-neutrino flux predictions. 
The unfolded neutrino spectrum (circles) is complementary to the
Frejus data \cite{Frejus_95} (squares) which are at lower energies. The black dashed lines
in this figure show the horizontal and vertical atmospheric neutrino flux. The
upper line represents the horizontal flux, while the prediction for the vertical flux
is given with the lower
line. The atmospheric flux spectrum above an energy of E~$>$~100~GeV is
parameterized according to Volkova \cite{Volkova_80}. Below this energy the
parameteri\-zation is given according to Honda et al. \cite{Honda_95}. The
reconstructed flux contains events from the lower hemisphere except events
very near to the horizon and is in good conformity with the atmospheric
prediction.
In addition to the unfolded flux, an upper limit on the extragalactic neutrino
signal of $2.6~\cdot~10^{-7}$~\funit\ is given. The limit clearly gives
restrictions on model~1 (StSa) \cite{Stecker_96}, assuming neutrino production in
$p\,\gamma$ interactions in AGN cores. This model with the parameterization as
given in~\cite{Stecker_96} can be excluded. Model~2 (MPR-max,
\cite{Mannheim_01}) represents the maximum neutrino flux from blazars in
photo-hadronic interactions and lies within the sensitivity range of AMANDA. In this
context, an upper bound on the
flux from these sources were estimated in \cite{Mannheim_01}, which is
indicated in the figure as
the shaded region (Model~3, MPR-bound). The horizontal line represents the
limit for sources that
are optically thick to $n\gamma$ interactions, $\tau_{n\gamma}>>1$, the lower
bound of the shaded region gives the bound for optically thin sources
$(\tau_{n\,\gamma}<1)$. In future analyses with a larger data set in AMANDA, it should be
possible to set limits lying within the shaded regions, so that the opacity of
the sources can be constrained.

\newpage

\setcounter{section}{0}
%
\title[Sensitivity of AMANDA-II to UHE Neutrinos]{Sensitivity of
  AMANDA-II to UHE Neutrinos}
\author[L. Gerhardt for the IceCube Collaboration] {L. Gerhardt$^a$ 
for the IceCube Collaboration \\
        (a) Department of Physics and Astronomy, University of California, Irvine, Irvine, CA USA \\
        }
\presenter{Presenter: Lisa Gerhardt (gerhardt@cosmic.ps.uci.edu), \  
usa-gerhardt-L-abs1-og25-oral}
\maketitle

\section*{\Large Sensitivity of AMANDA-II to UHE Neutrinos}

\vskip 0.1cm
{\large L. Gerhardt$^{\it a}$ for the IceCube Collaboration} \\
{\it (a) Department of Physics and Astronomy, University of California, Irvine, Irvine, CA USA}
\vskip 0.1cm
{\large Presenter: Lisa Gerhardt (gerhardt@cosmic.ps.uci.edu), \  
usa-gerhardt-L-abs1-og25-oral}

\begin{abstract}

The sensitivity of the AMANDA-II detector to ultra high energy (UHE)
neutrinos (energy greater than 10$^{6}$ GeV) is derived using data
collected during the year 2000. Due to absorption of UHE neutrinos in
the earth, the signal is concentrated at the horizon and has to be
separated from the background of large muon-bundles induced by cosmic ray air showers. This 
analysis 
leads to a sensitivity for an E$^{-2}$ all neutrino spectrum (assuming a 
1:1:1 flavor ratio) of 3.8 x 10$^{-7}$ cm$^{-2}$s$^{-1}$ sr$^{-1}$ GeV for an 
energy range between 1.8 x 10$^{5}$ GeV and 1.8 x 10$^{9}$
GeV. Sensitivites for five years of data taking and the
future IceCube array are given.

\end{abstract}
\section{Introduction}
AMANDA is a large volume neutrino telescope with the capability to search for neutrinos 
from astrophysical sources \cite{Casc-AMA}. In a previous publication \cite{UHE-AMA} it was shown 
that neutrino telescopes are able to search for UHE neutrinos (neutrinos with energy greater than 
10$^{6}$ GeV). UHE neutrinos are of interest because they are associated with the potential 
acceleration of hadrons by AGNs \cite{P95,St92,St96}, are produced by the decays of exotic objects 
such as topological defects \cite{Sigl98} or z-bursts \cite{Yosh98} and are guaranteed by-products of the
interaction of high energy cosmic rays with the cosmic microwave background \cite{Engel01}. 

Above 10$^{7}$ GeV the Earth is essentially opaque to 
neutrinos \cite{Klein}. This, combined with the limited overburden above AMANDA (approximately 
1.5 km), means that UHE neutrinos will be concentrated at the horizon. The background for this 
analysis consists of bundles of down-going, high energy muons from atmospheric showers. The muons 
from these bundles can spread over areas as large as 10$^{4}$ m$^{2}$. Separation of signal from the 
background takes advantage of the fact that signal events have a higher light density than background events, which 
causes multiple hits in multiple channels. Using this as well as the differences in geometrical acceptance and hit 
topology it is possible to remove almost all background while retaining a high sensitivity to signal. 
A previous 
analysis was performed using the inner ten strings of AMANDA (called AMANDA-B10) \cite{UHE-AMA}. 
This analysis uses 
all nineteen strings of AMANDA (called AMANDA-II, for description see \cite{Casc-AMA}).
Although the effective area of AMANDA-II is 
approximately the same as AMANDA-B10 for this analysis, the larger number of optical 
modules (OMs) offer improved background rejection leading to an improved sensitivity.

\section{Experimental and Simulated Data}
AMANDA-II collected 6.9 x 10$^{8}$ events between February and November of 2000, with an integrated 
lifetime of 173.5 days after retriggering and correcting for dead time and periods where the 
detector was unstable. Of this data 20\% was used to develop selection
criteria, while the rest, with a lifetime of 138.8 days, is set aside for the final analysis.
Two sets of cosmic ray air shower background events were generated
using CORSIKA \cite{CORSIKA}. One set uses composition and spectral
indices from \cite{Weibel}, i.e. the spectra follows approximately E$^{-3}$. 
In the other set, the statistical error and CPU time were reduced by biasing the Monte Carlo generation 
in both energy and composition (see \cite{UHE-AMA} for a full description). 
The UHE neutrinos were generated with energies between
10$^{3}$ GeV and 10$^{12}$ GeV using ANIS \cite{ANIS}. For more details on AMANDA simulation 
procedures see \cite{Casc-AMA,UHE-AMA}.

\begin{figure}
\begin{minipage}[t]{7.5cm}
\includegraphics[width=0.9\textwidth]{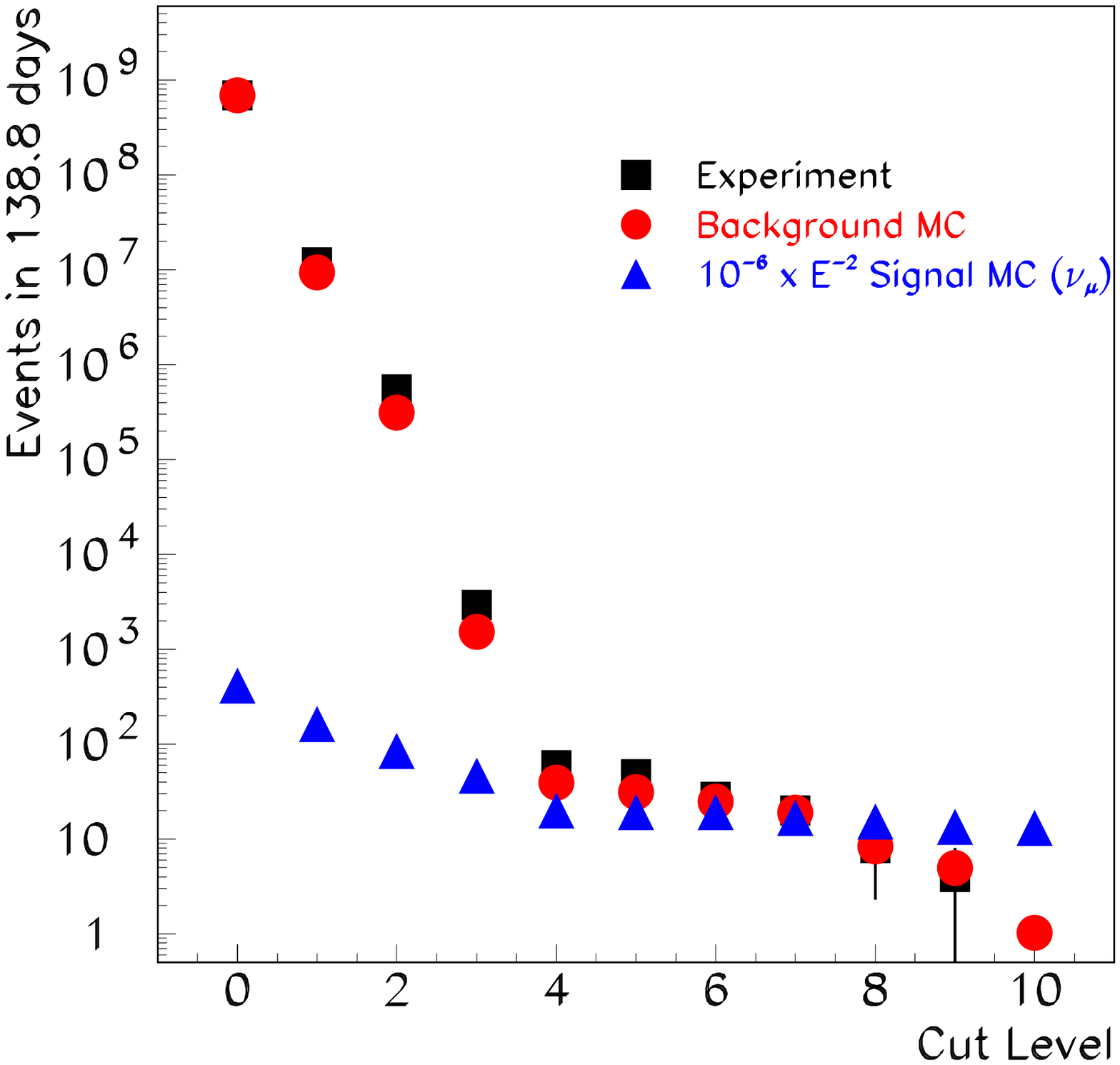}
\caption{\label{fig1c} The number of events passing the cuts as a function of cut 
number. The values for the experimental data are estimated from a 20\% 
subsample. Errors are statistical. The background MC has been scaled by a factor of 
1.24 so that the event rate agrees with the experiment at level 0. The signal MC 
is shown with a lower energy threshold of 10$^{5}$ GeV.}
\end{minipage} 
\hfill
\begin{minipage}[t]{7.5cm}
\includegraphics[width=0.9\textwidth]{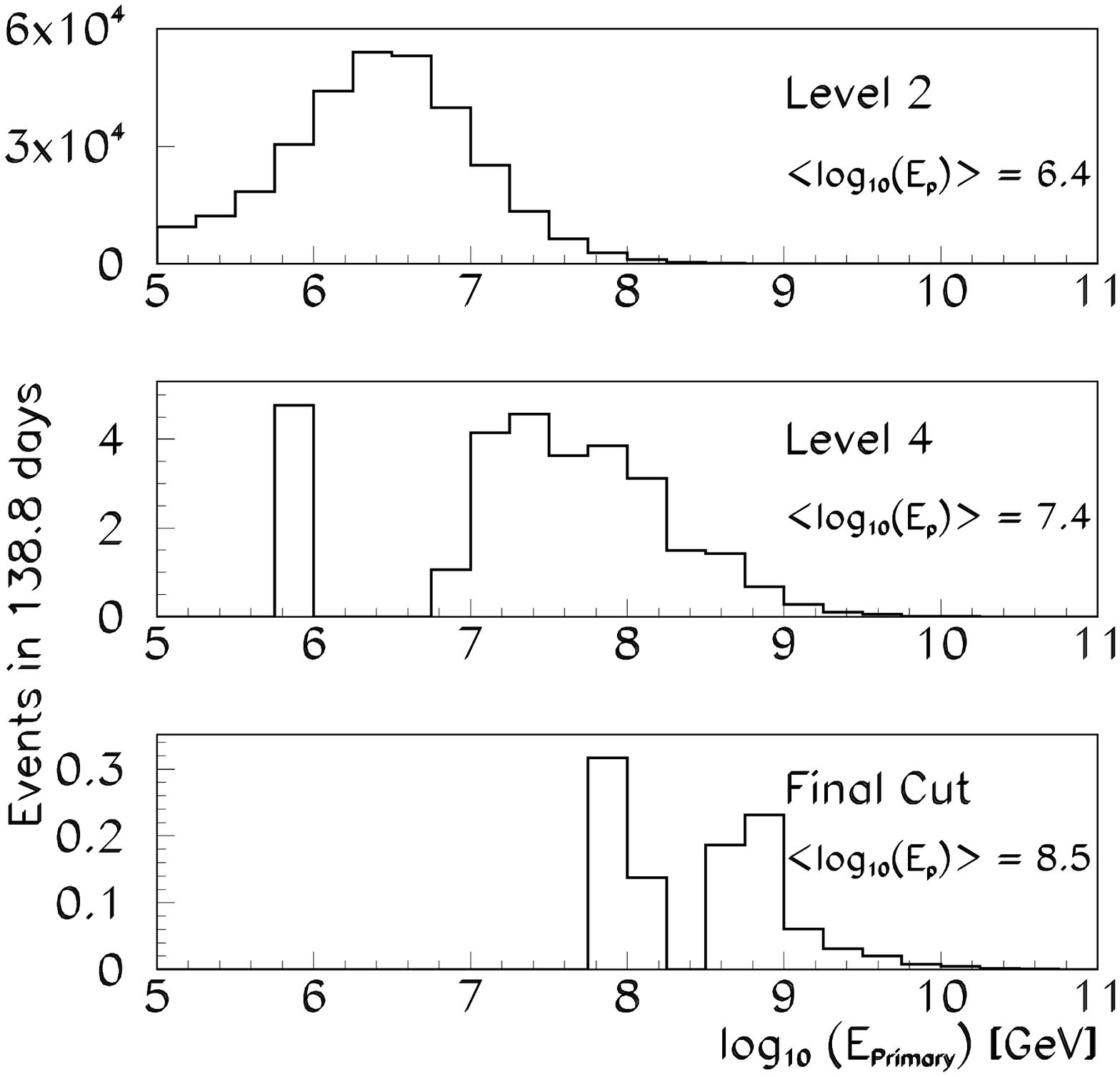}
\caption{\label{fig2c} The distribution of primary energy for simulated background at three 
different cut levels. The data shown is from the biased CORSIKA simulation and demonstrates the removal of lower 
energy background events by this analysis.}
\end{minipage}
\hfill
\end{figure}

\section{Method}
Twenty percent of the data from 2000 (randomly selected from February to November) was used to test 
the agreement with background MC. 
Following a blind analysis procedure this
20\% will be discarded and the developed selection criteria will be applied to the 
remaining 80\% of the data. Final cut values will be chosen by optimizing the model rejection factor 
\cite{MRF} for an E$^{-2}$ spectrum.

This analysis exploited the differences in light deposition caused by bundles of many low 
energy muons and single high energy muons. A muon bundle with the same total energy as a single high  
energy muon spreads its light over a larger volume, leading to a lower light density in the array. 
Both types of events have a large number of hit channels, but for the same number of hit OMs, 
the muon bundle has a lower total number of hits (NHITS). It also has a majority of OMs with a 
single hit, while the signal generates more multiple hits. The number of secondary hits is 
increased by the tendency of bright signals to produce afterpulses in the photomultiplier tube. The large amount of 
light deposited by high energy muons is also utilized in the reconstruction.  
The reconstruction algorithms used by the AMANDA Collaboration are optimized for the reconstruction of low 
energy muon bundles (from primaries with energies less than approximately 10$^{4}$ GeV), which makes them 
inaccurate for reconstructing the direction and energy of single high energy muons. However, loose cuts may be 
placed on the zenith angle based on the expectation that signal will come primarily from the horizontal direction, 
while background will come from the vertical, down-going direction. Single, high energy muons will also have 
distinct time residual distributions. The cylindrical geometry of the AMANDA-II array is also used to 
separate signal 
from background by estimating arrival direction. Down-going muon bundles will travel along the vertical strings of 
OMs in AMANDA-II. This, combined with asymmetries in the physical location of strings in the AMANDA-II 
array, pulls the 
center of gravity of hits away from the physical center of the array. Light from a single high energy muon will 
pass through a 
horizontal cross section of the array striking multiple strings, which pulls the center of gravity of hits closer 
to the physical center of the array.

Applying cuts on NHITS and the fraction of hit OMs with exactly one hit (F1H) reduced the 
data samples by a factor of 10$^{3}$ relative to retrigger level. At this point the data sets are 
split into a "high 
energy" and a "low energy" sample according to the energy deposited inside the array. A neural net trained on F1H, 
the closest distance between a reconstructed track and the detector center, and the radial distance from the center 
of the detector to the center of gravity of hits (RDCOG) served as an estimate of this energy selection value. 

The average energy of signal neutrinos in the "high energy" subset is 10$^{8}$ GeV. The energy deposited 
inside the array by these neutrinos is much greater than the energy deposited by a typical background 
event. This allows the application of simple selection criteria to separate signal from background 
events. Loose cuts on reconstruction variables, F1H and number of hit channels are sufficient to reduce the 
background expectation to less than 1 event for 138.8 days in this subset.

The "low energy" subsample consists of neutrinos with an average energy of less than 2 x 10$^{6}$ GeV. As the energy 
deposited inside the array by a typical background bundle of muons begins to approach the energy deposited by a 
single astrophysical neutrino, more refined selection criteria which depend on subtleties of the distribution of 
hit times must be utilized. The production of afterpulses by signal events, combined with the inaccurate 
reconstruction of signal direction means that UHE neutrino events have an excess of hits with a large time residual 
relative to the reconstructed track. A cut based on the timing of hits has been devised to take advantage of 
this.  Additionally, for the low energy subsample it is possible to 
take some advantage of differences in hit topology and multiplicity. Cutting on NHITS, the 
moment of inertia of the hits and the F1H of a subset of OMs helps to effectively separate 
signal from background. All these cuts, combined with cuts on reconstruction variables reduced the 
background by a factor of 10$^{8}$ relative to retrigger level. 

As can be seen from figures~\ref{fig1c} and ~\ref{fig2c}, this analysis is effective at removing lower energy 
background events while retaining higher energy signal events.

\section{Results and Outlook}
\begin{figure}
\begin{minipage}[t]{7.5cm}
\includegraphics[width=1.0\textwidth]{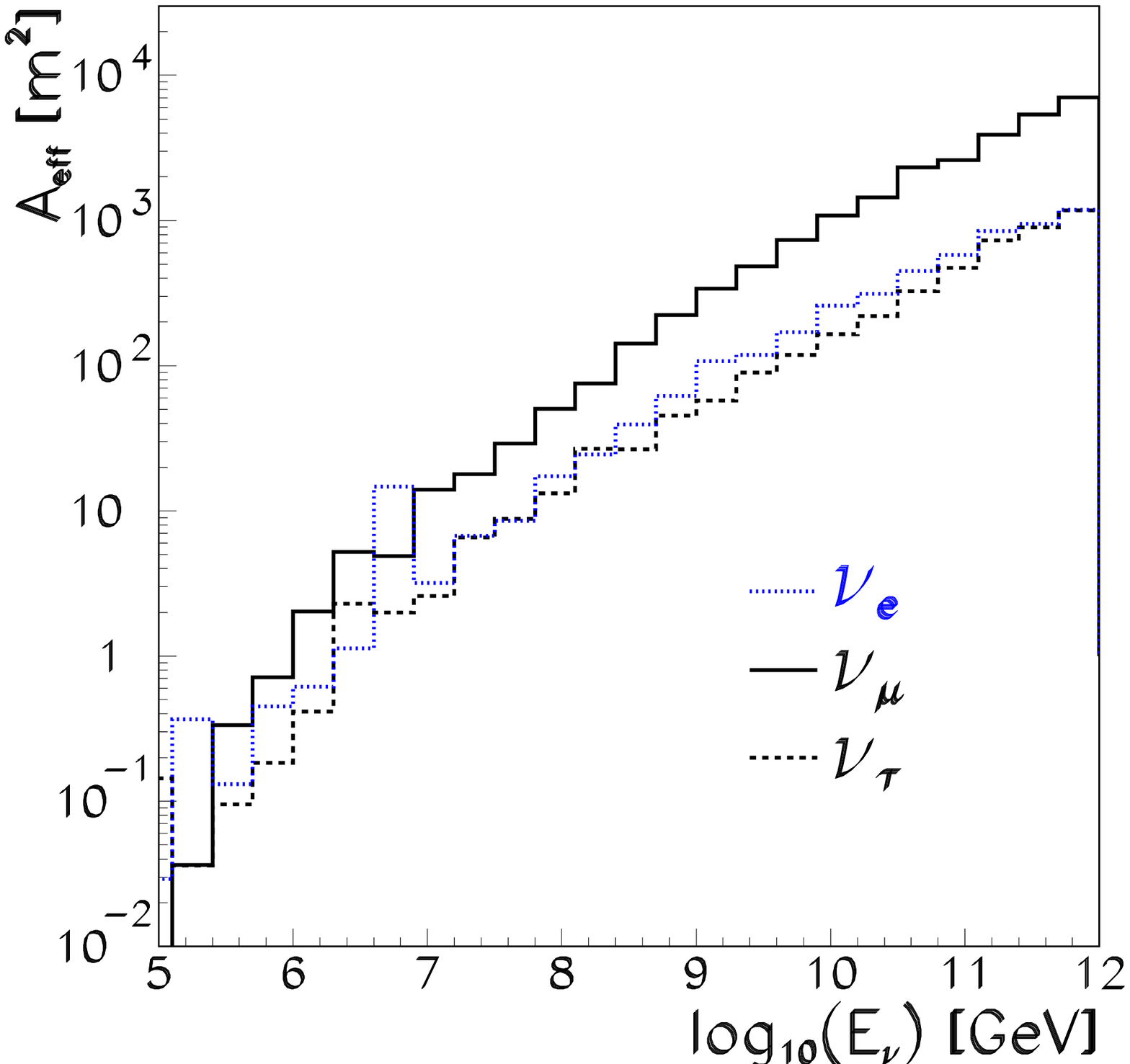}
\caption{\label{fig3c} Effective area averaged over all angles after all selection criteria have been 
applied as a function of neutrino energy. The peak in the electron neutrino effective area just below 10$^{7}$ GeV 
is due to the Glashow resonance.}
\label{F3}
\end{minipage} 
\hfill
\begin{minipage}[t]{7.5cm}
\includegraphics[width=1.0\textwidth]{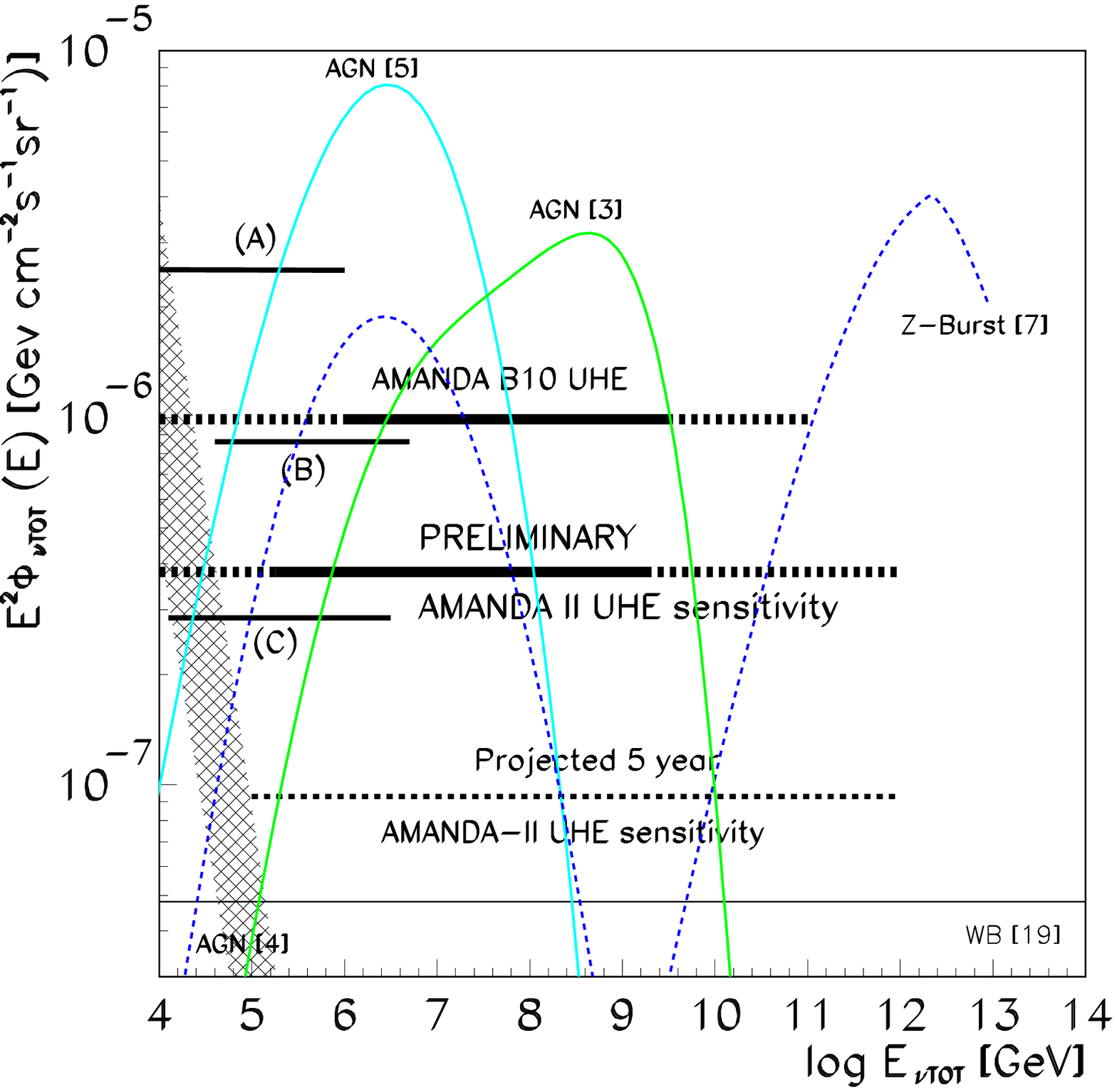}
\caption{\label {fig4} Limits on the flux of astrophysical neutrinos for an E$^{-2}$
spectrum. Shown are the results from the AMANDA B10 diffuse (A) \cite{Dif-AMA} and UHE analyses 
\cite{UHE-AMA}, the AMANDA-II cascade analysis (B) \cite{Casc-AMA}, the sensitivity of the 
three 
year 
AMANDA-II diffuse analysis (C) 
\cite{3Dif-AMA}, 
and the 
expected sensitivities for the AMANDA-II UHE analysis and five years of AMANDA-II.
Solid lines indicate the 90\% CL limit setting potential for an E$^{-2}$ spectrum. Also 
shown are 
the expected fluxes from a representative set of models.}
\label{F4}
\end{minipage}
\hfill
\end{figure}
Applying all the selection criteria leaves 1.0 \underline{+} 0.4 background MC event and 0 experimental events 
in the 20\% sample. The expected sensitivity 
\cite{Feldman} for an E$^{-2}$ all neutrino spectrum (assuming a 1:1:1 flavor ratio) is 
3.8 x 10$^{-7}$ cm$^{-2}$ s$^{-1}$ sr$^{-1}$ GeV with ninety percent of the events between 1.8 x 10$^{5}$ GeV and 
1.8 x 10$^{9}$ GeV (fig.~\ref{fig4}). This 
sensitivity is nearly a factor of two improvement over the previous limit set using 
AMANDA-B10. The expected neutrino effective area (fig.~\ref{fig3c}) is approximately the same as 
that of the previous analysis, but this analysis has increased background 
rejection which leads to an improved sensitivity.

Results from the analysis of the complete year 2000 data will be presented at 
the meeting. The AMANDA-II detector has been running since the beginning of 2000. Scaling this 
analysis to five years of AMANDA-II data results in an 
improvement of the sensitivity by a factor of 4. We expect additional 
improvements from the TWR system \cite{TWR} installed at the end of 2002. The TWR system provides 
additional information which will increase AMANDA-II's 
sensitivity to high energy events. For the future, the IceCube detector expects a 
sensitivity of 4.2 x 10$^{-9}$ cm$^{-2}$ s$^{-1}$ sr$^{-1}$ GeV (for 10$^{5}$ GeV to 10$^{8}$ GeV) for 
three years of 
operation \cite{ICECUBE}.

\setcounter{section}{0}
%
\newpage

\section*{\Large An investigation of seasonal variations in the atmospheric neutrino rate with the AMANDA-II neutrino telescope}

\vskip 0.05cm
{\large M.Ackermann$^{\it a}$, E. Bernardini$^{\it a}$ for the IceCube Collaboration} \\
{\it (a) DESY Zeuthen, Platanenallee 6, D-15738 Zeuthen, Germany}
\vskip 0.05cm
{\large Presenter: Markus Ackermann (markus.ackermann@desy.de), ger-ackermann-M-abs3-he22-poster }

\title[An investigation of seasonal variations in the atmospheric neutrino rate ...]
  {An investigation of seasonal variations in the atmospheric neutrino rate with the AMANDA-II neutrino telescope}
\author[M. Ackermann and E. Bernardini for the IceCube Collaboration] {M.Ackermann$^a$, E. Bernardini$^a$ for the IceCube Collaboration\\
        (a) DESY Zeuthen, Platanenallee 6, D-15738 Zeuthen, Germany
\presenter{Presenter: Markus Ackermann (markus.ackermann@desy.de), ger-ackermann-M-abs3-he22-poster}
 }
\maketitle

\begin{abstract}
Besides representing a source of background for the searches of astrophysical objects, atmospheric neutrinos are the most 
direct calibration source for neutrino telescopes. The characterization of this ``test beam'' has been, in the past, mostly based on the 
reconstruction of the energy spectrum and on flux measurements. In this work we investigate the amplitude and phase 
of possible seasonal variations in the event rate for the sample of 3329 neutrino candidates, detected with the 
AMANDA-II neutrino telescope in the years 2000-2003 (cfr. the AMANDA-II point source search, this conference). 
A mechanism that could produce such seasonal variations is the modulation of the target density for interactions
of cosmic rays in the atmosphere. 
Its effect on the atmospheric muon rate is known and measurements have been performed using several underground
detectors including AMANDA-II.
Its effect on the rate of atmospheric neutrinos at energies above a few hundred GeV has not been studied before. 
In this paper we report about a calculation of the seasonal variations expected using a global temperature model for the 
atmosphere. The results are compared to the event rate of the AMANDA-II neutrino sample.
\end{abstract}

\section{Introduction}

One of the major goals of the large-scale neutrino detectors, AMANDA and IceCube, is 
to identify cosmic sources of high-energy neutrinos ($>$100GeV). This search
is performed by reconstructing the direction of neutrino-induced muons using
the pattern of their Cherenkov light emission.  
Muons and neutrinos produced in the interaction of cosmic rays with the 
atmosphere form the dominant background for this analysis.
Both types of particle are generated in the decay of charged mesons 
($\pi^{\pm}$, $K^{\pm}$), which originate in the inelastic scattering
of cosmic ray primaries with nuclei of the atmosphere. 
The muons lose energy by electromagnetic interactions in the ice and rock
surrounding the detectors and can therefore not penetrate more than 
a few kilometers into dense materials. Consequently, this background can be readily 
removed by restricting the observation to particles travelling in upward 
direction in the ice. 
Atmospheric neutrinos however reach the detectors from all directions and can not
be distinguished from extra-terrestrial neutrinos. Therefore they remain as a 
residual background in the data sample, and it is essential to study the properties
of this background well to quantify correctly possible deficits and excesses that
would be interpreted as neutrino sources.

The search for variations of the atmospheric neutrino rate in time is an example for such a study. 
We report here on an investigation dedicated to seasonal variations
in the atmospheric neutrino rate in the AMANDA-II detector. 
Annual temperature fluctuations of the atmosphere could be 
responsible for such variations. 
We perform a calculation of the expected magnitude of oscillation 
in the AMANDA-II event rate caused by this effect. The results of these calculations
are compared with atmospheric neutrino data from the AMANDA-II neutrino telescope,
 recorded in 2000-2003 (the data sample used for the point source analysis \cite{psource}).

\section{Seasonal variations}

It has been shown that temperature variations in the atmosphere lead to changes in 
the intensity of the cosmic ray induced muon flux \cite{barrett}. This effect has been measured by
several experiments, among them MACRO \cite{macro1} and AMANDA-B10 \cite{bouchta}. 
Since neutrinos and muons are produced in the same 
decays one might expect a corresponding variation in the neutrino rate.
However, for several reasons the magnitude of these rate oscillations for neutrinos  
can not be derived directly from the muon rate changes using AMANDA-II or similar detectors:
\begin{enumerate}
\item The energy threshold for muons is higher than for neutrinos ($E_{thres}^{\mu} \approx$ 400 GeV,  
  $E_{thres}^{\nu} \approx$ 50 GeV) due to the deep underground location of the detector.
\item For kinematic reasons the atmospheric neutrinos result predominantly 
from kaon decay, the muons from pion decay \cite{gaissertalk}.
\item The muons originate from the local atmosphere above the detector while
   neutrinos are observed from interactions anywhere in the earth's atmosphere.
\end{enumerate}
For these reasons, the analytic high-energy approximations, as used 
in \cite{macro1} to calculate the variation in the muon flux, is not valid for neutrinos. 
At high energy (E $\gg$ 115 GeV for $\pi^{\pm}$,
E $\gg$ 850 GeV for $K^{\pm}$) meson interaction dominates over meson decay making the muon and
neutrino flux more sensitive to temperature variations \cite{gaisser}. 
However, for the low energy threshold of 
AMANDA-II, free decay plays an important role.
The meson flux reaches its maximum at an altitude of $X\approx$ 10 - 20 km, so temperature variations
at high altitudes have to be taken into consideration. The lack of global high altitude temperature data 
makes it necessary to use an atmospheric temperature model instead of measurements. 
There is a variety of such models available ranging from ground to an altitude of several hundred
kilometers (an overview can be found at \cite{nasa}). A numerical calculation is performed here based 
on the NRLMSISE-00 atmospheric model by \cite{atmmodel}.
 
The neutrino flux at a certain atmospheric depth $X$ at energy $E$ produced by $\pi^{\pm}$-decay
can be expressed using a derivation given by Gaisser \cite{gaisser}:
$$\Phi_{\nu}(E,X,\theta)= N_{\pi}\;\frac{\epsilon_{\pi}}{(1-r_{\pi}) \cos \theta} 
	                  \int_{E (1-r_{\pi})^{-1}}^{\infty}  \;
			  \frac{E'^{-2.7} e^{-X/\Lambda_{\pi}}}{X E'}
                          \int_{0}^{X} \left(\frac{X'}{X}\right)^{\frac{\epsilon_{\pi}}{E' \cos \theta}} 
                          \exp \left(\frac{X'}{\Lambda_{\pi}}-\frac{X'}{\Lambda_{n}}\right)
			   dX' \; \frac{dE'}{E'} $$ 

where $\Lambda_{\pi}$, $\Lambda_{n}$ are pion and nucleon attenuation lengths and  
$\epsilon_{\pi}$ is the critical energy distinguishing between decay and interaction dominated 
regimes. $\theta$ is the angle between cosmic ray primary and atmosphere normal,
$r_{\pi}= 1 - m_{\mu}^2/m_{\pi}^2$ and $N_{\pi}$ is a normalization constant. 
To compute the neutrino flux produced by kaon decay the constants 
$\Lambda_{\pi}$, $\epsilon_{\pi}$, $r_{\pi}$ and $N_{\pi}$ have to be
replaced by their kaon counterparts.
The temperature dependence in this representation is hidden in the critical 
energy $\epsilon_{\pi}=\epsilon_{\pi}(T(X))$ \cite{macro1,gaisser}. 
The temperature as a function of atmospheric depth $T(X)$ is provided by the NRLMSISE-00 model.

By integrating the equation above, 
one obtains the total flux above a threshold energy $E_{thres}$.
The expected event rate in AMANDA-II is obtained by weighting the energy integral with the effective area of the
experiment:

$$n_{\nu}(\theta)= \int_{E_{thres}}^{\infty} A_{eff}(E,\theta) \int_0^{\infty} \Phi_{\nu}(E,X,\theta) dE dX$$
 
\begin{figure}[h]
\begin{center}
\begin{tabular}{cc}
\includegraphics[width=0.44\textwidth,angle=0,clip]{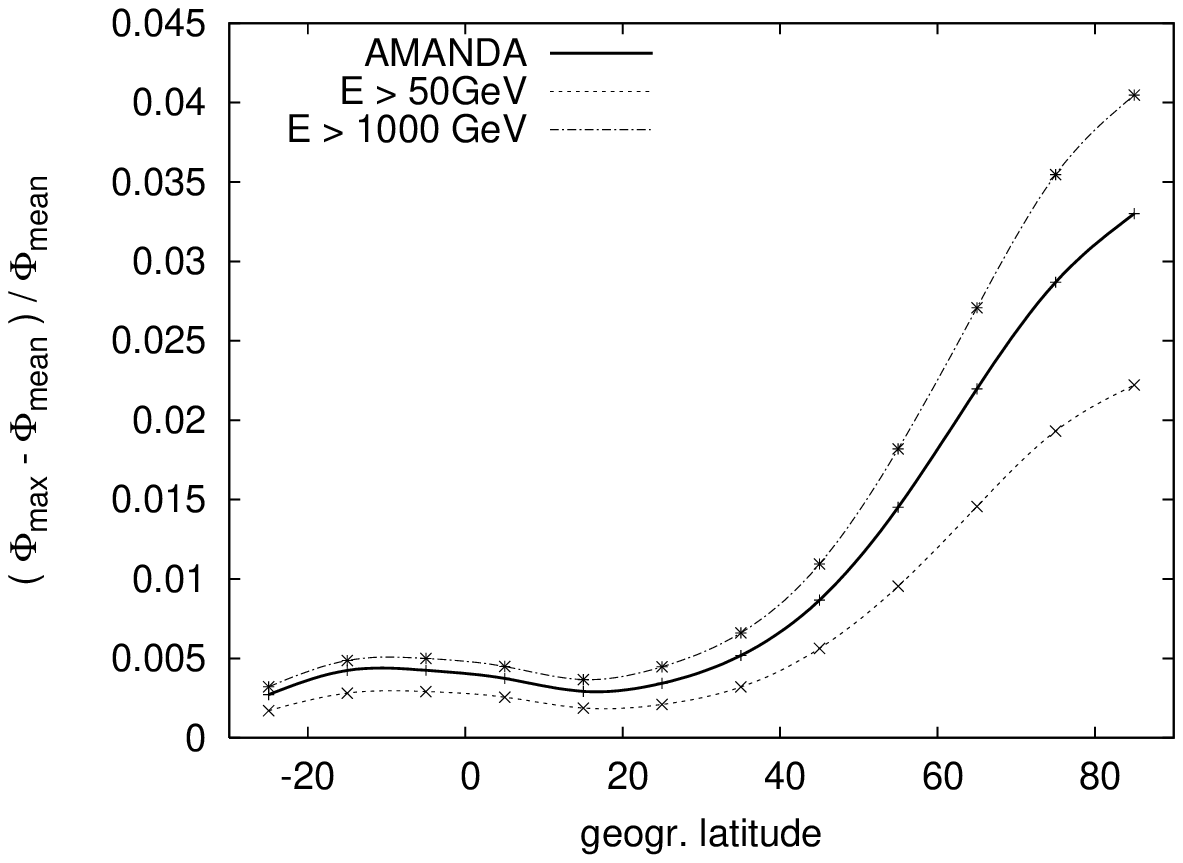} &
\includegraphics[width=0.44\textwidth,angle=0,clip]{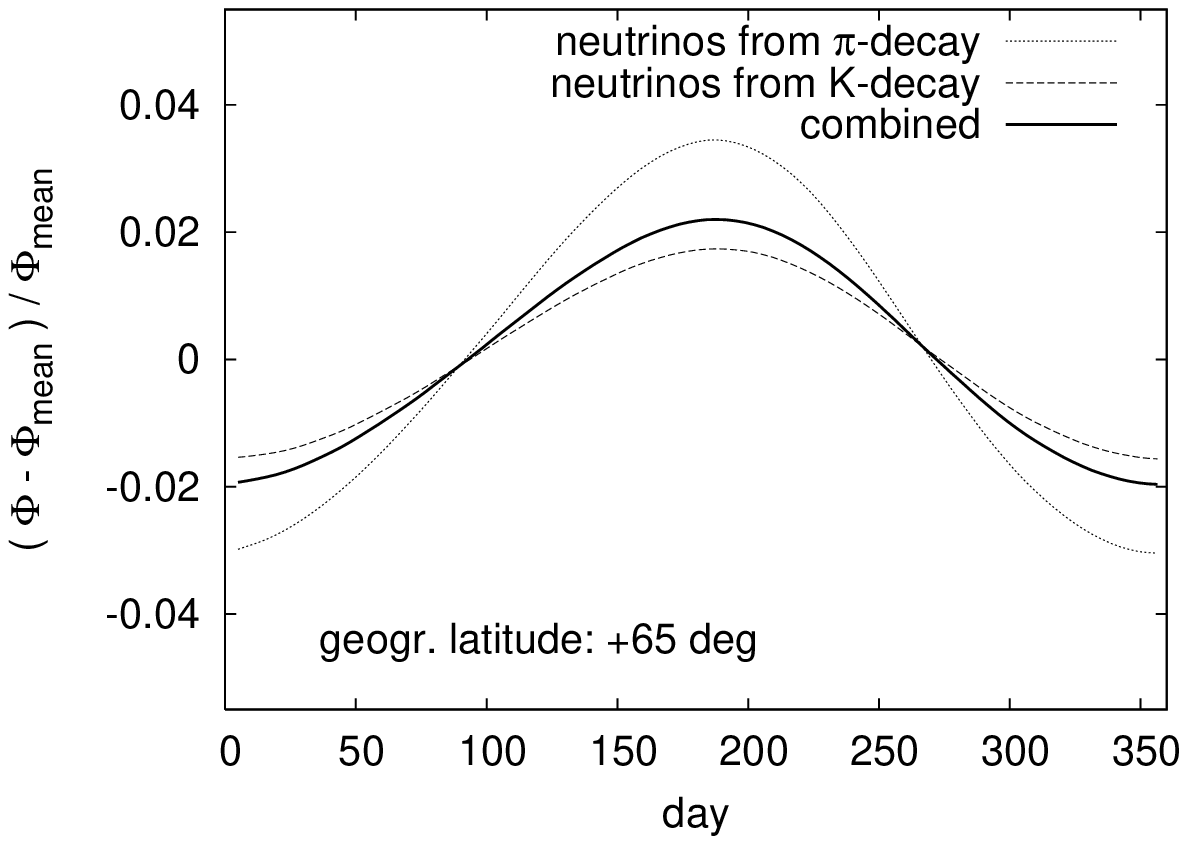}
\end{tabular}
\vskip -0.5cm
\caption{\label{fig1a} Left plot: Calculated geographical latitude dependence of the amplitude of the 
seasonal atmospheric neutrino flux variations relative to its annual average. The dotted line corresponds
to the flux integrated above 50 GeV, the dashed line to the flux above 1 TeV. 
The solid line is integrated above 50 GeV but the flux is weighted with a parametrization of the
AMANDA-II effective area. 
Right plot: Calculated time development of the flux variations at a latitude of $65^{\circ}$N 
(E $>$ 50 GeV, AMANDA-II effective area). 
The dotted line shows the variation of the $\pi$-decay component alone, the dashed line the K-decay component.
The solid line displays the combination of both components.
  }
\end{center}
\end{figure}

The integration of these equations can only be performed numerically; the technique used here is Monte Carlo 
integration.
The neutrino flux is calculated for a grid of $\theta$ ($\theta<60^{\circ}$) and $d$ (day of the year) values.  
In figure~\ref{fig1a} we show some results of this calculation for different threshold energies as well as with
an additional weight, accounting for the AMANDA-II effective area. 
The maximum relative neutrino flux deviation from the mean is shown as a function of 
geographical latitude in the left plot; the time development for a selected geographical latitude is illustrated on the right.
Notice that for neutrinos detected in AMANDA-II there are simple relations between geographical latitude $l$,
$\theta$ and the declination $\delta$, which are $\theta = 1/2 (90^{\circ}-l)$ and $\delta=90^{\circ}-\theta$. 

One can see that the expected variability of neutrino fluxes in AMANDA-II ranges 
between 3.5\% for particles from high latitudes and less than 0.5\% from low latitudes, where seasonal temperature 
changes become very small. The maximum flux for high latitudes is expected around day 190, the minimum flux
around day 360. Even for the highest latitudes the variation is considerably smaller than the $\approx 9\%
$ seasonal flux variation measured for muons in AMANDA-B10 \cite{bouchta}.

\section{Comparison to experimental data}

The AMANDA-II 2000-2003 point source sample provides a good set of events for investigating seasonal variations in the 
neutrino rate. While a small fraction of the events in this sample are expected to come from mis-reconstructed muons,
the majority forms the largest sample of high energy atmospheric neutrinos available in AMANDA-II.  
For this analysis the data is divided into 3 angular regions.

\begin{tabular}{cccl}
 declination $\delta$ & geogr. latitude $l$ &  \# events & description \\
$0^{\circ} < \delta \le 30^{\circ}$ &    $-90^{\circ} < l \le -30^{\circ} $  & 1227 & excluded from analysis, bin contaminated  \\
&&& with misreconstructed down-going muons \\
$30^{\circ} < \theta \le 60^{\circ}$ &   $-30^{\circ} < l \le +30^{\circ}  $  & 1492 & equatorial region \\ 
$60^{\circ} < \theta \le 90^{\circ}$ &   $+30^{\circ} < l \le +90^{\circ}  $  & 610 & northern hemisphere, high latitudes\\
\end{tabular}

\begin{figure}[h]
\begin{center}
\begin{tabular}{cc}
\includegraphics[width=0.44\textwidth,angle=0,clip]{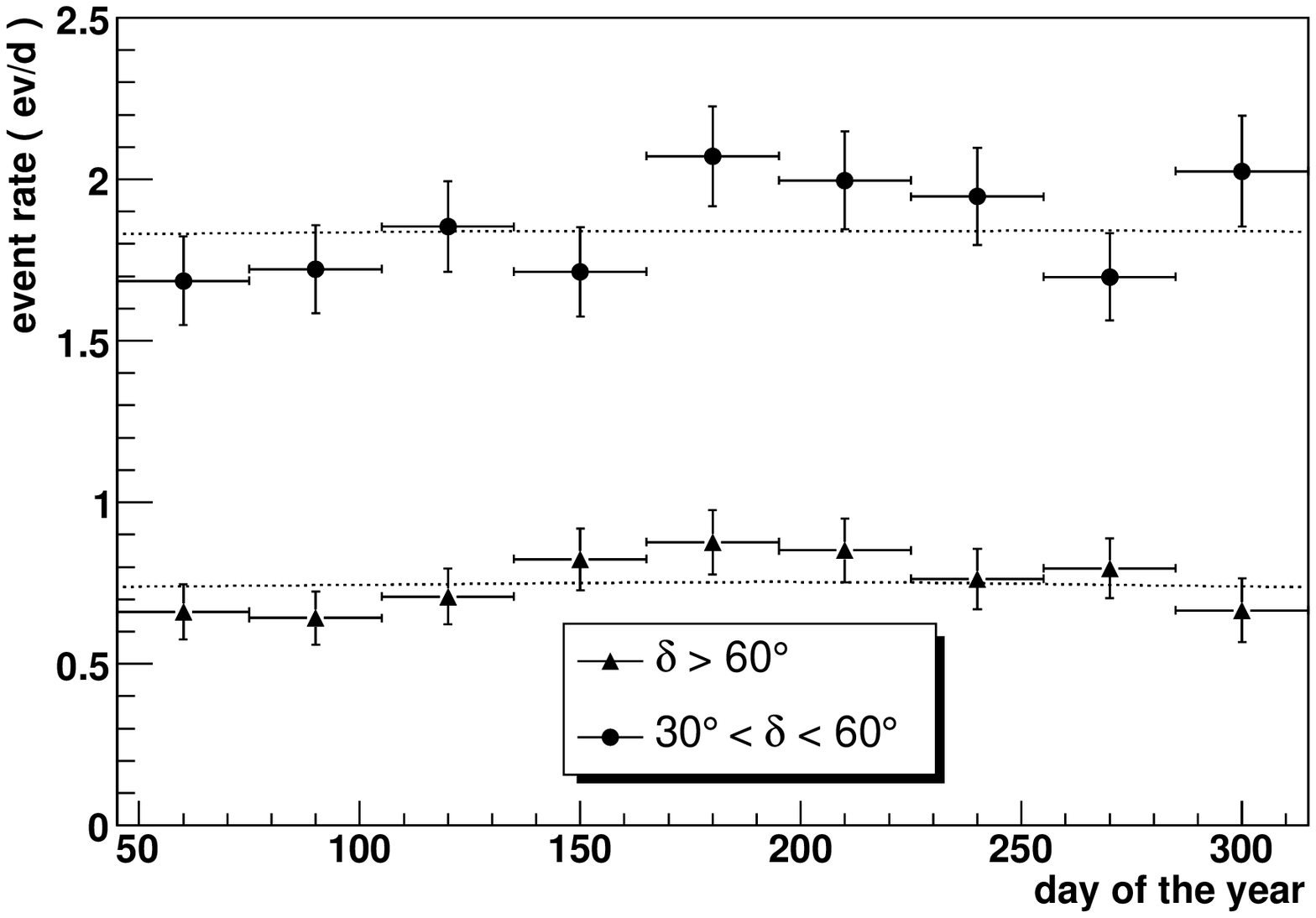} &
\includegraphics[width=0.44\textwidth,angle=0,clip]{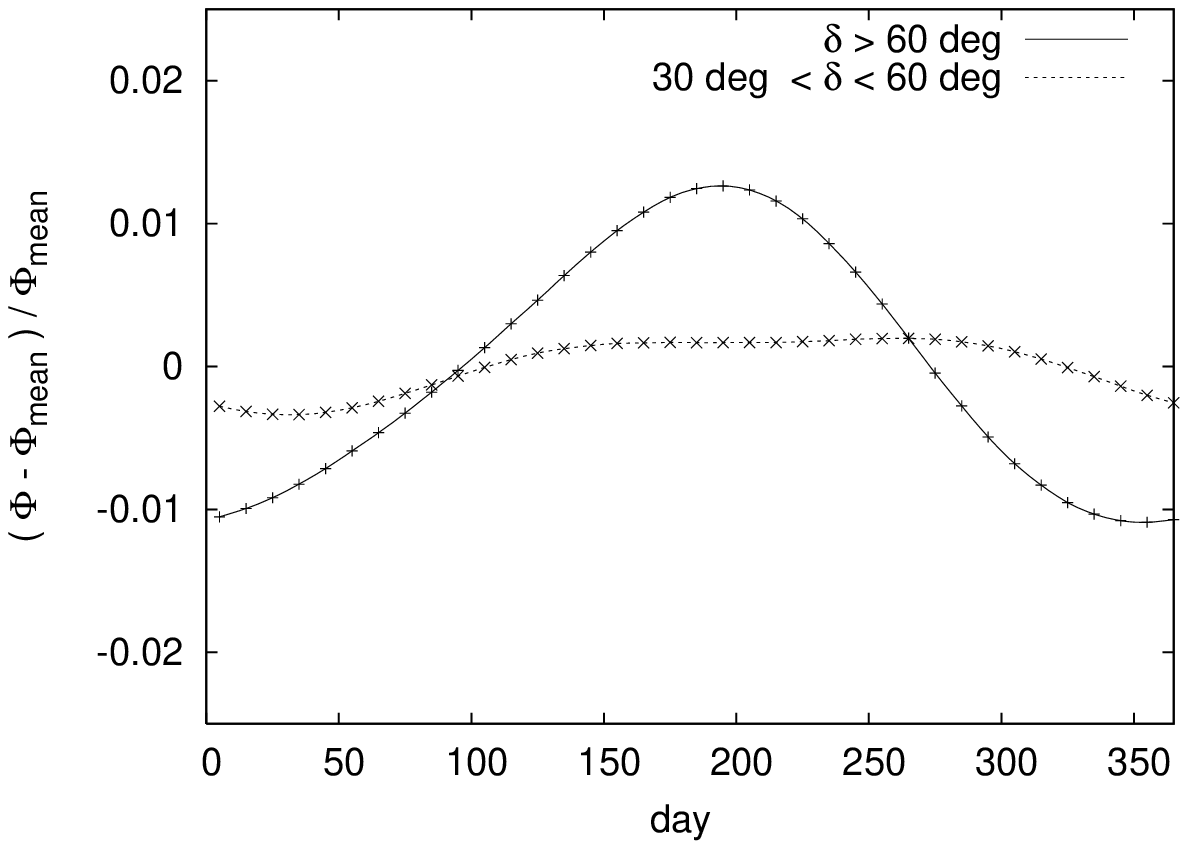} \\
\end{tabular}
\vskip -0.6cm
\caption{\label{fig2} Left plot: Event rates in the AMANDA-II neutrino telescope vs. time coming from
geographical latitudes above $30^{\circ}N$ ($\delta > 60^{\circ}$) and between $30^{\circ}S$ and
 $30^{\circ}N$ ($30^{\circ} < \delta \le 60^{\circ}$). The dotted lines correspond to fits with a 
 constant event rate plus the calculated relative modulation. Right plot: Relative seasonal variations
 of the atmospheric neutrino flux expected for these latitude bins from the numerical calculations 
 described above.}
 
\end{center}
\end{figure}

The small number of angular bins is due to the limited statistics of 3329 events in the AMANDA-II 
atmospheric neutrino sample. Figure \ref{fig2} (right) shows the annual 
expected relative flux oscillation, from the calculation described above, 
for the equatorial and high latitudes region.
In the high latitude region, $\Delta\Phi_{max}/\Phi$ is approximately 1.2\%
while for the equatorial region it is below 0.5\%. So, in both cases the seasonal variations 
should be well hidden within the statistical error of the sample. 
Therefore we can only test if we find data rate variations in AMANDA-II which are incompatible with such a small modulation.  

Figure \ref{fig2} (left) shows the AMANDA-II event rates in 30 day bins for the two latitude regions (with the years 2000-2003 
superimposed in the same bin). These event rates have been corrected for dead-time and 
down-time of the detector. 
The distribution is fitted with the calculated intensity variations on top of a constant function.
The $\chi^2$ for the high latitude bin is  $\chi^2/n_{free}=6.6/8$, the value for the equatorial bin is
$\chi^2/n_{free}=8.8/8$. The distributions are compatible with the flux variations calculated.
 
%


\section{Conclusions}

For the first time the expected amplitude of seasonal variations in the
atmospheric neutrino rates due to temperature fluctuations was calculated for a 
high energy neutrino detector. The calculations result in a variation ranging between
0.5\% and 3\% depending on the geographical latitude, which is too small to be resolved within 
the limited statistics of high energy atmospheric neutrinos from the AMANDA-II detector. 
IceCube and other km$^2$-detectors will provide samples with hundreds of thousands of atmospheric
neutrinos \cite{halzen}, allowing precision measurements of fluxes. Atmospheric neutrino rate 
modulations on the 1\% level will be measurable with these detectors.



\def\cere{\v{C}erenkov }
\def\numu{${\nu}_{\mu}$}
\def\nue{${\nu}_{e}$}
\def\nutau{${\nu}_{\tau}$}
\def\nuall{${\nu}_{e, \mu, \tau}$}
\def\nuetau{${\nu}_{e, \tau}$}
\setcounter{section}{0}
%
\newpage

\section*{\Large Search for high energy neutrino point sources in the northern hemisphere
  with the AMANDA-II neutrino telescope}

\vskip 0.05cm
{\large M. Ackermann$^{\it a}$, E. Bernardini$^{\it a}$, 
        T. Hauschildt$^{\it b}$ for the IceCube Collaboration} \\
{\it (a) DESY Zeuthen, Platanenallee 6, D-15738 Zeuthen, Germany\\
(b) Bartol Research Institute University of Delaware, 217
            Sharp Lab Newark, DE 19716, USA}
\vskip 0.05cm
{\large Presenter: Markus Ackermann (markus.ackermann@desy.de), ger-ackermann-M-abs3-he22-poster }

\author[M. Ackermann et al.] {M.Ackermann$^a$, E. Bernardini$^a$ for the IceCube Collaboration\\
        (a) DESY Zeuthen, Platanenallee 6, D-15738 Zeuthen, Germany
\presenter{Presenter: Markus Ackermann (markus.ackermann@desy.de),  ger-ackermann-M-abs2-og25-poster}
 }

\title[Search for high energy neutrino point sources in the northern hemisphere
  with the AMANDA-II neutrino telescope]
  {Search for high energy neutrino point sources in the northern hemisphere
  with the AMANDA-II neutrino telescope}

\author[M. Ackermann for the IceCube Collaboration]{M. Ackermann$^a$, E. Bernardini$^a$, 
        T. Hauschildt$^b$ for the IceCube Collaboration \\
        (a) DESY Zeuthen, Platanenallee 6
            D-15738 Zeuthen, Germany\\
        (b) Bartol Research Institute University of Delaware, 217
            Sharp Lab Newark, DE 19716, USA
}
\presenter{Presenter: M. Ackermann (Markus.Ackermann@desy.de), ger-ackermann-M-abs2-og25-poster}
\vskip -.5cm
\maketitle
\begin{abstract}
In this paper we report the most recent
survey of the northern sky to search
for neutrino point sources using the AMANDA-II telescope.
A search for astrophysical neutrinos of energies above a few tens
of GeV was performed on the data collected between the years 2000 and
2003 for a total live-time of 807 days.
Thanks to a higher reconstruction accuracy and background 
rejection power compared to past analyses, together with a longer
exposure time, a noticeable improvement has been achieved in the
sensitivity of the telescope.
The sensitivity to individual point sources, assuming a
signal energy spectrum proportional to d$\Phi$/dE $\sim E^{-\gamma}$ with
a spectral index of 2, is
E$^{2}\cdot$d$\Phi$/dE$\leqslant 6\cdot 10^{-8}$GeVcm$^{-2}$s$^{-1}$, weakly dependent on declination.
We have obtained a
 large sample of neutrino candidates with high reconstructed track
quality, consisting of 3329 selected up-going events.
We searched this sample for a signal from point sources.
Individual potential neutrino sources belonging to a catalogue of 33
preselected objects were scanned together with the
complete northern sky.  
We report the outcomes of the individual observations and the
significance map of the northern sky.
\end{abstract}

\section{Introduction}
\label{Sec:Intro}
The search for high energy extraterrestrial neutrinos is the major
focus of research of the Antarctic Muon And Neutrino
Detector Array AMANDA~\cite{amanda2}. The goal is the understanding of the
origin, propagation and nature of cosmic rays. 
The elusive nature of neutrinos makes them rather unique astronomical
messengers: neutrinos can escape from dense matter regions and
propagate freely over cosmological distances. Their observation would
also provide an incontrovertible signature of a hadronic
component in the flux of accelerated particles.  
Any source that accelerates charged hadrons to high energy is a likely
source of neutrinos: high energy particles will interact with other
nuclei or the ambient photon fields producing hadronic showers. In
these scenarios, high energy photons and neutrinos are expected to be
produced simultaneously. 
The search for high energy cosmic neutrinos reported in this paper
strongly focuses on identified sources of high energy gamma-rays.

Searches for astrophysical sources of neutrinos have
to cope with the backgrounds from
the interaction of cosmic rays with the Earth's atmosphere.
This results in a background of downward-going muons and a more uniform
background of neutrinos from mesons decay.  Downward-going muons are
rejected by selecting only events that are reconstructed as
upward-going, 
yet an
indistinguishable background remains, composed of atmospheric neutrino
induced muons and mis-reconstructed downward-going muons.  Both
sources of background are equivalent within the scope of this work and
are treated identically. 
The final event sample was selected 
in a blind approach to avoid the enhancement of apparent excesses 
in the data or
the introduction of biases that cannot be
statistically described. This was accomplished by 
randomizing the events in right ascension. 

\section{Event reconstruction and selection}
\label{Sec:Rec}
The major goal of this analysis was the selection of a high statistics sample
of high energy events which would be searched for evidence of steady
and transient point sources in the northern sky. 
Event reconstruction and selection were therefore optimized
to provide tracks with good angular resolution in a wide energy range.
The analyzed data were collected with the
AMANDA-II detector between the years 2000 and 2003. Periods corresponding
to the detector maintenance activities (roughly from November to
February)  
have not been used. The total live-time, after data
quality selection, is 807 days. 
Details of the pre-processing techniques (hits and Optical Modules
selection) and of the reconstruction algorithms can be found
in~\cite{reco}. 

\begin{figure}[!thb]
\begin{center}
\includegraphics[width=8cm,height=4.5cm]{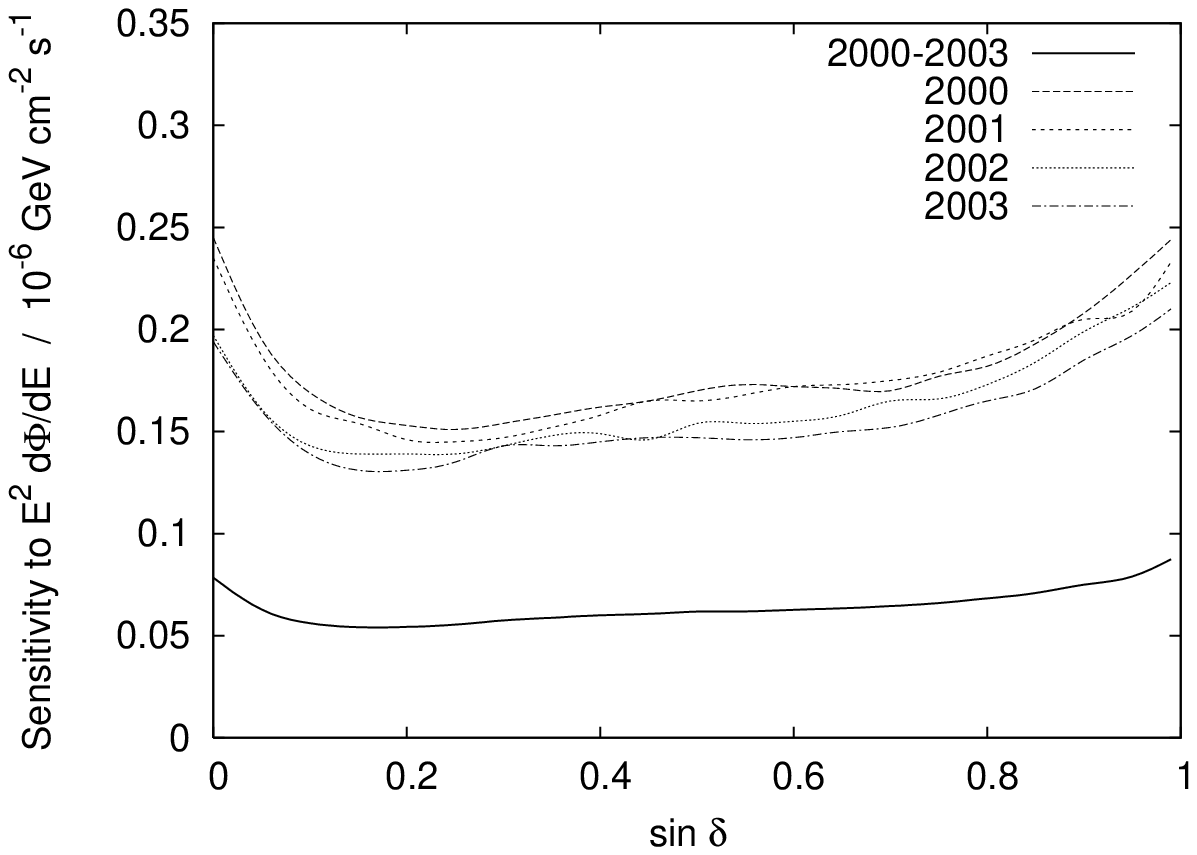}\hspace*{-0.5cm}\includegraphics[width=8cm,height=4.5cm]{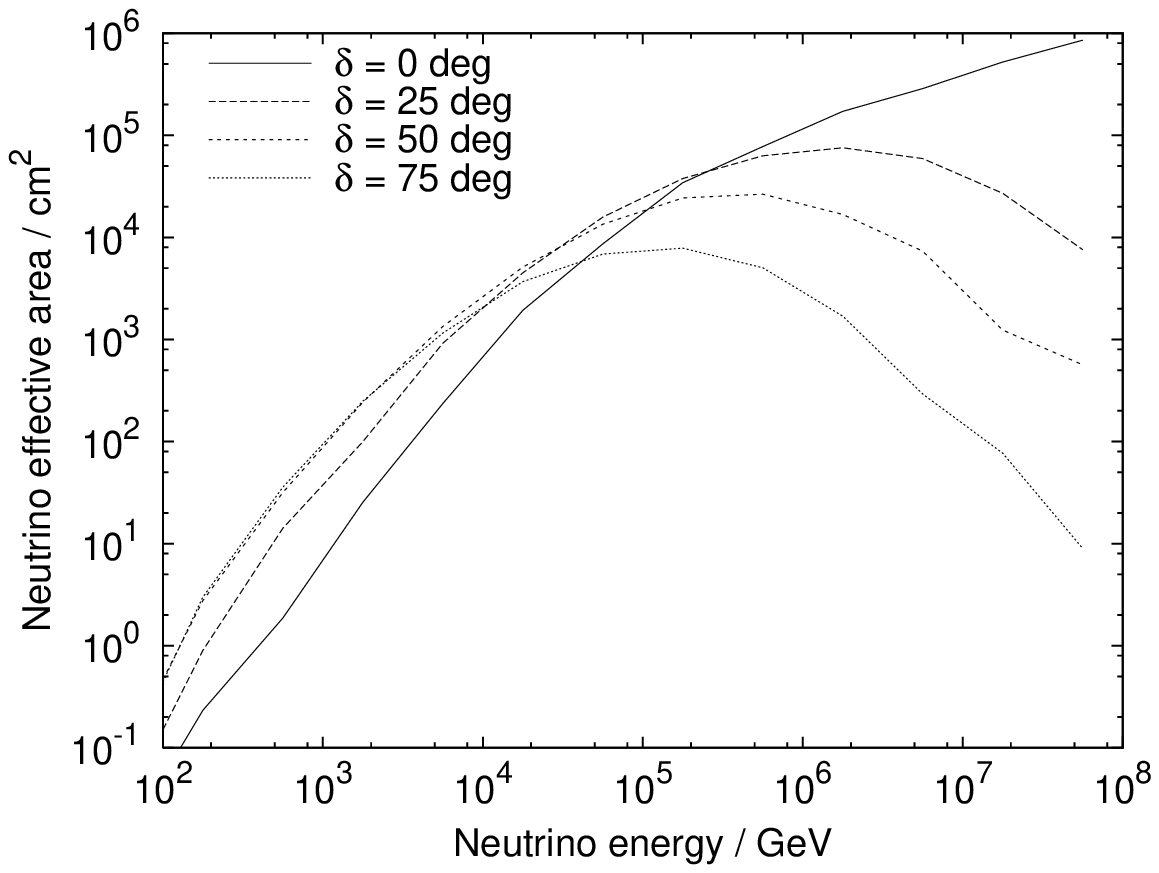}
\end{center}
\vskip -0.6cm
\caption{Left: Sensitivity as a function of declination ($\delta$), for a
signal spectral index of 2. The results for individual years and for the
combined data sample are shown. 
Right: Neutrino effective area as a function of the neutrino
energy.}  
\label{Fig:Sens}
\end{figure}

\begin{figure}[thb]
\begin{center}
\includegraphics[width=8cm]{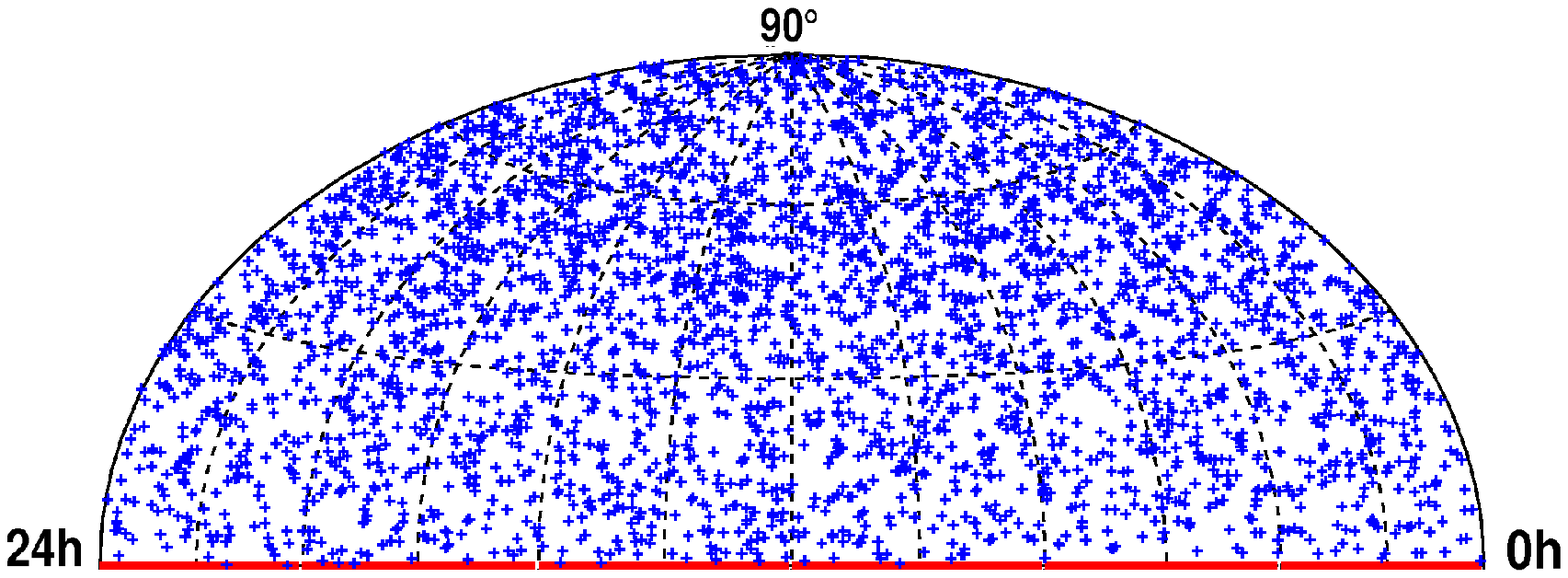}\hspace*{-0.5cm}
\includegraphics[width=8cm]{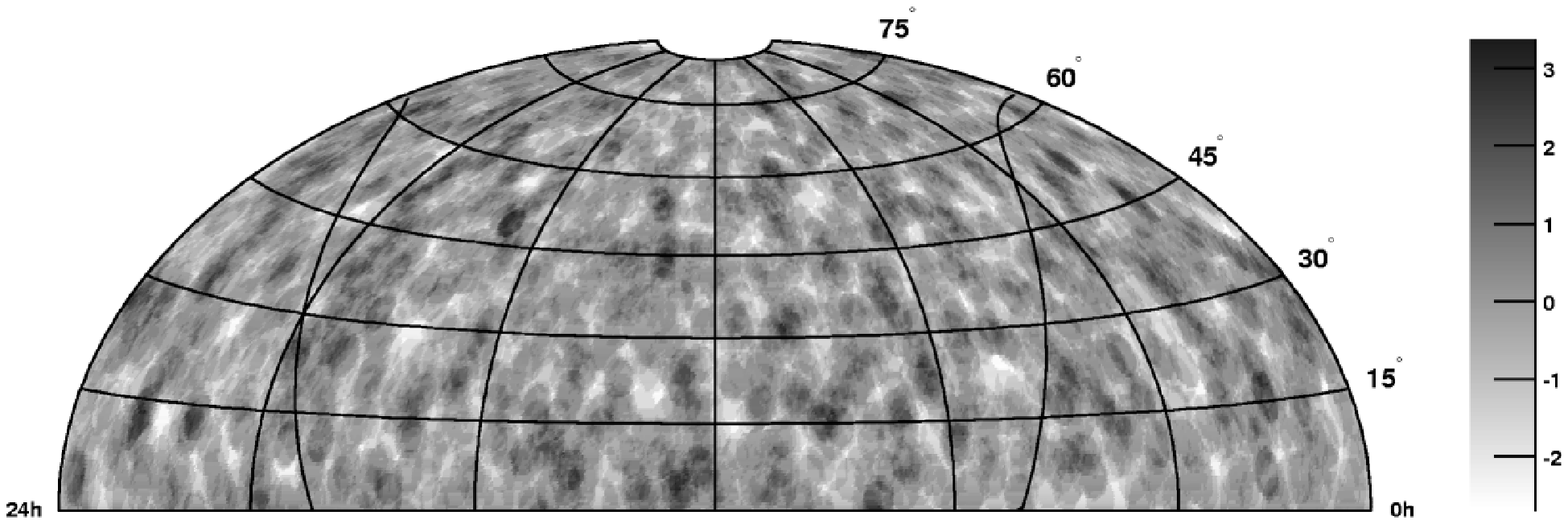}
\end{center}
\vskip -0.6cm
\caption{Left: Sky-plot (in celestial coordinates) of the selected 3329
up-going neutrino candidate events. 
Right: Significance map from a scan of the northern
sky to search for event clusters. The significance is positive for
excesses and negative for deficits of events (compared to the expected
background).}
\label{Fig:Sky}
\end{figure}

Neutrino induced up-going tracks were selected by
imposing track quality requirements.
Event selection criteria were chosen to achieve the best average flux
upper limit (``sensitivity''~\cite{mrf}) and were optimized for each
declination 
band independently. Selection criteria included:
a parameter describing the hit distribution along the track,
the fit likelihood (from two independent track
reconstruction procedures) and the event-based angular
resolution~\cite{neuphd}.   
The search bin radius in the sky was
an additional free parameter. 
The event selection depends also on energy, 
due to the energy dependence in the light deposit in the 
array and a varying detection efficiency. 
We therefore considered two extreme spectral indexes as 
reference: $\gamma$=2 and $\gamma$=3.
The effects of the different signal spectra on the event cut optimization
were  investigated separately and the results were combined in the final
event selection, to achieve the best performance for both spectra
simultaneously. 
Figure~\ref{Fig:Sens}  
shows the resulting sensitivity and effective area as a function of
declination. An overall improvement of
about a factor three was obtained compared to the baseline sensitivity of
the fully deployed AMANDA-II detector after 197 days of
exposure~\cite{2000ps}.
\vskip -0.6 cm
A final sample of 3369 events was selected, of which 3329 are
up-going. 
The corresponding directions are shown in
Fig.~\ref{Fig:Sky} (left). A relatively uniform coverage of the 
northern sky is obtained. 
Due to the event selection optimization allowing wider spectral
scenarios compared to~\cite{2000ps,0002ps}, this data sample contains
a certain contribution from lower energy events. 
\section{Search for point sources in the northern sky}
\label{Sec:SteadyPoint}
\label{Sec:Pnt}
\begin{table}[t]
 \begin{center}
 \begin{footnotesize}
  \begin{tabular}{lccccc|lccccc}\hline\hline
      Candidate & $\delta$($^\circ$) & $\alpha$(h)    &
                $n_{\mathrm{obs}}$ & $n_{b}$          &
                $\Phi_{\nu}^{\mathrm{lim}}$           &
      Candidate & $\delta$($^\circ$) & $\alpha$(h)    &
                $n_{\mathrm{obs}}$ & $n_{b}$          &
                $\Phi_{\nu}^{\mathrm{lim}}$           \\\hline
\multicolumn{12}{c}{ \emph{TeV Blazars} } \\
      Markarian 421  & 38.2 & 11.07 & 6 & 5.6 & 0.68&
      1ES 2344+514   & 51.7 & 23.78 & 3 & 4.9 & 0.38\\
      Markarian 501  & 39.8 & 16.90 & 5 & 5.0 & 0.61&
      1ES 1959+650   & 65.1 & 20.00 & 5 & 3.7 & 1.0 \\ 
      1ES 1426+428   & 42.7 & 14.48 & 4 & 4.3 & 0.54&
		     &&&&&\\
\multicolumn{12}{c}{ \emph{GeV Blazars} } \\
      QSO 0528+134   & 13.4 &  5.52 & 4 & 5.0 & 0.39&
      QSO 0219+428   & 42.9 &  2.38 & 4 & 4.3 & 0.54\\
      QSO 0235+164   & 16.6 &  2.62 & 6 & 5.0 & 0.70&
      QSO 0954+556   & 55.0 &  9.87 & 2 & 5.2 & 0.22\\
      QSO 1611+343   & 34.4 & 16.24 & 5 & 5.2 & 0.56&
      QSO 0716+714   & 71.3 &  7.36 & 1 & 3.3 & 0.30\\  
      QSO 1633+382   & 38.2 & 16.59 & 4 & 5.6 & 0.37&
		     &&&&&\\
  \multicolumn{12}{c}{ \emph{Microquasars} } \\
      SS433          &  5.0 & 19.20 & 2 & 4.5 & 0.21&
      Cygnus X3      & 41.0 & 20.54 & 6 & 5.0 & 0.77\\
      GRS 1915+105   & 10.9 & 19.25 & 6 & 4.8 & 0.71&
      XTE J1118+480  & 48.0 & 11.30 & 2 & 5.4 & 0.20\\
      GRO J0422+32   & 32.9 &  4.36 & 5 & 5.1 & 0.59&
      CI Cam         & 56.0 &  4.33 & 5 & 5.1 & 0.66\\
      Cygnus X1      & 35.2 & 19.97 & 4 & 5.2 & 0.40&
      LS I +61 303   & 61.2 &  2.68 & 3 & 3.7 & 0.60\\
   \multicolumn{12}{c}{ \emph{SNR \& Pulsars} }\\
      SGR 1900+14    &  9.3 & 19.12 & 3 & 4.3 & 0.35&
      Crab Nebula    & 22.0 &  5.58 &10 & 5.4 & 1.3\\
      Geminga        & 17.9 &  6.57 & 3 & 5.2 & 0.29& 
      Cassiopeia A   & 58.8 & 23.39 & 4 & 4.6 & 0.57\\
  \multicolumn{12}{c}{ \emph{Miscellaneous} }\\
      3EG J0450+1105 & 11.4 &  4.82 & 6 & 4.7 & 0.72&
      J2032+4131     & 41.5 & 20.54 & 6 & 5.3 & 0.74\\
      M 87           & 12.4 & 12.51 & 4 & 4.9 & 0.39&
      NGC 1275       & 41.5 &  3.33 & 4 & 5.3 & 0.41\\
      UHE CR Doublet & 20.4 &  1.28 & 3 & 5.1 & 0.30&
      UHE CR Triplet & 56.9 & 11.32 & 6 & 4.7 & 0.95\\    
      AO 0535+26     & 26.3 &  5.65 & 5 & 5.0 & 0.57&
      PSR J0205+6449 & 64.8 &  2.09 & 1 & 3.7 & 0.24\\
      PSR 1951+32    & 32.9 & 19.88 & 2 & 5.1 & 0.21&
		     &&&&&\\
       \hline\hline
  \end{tabular}
 \end{footnotesize}
\vskip -0.2cm
  \caption{\label{Tab:33}
                Results from the search for neutrinos from selected objects.
                $\delta$ is the declination in degrees, $\alpha$ the
		right ascension in hours,
                $n_{obs}$ is the number of observed events and $n_{b}$
                the expected 
                background. $\Phi_{\nu}^{\mathrm{lim}}$ is the
                90\% CL upper limits in units of $10^{-8}
                \mathrm{cm}^{-2}\mathrm{s}^{-1}$ 
                for a spectral index of 2 
                and integrated above 10 GeV. 
                These results are preliminary (the systematic errors
                are under assessment).} 
 \end{center}
  \end{table}
A search for point sources of neutrinos in the sample of 3329 up-going neutrino
candidates was performed by looking for 
excesses of events from the directions of individual known high-energy
gamma emitting objects and by a survey of
the full northern sky.
In both surveys we used circular search bins, with a size defined by
the bin radius optimized together with the event selection for optimal
sensitivity. The radius
depends on declination and varies
between 2.25$^{\circ}$ and
3.75$^{\circ}$. The number of events in each 
declination band is a few hundred and   
the statistical uncertainty in the background
in any given search bin is below 10\%.

A  sample of 33 candidate neutrino sources have been tested for
an excess (or deficit) of events. The investigated sources include
galactic and extragalactic objects and their corresponding locations are
listed in Tab.~\ref{Tab:33}. The directions of two cosmic rays
multiplets (a triplet 
and the highest energy doublet~\cite{agasa}) were also tested. 
The background is estimated by averaging in right ascension 
the event density as a function of declination.
A toy Monte Carlo, simulating equivalent tests using sets of 
events with randomized right ascension values, was
used to evaluate the significance of the observations (which 
expresses the probability of a background
fluctuation in units of standard deviations).
All the observations are compatible with the expected background.
The highest excess corresponds to the direction of the
Crab Nebula, with 10 observed events compared to an average of 5.4 
expected background (about 1.7 $\sigma$). The probability that a background
fluctuation produces this  or a larger deviation in any of the
33 search bins is 64\%, taking into account 
the trial factor (due to the multiplicity
of the directions examined and the correlation between
overlapping search bins). 

A full scan of the northern sky was also performed
to look for any localized event
cluster. We used overlapping search bins with optimal radius and 
centered on a grid with a spacing of
$0.5^{\circ}$. The search was extended up to $85^{\circ}$ in
declination\footnote{For a telescope located at the South Pole 
the zenith angle of a sources is fixed. This causes a 
sky coverage which is constant in time and equal for all directions. 
A simple integration in right ascension of the event density at
different declinations 
allows a measurement of the background without time-dependent
corrections. However, the limited statistics in the polar bin 
prevents an accurate estimation of the background.}.
Figure~\ref{Fig:Sky} shows a sky map of the 3329 neutrino
events and a map of significances from the northern sky cluster search.
All the observations are compatible with the background hypothesis. 
The highest excess corresponds to a significance of about 3.4
$\sigma$. The probability to observe this  or a higher excess, taking into
account the trial factor, is 92\%. 

\section{Summary and outlook}
We performed a search for a signal from point sources 
of neutrinos in the northern sky with data from the AMANDA-II neutrino 
telescope. Improved event
reconstruction and selection techniques have been applied 
to the data collected between the years 2000 and 2003. 
Special emphasis has been put on the
energy spectrum of the Monte Carlo events passing the selection cuts, 
to be sensitive to the largest variety of possible 
signal energy distributions. 
The achieved sensitivity to point sources is 
the most relevant numerical outcome of this analysis, and is equal to 
E$^{2}\cdot$d$\Phi$/dE$\leqslant 6\cdot 10^{-8}$GeVcm$^{-2}$s$^{-1}$, after
807 days of exposure and assuming a signal spectral index of 2.
The sensitivity is weakly dependent on
declination.  
We have obtained a large sample of
neutrinos with high energies, consisting of 3329
selected up-going events. 
No statistically significant excess
has been observed in the search for a signal from either candidate sources
from a catalogue of pre-selected objects or in the full northern
sky. 
We are currently extending this analysis to the data collected in the
year 2004.
An investigation of the possible sources of systematic uncertainties
is also in progress and the  upper limits reported here will be updated
to account for the systematic error.

Three other contributions to this conference present preliminary
results on searches with the four years sample of 3329 events 
for a variable signal from candidate neutrino sources~\cite{ps-var}, 
for a cumulative excess for classes
of objects from predefined source catalogues (source stacking
analysis)~\cite{ps-stck} and for a neutrino signal
from the galactic plane~\cite{ps-gal}.


%

\setcounter{section}{0}
%
\newpage

\section*{\Large Multiwavelength comparison of selected neutrino point source candidates}

\vskip 0.05cm
{\large M. Ackermann$^{\it a}$, E. Bernardini$^{\it a}$, 
        T. Hauschildt$^{\it b}$, E. Resconi$^a$ 
for the IceCube Collaboration} \\
{\it (a) DESY Zeuthen, Platanenallee 6, D-15738 Zeuthen, Germany\\
(b) Bartol Research Institute University of Delaware, 217
            Sharp Lab Newark, DE 19716, USA}
\vskip 0.05cm
{\large Presenter: M. Ackermann (Markus.Ackermann@ifh.de), ger-ackermann-M-abs1-og25-oral}

\title[Multiwavelength comparison of selected neutrino point source candidates]{Multiwavelength comparison of selected neutrino point source candidates}
\author[M. Ackermann et al.] {M. Ackermann$^a$, E. Bernardini$^a$, T. Hauschildt$^b$, E. Resconi$^a$ for the IceCube Collaboration\\
(a) DESY Zeuthen, Platanenallee 6, D-15738 Zeuthen, Germany.\\
(b) Bartol Research Institute University of Delaware, 217 Sharp Lab Newark, DE 19716, USA.
}
 
\presenter{Presenter: M. Ackermann (Markus.Ackermann@ifh.de), ger-ackermann-M-abs1-og25-oral }

\maketitle
\begin{abstract}
In this paper, we report the first results of an analysis of AMANDA-II
data to search for time-variable neutrino point sources.  A large
sample of 3329 neutrino candidate events from the northern hemisphere
was analyzed.  The investigation is based on the observation that many
cosmic sources have violent variations in their electromagnetic
emission. We have tested the hypothesis that neutrino production in
such sources is correlated with the electromagnetic activity. Using an
independent approach, we have also searched for occasional neutrino
flares using a sliding-window technique. Such flares might be
detectable with a dedicated time variability investigation and under
favorable conditions of signal enhancement and duration. The two
search methods will be described and the results reported.

\end{abstract}

\vspace{-0.20in}
\section{Introduction}
TeV neutrino candidate sources often show large and violent variations
in the electromagnetic emission. Under the assumption that neutrino
emission shows a similar variability, a set of methods that test the
flare behavior of a source have been developed. Under favorable
conditions of signal enhancement and period duration, such flares
might be detectable with a dedicated time-variability investigation
and still not be evident in the time-integrated point-source search
\cite{4y}.\\
The driving criteria used for the selection of the sources considered in this analysis are:
the source presents an evident variable character in one or more wavelengths,  
flares are plausible in the period of interest for this analysis (2000-2003) and  
the total time  of the flare periods is long enough in order to allow a reasonable detection probability.
Three categories of sources have been selected: blazars, microquasars and variable sources from the EGRET catalog.\\
Two different methods have been developed in order to 
search for a variable neutrino signal from these families of sources: 
(A) a multiwavelength comparison when the source presents a resolved variability in 
one or more wavelengths, (B) a search for neutrino flare with a sliding-time window
when the variable character of the source is evident but electromagnetic observations are limited.\\
The test data sample is provided by the time-integrated point
source search \cite{4y}. In the multiwavelength method (A) 
an appropriate re-optimization for shorter live-times is performed.
The complete catalog of sources used in this approach with the results obtained are reported in Table 1.
The analysis has been performed following the principle of "blind analysis" to avoid the introduction of biases
 that cannot be statistically quantified. Details about this topic are discussed in \cite{4y}.

\vspace{-0.1in}
\section{Method A: Multiwavelength Comparison} 
A multiwavelength comparison has been developed in order to analyze TeV blazars and microquasars in the 
scenario of non-steady-state neutrino emission. 
Details of the method that are specific to these families of sources are reported below.\\

\begin{figure}[ht]
\begin{center}
\includegraphics[height=1.8in]{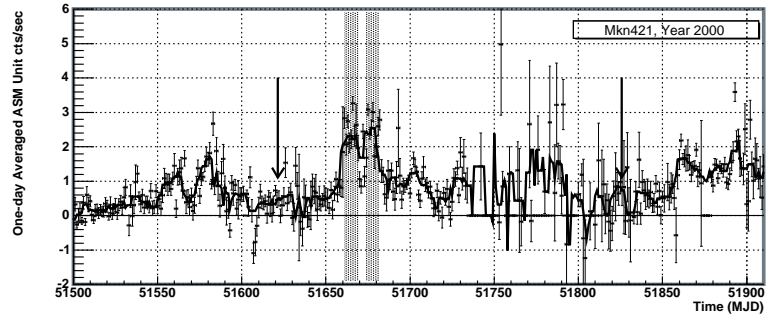}
\caption{\label{fig1b}  
The periods of interest for TeV blazars have been selected based on 
the X-ray light curves provided by ASM/RXTE \cite{ASM} (points with error bars).
A median filter is first applied to the light curves in order to make  the trend of the curve more evident
(continues line). 
The arrows represent the arrival time of the
neutrino events. The shaded regions are the periods of high activity selected.
For space reason, only the result of Mkn421 and the year 2000 is shown, however 
the optimization runs on the entire 4 years of data. 
 }
\end{center}
\end{figure}

{\bf TeV blazars} show dramatic variability correlated between multiple wavelengths of
the electromagnetic spectrum; correlations between TeV $\gamma$-rays
and X-rays occur on time-scales of hours or less.
This correlated variability is often interpreted as a strong argument in favor of
pure electromagnetic models (leptonic models) in which the same
population of ultra-relativistic electrons is responsible for
production of both X-rays and TeV $\gamma$-rays.
In fact, these observations do not rule out models involving
the acceleration and interaction of protons (hadronic models see e.g. \cite{Aharonian}).
In hadronic models, pions produced by $p\gamma$ or $pp$ interaction result in the simultaneous
emission of  $\gamma$-rays and neutrinos. 
Imaging atmospheric Cherenkov telescopes have detected various 
$\gamma$-ray flares in the energy region GeV to TeV. 
If it were not for large gaps in time between the measurements of these
flaring periods, the measured on-times of the $\gamma$-rays would clearly
define the high state of activity of the sources. On the contrary,
the use of the $\gamma$-ray measurements in this context is quite
limited.
TeV and X-ray flares are, with few exceptions, well correlated (see e.g.
\cite{Kraw}), and all-sky X-ray measurements guarantee a quasi-continuous data
record. Therefore a reasonable strategy, which we  used, is to
select the periods of interest on the basis of the X-ray light curves
provided by ASM/RXTE \cite{ASM}, see Fig.~\ref{fig1b}  We have looked for an excess of events in the
on-source direction by comparing the integrated number of neutrino counts
versus the estimated atmospheric neutrino background for the selected time
periods. The optimization was performed 
over the entire 4 years of data. Quantitative results are reported in Tab. 1.\\

\vspace{-10pt}

\vspace{-10pt}
{\bf Microquasars} are galactic X-ray binary (XRB) systems, which exhibit relativistic radio jets 
\cite{DiStefano}.
Microquasars are associated with several classes of XRBs and  variable time behavior. 
The observed radiation from microquasar jets, typically in the radio,  in some cases also in the IR band, 
is consistent with non thermal synchrotron radiation emitted by a population of relativistic, shock-accelerated electrons.
The composition of microquasar jets is still an open issue.
An indication of e-p jets is the presence of Doppler-shifted spectral lines
proving the presence of nuclei in the jets of this source.
Neutrino bursts are predicted in such sources. The duration should be of the order of the ejection time of the blob.
It should precede the associated radio outburst
by several hours. 
Radio data are used as identification time of the emission 
of the jets and of the emission of the possible neutrino production. 
Following these observations we have decided to include one microquasar (Cygnus X-3)
in the search for neutrino bursts. In this case,  periods of high activity are selected 
based on the radio light curve from the Ryle telescope,  
which has a nearly constant record of the source  
(data available courtesy from Guy Pooley). 
A window of three days is added prior to this period in order to take  into 
account  that the neutrino signal should precede the radio emission. 

\section{Method B: Flare Search}
In this section a search method sensitive to occasional neutrino flares, whose times and
durations are unknown,
is described.
Observations of strong variability in the electromagnetic emission exist for various
 TeV neutrino candidate sources. 
However, often there is no continuous observation of the flux (e.g. the EGRET sources) 
and/or no prediction for a time correlation between the photon and neutrino
emission.\\
The data selection has been optimized with the help of toy Monte Carlo events.
The data sample produced by the time-integrated point source search \cite{4y} 
without any re-optimization gives the highest detection probability.
A method based on a sliding window of fixed duration has been investigated, where
the length of the window is a parameter to be optimized.
As the flare duration is not known 
the detection probability in dependence of the time window length has to be calculated.
Note that short flares will be not detectable if the peak to valley
ratio is much higher than
the one observed in $\gamma$-rays. 
The signal contribution in very long windows is limited by the number of 
events observed in the time-integrated search.
The optimum choice is a 20 days window for galactic objects and a 40 days window for
extra-galactic ones.
The three categories of sources selected  
are a subset of the standard list of TeV neutrino candidate sources 
used in the time-integrated point source analysis \cite{4y}.
Moreover, we have included in our analysis 
three sources from the EGRET catalog which show 
extraordinarily large variations in the MeV $\gamma$-ray flux.

\begin{figure}[ht]
\begin{center}
\includegraphics*[width=1.0\textwidth,angle=0,clip]{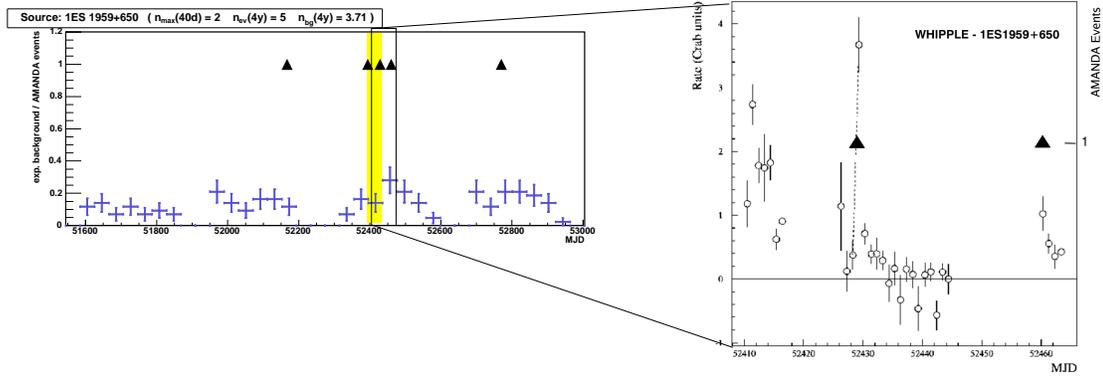}
\caption{\label{fig2b}  
Left: AMANDA-II neutrino candidates within $2.25^o$ from the direction of the blazar 1ES 1959+650. 
The triangles indicate the arrival time
of the observed events; the crosses refer to the background events in the 
40-days windows.  
The window showing the highest multiplicity is highlighted. Right: Zoom-in of the time-window MJD 52410-52460.
The arrival time of two out of the five AMANDA-II events is compared with the Whipple light curve from 
\cite{whipple}.}
\end{center}
\end{figure}

\vspace{-20pt}
\section{Results and Discussion}
The results obtained with the two methods discussed in section 2 and section 3 are reported in Tab.~\ref{table}. 
In both cases, no statistically significant excess of events over the background expected 
has been observed.

\begin{table}[ht]  

\begin{center}
\begin{tabular}{c|ccc|ccc} \hline \hline 
~Source~ &             & ~Multiwavelength~ & & &~Sliding Window~ &       \\
         & $n_{obs}$  & $n_{bg}$  & $n_{obs}/n_{bg}$ & $n_{obs}$  & $n_{bg}$  & $n_{doublets}$ \\
	 \hline
Mkn421 &  7 & 9.44 & 0/1.63 & 6 & 5.58 & 0\\
1ES 1959+650 & 5 & 4.67 & 2/1.57 & 5 & 3.71 & 1\\
QSO 0235+164 &&&& 6 & 5.04 & 1\\
QSO 0528+134 &&&& 4 & 4.98 & 0\\
\hline
Cygnus X-3 & 13 & 9.86 & 2/1.39 & 6 & 5.04 & 0\\
Cygnus X-1 &&&& 4& 5.21 & 0\\
GRS 1915+105 &&&& 6 & 4.76 & 1\\
GRO J0422+32 &&&&  5 & 5.12 & 0\\
\hline
3EG J0450+1105 &&&& 6 & 4.67 & 1\\
3EG J1227+4302 &&&&  6 & 4.37 & 1\\
3EG J1828+1928 &&&& 3 & 3.32 & 0\\
3EG J1928+1733 &&&& 7 & 5.01 & 1\\

\hline
\hline

\end{tabular}
\caption{\label{table} Results from the search of neutrinos from selected variable sources. $n_{obs}$ is the number of observed 
events in 
 four years and $n_{bg}$ is the expected, corresponding background.
$n_{obs}/n_{bg}$ are the event numbers observed during to the periods 
of high X-ray activity for the two blazars and in the radio band for the micro-quasar. $n_{doublets}$ corresponds to the 
 cluster of two events within the time window (20/40 days).  }
\end{center}
\end{table}

Although the results obtained are not significant, 
the time structure of the neutrino candidates from the direction of the blazar 1ES 1959+650  (within $2.25^o$)
merits a dedicated discussion.
The sample is composed of five events in the entire four-years period.
They have been identified by the blind
analysis as described previously (see Fig.~\ref{fig2b}).
The following has been noticed:
\begin{itemize}
  \item Three events out of the five fall within 66 days (MJD 52394.0, 52429.0, 52460.3). 
  \item The period of time in which these three events fall is 
  partially overlapping with a period of exceptional high activity of the source. 
  The activity of the
  source has been monitored by a multiwavelength campaign during the period MJD 52410-52500. 
  The detailed analysis of the campaign was reported earlier \cite{Kraw}. 
  \item In the paper cited, the detection of a $\gamma$-ray flare without its X-ray counterpart
  is reported (MJD 52429.3). This event has been defined as the first unambiguous 
  example of an ``orphan'' $\gamma$-ray flare from a blazar. 
  The main conclusion from the observation of the "orphan" flare is that it cannot be explained with a 
  conventional one-zone synchrotron self-Compton model. Several authors interpret such "orphan" as indicators  of 
hadronic processes occurring in the blazar jet.  
High energy neutrinos are expected in this case. 
  One of the five events  of our sample (MJD 52429.0)
  is within few hours from  the "orphan" flare.  
\end{itemize}
On the base of these observations different authors discussed 
the possible neutrino emission during the "orphan" flare 
(see \cite{Halzen}). 
The IceCube collaboration is currently working on methods to search for neutrinos connected
 to phenomena similar to the case of 1ES 1959+650.  To achieve that, a
close collaboration with the $\gamma$-ray community is essential. So, we encourage
 extensive multidisciplinary investigations and 
extended $\gamma$-ray monitoring of this and similar sources.


%
\setcounter{section}{0}
%
\newpage

\section*{\Large  A source stacking analysis of AGN as neutrino point source candidates  with AMANDA}

\vskip 0.05cm
{\large A.~Gro{\ss}$^{\it a}$ and T.~Messarius$^{\it a}$ for the IceCube Collaboration}\\ 
        
{\it (a) Institute for Physics, University of Dortmund, D-44221 Dortmund, 
            Germany}
\vskip 0.05cm
{\large Presenter: T.~Messarius (timo@physik.uni-dortmund.de), \  
ger-messarius-A-abs2-og25-poster}

\title[AGN source stacking with AMANDA]{ A source stacking analysis of AGN as neutrino point source candidates  with AMANDA}
\author[A.~Gro{\ss} and T.~Messarius for the IceCube Collaboration] {A.~Gro{\ss}$^a$ and T.~Messarius$^a$ for the IceCube Collaboration\\ 
        (a) Institute for Physics, University of Dortmund, D-44221 Dortmund, 
            Germany 
        }
\presenter{Presenter: T.~Messarius (timo@physik.uni-dortmund.de), \  
ger-messarius-A-abs2-og25-poster}

\maketitle

\begin{abstract}

Source stacking methods have been applied in $\gamma$-astronomy
and in optical astronomy to detect generic point sources at the sensitivity
limit of the telescopes. In such an analysis, the cumulative signal and
background of several selected sources of the same class is evaluated. 
In this report we introduce a systematic classification of AGN into several
categories that are each considered as a TeV neutrino source candidate.
Within each of these AGN categories the individual sources are stacked and 
tested for a cumulative signal using the AMANDA-II data.
\end{abstract}

\section{Introduction}
Active Galactic Nuclei (AGN) are known high luminosity photon emitters 
reaching photon energies up to some ten TeV. Additionally they are candidates
for the production of high energy charged particles and neutrinos. A detection
of TeV neutrinos 
($\nu$'s) from AGN would provide invaluable insight into their nature and their
contribution to the measured flux of high energy charged particles in cosmic
rays. 

AMANDA-II is a neutrino telescope operating at the geographic South
Pole~\cite{AMANDA}. 
In AMANDA-II point source analyses~\cite{PSA_Tonio,
  PSA_Paolo,PSA_Zeuthen_ICRC}, no statistically significant signal from 
any point source has been detected yet. The most sensitive analysis is based on
data collected in 2000-2003~\cite{PSA_Zeuthen_ICRC}.  A further increase in
sensitivity  may be achieved by evaluating the same data set for the
cumulative signal of several AGN.
A source stacking analysis has been developed using individual classes of
AGN that have been systematically categorized~\cite{stacking_paper}.

For each AGN type, the number of sources is optimized for the analysis with
AMANDA-II. 
The samples are analyzed for a cumulative neutrino flux using the data set of
the point source analysis~\cite{PSA_Zeuthen_ICRC}.

\section{AGN classification and selection scheme}
With the aid of a systematic classification of AGN, we define categories of
AGN for which individual AGN in each category are stacked.
A large diversity of AGN types (like Seyfert galaxies, radio galaxies or
quasars) has been observed and has historically been classified and named due
to their appearance in  telescopes sensitive to various wavelengths at Earth. 
The differences between the
various AGN types may partially be explained by a unified geometrically
axisymmetric model~\cite{urry}.

\begin{figure}[h]
\begin{center}
\includegraphics*[width=0.9\textwidth,angle=0,clip]{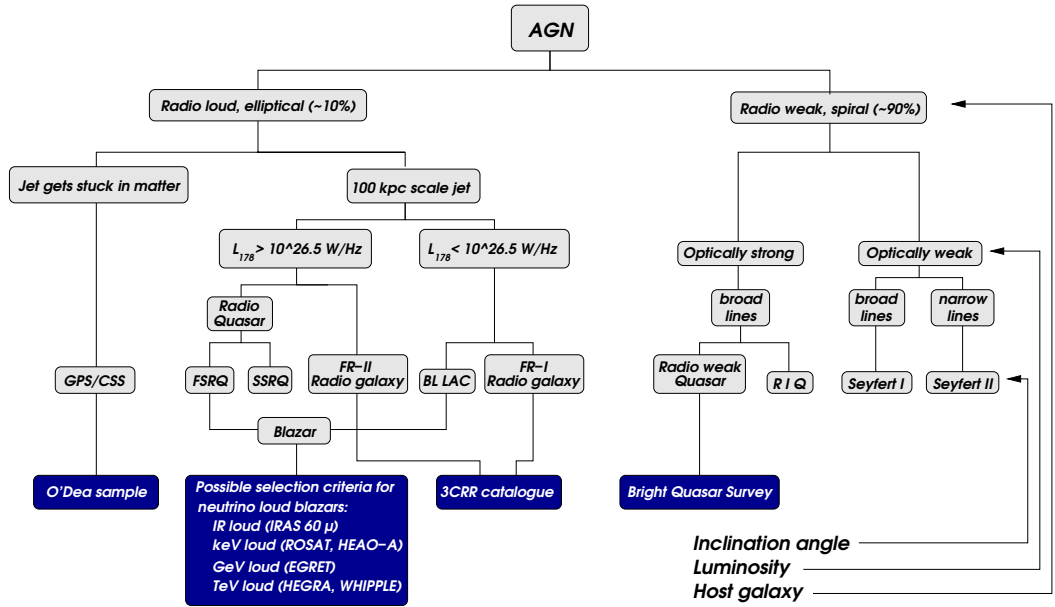}
\caption{\label{tree} 
AGN classif\/ication according to host galaxy, luminosity and
inclination angle. In the scheme, SSRQ and FSRQ stand for steep spectrum radio
quasar resp.\ flat spectrum radio quasar. Radio intermediate quasars are
labeled RIQ. The GHz Peaked Sources and the Compact Steep Spectrum sources are
represented by GPS/CSS.
}
\end{center}
\end{figure}

\begin{figure}
\begin{center}
\begin{tabular}{cc}
\includegraphics*[width=0.32\textwidth,angle=0,clip]{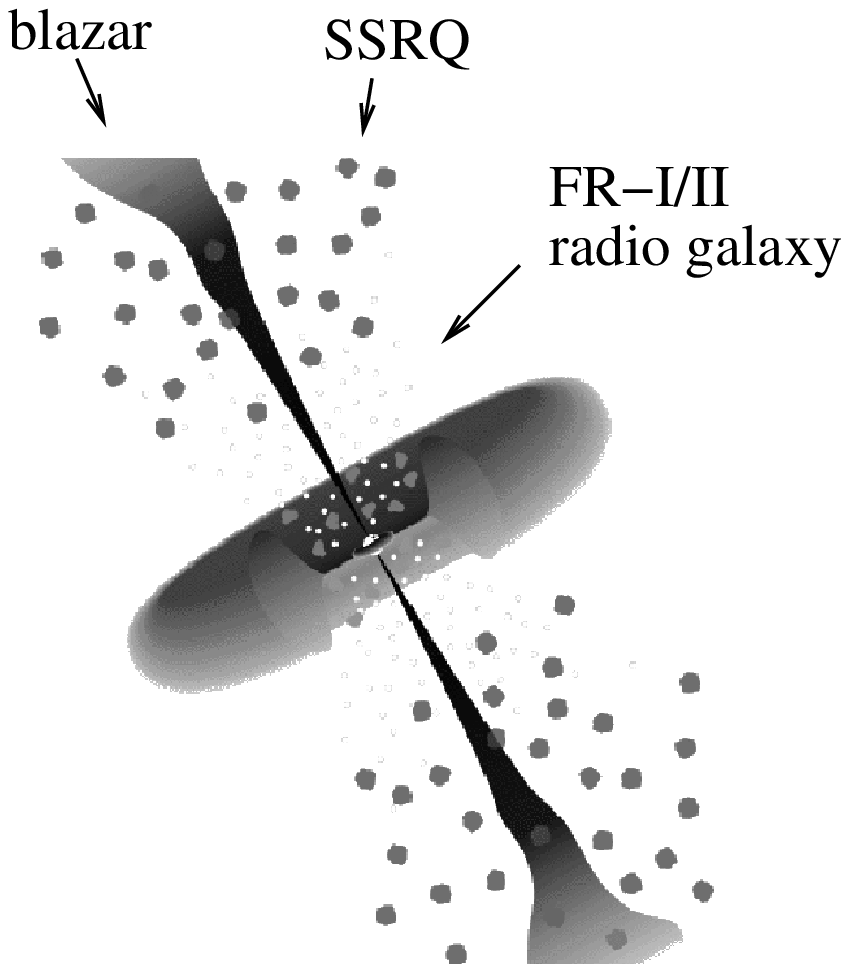}&
\includegraphics*[width=0.5\textwidth,angle=0,clip]{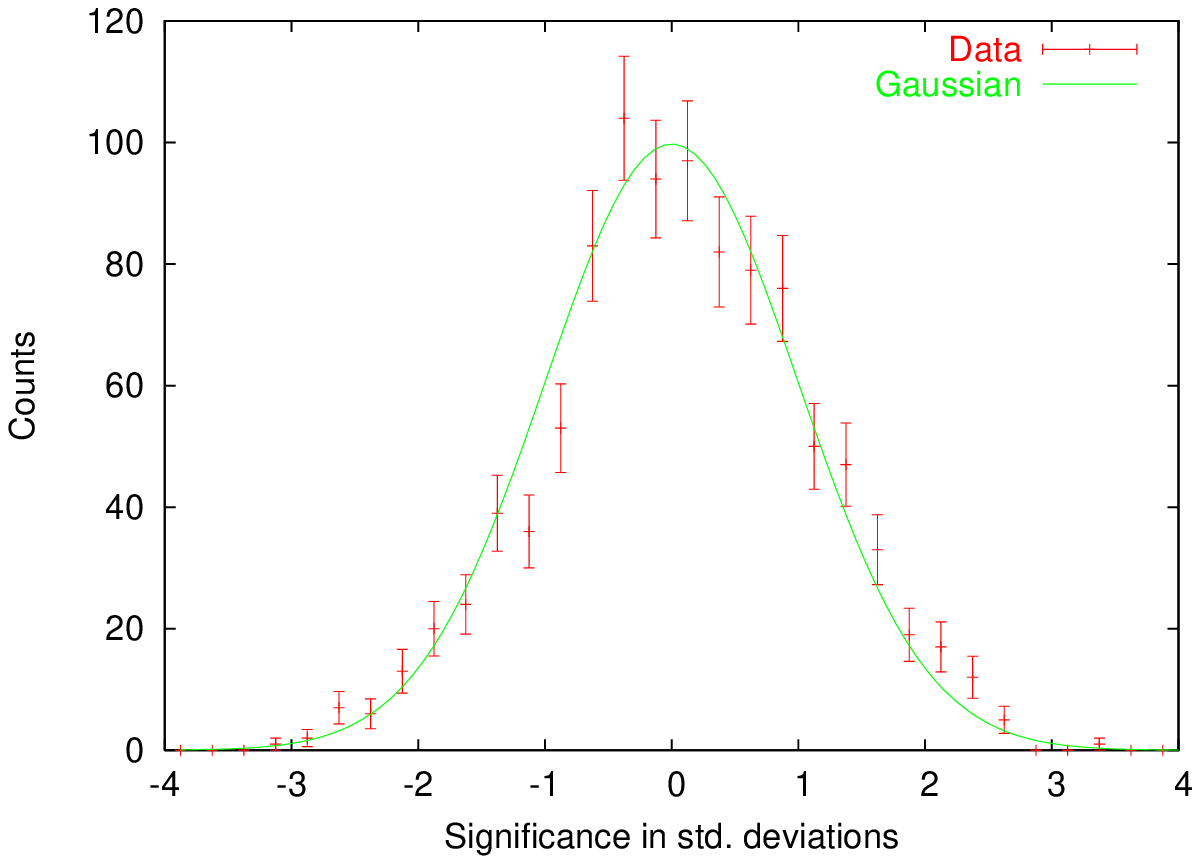}
\end{tabular}
\caption{\label{scheme} On the left: Scheme of an AGN with a black hole in the center and an accretion disc
perpendicular to the direction of two jets along its rotation axis.
On the right: Significance distribution for hypothetical source
samples with random positions.}
\end{center}
\end{figure}

In Fig.~\ref{tree}, we present an AGN classification scheme depending on host
 galaxy, luminosity and angle of observation.
 A schematic view of an AGN is shown in Fig.~\ref{scheme}.
AGN are generally divided into radio-loud and radio-weak sources as it is
 indicated in the first branching of the classification scheme.
Radio-loud AGN are mostly found in elliptical galaxies, while radio-weak ones
 are mostly located in spiral galaxies~\cite{dowd}. Radio-loud AGN with large
 scale jets show  different morphologies depending on the luminosity at $178$
 MHz~\cite{FR}.  

The catalogs which are used for the selection of generic source classes are
listed at the bottom of Fig.~\ref{tree} (black squares). 
A list of compact radio-loud objects, possibly young AGN where the jets are
stopped in dense matter, is given by a sample defined in~\cite{odea}.
Blazars are radio-loud AGN observed in jet direction, including 
subclasses of different luminosity at low radio frequencies, Flat Spectrum
Radio Quasars (FSRQ) and BL Lac objects. 
Blazar emission is dominated by relativistic beaming effects and characterized
by a flat radio spectrum. 
Radio galaxies, which are radio-loud AGN observed at a high inclination angle,
are selected by their radio flux at $178$ MHz as provided by the 3CR
catalog. The radio galaxies are further divided into
FR-I and FR-II radio galaxies, depending on different
jet morphologies which are correlated with the luminosity at $178$
MHz~\cite{FR}. 
A sample of radio-weak quasars is defined by the Bright Quasar Survey,
selected at optical and ultraviolet frequencies.

Further details of the classification can be found in~\cite{stacking_paper}. 

In most theoretical models for $\nu$ production in AGN, the dominant
$\nu$ production mechanism is the decay of  charged pions. The pions are
decay products of the $\Delta$ resonance produced by $pp$ or $p\gamma$
interactions of high energy protons. In AGN, the jets and the accretion disk
are possible acceleration sites for protons.

In this scenario, the production of $\nu$'s is always coincident with the
production of a similar number of $\gamma$'s from the decay of neutral pions.  
For an observer at the Earth, the different optical depths for the propagation
of neutrinos and photons may affect the spectra. The high cross
sections of photons, e.g.\ for pair creation,  mean a higher probability that photons will
 interact or be absorbed in the source region or in the
interstellar medium. Then, the energy of the photons 
gets redistributed on several photons of lower energy and 
the $\gamma$ flux is shifted to lower energies than the $\nu$ flux. 

The $\nu$ production scenario motivates a selection of candidate sources from
the various AGN classes according to their photon flux. 
The possible cascading of photons suggests considering also
photon energies which are lower than  the energy threshold of AMANDA-II for
neutrinos, which is at about $50$ GeV. 
For blazars, different selections are made according to their measured flux at
IR, keV, GeV and TeV energies. Radio-weak quasars are selected according to
their photon flux at $60\, \mu$m. For FR-I and FR-II radio galaxies, the radio
flux at $178$ MHz is used as the selection criterion. The compact objects (GPS
and CSS) 
are sorted according to their optical strength. 

The number of sources in each generic sample is determined by 
the assumption, that the hypothetical  $\nu$ flux at
TeV energies is proportional to the photon flux
at the selection energy. The normalization of the hypothetical signal is
chosen in a way that the signal of the strongest source of each sample 
does not exceed the
sensitivity of the point source analysis. 

For most AGN categories, we found the optimum number of sources to include for
the AMANDA-II analysis was about ten.
The exceptions are FR-I  and FR-II radio 
galaxies where specific distributions of the candidate sources constraint
source stacking. FR-I galaxies are strongly dominated by the local source M87,
while many FR-II galaxies show a comparable $\gamma$ flux.
Thus, for FR-I galaxies, the analysis of only M87 turns out to be most
sensitive. 
In contrast, there is no optimum number of FR-II galaxies above which the
expected sensitivity is decreasing. In this case the techniques of point
source analysis become less suitable to analyze the signal than the search for
an isotropic flux. 

Even though in these two cases source stacking seems to be less sensitive to
the hypotheses than other analyses, we look also at samples of these
sources. For FR-II galaxies, this takes into account that
point source methods are complementary to diffuse methods. For FR-I
galaxies  the possibility that M87 may be an exception from a
general correlation motivates the consideration of an additional sample 
of FR-I galaxies where
M87 is excluded. For a source listing of all samples see~\cite{stacking_paper}.

\section{Data analysis}

For the samples of the AGN categories as selected above, the cumulative signal was 
evaluated with a data sample of events collected with AMAN\-DA-II in the
year 2000~\cite{PSA_Tonio}. None of the AGN categories had a statistically significant 
deviation from the background expectation~\cite{moriond,stacking_paper}. 
Here, we present an application of the method to a data set collected with
AMANDA-II in 2000-2003~\cite{PSA_Zeuthen_ICRC}. 
All optimization steps for the stacking results were redone, resulting in
no significant change in the number of sources to include. The optimum bin
size  was found at a radii of $2.3-2.5^\circ$, which is about $15 \%$ smaller
than for the previous sample.

For the evaluation of a possible signal, the number of background events
expected in a circular search bin around each 
source is assumed to be proportional to the event density in the zenith band
of that source. Events
which are in overlapping bins contribute only once to the cumulative
signal. The background expectation for the stacked sources is  corrected for
the background expectation of the overlap regions. In this way, a double count
of statistical fluctuations in  overlapping bins is avoided.
The analysis was tested in various ways. In
total, $1000$ test data sets with randomized right ascension (RA) for all
events were generated. The resulting distribution of on-source counts is in
agreement with the expectation from Poisson statistics. Additionally, a
collection of samples containing $10$ hypothetical sources with random source
positions was analyzed, using the final data set (without 
randomization of RA). Also in this case, no deviations from the expectations
were found. The significance distribution for these test data sets is shown in
Fig.~\ref{scheme}(on the right).

\begin{floatingtable}{
\small
\begin{tabular}{|c|c|c|c|c|}
\hline
AGN category & $N_{src}$ & $N_{\nu}^{obs}$ & $N_{\nu}^{bg}$ & $f_{sens}$  \\
\hline
GeV blazars & 8 & 12 & 20.5 &1.5 \\
unid.\ GeV sources& 22 & 62  & 60.1&2.6\\
IR blazars & 11 & 30 & 34.1 &2.0\\
keV blazars (HEAO-A) & 3 & 7 & 11.0&1.4\\
keV blazars (ROSAT)  & 8 & 19 & 25.8 &1.8\\
TEV blazars & 5 & 14 & 18.3&1.5\\
GPS and CSS & 8 & 16 & 22.7&1.7\\
FR-I galaxies & 1 & 2 & 2.5&0.7\\
FR-I without M87 & 17 & 28 & 45.0&2.4\\
FR-II galaxies& 17 & 58 & 53.8&2.6\\
radio-weak quasars& 11 & 29 & 32.5&2.1\\
\hline
\end{tabular}
}
\caption{Results of the stacking analysis for each AGN category:
the number of included sources is given by $n_{src}$, the number of
  expected events is listed under $n_{bg}$ and  the number of observed
  events is given by $n_\nu$. The sensitivity of the analysis is listed under
  $f_{sens}$ in units of $10^{-8} \mbox{cm}^{-2}
\mbox{s}^{-1}$ for the integral flux above $10$ GeV.}
\label{resulttable}
\end{floatingtable}
The evaluation of the signal yields no excess for any of the considered
AGN categories. The observed and expected event counts for the selected AGN samples
are listed in
Table~\ref{resulttable} together with the sensitivity to a flux from
the corresponding AGN category. The sensitivities are given at $90\%$
C.L. and do not include systematic errors. Limits on the flux from the
considered AGN categories will be given after the investigation of systematic
errors.

With help of the stacking analysis, the sensitivity per generic source has
been lowered by a factor of $3$ for a typical AGN category. No evidence of a signal
has been found.


\newpage

\setcounter{section}{0}
\section*{\Large A Search for High-energy Muon Neutrinos from the Galactic Plane with AMANDA-II}

\vskip 0.05cm
{\large J.L. Kelley$^a$ for the IceCube Collaboration} \\
{  \it (a) Physics Department, University of Wisconsin, Madison, WI 53706 USA}
\vskip 0.05cm
{\large Presenter: J.L. Kelley (jkelley@icecube.wisc.edu), \  
usa-kelley-j-abs1-og25-oral }

\title[A search for neutrinos from the galactic plane with AMANDA-II ...]{A
  Search for High-energy Muon Neutrinos from the Galactic Plane with AMANDA-II}
\author[J.L. Kelley et al.] {J.L. Kelley$^a$ for the IceCube Collaboration \\
  (a) Physics Department, University of Wisconsin, Madison, WI 53706 USA
}
\presenter{Presenter: J.L. Kelley (jkelley@icecube.wisc.edu), \  
usa-kelley-j-abs1-og25-oral}

\maketitle


\begin{abstract}

Interactions of cosmic rays with the galactic interstellar medium produce
high-energy neutrinos through the decay of charged pions and kaons.  We
report on a search with the AMANDA-II detector for muon neutrinos from the
region of the galactic plane below the horizon from the South Pole
($33^{\circ} <$ galactic longitude $< 213^{\circ}$).  Data from 2000 to
2003 were used for the search, representing a total of 807 days of livetime
and 3329 candidate muon neutrino events.  No excess of events was
observed. For a spectrum of $E^{-2.7}$ and Gaussian spatial distribution
($\sigma = 2.1^\circ$) around the galactic equator, we calculate a flux
limit of $4.8\times
10^{-4}\ \mathrm{GeV}^{-1}\ \mathrm{cm}^{-2}\ \mathrm{s}^{-1}\ \mathrm{sr}^{-1}$ in 
the energy range from 0.2 to 40 TeV.

\end{abstract}


\section{Introduction}
\label{Introduction}

High-energy neutrinos are produced in the disk of the Galaxy as cosmic rays
interact with the interstellar medium (ISM), creating charged pions and
kaons.  Because of the low density of the ISM, the particles produced
typically decay before interacting again, and the energy spectrum of the
neutrinos follows the primary cosmic ray spectrum of $E^{-2.7}$.  Most
models of this emission predict a flux that is proportional to the column
density of the ISM, and thus highest towards the Galactic Center
\cite{Berezinsky93}, \cite{Ingelman96}.

The AMANDA-II detector, a subdetector of the IceCube experiment, is an
array of 677 optical modules buried in the ice at the geographic South Pole
which detects the \v{C}erenkov radiation from charged particles produced in
neutrino interactions with matter \cite{Andres01}.  In particular, muons
produced in charged-current $\nu_{\mu}$ and $\bar{\nu}_{\mu}$ interactions
deposit light in the detector with a track-like topology, allowing us
to use directional reconstruction to reject the large background of
down-going atmospheric muon events.  Up-going atmospheric neutrinos are the
primary remaining background for this search.  Because we restrict
ourselves to events originating below the horizon, we are not sensitive to
the region near the Galactic Center; rather, we perform a search for
neutrinos from the outer region of the galactic plane, $33^\circ < $ galactic
longitude $< 213^\circ$.  Using the parametrization
in ref. \cite{Ingelman96} with an average ISM column density in this region
of $0.8\times10^{22}\ \mathrm{cm^{-2}}$, we expect at Earth an average
$\nu_{\mu} + \bar{\nu}_{\mu}$ flux of $3.9 \times
10^{-6}\ \mathrm{GeV}^{-1}\ \mathrm{cm}^{-2}\ \mathrm{s}^{-1}\ \mathrm{sr}^{-1}$.


\section{Signal Hypothesis and Simulation}

The actual distribution of the ISM in the galactic plane is quite irregular
\cite{Nakanishi03}, so we use a simplified signal hypothesis.  Because the
ISM column density in the outer Galaxy does not vary too much (see e.g. the
map by Bloemen in ref. \cite{Berezinsky93}), we model the neutrino flux as
isotropic in galactic longitude.  The expected profile in galactic latitude
has not been discussed in detail in the literature, although we can study
models of $\gamma$-ray emission as a guide.  A recent model by Strong
\textit{et al.}  of the $\gamma$-ray emission from $\pi^0$ decay in a
somewhat lower energy range (4-10 GeV) has an approximately Gaussian
profile with $\sigma \approx 2.1^\circ$ around the galactic equator
\cite{Strong04}.

For our initial signal hypothesis, we have simulated a line source from the
galactic equator that is isotropic in galactic longitude.  As discussed
later, we also use two other spatial profiles: a diffuse flux near the
galactic equator, and a Gaussian with $\sigma = 2.1^\circ$.  The spectral
slope is assumed to be -2.7, but other values ranging from -2.4 to -2.9 are
also tested (for specific models, see e.g. \cite{Strong04}).  We do not model
the change of slope at the knee of the cosmic ray spectrum, since the
resulting difference in number of events is negligible.  Possible point
sources in the galactic plane have also not been considered.  To produce
the signal Monte Carlo (MC), we use a reweighting method to transform an
isotropic distribution of simulated events \cite{Ahrens03} to a line source
originating from the galactic equator.  The absolute normalization of the
simulated signal flux is adjusted after normalizing the atmospheric neutrino
MC to the data sample.


\section{Data Sample}

The data sample used for this analysis consists of 3329 candidate muon
neutrino events collected from 2000 to 2003, representing 807 days of
livetime.  The event selection involved a number of quality criteria to
reject mis-reconstructed down-going muon events, and was optimized for a
broad sensitivity to an $E^{-2}$ to $E^{-3}$ spectrum.  This sample was
originally used for a point-source neutrino search, and details of the data
selection procedure are presented elsewhere in these proceedings
\cite{Ackermann05}.  During the design and optimization of event selection
criteria, the right ascension of the data events was scrambled in
accordance with our blind analysis procedures. 


\section{Background Estimation and Optimization of Selection Criteria}

\begin{figure}[ht]
\centering
\includegraphics[scale=0.42]{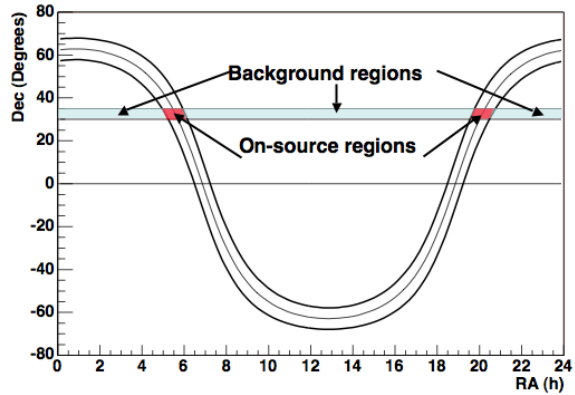}
\caption{Regions of the sky used for on-source event counting
  and background estimation for a particular declination slice.}
\label{fig_signal_bg}
\end{figure}

Because of our isotropic detector response in right ascension, we can use the
data to estimate the background in a point-source search by counting
events in a declination band around the sky.  For this analysis, however,
the source is extended across a large range of declinations, requiring a
modification of this technique.  We define the \textit{on-source region} as
the band of sky within $B$ degrees of the galactic equator.  The on-source
region is first divided into slices of equal declination $5^\circ$ wide, and
the background is estimated by counting the number of events in the
declination band outside the on-source region and scaling by the solid
angle ratio (see fig.~\ref{fig_signal_bg}). The total number of on-source and background events is then
calculated by summing over the declination bands.

The direction of the events is the most useful parameter in distinguishing a
galactic signal, and we optimize our sensitivity by varying the size of the
on-source region.  The region size is chosen to minimize the model
rejection potential \cite{Hill03}, the ratio of the average event upper
limit at the 90\% confidence level to the number of expected signal events
at a given reference flux.

\begin{figure}[ht]
\centering
\begin{minipage}[t]{7.2cm}
\includegraphics[width=0.9\textwidth]{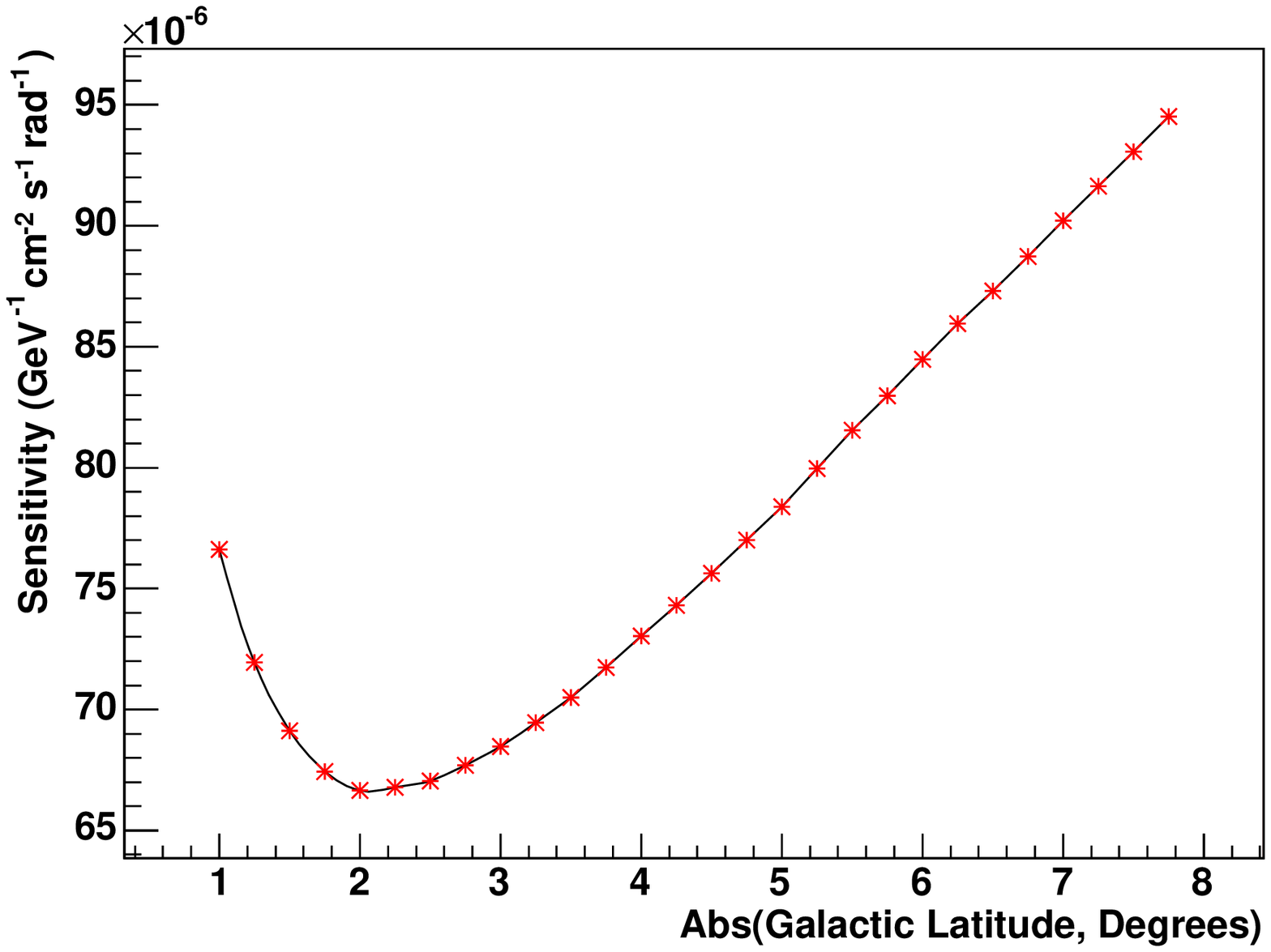}
\caption{Sensitivity to a line source as a function of on-source region size.}
\label{fig_sens_line}
\end{minipage}
\hfill
\begin{minipage}[t]{7.2cm}
\includegraphics[width=0.9\textwidth]{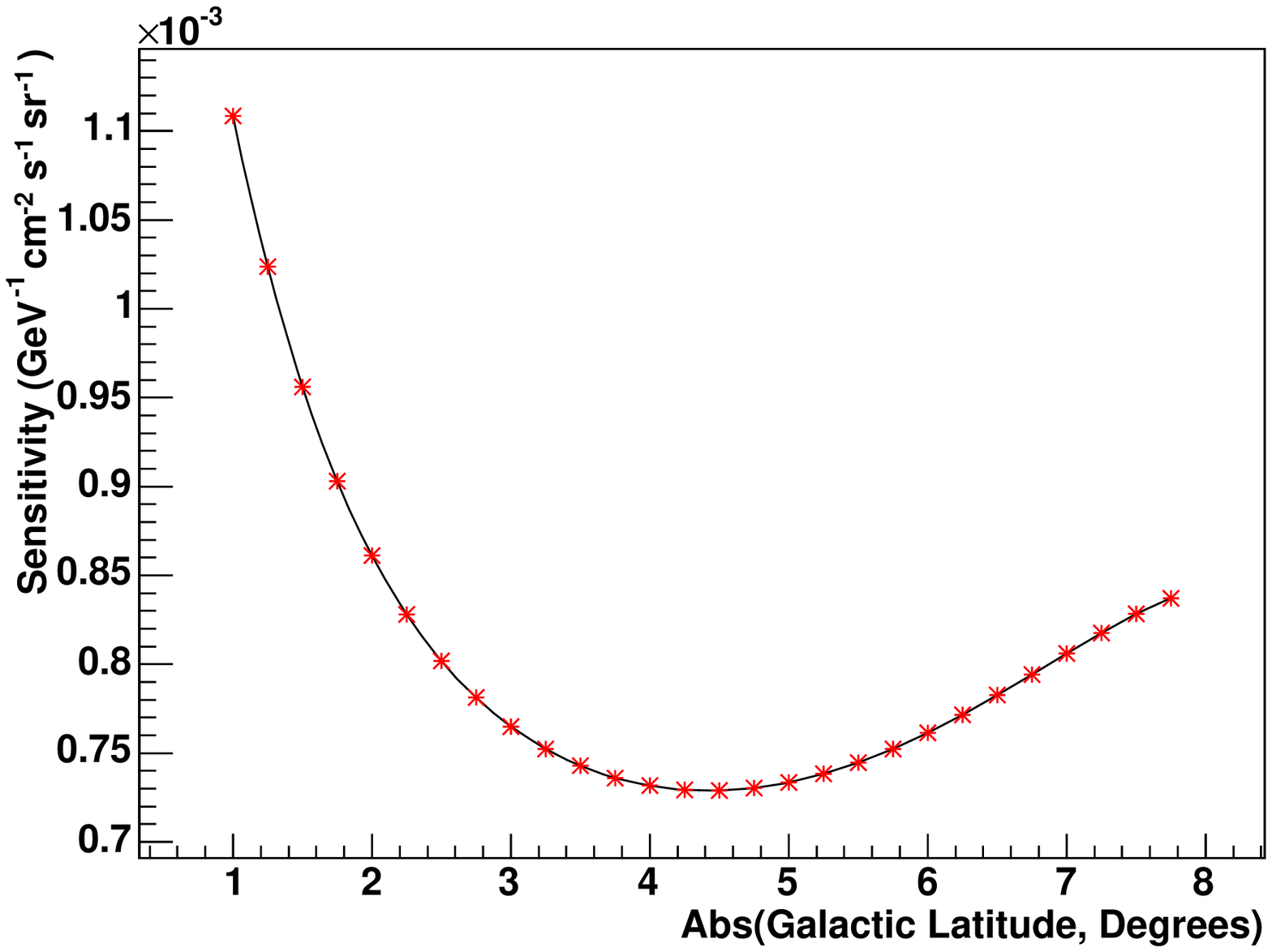}
\caption{Sensitivity to a Gaussian source ($\sigma = 2.1^\circ$) as a
  function of on-source region size.}
\label{fig_sens_gauss}
\end{minipage}
\hfill
\end{figure}

The optimal on-source region size for a line source was found to be
$B=2^\circ$ (see fig.~\ref{fig_sens_line}).  This optimization is, however, a bit artificial since it is
primarily determined by our line spread function.  To obtain a value for
more realistic flux distributions, we use an analytical method to estimate
the sensitivity to a diffuse flux in the on-source region, as well as to a
Gaussian signal profile of a given width.  Our sensitivity to another
signal profile is the flux level at which the total number of events in the
on-source region is equal to that of the original line-source flux.  This
approximation is valid as long as the zenith angle does not vary too much
over the on-source region.

First, we convert the line-flux sensitivity $\Phi_{line}$ (angular units of
$\mathrm{rad}^{-1}$) to a diffuse-flux sensitivity $\Phi_{diff}$ (angular
units of $\mathrm{sr}^{-1}$).  We integrate the line flux over $\pi$
radians of galactic longitude, divide by the solid angle $\Omega_{gal}$ of
the on-source region, and include an efficiency factor $\eta$ as the
fraction of signal events in that region:

\begin{equation}
\label{sens_diff}
\Phi_{diff} = \eta\ \pi\ \Phi_{line}\ /\ \Omega_{gal}\ .
\end{equation}

A similar procedure can be used to estimate the sensitivity to a Gaussian
signal profile.  The convolution of a Gaussian signal of width $\sigma_{sig}$ with
the line spread function (also approximated as a Gaussian, of width
$\sigma_{lsf} \approx 1.5^\circ$) results in a wider Gaussian.  As before,
by integrating to equalize the number of events in the angular region, we
solve for the Gaussian peak sensitivity $\Phi_{peak}$ in terms of the line
source sensitivity $\Phi_{line}$:

\begin{equation}
\label{sens_gauss}
\Phi_{peak}(B) = \frac{\Phi_{line}(B)}{\sqrt{2\pi(\sigma_{lsf}^2 +
    \sigma_{sig}^2)}} \ \operatorname{erf}(B / \sqrt{2}\sigma_{lsf})
    \ /\ \operatorname{erf}(B / \sqrt{2 (\sigma_{lsf}^2 +
      \sigma_{sig}^2)})\ .
\end{equation}

Using the relationship between the line-flux region size $B$ and the
sensitivity $\Phi_{line}$, we can reoptimize for the wider Gaussian signal
profile.  For a Gaussian with $\sigma_{sig} = 2.1^\circ$, we find an
optimal on-source region of $B=4.4^\circ$ (see fig.~\ref{fig_sens_gauss}).

\section{Results}

After unblinding the data, we observe no excess of events.  We first
calculate a limit on the line source flux at the 90\% confidence level, and
then calculate corresponding limits for diffuse and Gaussian spatial profiles
using the analytical expressions above (eqns. \ref{sens_diff} 
and \ref{sens_gauss}).  These results are presented in table
\ref{tbl_results} for a signal spectrum of $E^{-2.7}$.  The energy range of
these limits, incorporating the central 90\% of the signal spectrum after
all selection criteria, is 0.2 to 40 TeV.  The calculation has been
repeated using spectral slopes from -2.4 to -2.9, resulting in
diffuse-flux limits ranging from $5.3\times10^{-5}$ to
$3.1\times10^{-3}\ \mathrm{GeV}^{-1} \mathrm{cm}^{-2} \mathrm{s}^{-1}
\mathrm{sr}^{-1}$.

Because the signal flux is normalized using atmospheric neutrino MC, the
largest systematic error is the uncertainty of $\sim$30\% on the absolute
atmospheric neutrino flux, and this has been incorporated into the limits
\cite{Conrad03},\cite{Hill03b}.   Possible unquantified sources of error are 
variations in the width of the Gaussian signal profile, and the offset of
the peak flux from the galactic equator.

\begin{table}[ht]
\begin{center}
\begin{tabular}{|c|c|c|c|c|c|c|c|} 
\hline \hline
On-source & On-source & Expected & Event & Line source limit & Diffuse limit & Gaussian limit \\
region & events & background & upper limit & & & \\
\hline \hline
$\pm2.0^\circ$    &   128    &   129.4   &  19.8 & $6.3 \times 10^{-5}$ &  
$6.6 \times 10^{-4}$ &   --  \\ 
\hline
$\pm4.4^\circ$    &   272    &   283.3   & 20.0 & -- & 
-- &  $4.8\times 10^{-4}$ \\
\hline
\end{tabular}
\caption{\label{tbl_results} Preliminary limits on the
  $\nu_{\mu}+\bar{\nu}_{\mu}$ flux at Earth from the outer galactic plane,
  for an $E^{-2.7}$ spectrum (systematic errors included). 
  Units on the line source limit are $\mathrm{GeV}^{-1} \mathrm{cm}^{-2}
  \mathrm{s}^{-1} \mathrm{rad}^{-1}$; units on the other two limits are 
  $\mathrm{GeV}^{-1} \mathrm{cm}^{-2} \mathrm{s}^{-1} \mathrm{sr}^{-1}$.}
\end{center}
\end{table}


\section{Conclusions}

Comparing the limit for a Gaussian flux profile in table \ref{tbl_results}
with the model prediction in section \ref{Introduction}, we find that the
sensitivity of this analysis is approximately two orders of magnitude above
the predicted flux.  IceCube, the $\mathrm{km}^3$-scale successor to
AMANDA-II, will have a larger effective area and better angular resolution
\cite{Ahrens04}.  Only for the most optimistic case, in which the source
profile is nearly a line, will the increase in angular resolution allow a
significantly smaller on-source window.  For five years of data from the
complete detector, the total improvement in sensitivity is just over one
order of magnitude.  Other approaches may be more sensitive -- for example,
we can focus only on dense areas of the ISM, such as the Cygnus region.
Also, recent calculations by Candia suggest that IceCube may be sensitive
to the flux from the Galactic Center using cascades from down-going
neutrinos \cite{Candia05}.  Detection of specific sources within the plane
may well precede discovery of a truly diffuse flux from the galactic disk.


%
\newpage

\setcounter{section}{0}

\section*{\Large The Search for Neutrinos from Gamma-Ray Bursts with AMANDA}

\vskip 0.05cm
{\large K. Kuehn$^a$, for the IceCube Collaboration and the IPN Collaboration}\\
{\it  (a) Department of Physics and Astronomy, University of California
	Irvine, CA 92697-4575 USA} 
\vskip 0.05cm
{\large Presenter: K.Kuehn (kuehn@HEP.ps.uci.edu), usa-kuehn-K-abs1-og25-oral }

\title[The Search for Neutrinos from GRBs with AMANDA]{The Search for Neutrinos from Gamma-Ray Bursts with AMANDA}
\author[K. Kuehn for the IceCube Collaboration] {K. Kuehn$^a$, for the IceCube Collaboration and the IPN Collaboration\\
        (a) Department of Physics and Astronomy, University of California
		Irvine, CA 92697-4575 USA\\ 
        }
\presenter{Presenter: K.Kuehn (kuehn@HEP.ps.uci.edu), usa-kuehn-K-abs1-og25-oral}

\maketitle

\begin{abstract}
The Antarctic Muon and Neutrino Detector Array (AMANDA), located at the South Pole, has 
been searching the heavens for astrophysical neutrino sources (both discrete and diffuse) 
since 1997. The AMANDA telescope detects \v{C}erenkov radiation caused by high-energy 
neutrinos traveling through the nearby ice; here we describe AMANDA's technique to 
search for neutrinos from gamma-ray bursts, both concurrent with the photon emission and 
prior to it (during the "precursor" phase).  We present preliminary results from several 
years (1997-2003) of observations, and we also briefly discuss the current status and future 
potential of an expanded search for GRBs.
\end{abstract}

\section{Introduction}
Gamma-ray bursts (GRBs) are among the most energetic phenomena in the universe.  All are
characterized by prodigious gamma-ray emission, hypothesized to occur as a result of the
collapse of a massive star or the merger of compact objects.  Aspects of these theories
have been corroborated by recent observations \cite{Price,Lee}; however, many questions 
about the nature of GRBs still remain.  One of the promising techniques currently
available to answer such questions is to use underwater or under-ice detectors to
observe possible high-energy neutrinos from these sources \cite{WaxBah,Alvarez}.  The search 
for neutrino emission will help to test models of hadronic acceleration in the fireball
\cite{WaxBah,Alvarez,Dermer} or other GRB scenarios, and the search for precursor 
neutrinos may constrain models of GRB progenitors \cite{Razzaq,Meszar}.  AMANDA uses the ice 
at the South Pole to detect \v{C}erenkov radiation from neutrino-induced muons from both 
atmospheric interactions and astrophysical sources \cite{Ahrens}, including, potentially, 
GRBs.  In its initial configuration (AMANDA B-10, operational from 1997-1999), the detector 
consisted of an array of 302 photo-detectors housed in optical modules (OMs) beneath the 
surface of the ice cap; the upgraded configuration of 677 OMs (operational from 2000 to 
2004) was known as AMANDA-II.

\section{Observation and Analysis}
The AMANDA GRB search relies on spatial and temporal correlations with photon observations 
of other instruments, such as the Burst and Transient Source Experiment \cite{BATSE} or the 
satellites of the InterPlanetary Network \cite{Hurley}.  For each GRB, we search for 
coincident neutrino emission during the entire burst duration, plus the 10 seconds prior to 
the burst start (plus corrections associated with uncertainties in burst timing).  We also 
perform a search for precursor neutrino emission from 110 seconds prior to the burst trigger 
up to 10 seconds before the trigger.  To determine the expected background rate for each 
burst, a larger period of one hour and 50 minutes of data is analyzed -- from one hour 
before the burst to one hour after the burst.  The 10 minute period during and immediately 
surrounding the burst is excluded from the background region, to ensure that the data 
selection criteria are not chosen in a biased fashion (a "blind analysis").  In addition, 
the event count per 10 second time bin during the background period is compared to the 
expected (temporally uncorrelated) distribution of background events.  This test determines 
if there are significant fluctuations in data rate due to any intrinsic instability in the 
detector which could be misinterpreted as a signal event.  All data included in the GRB 
searches satisfy these criteria established for stable detector operation.

\begin{figure}[htb]
\begin{center}
\includegraphics*[width=0.45\textwidth,angle=0,clip]{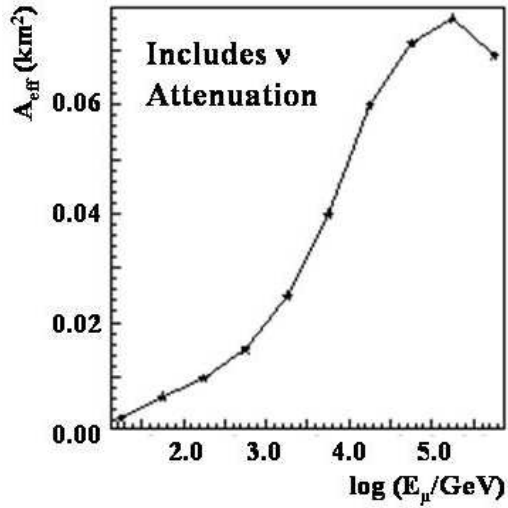}
\caption{\label{Fig2} Angle-averaged muon effective area as a function of muon energy, 
utilizing the data selection criteria from the year 2000 coincident GRB search.}
\end{center}
\end{figure}

The data selection criteria for the coincident and precursor searches are determined by 
minimizing the Model Rejection Factor \cite{Hill031}, which is defined as the 90$\%$ event 
upper limit derived from observed background events divided by the expected number of signal 
events determined from Monte Carlo simulations of the predicted neutrino flux.  For the 
coincident search, the predicted flux is derived from the Waxman-Bahcall model \cite{WaxBah} 
(corrected for neutrino oscillations), and for the precursor search, the predicted flux is 
derived from the model of Razzaque $\it{et}$ $\it{al.}$ \cite{Razzaq}.  In addition to 
temporal coincidence, several other selection criteria were used in these analyses, 
including: the angular mismatch between the burst position and the reconstructed event track 
(based upon a maximum-likelihood pattern recognition algorithm applied to the photon arrival 
time at each OM), the angular resolution of the reconstructed event track, and the 
uniformity of the spatial distribution of the hit OMs.  The detector's effective area 
($A_{eff}$) is determined after all selection criteria are applied, and provides a measure 
of the detector's sensitivity to neutrino-induced muons passing through (or nearby) the 
detector.  Though the selection criteria for the coincident and precursor searches differ 
slightly, both require only modest background rejection, giving AMANDA-II 
an $A_{eff}$ larger than any other currently-operating neutrino detector (Figure \ref{Fig2}).

\section{Results and Discussion}
\begin{figure}
\begin{minipage}[t]{7.5cm}
\includegraphics[width=0.9\textwidth,angle=0,clip]{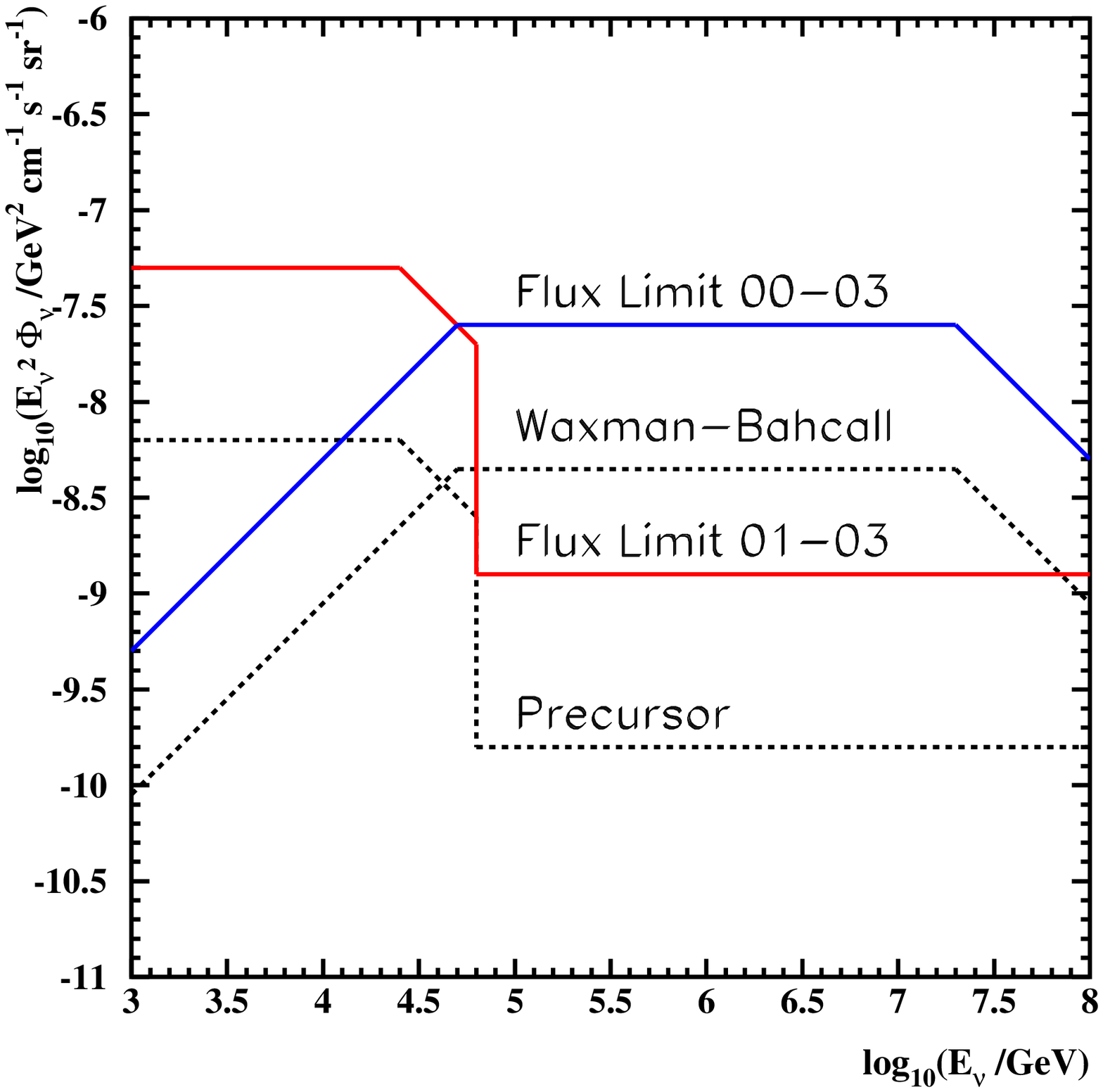}
\caption{\label {Fig3}Flux Limits for AMANDA-II observations (solid lines) and flux 
predictions for precursor and coincident spectra (dashed lines)}
\end{minipage}
\hfill
\begin{minipage}[t]{7.5cm}
\includegraphics[width=1.1\textwidth,angle=0,clip]{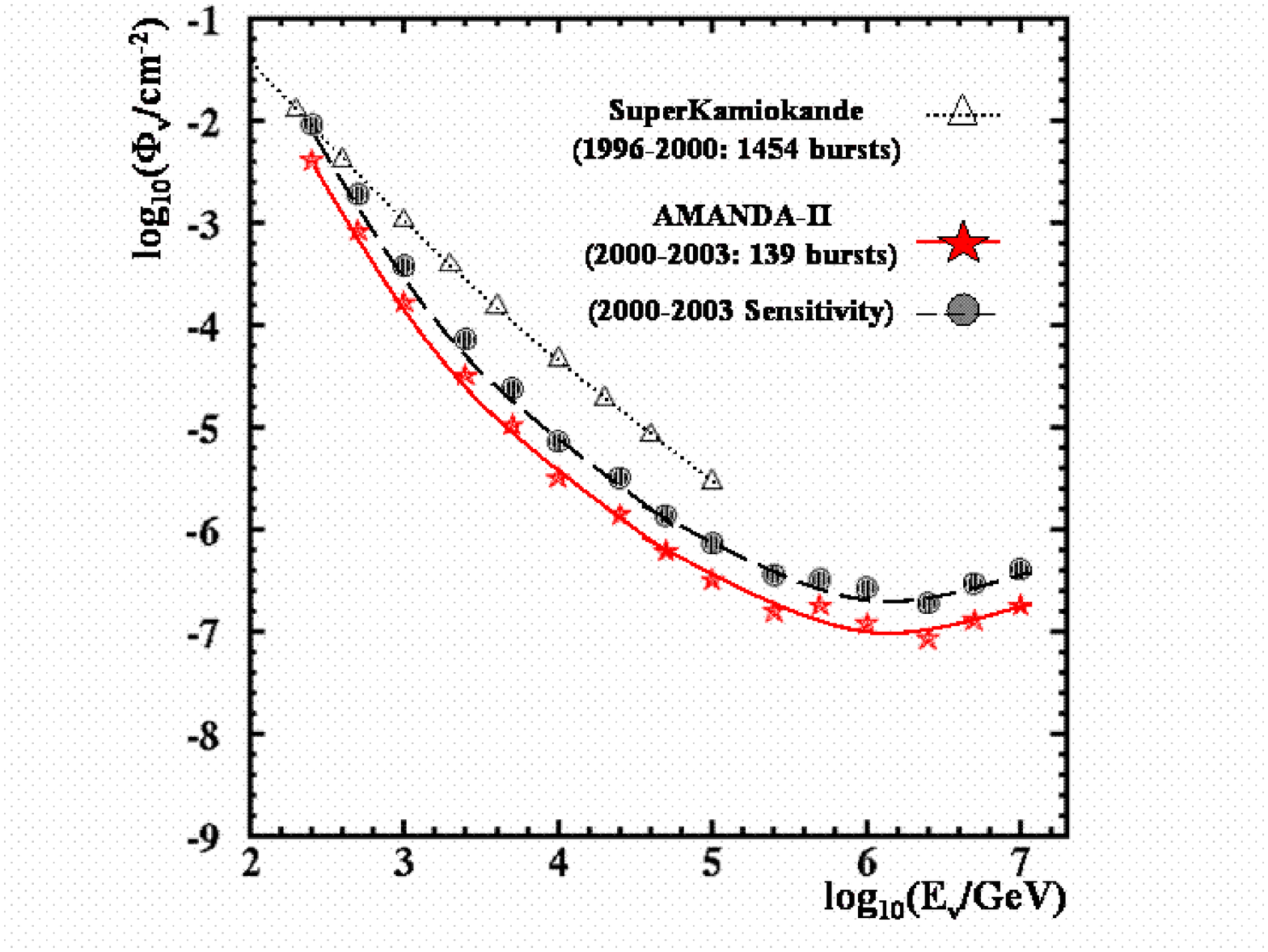}
\caption{\label {Fig4}Green's Function Fluence Limit for AMANDA-II observations of BATSE and 
IPN triggered bursts.  Note that the Super-Kamiokande results also include non-triggered 
BATSE bursts.}
\end{minipage}
\hfill
\end{figure}

AMANDA data from 1997-2000 were searched for emission coincident with 312 bursts detected 
by BATSE's online triggering system.  As reported elsewhere \cite{ICRC}, zero events were 
observed, which results in an observed upper limit on the muon neutrino flux of 4 $\times$ 
10$^{-8}$ GeV cm$^{-2}$ s$^{-1}$ sr$^{-1}$ (assuming a Waxman-Bahcall energy spectrum).  
Since the time of these observations, the coincident analysis has been expanded to include 
bursts detected by other satellites of the InterPlanetary Network.  This search for 139 
BATSE + IPN bursts from 2000-2003 also resulted in zero observed events, leading to an even 
more stringent upper limit of 3 $\times$  10$^{-8}$ GeV cm$^{-2}$ s$^{-1}$ sr$^{-1}$. 
These observations specifically exclude significant neutrino emission from GRB030329 
(for a more detailed independent analysis of this unique burst, see \cite{Mike}).
\footnote{Not included in this flux limit are the non-triggered BATSE bursts discovered in 
offline searches of BATSE's archival data \cite{Stern,Schmidt}; we searched for coincident 
emission from twenty-six such bursts and observed zero neutrinos.}  Additionally, we 
searched the 2001-2003 subset of bursts for a precursor neutrino signal; no events were 
observed.  Therefore, an upper limit of 5 $\times$ 10$^{-8}$ GeV cm$^{-2}$ s$^{-1}$ 
sr$^{-1}$ is derived for the precursor neutrino flux predicted by Razzaque $\it{et}$ 
$\it{al.}$.  The observations are summarized in Table \ref{Table1} and the flux upper limits 
are shown in Figure \ref{Fig3}.

\begin{table}[h]   
\caption{\label{Table1} Gamma-Ray Bursts Included in the AMANDA Observations}
\begin{center}
\begin{tabular}{||c|c|c|c|c|c|c||} \hline \hline

~Year~ & ~$N_{Bursts}$~  & ~$N_{BG}$~ & $N_{0bs}$ & Event U.L. & MRF & MRF (Sensitivity) \\
                             \hline \hline

2000        & 88 &  1.02  & 0 & 1.61 & 9 & 13 \\
                             \hline

2001        &  15  & 0.05 & 0 & 2.38 & 64 & 66 \\

Precursor   & 15 & 0.06 & 0 & 2.39 & & \\
                             \hline
2002        & 17 & 0.08 & 0 & 2.36 & 54 & 54 \\

Precursor   & 17 & 0.06 & 0 & 2.38 & & \\
                             \hline

2003        & 19 & 0.10 & 0  & 2.34 & 52 & 54 \\

Precursor   & 18 & 0.06 & 0  & 2.38 & & \\
                             \hline
00-03   & 139 & 1.25  & 0 & 1.47 & 5 & 10 \\

Precursor   & 50 & 0.16 & 0 & 2.28 & & \\
                             \hline

\end{tabular}
\end{center}
\end{table}

Though specific neutrino energy spectra have been assumed thus far, the results of
these analyses can be applied to other energy spectra as well, by using the Green's Function 
fluence method as presented by the Super-Kamiokande Collaboration \cite{SuperK}.  By 
folding the energy-dependent sensitivity of the detector into a desired theoretical 
spectrum, one can straightforwardly derive a flux limit for that specific spectrum.  
The Green's Function fluence limit for AMANDA-II (Figure \ref{Fig4}) extends several orders 
of magnitude beyond the energy range of the Super-Kamiokande limit.  It is also 
significantly (up to a factor of 10) lower compared to the Super-Kamiokande results in the 
overlapping energy region, primarily due to the much larger effective area of AMANDA-II.

\section{Conclusions}

AMANDA has searched for neutrino emission from nearly 500 GRBs based on temporal and
spatial coincidence with photon detections from numerous other observatories.  Thus  
far, zero neutrino events have been observed in correlation with these bursts.  These 
results lead to upper limits on the fluxes that are approaching the predictions for 
several canonical GRB models.  However, it has been observed that the individual bursts vary 
significantly in their expected neutrino spectra; therefore we are also constructing an 
analysis procedure based on more detailed models that incorporate individual burst 
parameters \cite{Mike}.  AMANDA is continuing its search for neutrino emission from various 
sources; even in the absence of a detection, the final results from AMANDA's observations 
from 1997-2004 should result in an improvement of the flux upper limits particularly for 
GRBs.  In February of 2005, AMANDA's successor experiment known as IceCube \cite{IceCub} 
began operation, and its increased collecting area will allow it to swiftly improve upon the 
limits of its predecessor.  Furthermore, the Swift satellite, operational since December 
2004, is expected to localize as many as 100 additional GRBs per year \cite{Swift}, which 
will provide a significantly larger dataset for the continuing search for neutrinos from 
gamma-ray bursts.

\section{Acknowledgements}

K.Kuehn would like to thank E. Waxman and S. Razzaque for productive and thought-provoking discussions, as well as S. 
Desai for detailed discussions regarding the Super-Kamiokande GRB analysis.

\newpage

\setcounter{section}{0}
%

\section*{\Large Probing for Leptonic Signatures from GRB030329 with AMANDA-II}

\vskip 0.05cm
{\large Michael Stamatikos$^{\it a}$ for the IceCube Collaboration, 
Jenny Kurtzweil$^{\it a}$ and Melanie J. Clarke$^{\it a}$}\\
{\it (a) Department of Physics, University of Wisconsin, Madison, 1150 University Avenue, Madison, WI 53706, USA}
\vskip 0.05cm
{\large Presenter: M. Stamatikos (michael.stamatikos@icecube.wisc.edu), \
usa-stamatikos-M-abs1-og24-oral }

\title[Probing for Leptonic Signatures from GRB030329 with AMANDA-II]{Probing for Leptonic Signatures from GRB030329 with AMANDA-II}

\author[M. Stamatikos et al.] {Michael Stamatikos$^a$ for the IceCube Collaboration, Jenny Kurtzweil$^a$ and Melanie J. Clarke$^a$
        \newauthor
            \\
        (a) Department of Physics, University of Wisconsin, Madison, 1150 University Avenue, Madison, WI 53706, USA \\
        }

\presenter{Presenter: M. Stamatikos (michael.stamatikos@icecube.wisc.edu), \
usa-stamatikos-M-abs1-og24-oral}

\maketitle

\begin{abstract}

The discovery of high-energy (TeV-PeV) neutrinos from gamma-ray bursts (GRBs) would shed light on their intrinsic microphysics by confirming hadronic acceleration in the relativistic jet; possibly revealing an acceleration mechanism for the highest energy cosmic rays. We describe an analysis featuring three models based upon confronting the fireball phenomenology with ground-based and satellite observations of GRB030329, which triggered the High Energy Transient Explorer (HETE-II). Contrary to previous diffuse searches, the expected \emph{discrete} muon neutrino energy spectra for models 1 and 2, based upon an isotropic and beamed emission geometry, respectively, are directly derived from the fireball description of the prompt $\gamma$-ray photon energy spectrum, whose spectral fit parameters are characterized by the Band function, and the spectroscopically observed redshift, based upon the associated optical transient (OT) afterglow. For comparison, we also consider a model (3) based upon \emph{averaged} burst parameters and isotropic emission. Strict spatial and temporal constraints (based upon electromagnetic observations), in conjunction with a single, robust selection criterion (optimized for discovery) have been leveraged to realize a nearly background-free search, with nominal signal loss, using archived data from the Antarctic Muon and Neutrino Detector Array (AMANDA-II). Our preliminary results are consistent with a null signal detection, with a peak muon neutrino effective area of $\sim80$ m$^2$ at $\sim 2$ PeV and a flux upper limit of $\sim0.150\:\text{GeV}/\text{cm$^{2}/$s}$ for model 1. Predictions for IceCube, AMANDA's kilometer scale successor, are compared with those found in the literature. Implications for correlative searches are discussed.

\end{abstract}

\section{Introduction}

Neutrino astronomy may provide us with a new glimpse at the internal processes of gamma-ray bursts (GRBs). The Antarctic Muon and Neutrino Detector Array (AMANDA), which has been calibrated upon atmospheric neutrinos, has demonstrated the viability of high energy neutrino astronomy by using the ice at the geographic South Pole as a Cherenkov medium. Canonical fireball phenomenology \cite{Piran:2004}, in the context of hadronic acceleration, predicts correlated MeV-EeV neutrinos from GRBs via various hadronic interactions \cite{Stamatikos:2004}. Ideal for detection are TeV-PeV neutrinos in coincidence with prompt $\gamma$-ray emission, resulting in a nearly background-free search. A detailed description of the modeling techniques and an ongoing analysis featuring correlated neutrino searches of individual GRBs from the Burst and Transient Source Experiment (BATSE) are described elsewhere \cite{Stamatikos:2005,Stamatikos:2004}. Here, we report on a complementary search for correlated leptonic ($\nu_{\mu},\bar{\nu}_{\mu}$) emission, using models based upon the unique \emph{discrete} electromagnetic characteristics and emission geometry of GRB030329, gleaned directly from satellite and ground-based observations. This represents a novel departure from searches~\cite{Hardtke:2003} based upon a diffuse formulation \cite{WB:1999}, which utilize averaged burst parameters.

\section{GRB030329: Electromagnetic Emission \& The GRB-Neutrino Connection}

On March 29, 2003, at $11^{\text{h}}37^{\text{m}}14.^{\text{s}}67$ (UTC), HETE-II was triggered by GRB030329 (H2652), a watershed transient which confirmed the connection between a core collapse Type Ic supernova and long duration GRB.  Electromagnetic investigations of the prompt $\gamma$-ray and multi-wavelength afterglow emissions associated with GRB030329 abound in the literature (see table~\ref{GRB030329_electromagnetic_properties}), making it a perfect specimen for study. Via forward folding deconvolution, the photon energy spectrum, was fit to the \emph{Band function} \cite{Band:1993}, an empirically derived power law with smooth transition. For spectral indices $\alpha > -2$ and $\beta < -2$, the characteristic peak energy is defined as $E_{p}=[(2+\alpha)\epsilon_{\gamma}^{b}](\alpha-\beta)^{-1}$, where $\epsilon_{\gamma}^{b}$ is known as the photon break energy \cite{Band:1993}. Hence, using standard error propagation, we find that $\epsilon_{\gamma}^{b}=115.6\pm9.9$ keV. Doppler spectroscopy of the OT afterglow revealed a redshift measurement, which, under an assumed $\Lambda_{CDM}$ cosmology\footnote{$\Lambda_{CDM}$ cosmology: \emph{$H_{o}=72\pm5\:\text{km}/\text{Mpc$/$s},\: \Omega_{\text{m}}=0.29\pm0.07,\:\Omega_{\Lambda}=0.73\pm0.07$ Spergel et al., ApJS 148, 175-194 (2003) }} placed GRB030329 at a luminosity distance of $2.44^{+0.20}_{-0.18}\times10^{27}$ cm. Coupled with the peak energy flux, this implies an intrinsic peak isotropic luminosity of $L_{\gamma}^{iso}\approx 5.24^{+0.86}_{-0.77}\times10^{50}\:\text{ergs}/\text{s}$ in the 30-400 keV energy band pass. Evidence for anisotropic emission, in the form of a two component break in the afterglow spectrum, was revealed by radio calorimetry and is consistent with collimated prompt emission within a jet of opening half angle $\theta_{jet}$. This requires a beaming fraction correction, which reduces the intrinsic peak luminosity to $L_{\gamma}^{jet}=L_{\gamma}^{iso}(1-\cos\theta_{jet})\approx 1.99^{+0.33}_{-0.29}\times10^{48}\:\text{ergs}/\text{s}$. Extended calorimetry provided an estimate for the fractions of shock energy imparted to the electrons ($\epsilon_{e}$) and magnetic field ($\epsilon_{B}$). Table~\ref{GRB030329_electromagnetic_properties} summarizes the observed electromagnetic properties used in this analysis.
\vskip -0.4 cm
\begin{table} [h]
\begin{center}
\begin{minipage}[c]{1.00\textwidth}
\caption [l] {Electromagnetic Properties of GRB030329: Prompt $\gamma$-ray and Multi-wavelength Afterglow Emission}
\begin{tabular}{lll}
\hline
\small{\textbf{Parameter(s)}} & \small{\textbf{Value}} & \small{\textbf{Reference}}\\
\hline
\scriptsize{Positional Localization $\left(\alpha_{\text{J2000}},\delta_{\text{J2000}},\sigma_{R}\right)$} &  \scriptsize{$161.2081646^{\circ},21.5215106^{\circ},3.0\times10^{-7}\:^{\circ}$} & \scriptsize{Taylor et al., GCN Report 2129} \\
\scriptsize{Trigger (T) \& Duration (T$_{90}$) [30-400 keV]} & \scriptsize{$41,834.67\:\text{SOD},22.8\pm0.5\:\text{s}$} & \scriptsize{Vanderspek et al., ApJ 617: 1251-1257 (2004)} \\
\scriptsize{Energy Fluence ($F_{\gamma}$) [2-400 keV]} & \scriptsize{$1.630^{+0.014}_{-0.013}\times10^{-4}\:\text{ergs}/\text{cm$^2$}$} & \scriptsize{T. Sakamoto et al., astro-ph/0409128} \\
\scriptsize{Band Parameters $\left(\alpha,\:\beta,\:E_{p}\right)$ [2-400 keV]} & \scriptsize{$-1.32\pm0.02$,$-2.44\pm0.08$, $70.2\pm2.3$} & \scriptsize{Vanderspek et al., ApJ 617: 1251-1257 (2004)} \\
\scriptsize{Peak Energy Flux ($\Phi_{\gamma}^{\text{Peak}}$) [30-400 keV]} & \scriptsize{$\sim7\times10^{-6}\:\text{ergs}/\text{cm$^{2}$/s}$} & \scriptsize{Vanderspek et al., GCN Report 1997} \\
\scriptsize{Spectroscopic Redshift (z)} & \scriptsize{$0.168541\pm0.000004$} & \scriptsize{Bloom et al., GCN Report 2212} \\
\scriptsize{Jet Opening Half Angle ($\theta_{jet}$)} & \scriptsize{$\sim5^{\circ}\approx0.09$ rad} & \scriptsize{Berger et al., Nature 426, 154-157 (2003)} \\
\scriptsize{Electron \& Magnetic Field Energy Fractions } & \scriptsize{$\epsilon_{e}\approx0.19$, $\epsilon_{B}\approx0.042$} & \scriptsize{Frail et al., ApJ 619, 994-998 (2005)} \\
\hline
\end{tabular}
\label{GRB030329_electromagnetic_properties}
\end{minipage}
\end{center}
\end{table}

The generic mechanism responsible for the super-Eddington luminosities associated with GRBs is the dissipation, via shocks, of highly relativistic kinetic energy, acquired by electrons and positrons Fermi accelerated in an optically thick, relativistically expanding plasma, commonly referred to as a \emph{fireball}. The acceleration of electrons in the intense magnetic field of the fireball leads to the emission of prompt non-thermal $\gamma$-rays via synchrotron radiation. The temporal variability ($t_{v}\sim 10$ ms) in the light curves imply compact sources. In order to ensure a transparent optical depth to photons of energy $\epsilon_{\gamma}^{max}\approx100\:\text{MeV}$, a minimum bulk Lorentz boost factor ($\Gamma$) was assigned (see equation~\ref{Gamma}). Hadronic acceleration within the ambient photon field produces TeV-PeV leptons, whose spectra $(dN_{\nu_{\mu}}/d\epsilon_{\nu_{\mu}}\equiv\Phi_{\nu_{\mu}})$ trace the photon spectra, via the following photomeson interaction: $p+\gamma \rightarrow \Delta ^{+} \rightarrow \pi ^{+} + [n] \rightarrow \mu^{+}+\nu_{\mu} \rightarrow e^{+} + \nu_e + \bar{\nu}_{\mu}+\nu_{\mu}$, as illustrated in  figures~\ref{neutrino_spectrum_and_response} and \ref{equations} \cite{Stamatikos:2005,Guetta:2004}. Hence, these $\nu$s are expected to be spatially and temporally correlated with prompt $\gamma$-ray emission.

\section{Neutrino Astronomy with AMANDA-II: Analysis, Results \& Discussion}

AMANDA-II is comprised of 677 optical modules buried at depths between 1500-2000 m. The background consists of cosmic ray induced \emph{down-going} atmospheric muons, detected at a rate of $\sim100$ Hz, with a perturbation of atmospheric neutrinos, detected at a rate of $\sim10^{-4}$ Hz. The astrophysical neutrino signal, detected via charged current interactions such as: $\nu_{\mu}+N\rightarrow\mu^{\pm}+X$, is isolated by utilizing topologically identified \emph{up-going} muon events, which are reconstructed by a maximum likelihood method. On-source, off-time data were used to estimate the stability of the background rate in order to maintain blindness, facilitating an unbiased analysis. After data filtering (see \cite{Ackermann:2005}), the total off-time background, excluding a 10 minute blinded window centered upon the trigger time, was consistent with a Gaussian fit and accrued $24,972\pm158$ events over a 57,328.04 second interval, resulting in a rate of $0.436\pm0.003$ Hz. Based upon a visual inspection of the light curve, a conservative search window of 40 seconds (beginning at $T$) was chosen. Hence, $17.44\pm0.12$ background events ($n_{b}$) were expected on-time in AMANDA-II prior to any quality selection. Signal neutrino spectra were simulated for three models (see figure 26) by propagating a total of $\sim440,000$ muon neutrinos ($\nu_{\mu}+\bar{\nu}_{\mu}$) from an error box in the sky defined by the spatial localization of the radio afterglow (see table~\ref{GRB030329_electromagnetic_properties}). Event quality selection was optimized for the best limit setting and discovery potential by minimizing the model rejection factor (MRF) \cite{Hill:2003} and the model discovery factor (MDF) \cite{Hill:2005}, respectively. Although multiple observables were investigated, a single criterion emerged, based upon the maximum size of the search bin radius ($\Psi$), i.e. the space angle between the reconstructed muon trajectory and the GRB's position.

\begin{figure}[t]
\centering
\begin{minipage}[c]{0.46\textwidth}
\includegraphics[width=1.00\textwidth]{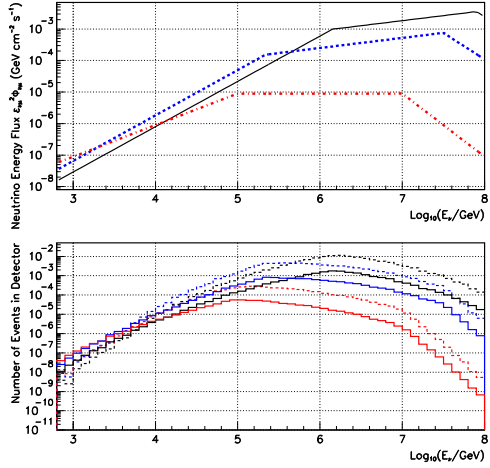}
\end{minipage}
\begin{minipage}[c]{0.46\textwidth}
\tiny{
\begin{equation}
\epsilon_{\nu_{\mu}}^{2}\Phi_{\nu_{\mu}}\approx A_{\nu_{\mu}}\times
\begin{cases}
\left(\frac{\epsilon_{\nu_{\mu}}}{\epsilon_{\nu}^{b}}\right)^{-\beta-1} & \epsilon_{\nu_{\mu}}<\epsilon_{\nu}^{b} \\
\left(\frac{\epsilon_{\nu_{\mu}}}{\epsilon_{\nu}^{b}}\right)^{-\alpha-1} & \epsilon_{\nu}^{b}<\epsilon_{\nu_{\mu}}<\epsilon_{\pi}^{b} \\
\left(\frac{\epsilon_{\nu_{\mu}}}{\epsilon_{\nu}^{b}}\right)^{-\alpha-1}\left(\frac{\epsilon_{\nu_{\mu}}} {\epsilon_{\pi}^{b}}\right)^{-2} & \epsilon_{\nu_{\mu}}>\epsilon_{\pi}^{b}
\end{cases}
\label{neutrino_energy_spectrum}
\end{equation}
}
\tiny{
\begin{equation}
A_{\nu_{\mu}}\approx\frac{F_{\gamma} f_{\pi}}{8\epsilon_{e}\ln(10)T_{90}}\approx9.86\times10^{-4}\:\text{GeV}/\text{cm$^{2}/$s}
\label{neutrino_normalization}
\end{equation}
}
\tiny{
\begin{equation}
f_{\pi}\simeq 0.2\times \frac{L_{\gamma,52}}{\Gamma_{2.5}^{4}t_{v,-2}\epsilon_{\gamma, MeV}^{b}(1+z)}\approx0.77
\label{proton_efficiency}
\end{equation}
}
\tiny{
\begin{equation}
\Gamma\gtrsim276\left[L_{\gamma,52}t_{v,-2}^{-1}\epsilon_{\gamma,MeV}^{max}(1+z)\right]^{\frac{1}{6}}\approx178
\label{Gamma}
\end{equation}
}
\tiny{
\begin{equation}
\epsilon_{\nu}^{b}\approx\left[\dfrac{7\times10^{5}}{(1+z)^{2}} \dfrac{\Gamma_{2.5}^{2}}{\epsilon_{\gamma,\text{MeV}}^{b}}\right]\:\text{GeV}\approx1.40\times10^{6}\:\text{GeV}
\label{neutrino_break_energy}
\end{equation}
}
\tiny{
\begin{equation}
\epsilon_{\pi}^{b}\approx\left[\dfrac{10^{8}}{(1+z)}\epsilon_{e}^{\frac{1}{2}}\epsilon_{B}^{-\frac{1}{2}} (L_{\gamma,52})^{-\frac{1}{2}}\Gamma_{2.5}^{4}t_{v,-2}\right]\:\text{GeV}\approx7.98\times10^{7}\:\text{GeV}
\label{synchrotron_break_energy}
\end{equation}
}
\end{minipage}
\begin{minipage}[t]{0.46\textwidth}
\caption{Upper panel - Prompt neutrino energy flux for models 1 (solid black), 2 (dashed blue) and 3 (dot-dashed red), based upon equation~\ref{neutrino_energy_spectrum}. Lower panel - Detector response for models 1 (black), 2 (blue) and 3 (red) for AMANDA-II (solid) and IceCube (dashed). The effects of neutrino flavor oscillations have been included.}
\label{neutrino_spectrum_and_response}
\end{minipage}
\hspace{0.5cm}
\begin{minipage}[t]{0.46\textwidth}
\caption{Parameterization \cite{Stamatikos:2005,Guetta:2004} of the neutrino energy spectrum. Where $L_{\gamma}\equiv L_{\gamma,52}\cdot10^{52}\:\text{ergs/s}$, $\Gamma\equiv\Gamma_{2.5}\cdot10^{2.5}$, $t_{v}\equiv t_{v,-2}\cdot 10\:\text{ms}$, $\epsilon_{\gamma}^{b}\equiv\epsilon_{\gamma,\text{MeV}}^{b}\cdot1\:\text{MeV}$, and $\epsilon_{\gamma}\equiv\epsilon_{\gamma,MeV}^{max}\cdot100\:\text{MeV}$. Values are given for model 1. Note the explicit dependence on discrete $\gamma$-ray photon observables.}
\label{equations}
\end{minipage}
\end{figure}

\begin{figure}[t]
\centering
\begin{minipage}[c]{0.46\textwidth}
\includegraphics[width=1.00\textwidth,height=5.5cm]{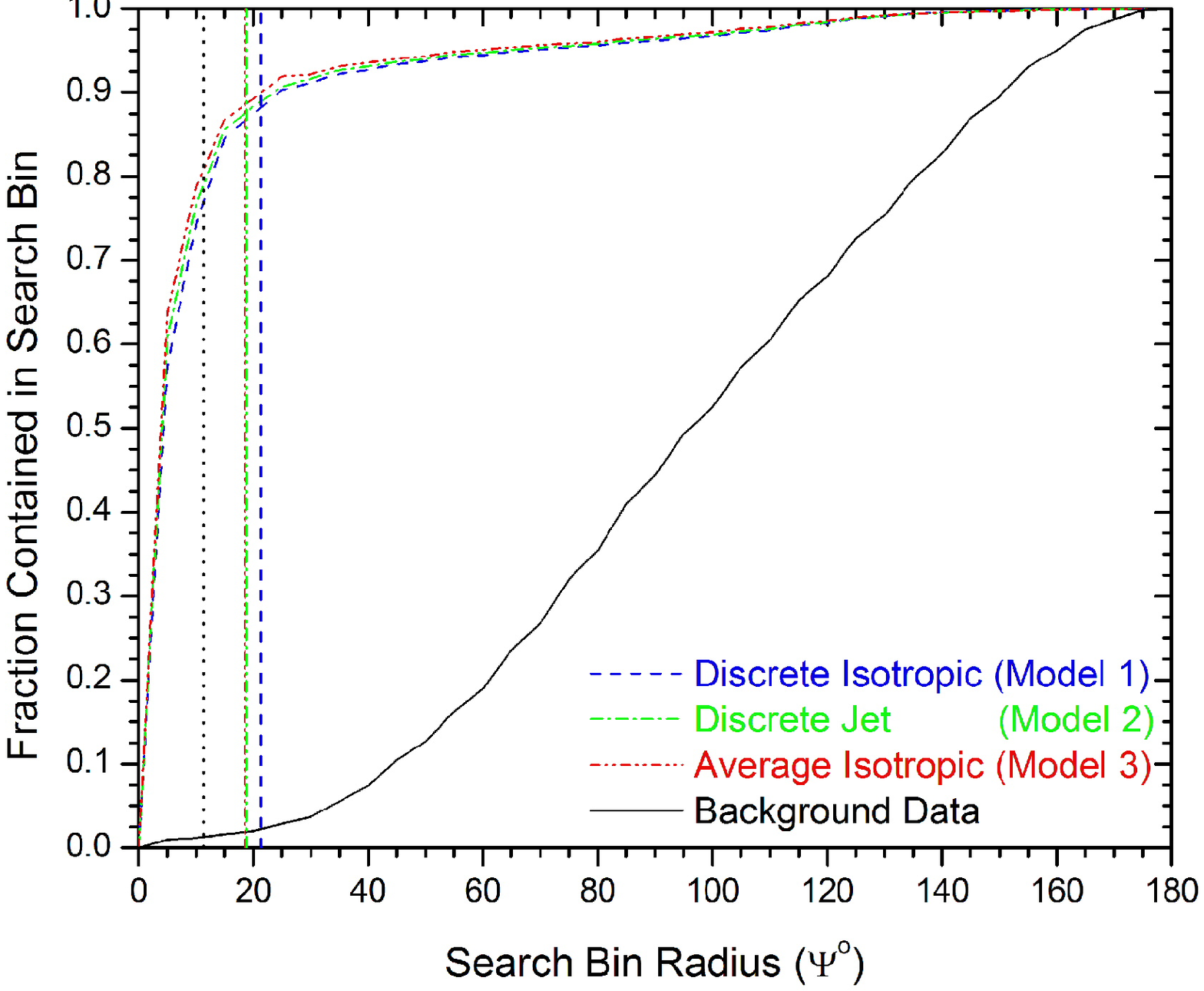}
\end{minipage}
\begin{minipage}[c]{0.46\textwidth}
\includegraphics[width=1.00\textwidth,height=5.5cm]{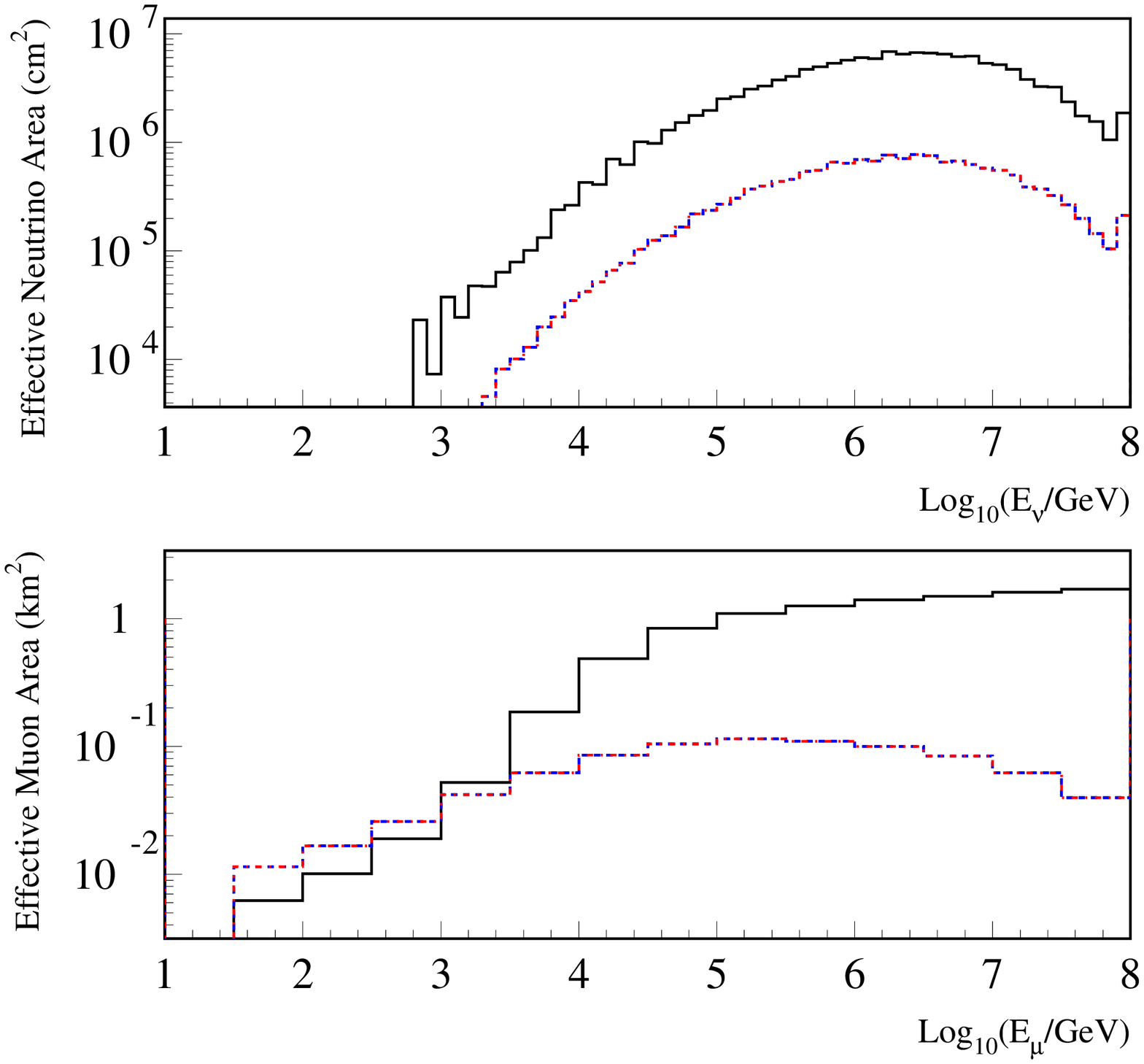}
\end{minipage}
\begin{minipage}[t]{0.46\textwidth}
\caption{AMANDA-II signal efficiency/background rejection for models 1-3 using MRF (vertical dashed blue, dashed-dot green and dashed-dot-dot red) and MDF (vertical black dotted) optimizations. Both MRF and MDF selections reject $\sim99\%$ of the background while retaining $\sim86\%$ and $\sim77\%$ of the signal, respectively.}
\label{frac_v_psi}
\end{minipage}
\hspace{0.5cm}
\begin{minipage}[t]{0.46\textwidth}
\caption{Upper panel - the effective muon neutrino area. Lower panel - the effective muon area for energy at closest approach to the detector. MDF optimized AMANDA-II results for models 1 (dashed black), 2 (dashed blue) and 3 (dot-dashed red) and predicted IceCube (solid black) curves are illustrated for $\delta_{\text{J2000}}\approx22^{\circ}$.}
\label{neutrino_effective_area}
\end{minipage}
\end{figure}

Our search, optimized for $5\sigma$ discovery (requiring 4 events within $\Psi\leq11.3^{\circ}$), is consistent with a null signal. Upper limits, summarized with our results in table~\ref{results}, do not constrain the models tested using AMANDA-II. Effective neutrino and muon areas are given in figure~\ref{neutrino_effective_area}. The number of expected signal events in IceCube ($N_{s}$) for model 1 is consistent with \cite{Razzaque:2004}, when neutrino oscillations are considered. For model 3, our results for $N_{s}$ are in agreement with \cite{Guetta:2004} and \cite{Ahrens:2004}, when one adjusts for the assumptions of \cite{WB:1999}. Selection based upon $\Psi$ was robust across the models tested in AMANDA-II, as illustrated in figure~\ref{frac_v_psi}. However, the MRF/MDF and hence the limit setting/discovery potential was strongly model dependent, varying by over an order of magnitude for models 1 and 3. Furthermore, using the same theoretical framework, the response of AMANDA-II and IceCube to spectra based upon discrete and average parameters are discrepant in mean neutrino energy and event rate by over an order of magnitude as illustrated in figure~\ref{neutrino_spectrum_and_response} and table~\ref{results}. Such variance in detector response unequivocally demonstrates the value of a discrete modeling approach when making correlative neutrino observations of individual GRBs, especially in the context of inferred astrophysical constraints, in agreement with \cite{Guetta:2004}.
\vskip -.5 cm
\begin{table} [h]
\begin{center}
\begin{minipage}[c]{1.00\textwidth}
\caption [c] {Results Summary. Primes indicate value after selection. Superscripts indicate A=MRF and B=MDF optimization.}
\begin{tabular}{ccccccc}
\hline
\small{\textbf{Model}} & \small{\textbf{$\Psi^{A},\Psi^{B}\:^{\circ}$}} & \small{\textbf{$n_{b},n_{b}^{A\prime},n_{b}^{B\prime}$}} & \small{\text{\textbf{$N_{s}$}},\:\textbf{$n_{s}$},\:\textbf{$n_{s}^{B\prime}$}} & \small{\textbf{$n_{obs},n_{obs}^{B\prime}$}} & \small{\textbf{MRF, MDF}} & \small{Sensitivity$^{B}$, Limit$^{B}$ $\frac{\text{GeV}}{\text{cm$^{2}\cdot$s}}$} \\
\hline
\scriptsize{1} & \scriptsize{21.3, 11.3} & \scriptsize{17.44, 0.23, 0.06} & \scriptsize{0.1308, 0.0202, 0.0156} & \scriptsize{15, 0} & \scriptsize{152, 424} & \scriptsize{0.157, 0.150} \\
\scriptsize{2} & \scriptsize{18.8, 11.3} & \scriptsize{17.44, 0.17, 0.06} & \scriptsize{0.0691, 0.0116, 0.0092} & \scriptsize{15, 0} & \scriptsize{256, 716} & \scriptsize{0.041, 0.039} \\
\scriptsize{3} & \scriptsize{18.5, 11.3} & \scriptsize{17.44, 0.17, 0.06} & \scriptsize{0.0038, 0.0008, 0.0006} & \scriptsize{15, 0} & \scriptsize{3864, 10794} & \scriptsize{0.036, 0.035} \\
\hline
\end{tabular}
\label{results}
\end{minipage}
\end{center}
\end{table}

\newpage
\clearpage
\setcounter{section}{0}
\section*{\Large Neutrino-Induced
  Cascades From GRBs With AMANDA-II}

\vskip 0.05cm
{\large B.~Hughey$^{\it a}$,
  I.~Taboada$^{\it b}$ for the IceCube Collaboration}\\
{\it 
       (a) Physics Dept. University of Wisconsin. Madison, WI 53706, USA \\
       (b) Physics Dept. University of California. Berkeley, CA 94720, USA\\
        }

\vskip 0.05cm
{\large  Presenter: B.~Hughey (brennan.hughey@icecube.wisc.edu), \
  usa-hughey-B-abs1-og25-oral}

\title[$\nu$-induced cascades from GRBs with AMANDA-II]{Neutrino-Induced
  Cascades From GRBs With AMANDA-II}
\author[B.~Hughey, I.~Taboada for the IceCube Collaboration] {B.~Hughey$^a$,
  I.~Taboada$^b$ for the IceCube Collaboration\\
        (a) Physics Dept. University of Wisconsin. Madison, WI 53706, USA \\
        (b) Physics Dept. University of California. Berkeley, CA 94720, USA\\
        }
\presenter{Presenter: B.~Hughey (brennan.hughey@icecube.wisc.edu), \
  usa-hughey-B-abs1-og25-oral}

\maketitle

\begin{abstract}
Using AMANDA-II we have performed a search for $\nu$-induced
cascades in coincidence with 73 bursts reported by BATSE in 2000.
Background is greatly suppressed by the BATSE temporal constraint. No
evidence of neutrinos was found. We set a limit on a WB-like spectrum, 
$A_{90}^{\rm all\;
  flavors}$~=~9.5$\times$10$^{-7}$~GeV~cm$^{-2}$~s$^{-1}$~sr$^{-1}$. 
The determination of systematic uncertainties is in progress, and the limit
will be somewhat weakened once these uncertainties are taken into account.
We are also conducting a rolling time-window search for $\nu$-induced
cascades consistent with a GRB signal in 2001. The data set is searched for a
statistically significant cluster of signal-like events within a 1~s or 100~s
time window.  The non-triggered search has the
potential to discover phenomena, including gamma-ray dark choked bursts, which
did not trigger gamma-ray detectors.
%
%
\end{abstract}

\section{Introduction}

Gamma Ray Bursts are among the most energetic processes in the universe. High
energy neutrinos ($\approx 10^{14}$~eV) are thought 
to be produced via the process $p+\gamma \rightarrow \Delta^+ \rightarrow \pi^+
[+n] \rightarrow \mu^+ + \nu_\mu \rightarrow e^+ + \bar{\nu}_e +
\nu_\mu$. Neutrino oscillations result in a flavor flux ratio, 
$\phi_{\nu_e+\bar{\nu}_e}$:$\phi_{\nu_\mu+\bar{\nu}_\mu}$:$\phi_{\nu_\tau+\bar{\nu}_\tau}$,
equal to 1:1:1 at Earth\footnote{But the ratio $\phi_\nu$:$\phi_{\bar{\nu}}$ is
  not 1:1.}.
AMANDA-II, a sub-detector of IceCube, was commissioned in 2000 with a total of
677 optical modules arranged on 19 strings, at depths between 1500~m and
2000~m below the surface of the ice at the South Pole. Each OM contains a
20~cm photo-multiplier tube in a pressure vessel. AMANDA-II uses polar ice as a
Cherenkov medium. Searches for $\nu$-induced muons with
AMANDA-II~\cite{ama:nature-pub,nu2004:ama} in coincidence with bursts 
reported by satellites have been done for 1997-2003~\cite{icrc03:grb}. These
searches take advantage of the spatial and temporal  localization of the
bursts to reduce background, but are restricted to bursts with positive
declination because AMANDA-II, located at the South Pole, relies on the use of
the Earth to filter out all non-neutrino particles from the northern
hemisphere.
The cascade channel is complementary to the muon channel. AMANDA-II is uniformly
sensitive to cascades from all directions, so objects at any declination can
be studied. Further, GRBs without directional information can be used as no
correlation to the cascade direction is required. Even though the detector's
effective volume is smaller for cascades than for muons, more bursts can be
studied with the cascade channel.
Isolated cascades are produced by several interactions: $\nu_e N$
charged current, $\nu_x N$ neutral current, $\bar{\nu}_e e^-$ at 6.3 PeV
(Glashow resonance) and $\nu_\tau N$ charged current in the case when the
$\tau$ travels a short distance before decaying and the decay cascade overlaps
the $\nu_\tau N$ hadronic cascade. A 100~TeV $\tau$  will travel $O(5~m)$
before decaying. As a comparison, a 100 TeV electromagnetic cascade is
$\approx$8.5~m long in ice.
 
We present two analyses that search for $\nu$-induced
cascades in coincidence with GRBs. For the first analysis, hereafter referred
to as the \textit{Rolling} analysis, we do not use any correlation with bursts reported by satellites. Instead
two time windows, 1~s and 100~s, are rolled along the data taken by
AMANDA-II in the year 2001, to search for statistical excess. This technique has
the advantage of being sensitive to bursts that were not reported by 
satellites. The second analysis, hereafter referred to as the
\textit{Temporal} analysis, uses the temporal, but not the spatial,
correlation with bursts reported by BATSE~\cite{batse} in the year 2000. Using
this correlation reduces the background significantly.

\section{Simulation and Reconstruction}

For both analyses neutrino-induced cascades for all three neutrino
flavors were simulated with \texttt{ANIS}~\cite{anis} from 100~GeV to 100~PeV
following an $E^{-1}$ spectrum. This simulation was then re-weighted to 
follow the flux predicted by the Waxman-Bahcall model~\cite{wb}. This spectrum
is derived from average burst 
characteristics, and thus it is adequate to describe a large number of bursts
simultaneously. Individual burst spectra may deviate significantly from the WB
spectrum. We use a break energy, $E_B$=100~TeV and a synchrotron energy,
$E_s$=10~PeV. For the Rolling analysis signal
simulation was verified with \texttt{TEA}~\cite{tea} which produces a
Waxman-Bahcall type broken power law spectrum directly. The outputs of
\texttt{ANIS} and \texttt{TEA} were found to be consistent. In both
the Rolling and Temporal analyses, background muon events were simulated
with \texttt{CORSIKA}~\cite{corsika1}. Muons were propagated through ice using
\texttt{MMC}~\cite{MMC} and detector response was simulated with
\texttt{AMASIM}~\cite{amasim} for both signal and background simulation.   

For both analyses, data were reconstructed with 2 different hypotheses: a
cascade hypothesis and a muon hypothesis. Muon and cascade reconstruction
methods are described in
refs.~\cite{ama:mu-reco,kowalski:workshop,ama:b10casc-pub}. We obtain a cascade
vertex resolution of about 6~m in the x,y coordinates and slightly better in
the z coordinate. We obtain a cascade energy resolution of $\log_{10}E_{\rm
  true}/E_{\rm reco} \approx$~0.15. The Rolling analysis reconstructs the 
position of cascades while for the Temporal analysis both the position and the
energy of the cascade is reconstructed. The angular resolution of the muon
reconstruction is about $5^\circ$ \footnote{Better angular resolution is
achieved by analyses that focus on the muon channel.}.

\section{Rolling Time Window Analysis}

The Rolling analysis currently uses data from the year 2001.  We scan the
entire data sample for a clustering of events which survive cuts and are not
consistent with the expected background.  Therefore, it has the potential to
detect signals which are not coincident with prior gamma-ray detections.
These sources include gamma-ray dark neutrino sources, such as choked
GRBs~\cite{choke-bursts} as well as conventional GRBs not detected by the 
Third Interplanetary Network (IPN3)~\cite{ipn3}. 
The live-time of this analysis is $\approx$233 days.  
Two separate rolling searches are performed, with time window lengths of 1
and 100 seconds.  These lengths were chosen, based on the bimodal plot of GRB
durations produced by BATSE~\cite{bimodal}, to contain the majority of signal
from short and long bursts, respectively, while still being short enough to
keep out extra background events.
Since there is no temporal or spatial coincidence to aid in background
rejection, the use of cuts to reduce the background of atmospheric muons
becomes very important.  Cuts based on both topology and number of hits in
optical modules (which is indirectly tied to event energy) are utilized. 
After an initial filter is applied to take only high energy events, a final
cut is made using a support vector machine (SVM)~\cite{svm}. 
Background muon Monte Carlo was found to be in good agreement with
experimental data in all 6 variables used in the SVM. A sample of experimental
data taken from 5 runs distributed throughout the year was used as background
in the SVM. The SVM cut was optimized independently for the 1 second and 100
second searches.
Since the background is of stochastic nature, Poisson statistics can be
used to estimate the the statistical significance of a cluster of
events. Preliminary calculations result in a sensitivity of
$2.7\times$10$^{-6}$~GeV~cm$^{-2}$~s$^{-1}$~sr$^{-1}$ for a time-averaged
diffuse neutrino flux of all flavors, with energy spectrum according to the
Waxman-Bahcall model. This sensitivity assumes 425 bursts during the live-time
of this analysis based on the average rate of GRB detection by the BATSE
experiment and does not account for the unknown number of bursts with weaker
or non-existent gamma-ray signals.  Final results are not yet available at the
time of writing. 

\section{Temporal Analysis: Bursts reported by BATSE in 2000}

\begin{figure}[ht]
\begin{center}
\mbox{
\includegraphics*[width=0.36\textwidth]{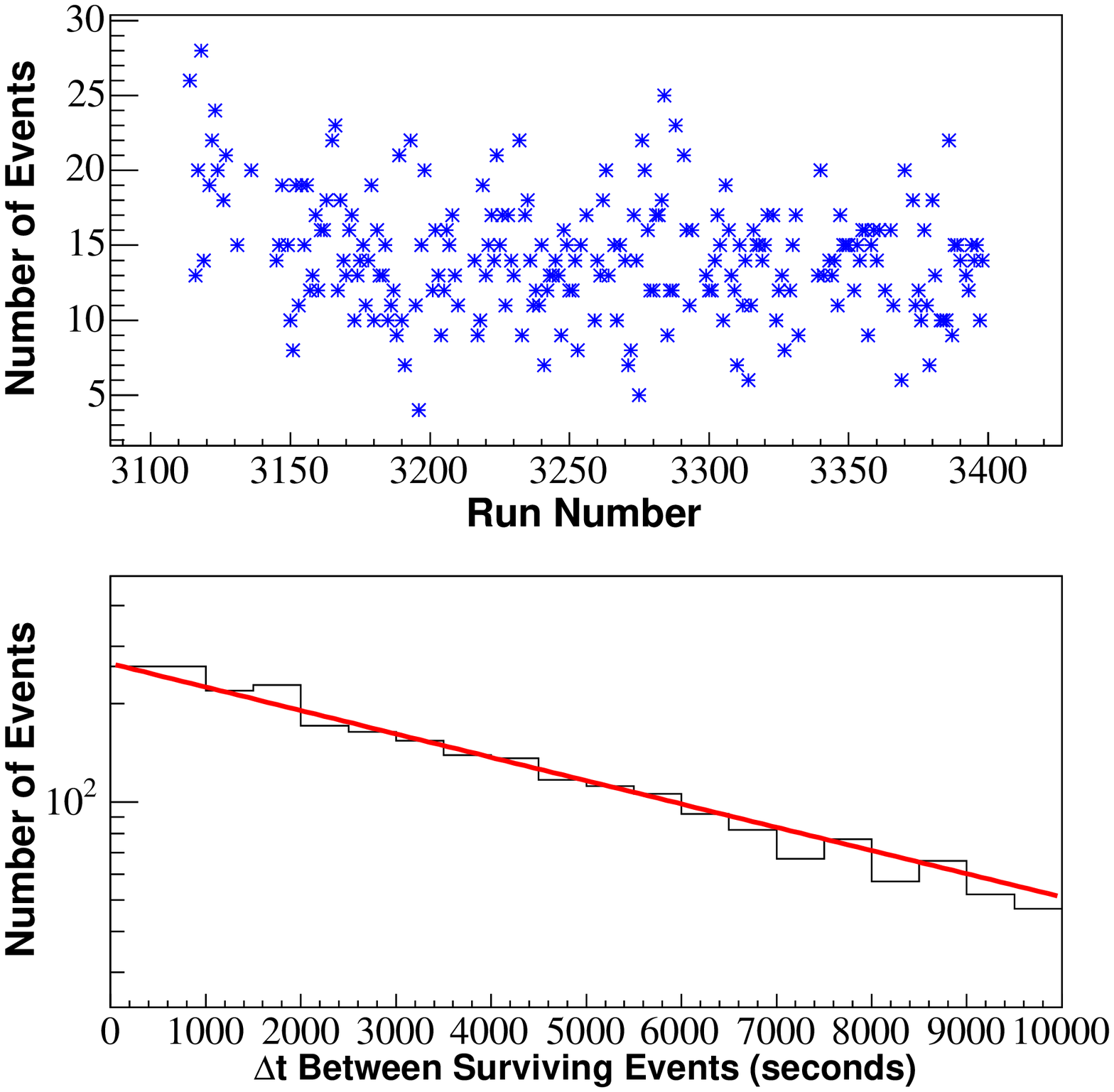}
\includegraphics*[width=0.37\textwidth]{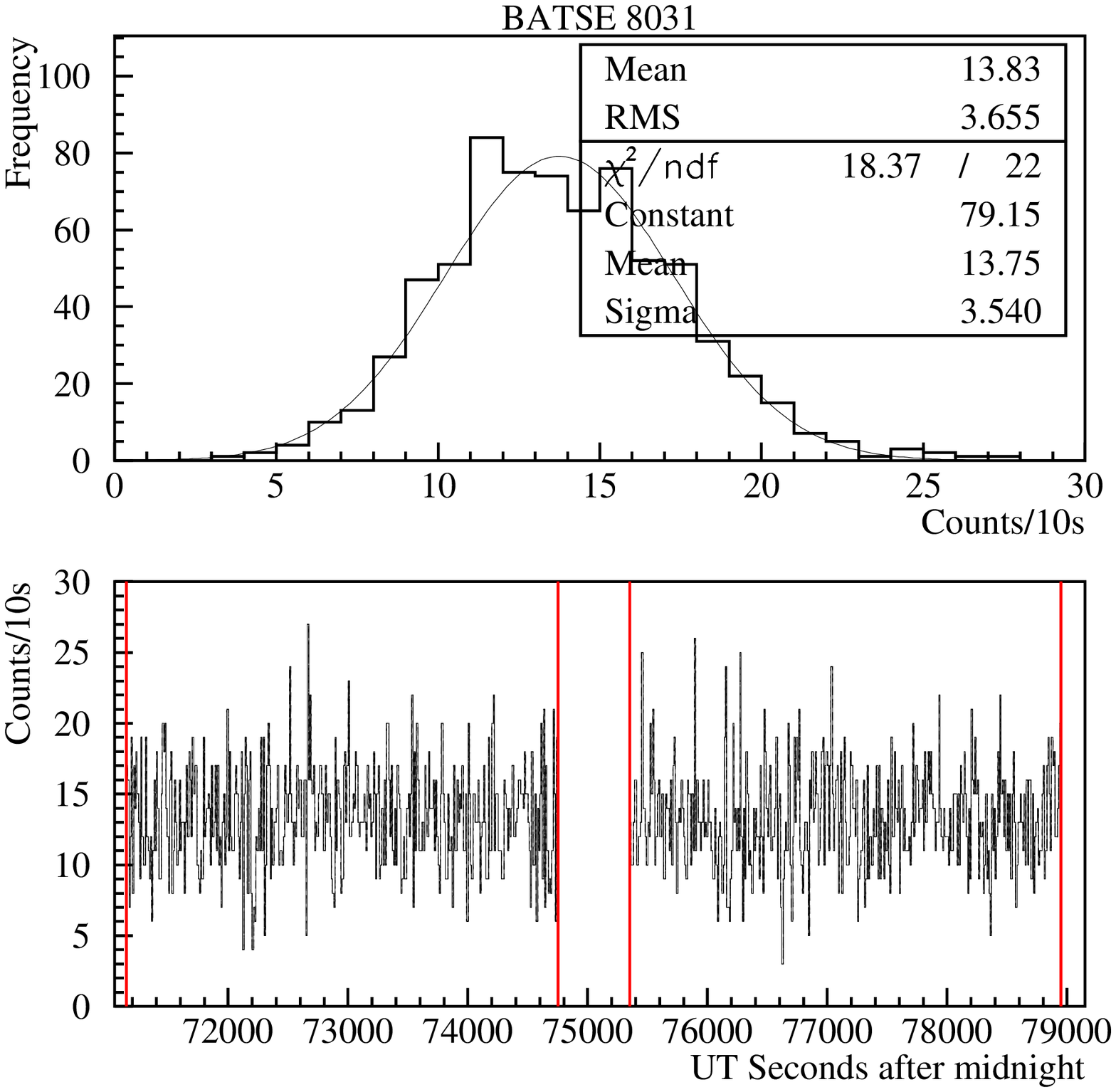}}
\caption{\label{fig:stability} Left - Rolling Analysis:
Plots showing background stability. The upper plot shows background counts per
day after the SVN cut and the lower plot shows time between
events surviving cuts.  The line 
is the prediction assuming Poissonian statistics. Right - Temporal
Analysis: Stability plots for BATSE-8031. The upper panel shows
the frequency of events/10s that pass the filter in the off-time window. The
lower panel shows the number of events/10s that pass the filter versus time in
seconds after midnight (UTC). The vertical lines indicate the off-time
period. The on-time period is analyzed according to our blindness
procedures.}
\end{center}
\end{figure}

AMANDA-II began normal operation Feb. 13, 2000. The last BATSE burst was
reported May 26, 2000. In this period 76 bursts were reported. Since the GRB
start time and duration\footnote{We use $T90$, the time over which a burst 
emits from 5\% to 95\% measured fluence, as the duration}, are well known, the
separation of $\nu$-induced cascade signal from the down-going muon
background is simplified. We use three selection criteria based
on the two reconstruction hypotheses to discard the down-going muon background
and keep the neutrino-induced cascade signal.
A total of $\approx$7800~s per burst were studied. A period of 600 s
(\textit{on-time} window) centered  at the start time of the GRB was initially
set aside in accordance with our blind analysis procedures. Two periods of
data of 1 hour duration (\textit{off-time} window) just 
before and after the on-time window are also studied. We optimize the
selection criteria using the off-time window and signal simulation. Thus the
background is experimentally measured. We only examined the
fraction of the on-time window corresponding to the duration of each
burst. Keeping the rest of the on-time window blind 
allows for other future searches, e.g. precursor neutrinos.
We determined the detector stability using the off-time window
experimental data. Only GRBs for which the detector is found to be stable in
the off-time windows were used. Of the 76 bursts reported
by BATSE in coincidence with AMANDA-II, for two bursts there are gaps in the
AMANDA-II data and for one burst, AMANDA-II data was found to be unstable. Figure
\ref{fig:stability} shows a sample of the plots used to determine the
stability.
We applied the selection criteria in two steps, a filter and final
selection. The filter rejects down-going muons, $\theta_\mu > 
70^\circ$, and keeps events that are cascade-like, $L_{mpe} < 7.8$. The
parameter $L_{mpe}$ is the reduced likelihood of the cascade vertex
reconstruction and has smaller values for cascade-like events.
The final selection criteria are $L_{mpe}<6.9$ and $E>40$~TeV, where $E$ is
the reconstructed cascade energy. One event in the off-time window remains
after all cuts. This is equivalent to a background of
$n_b=0.0054^{+0.013}_{-0.005}$~(stat) in the on-time window.
After un-blinding the on-time window, no events were found. To set a
limit, we assume a WB-like spectrum with $E_b=100$~TeV and $E_s=10$~PeV. We
assume neutrino flavor flux ratio of 1:1:1 and p-$\gamma$ neutrino
generation. The 90\% c.l. limit on the all-flavor flux factor is  
$9.5\times$10$^{-7}$~GeV~cm$^{-2}$~s$^{-1}$~sr$^{-1}$. The event upper limit
is 2.44. These limits have not yet been corrected for systematic
uncertainties. Once the systematic uncertainties have been taken into account
this limit will worsen slightly.

\section{Outlook and Conclusions}

Two methods for searching for neutrino-induced cascades from GRBs using 
AMANDA-II have been presented.  A Rolling Time Window search is being
conducted to search for a neutrino GRB signal at any time and from any
direction.  This method serves as a useful complement to satellite-coincident
GRB searches conducted with AMANDA-II. Its sensitivity to individual bursts
suffers from the lack of temporal constraints, but it has the potential to
observe neutrino signals from transients which would otherwise be missed.
Although this search is currently being conducted on the 2001 data set, it is
relatively straightforward to expand the search to data sets from later years.
This method can also be adapted to use the muon channel in addition to
cascades.

Temporal correlation with satellites was used to perform a search with very
low background. No evidence for neutrinos was found and we have set a limit
based on the WB flux. The 90\% c.l. limit on the all-flavor flux factor,
supposing 1:1:1 flavor flux ratio, is
$9.5\times$10$^{-7}$~GeV~cm$^{-2}$~s$^{-1}$~sr$^{-1}$. This value has not
yet been corrected for systematic uncertainties. Previous
searches by AMANDA-II, performed on a much larger set of
bursts~\cite{icrc03:grb}, have a significantly better sensitivity  
than what has been presented here. Given a large random set of bursts with
both positive and negative declination, we expect the cascade channel
sensitivity to be roughly half as sensitive as the muon channel. It is
expected that only a small fraction of all bursts will contribute
significantly an eventual observed neutrino flux. By monitoring both
hemispheres we increase the probability of discovery. The Temporal Analysis
can be expanded to include bursts reported by IPN3, Swift and by using newly
or soon to be deployed IceCube strings.




\newpage
\setcounter{section}{0}
\section*{\Large Search for neutralino dark matter with the AMANDA neutrino detector}

\vskip 0.05cm
{\large D. Hubert$^{\it a}$, A. Davour$^{\it b}$ and C. de los Heros$^{\it b}$ for the IceCube Collaboration}\\
{\it        (a) Vrije Universiteit Brussel, 
	    Dienst ELEM, B-1050 Brussels, Belgium \\
        (b) Division of High Energy Physics, 
	    Uppsala University, S-75121 Uppsala, Sweden \\
        }

\vskip 0.05cm
{\large  Presenter: D. Hubert (hubert@hep.iihe.ac.be), bel-hubert-D-abs1-he23-oral}

\title[Search for neutralino dark matter with the AMANDA neutrino detector]{Search for neutralino dark matter with the AMANDA neutrino detector}
\author[D. Hubert, A. Davour, C. de los Heros \textit{et al.}] {D. Hubert$^a$, A. Davour$^b$ and C. de los Heros$^b$ for the IceCube Collaboration\\
        (a) Vrije Universiteit Brussel, 
	    Dienst ELEM, B-1050 Brussels, Belgium \\
        (b) Division of High Energy Physics, 
	    Uppsala University, S-75121 Uppsala, Sweden \\
        }
\presenter{Presenter: D. Hubert (hubert@hep.iihe.ac.be), bel-hubert-D-abs1-he23-oral}

\maketitle

\begin{abstract}
Data taken with the AMANDA high energy neutrino telescope can be used in a search for an indirect dark matter signal.  This paper presents current results from searches for neutralinos accumulated in the Earth and the Sun, using data from 1997-1999 and 2001 respectively.  We also discuss future improvements for higher statistics data samples collected during recent years.
\end{abstract}

\vspace{-0.5pc}
\section{Introduction}
\vspace{-0.5pc}
Cosmological observations have long suggested the presence of non-baryonic dark matter on all distance scales.  The WMAP results \cite{wmap} confirmed our current understanding of the Universe, summarised in the concordance model.  In this model the Universe contains about 23\% non-baryonic cold dark matter, but nothing is predicted about the nature of this dark matter.

A massive, weakly interacting and stable particle appears in Minimally Supersymmetric extensions to the Standard Model that assume R-parity conservation.  Indeed, the supersymmetric partners of the electroweak neutral Standard Model bosons mix into an interesting dark matter candidate, the neutralino, whose mass is expected in the GeV-TeV range \cite{neutralinoDM}.

On their trajectory through the Universe these particles will scatter weakly on normal matter and lose energy.  Eventually, the dark matter particles will be trapped in the gravitational field of heavy celestial objects, like the Earth and the Sun \cite{trap}.
The particles accumulated in the center of these bodies can annihilate pairwise.  The neutrinos produced in the decays of the Standard Model annihilation products can then be detected with a high energy neutrino detector as an excess over the expected atmospheric neutrino flux.

In this paper we present the results of searches with the AMANDA detector for neutralino dark matter accumulated in the Earth (1997-1999 data set) and the Sun (2001 data set).  We also summarize current techniques that continue these efforts on higher statistics data samples accumulated during recent years.

The Antarctic Muon And Neutrino Detector Array \cite{amanda} at the South Pole uses the polar ice cap as a Cherenkov medium for the detection of relativistic charged leptons produced in high energy neutrino interactions with nuclei.  The detector was constructed between 1996 and 2000.  Now totaling 677 light sensitive devices distributed on 19 strings, the detector is triggered when at least 24 detector modules are hit within a sliding 2.5$~\mu$s window.  
Before 2000, the detector configuration consisted of between 10 and 13 strings and consequently a lower multiplicity trigger condition was able to cope with the high rate of events produced by cosmic ray interactions with the atmosphere.

Reconstruction of muons, with their long range, offers the angular resolution required to reject the atmospheric background and search for a neutralino-induced signal, which, due to the geographic location of AMANDA, yields vertical upward-going (Earth) or horizontal (Sun) tracks in the instrumented volume.  Indeed, it is possible to eliminate the dominant background, downward-going atmospheric muons.  However, upward-going atmospheric neutrinos will always contaminate the final, selected data sample.

\vspace{-0.5pc}
\section{Signal and background simulation}
\vspace{-0.5pc}
We have used the DARKSUSY program \cite{darksusy} to generate dark matter induced events for seven neutralino masses between 50$~$GeV and 5000$~$GeV, and two annihilation channels for each mass: the $W^+W^-$ channel produces a hard neutrino energy spectrum ($\tau^+\tau^-$ for a 50$~$GeV neutralino), while $b\bar{b}$ yields a soft spectrum.
The cosmic ray showers in the atmosphere, in which downward-going muons are created, are generated with CORSIKA \cite{corsika2} with a primary spectral index of $\gamma$=2.7 and energies between 600$~$GeV and 10$^{11}~$GeV.  The atmospheric neutrinos are produced with NUSIM \cite{nusim} with energies between 10$~$GeV and 10$^8~$GeV and zenith angles above$~$80$^\circ$.

\vspace{-0.5pc}
\section{\label{earth_section}Search for neutralino annihilations in the center of the Earth}
\vspace{-0.5pc}
\begin{figure}[t]
\vspace{-1.7pc}
\begin{center}
\includegraphics*[width=0.43\textwidth,height=5.6cm,angle=0,clip=true]{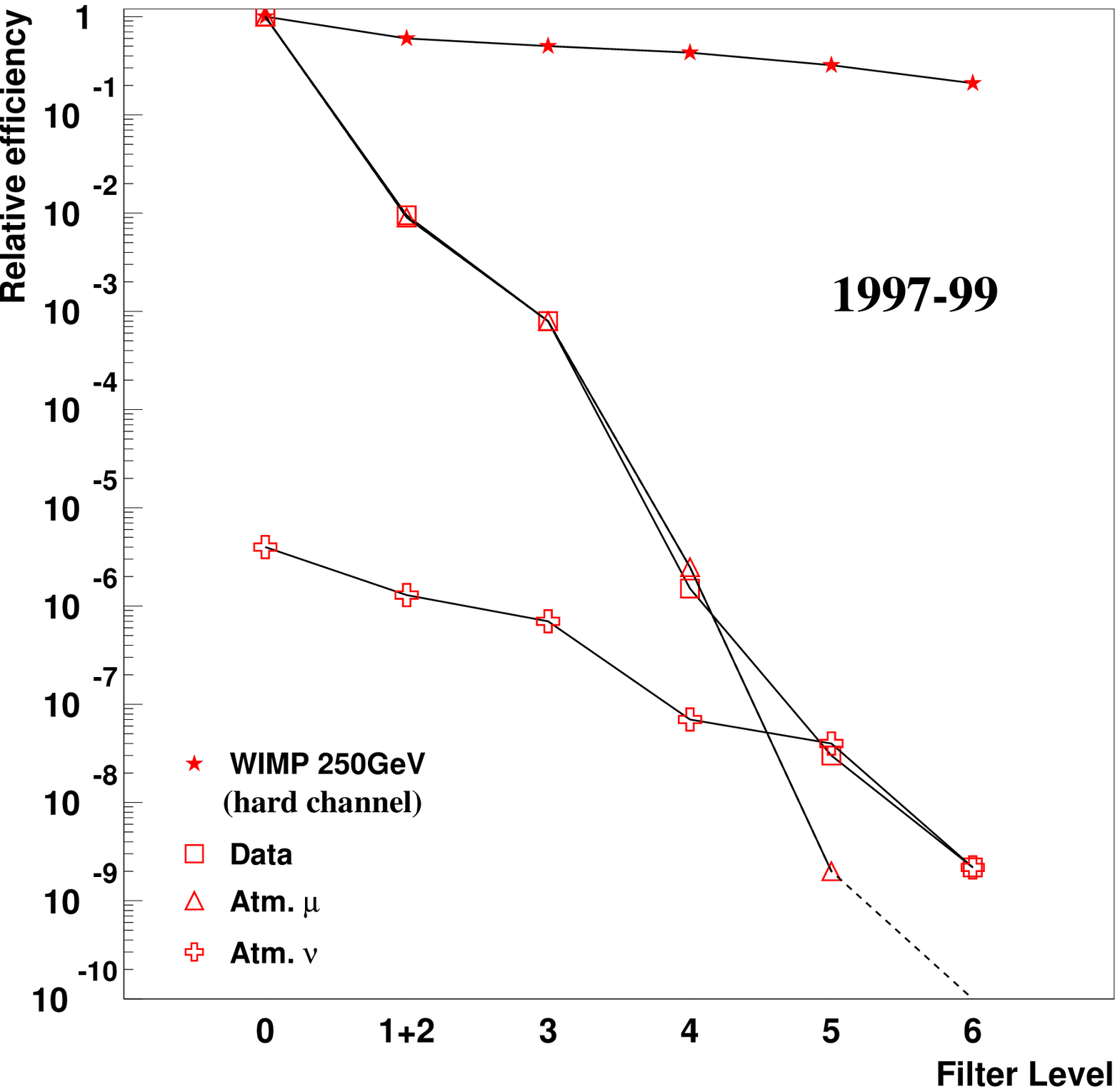}
\hfill
\includegraphics*[width=0.46\textwidth,height=5.6cm,angle=0,clip=true]{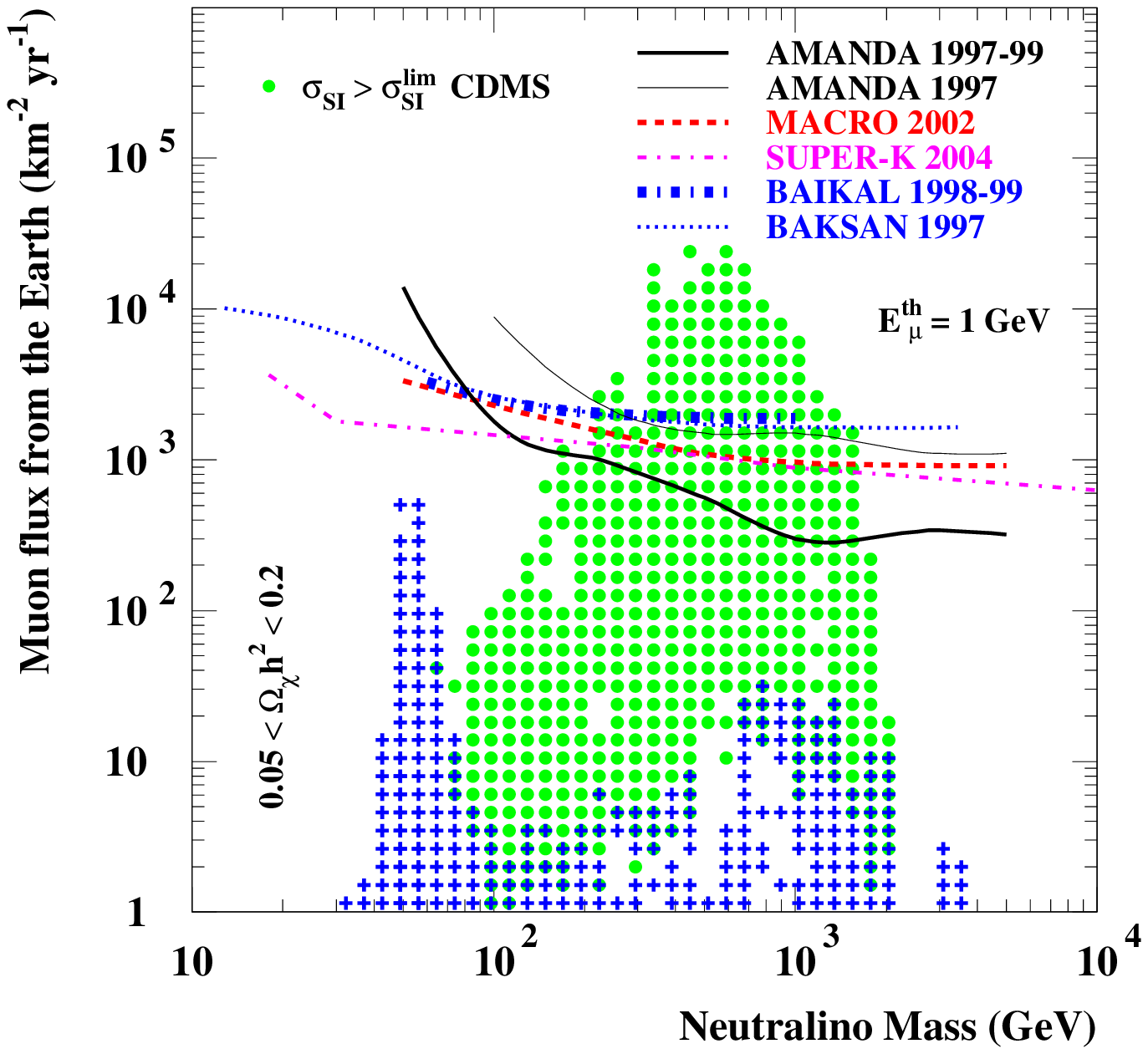}
\vspace{-1.5pc}
\caption{\label {earth} (a) Detection efficiencies relative to trigger level for the different filter levels in the terrestrial neutralino analysis ($m_{\chi}$=250$~$GeV, hard spectrum) for 1997-1999 data, neutralino signal, atmospheric muons and neutrinos.  (b) As a function of neutralino mass, the 90\% confidence level upper limit on the muon flux coming from hard neutralino annihilations in the center of the Earth compared to our results from 1997 data \cite{earth97} and other indirect experiments \cite{indirect}.  Markers show predictions for cosmologically relevant MSSM models, the dots represent parameter space excluded by CDMS \cite{cdms}.}
\end{center}
\vspace{-1.pc}
\end{figure}
A neutralino-induced signal from the center of the Earth was searched for in AMANDA data collected between 1997 and 1999, with a total effective livetime of 536.3 days.  To reduce the risk of experimenter bias, the complete data set of 5.0$\times$10$^9$ events was divided in a 20\% subsample, used for optimisation of the selection procedure, and a remaining 80\% sample, on which the selection was applied and final results calculated. Similarly, the sets of simulated events were divided in two samples: the first for use in the selection optimisation and the second for the selection efficiency calculations.  The simulated atmospheric muon sample contains 4.2$\times$10$^9$ triggered events (equivalent to an effective livetime of 649.6 days).  The sample of atmospheric$~$neutrinos totals 1.2$\times$10$^8$ events, which corresponds to 2.2$\times$10$^4$ triggers when scaled to the livetime of the analysis.

First, we try to suppress the dominant atmospheric muon background which is about 10$^6$ times more abundant than the atmospheric neutrino background.  This is partially done by selecting the events that are reconstructed as upward-going and that satisfy a cut correlated with reconstruction quality (``filter level 3'').  However, only a 10$^{-3}$ reduction of the atmospheric muons is obtained this way (Fig. \ref{earth}a) and more elaborate selection criteria are needed to reject downward-going muon tracks misreconstructed as upward-going.
Depending on the detector configuration and the neutralino model under study, the characteristics of the signal differ, which influences selection efficiencies significantly at this point.  Therefore, all further cuts are fine-tuned separately for each neutralino model and year of data taking.
At filter level 4, a neural network is trained using between 8 and 10 input observables, reaching another 10$^{-3}$ reduction.  Filter level 5 cuts sequentially on observables, with the goal of removing downward-going muons that resemble signal events.

At filter level 5 the data sample is dominated by atmospheric neutrinos (see Fig. \ref{earth}a).  With no significant excess of vertical tracks observed, the final selection on reconstructed zenith angle (filter level 6) was optimised for the average lowest possible 90\% confidence level upper limits on the muon flux.  From the number of observed events and the amount of (simulated) background in the final angular search bin, we infer the 90\% confidence level upper limit on the number of signal events.  Combined with the effective volume at the final cut level and the livetime of the collected data, this yields an upper limit on the neutrino-to-muon conversion rate, which can then be related to the muon flux \cite{newearth} (see Fig. \ref{earth}b).

\section{Search for neutralino annihilations in the Sun}
\begin{figure}[t] 
\begin{center}
\includegraphics*[width=0.43\textwidth,angle=0,clip]{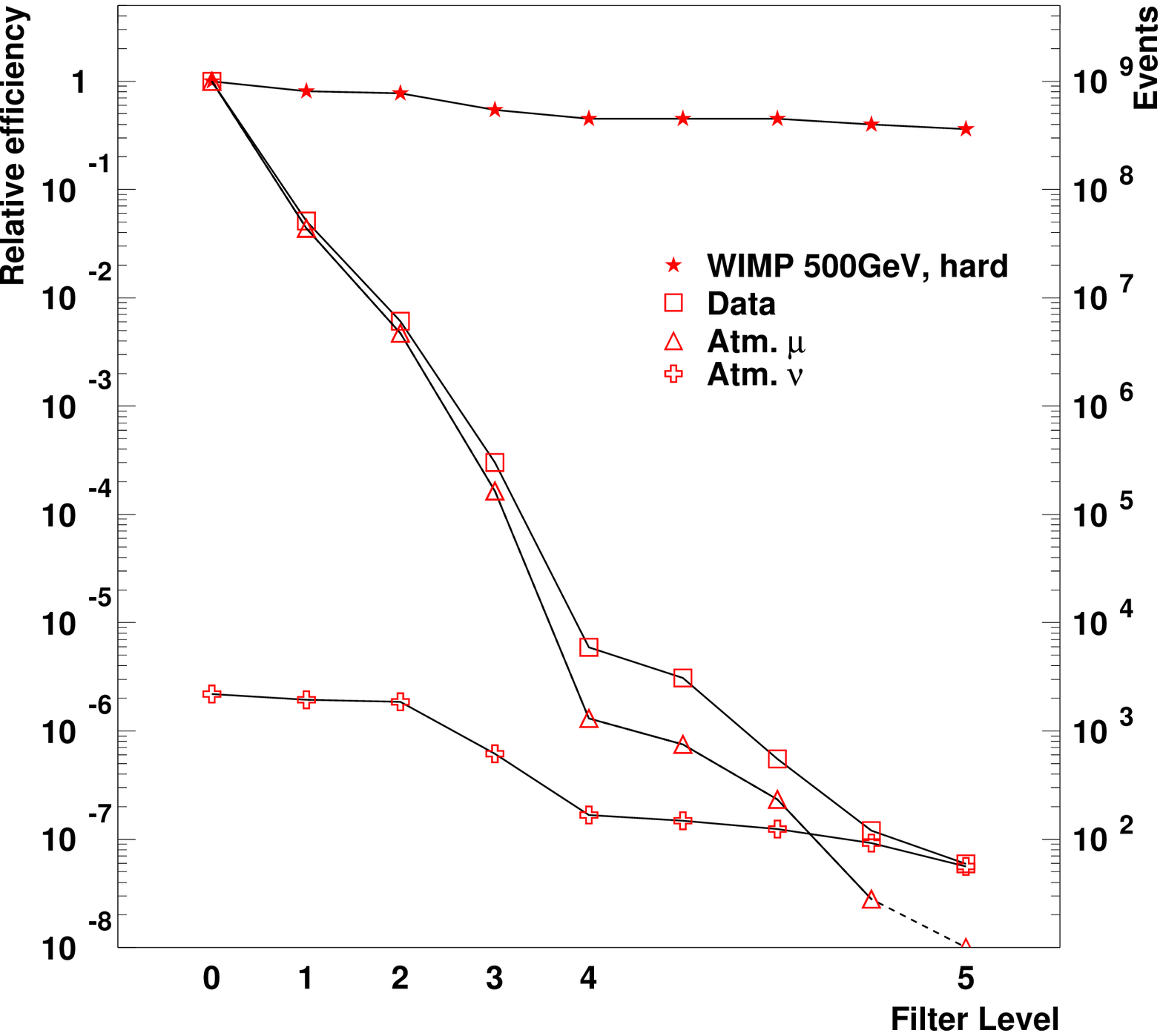}
\hfill
\includegraphics*[width=0.46\textwidth,angle=0,clip]{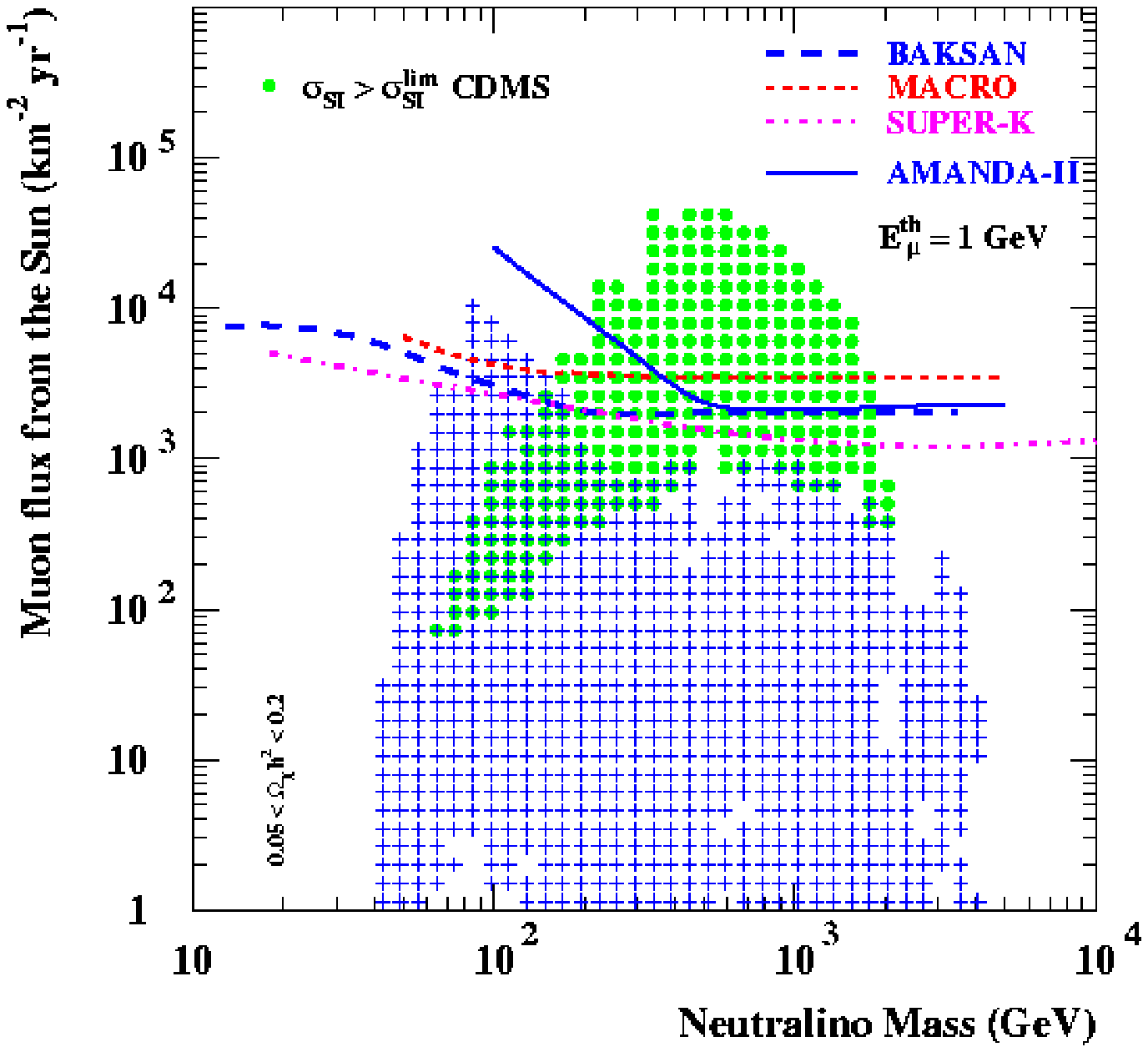}
\vspace{-1.5pc}
\caption{\label{sun} (a) Detection efficiencies relative to trigger level 
for the different filter levels in the solar neutralino analysis 
($m_{\chi}$=500$~$GeV, hard spectrum) for 2001 data, neutralino signal, 
atmospheric muons and neutrinos.  (b) As a function of neutralino mass, 
the 90\% confidence level upper limit on the muon flux coming from hard 
neutralino annihilations in the center of the Sun compared to other 
indirect experiments \cite{indirect}. Markers show predictions for 
cosmologically relevant MSSM models, the dots represent parameter space 
excluded by CDMS \cite{cdms}.}
\end{center}
\end{figure}
The AMANDA data used in the search for solar neutralinos consists of 8.7$\times$10$^8$ events, corresponding to 143.7 days of effective livetime, collected in 2001.  In contrast to the search in Section \ref{earth_section}, reducing the risk of experimenter bias in this analysis can be achieved by randomizing the azimuthal angles of the data.  The advantage of this procedure is that it allows the use of the full data set for cut optimisation.  The azimuthal angles are restored once the optimisation is finalised and results are calculated.  The simulated atmospheric background sample at trigger level totals 1.6$\times$10$^8$ muons (equivalent to 32.5 days of effective livetime) and$~$1.9$\times$10$^4~$neutrinos.

The solar neutralino analysis suffers the same backgrounds as the terrestrial neutralinos, but the signal is expected from a direction near the horizon, due to the trajectory of the Sun at the South Pole.  This analysis was only possible after completion of the AMANDA-II detector, whose 200 m diameter size provides enough lever arm for robust reconstruction of horizontal tracks. 

We adopted a similar analysis strategy as in Section \ref{earth_section}.  First, we select events with well-reconstructed horizontal tracks (filter level 1-3).  The remaining events are passed through a neural network that was trained separately for the neutralino models under study and used data as background (filter level 4).  Although a data reduction of $\sim$10$^{-5}$ compared to trigger level is achieved, the data sample is still dominated by misreconstructed downward-going muons.  As shown in Fig. \ref{sun}a, they are removed with extra cuts on observables related to reconstruction quality (filter level 5).

There was no sign of a significant excess of tracks from the direction of the Sun in the final data sample.  The expected background in the final search bin around the Sun was estimated from off-source data in the same declination band, which eliminates the effects of uncertainties in background simulation.  Combining this with the number of observed events, the effective volume and the detector livetime, we obtain 90\% confidence level limits on the muon flux coming from annihilations in the Sun for each considered neutralino mass \cite{newsun}, as shown in Fig. \ref{sun}b.

\vspace{-0.5pc}
\section{Discussion and outlook}
\vspace{-0.5pc}

Figures \ref{earth}b and \ref{sun}b present the AMANDA limits on the muon flux from neutralino annihilations into $W^+W^-$ (hard channel) in the Earth and the Sun respectively, together with the results from other indirect searches.  Limits have been rescaled to a common muon treshold of 1$~$GeV using the known energy spectrum of the neutralinos.  Also shown are the cosmologically relevant MSSM models allowed (crosses) and disfavoured (dots) by the direct search from CDMS \cite{cdms}.  Compared to our search for a terrestrial neutralino signal in 1997 AMANDA data \cite{earth97}, the limit has been improved by a factor which is more than that expected from additional statistics alone.  This is due mainly to the separate cut optimization for each neutralino mass, which exploits the characteristic muon energy spectrum of each model.

In 2001 an extra trigger was installed that lowered the energy threshold of the detector.  This trigger takes into account spatio-temporal correlations in the event hit pattern.  A preliminary analysis with data taken in 2001 and 2002 shows an improvement of a factor of about 5 in the effective volume in the search for 50$~$GeV neutralinos (soft annihilation channel) from the Earth with respect to the analysis presented in this note.  We are currently performing searches for a dark matter signal both from the Earth and the Sun with data taken from 2000 and later.  The increased detector exposure combined with improved reconstruction techniques and the new trigger setting will result in improved limits from these analyses (note that a 4-year exposure alone would already give an improvement of a factor of two).




\newpage
\setcounter{section}{0}
\section*{\Large Performance of AMANDA-II using Transient Waveform Recorders}

\vskip 0.05cm
{\large A. Silvestri$^a$ for the IceCube Collaboration}\\
{\it (a) Department of Physics and Astronomy, University of California, Irvine, CA 92697, USA}

\vskip 0.05cm
{\large  Presenter: A. Silvestri (silvestri@HEP.ps.uci.edu), usa-silvestri-A-abs1-og27-poster}

\title[Performance of AMANDA-II using Transient Waveform Recorders]{Performance of AMANDA-II using Transient Waveform Recorders}
\author[A. Silvestri for the IceCube Collaboration] {A. Silvestri$^a$ for the IceCube Collaboration\\
        (a) Department of Physics and Astronomy, University of California, Irvine, CA 92697, USA}
\presenter{Presenter: A. Silvestri (silvestri@HEP.ps.uci.edu), usa-silvestri-A-abs1-og27-poster}

\maketitle
\begin{abstract}
AMANDA-II data acquisition electronics was upgraded in January 2003 to 
readout the complete waveform from the buried PMTs using 
Transient Waveform Recorders (TWR).
We perform the same atmospheric neutrino analysis on data collected in
2003 by the TWR and standard AMANDA data acquisition 
system ({$\mu$-{\tt DAQ}}). Good agreement in event rate and angular distribution
verify the baseline performance of the TWR system.
\end{abstract}

\section{Description of {$\mu$-{\tt DAQ}} and {\tt TWR-DAQ} Systems in AMANDA}
\indent The Antarctic Muon And Neutrino Detector Array (AMANDA) is the first
neutrino telescope constructed in transparent ice,
and deployed between 1500 m and 2000 m beneath the surface of the
ice at the geographic South Pole in Antarctica.
AMANDA-II~\cite{pointsource} was completed in February 2000 and has taken
data routinely since that time~\cite{icrc2001}.
It is designed to search for neutrinos that originate in the most
violent phenomena in the observable universe.
AMANDA has searched for point sources in the entire northern sky and
for diffuse sources of high energy neutrinos of cosmic origin.\par 
Since 1997 the detector is taking data using a data acquisition
electronics based on Time to Digital Converters (TDC) which measures
the arrival time of the PMT pulses, and Analog to Digital Converters
(ADC), which record the maximum value of the PMT pulse amplitude.
Limitation of this system are the TDC with a maximum of 8
leading edges per trigger, and the peak ADC that records only one 
amplitude per event.
The limitation is particularly acute for bright, high energy events,
because important information on afterpulses is lost.
The data acquisition electronics of the AMANDA-II detector was
upgraded in 2003 to readout the complete waveform of the
photomultiplier tubes (PMTs) using Transient Waveform Recorders
(TWR)~\cite{icrc2003}. Afterpulse information is crucial to extend the
dynamic range of Number of photo-electron (Npe) measurement up to 5000
photons~\cite{icrc2003}.\par
In order to distinguish the two data acquisition
electronics of the AMANDA detector, we call {$\mu$-{\tt DAQ}} the
original system operating since 2000, and {\tt TWR-DAQ} the upgraded
system using waveforms since 2003.
The decision was made to run the two systems in parallel until the
{\tt TWR-DAQ} was proven to work as satisfactorily.
To compare the two systems the data from 2003 has been
analyzed with both read-out systems.
Additional information on the {\tt TWR-DAQ} hardware can be found
in~\cite{timo}.\par
Extending the analysis tools to include TWR data required several new
developments: (1) The {\tt TWR-DAQ} measures the integrated charge $\cal{Q}$
of the PMT pulses. 
In contrast, the {$\mu$-{\tt DAQ}} only measures the maximum amplitude
of the PMT waveform in a 2 $\mu$s window. (2) We account for various
time offsets and we extract a timing resolution of few nanoseconds.
The performance of the {\tt TWR-DAQ} is tested by comparing the
results for the absolute rate of atmospheric neutrinos and the
$cos(\theta)$ distribution with the results
from the standard {$\mu$-{\tt DAQ}}.

\section{Data Processing and Comparison of {$\mu$-{\tt DAQ}} and {\tt TWR-DAQ} Systems}
\label{processing}
The {\tt TWR-DAQ} data volume reaches 15TB per year 
compared to $\sim$ 1TB of the {$\mu$-{\tt DAQ}} system.
The data information of the two systems has been merged
according to GPS time and the fraction of overlapping PMTs
participating in the event in both systems.
For this analysis we restricted the capabilities of the {\tt TWR-DAQ}
system to mimic the features of the {$\mu$-{\tt DAQ}} system as close
as possible. 
The timing and amplitude information extracted from waveforms has been
used as input parameters to perform PMT-pulse cleaning, TOT
(Time-Over-Threshold) and cross-talk cleaning.\par
Due to small differences in threshold values, the {\tt TWR-DAQ}
collects $\sim$ 80\% of pulses observed in the {$\mu$-{\tt DAQ}}
system.\footnote{The pulse detection efficiency 
was increased in 2004 and 2005 seasons.}
A software re-trigger was applied at lower majority ($\cal{M} =$ 19) to obtain
similar rates after PMT-pulse cleaning procedures.
TOT-cleaning procedures were tested using {$\mu$-{\tt DAQ}} data,
which exclude anomalous PMT-pulses which are too short or too long.
The same TOT-cleaning was performed
on the {\tt TWR-DAQ} data, which causes a slight loss in
efficiency due to the higher threshold.
Cross-talk cleaning procedures were also tested using {$\mu$-{\tt DAQ}} data,
and cross-talk tables were designed according the ADC and TOT response
observed in {$\mu$-{\tt DAQ}} data.
We decided to apply the same cross-talk cleaning procedures to
perform a closer comparison, however small differences are expected since
the {$\mu$-{\tt DAQ}} data contains differences TOT values,
and peak ADC are replaced by the integrated charge $\cal{Q}$.\par
Timing and amplitude calibration have been performed using a threshold
algorithm which captures the leading and trailing edge of single PMT
pulses, and calculate the Npe from integrated charge.
Time offsets are calculated and have been applied as
$t_{offset} = t_{cable} - t_{module} - t_{delay} - t_{\cal{Q}}$, where
$t_{cable}$, $t_{module}$, $t_{delay}$ and $t_{\cal{Q}}$ account for
cabling offset, TWR module hardware clock, stop delay of trigger
response and amplitude-timing corrections, respectively.
Amplitude-timing corrections are performed by fitting the leading edge
versus $1/\sqrt{\cal{Q}}$, where $\cal{Q}$ is the integrated charge rather than
peak ADC measurement.
These measurements are then compared to the standard timing calibration
of the {$\mu$-{\tt DAQ}}.\par
\begin{figure}[h]
\begin{center}
\includegraphics*[width=0.49\textwidth,angle=0,clip]{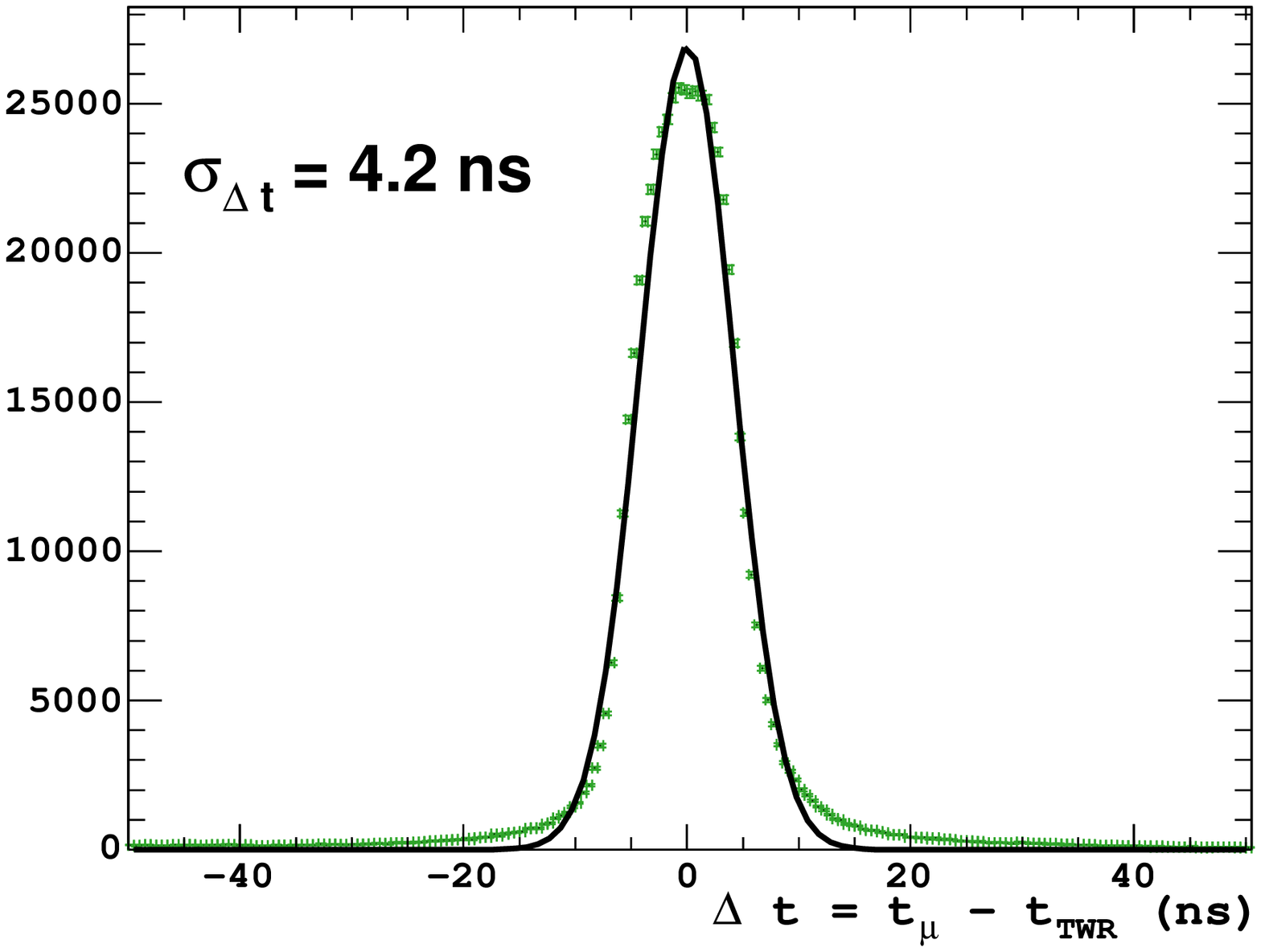}
\includegraphics*[width=0.49\textwidth,angle=0,clip]{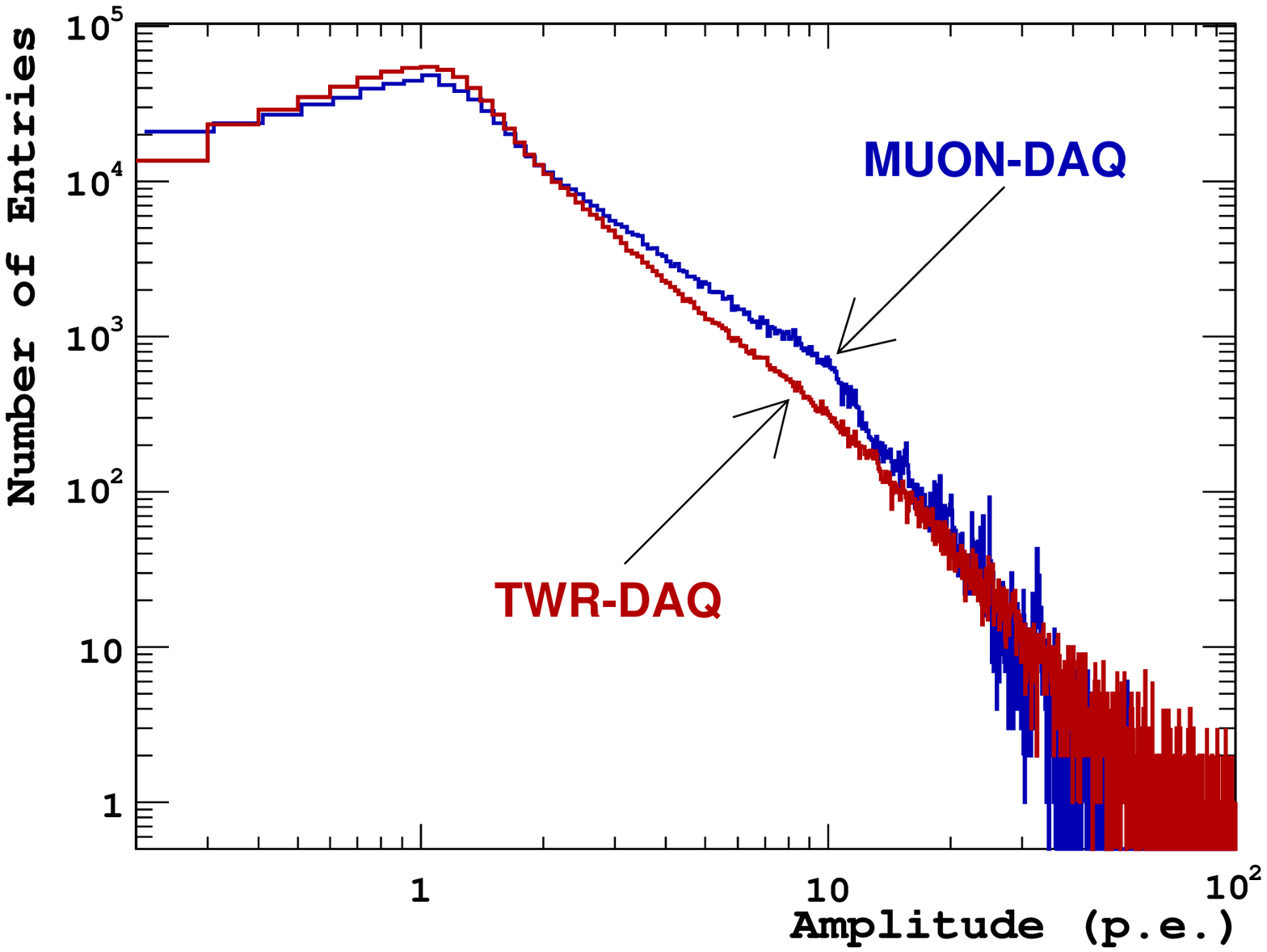}
\caption{\label{fig1d} 
(Left) Time difference between the PMT pulses recored by the two
  acquisition systems $\Delta t = t_{\mu} - t_{TWR}$. (Right)
  Calibrated amplitude normalized to 1 photo-electron (p.e.)
  value for {$\mu$-{\tt DAQ}} and {\tt TWR-DAQ} data. See text for details}
\end{center}
\end{figure}
A Gaussian fit of the distribution for 
$\Delta t = t_{\mu} - t_{TWR}$ yields $\sigma_{\Delta t} = 4.2$ ns
(Figure \ref{fig1d} (left)), which is dominated by the systematic error
of the time jitter between independent flash ADC clocks of the TWR
system. The {\tt TWR-DAQ} timing calculations are relative to the
values measured by the {$\mu$-{\tt DAQ}}.
The timing of the {\tt TWR-DAQ} system includes two sources of jitter,
both are related to a uniform window that is 10 ns in duration. This
account for the $\sigma_{\Delta t} \sim 4$ ns.
Amplitudes are also calibrated by extracting the number of
photo-electron (Npe)
detected from peak ADC of the {$\mu$-{\tt DAQ}} and charge $\cal{Q}$
of the TWR system and normalized to 1pe amplitude. 
By integrating the charge from pulse in the waveform, the dynamic
range of the {\tt TWR-DAQ} extends to Npe $\sim 100$, to be compared to
Npe $\sim 30$ of the {$\mu$-{\tt DAQ}}.
Figure
\ref{fig1d} (right) shows the reconstructed amplitude of the 
{\tt TWR-DAQ} compared to the {$\mu$-{\tt DAQ}}, which indicates a stable
power law distribution extending up to 100 Npe,
while the {$\mu$-{\tt DAQ}} system shows a ``knee'' around 10 Npe.
The knee is due to the amplitude saturation of the channels with
optical fibers, approximately 40\% of the AMANDA
readout.\footnote{High voltage values have been lowered in
January 2005 to increase linear dynamic range of optical channels.}
After cleaning, the muon track is reconstructed from the remaining
information. Details on the reconstruction techniques can be found
in~\cite{amandareco}.
Table \ref{rates} summarizes the passing rates from the raw data level 
to the final sample of atmospheric neutrinos.
\begin{table}[h]
\begin{center}
\begin{tabular}{c|c|c} 
~Selection~&~{\tt TWR-DAQ}~ & ~{$\mu$-{\tt DAQ}}~\\
                             \hline
Raw Sample    &  $1.86\times10^{9}$ & $1.86\times10^{9}$\\
Level-1       & $1.25\times10^{8}$ &$1.25\times10^{8}$   \\
Level-3       & $2.56\times10^{6}$& $1.99\times10^{6}$   \\
Final Sample  & 1112& 1026\\
\end{tabular}
\end{center}
\caption{\label{rates} Passing rates for increasing cut selection
  level for the {\tt TWR-DAQ} and {$\mu$-{\tt DAQ}} data analysis.}
\end{table}
\begin{figure}[h]
\begin{center}
\includegraphics*[width=0.49\textwidth,angle=0,clip]{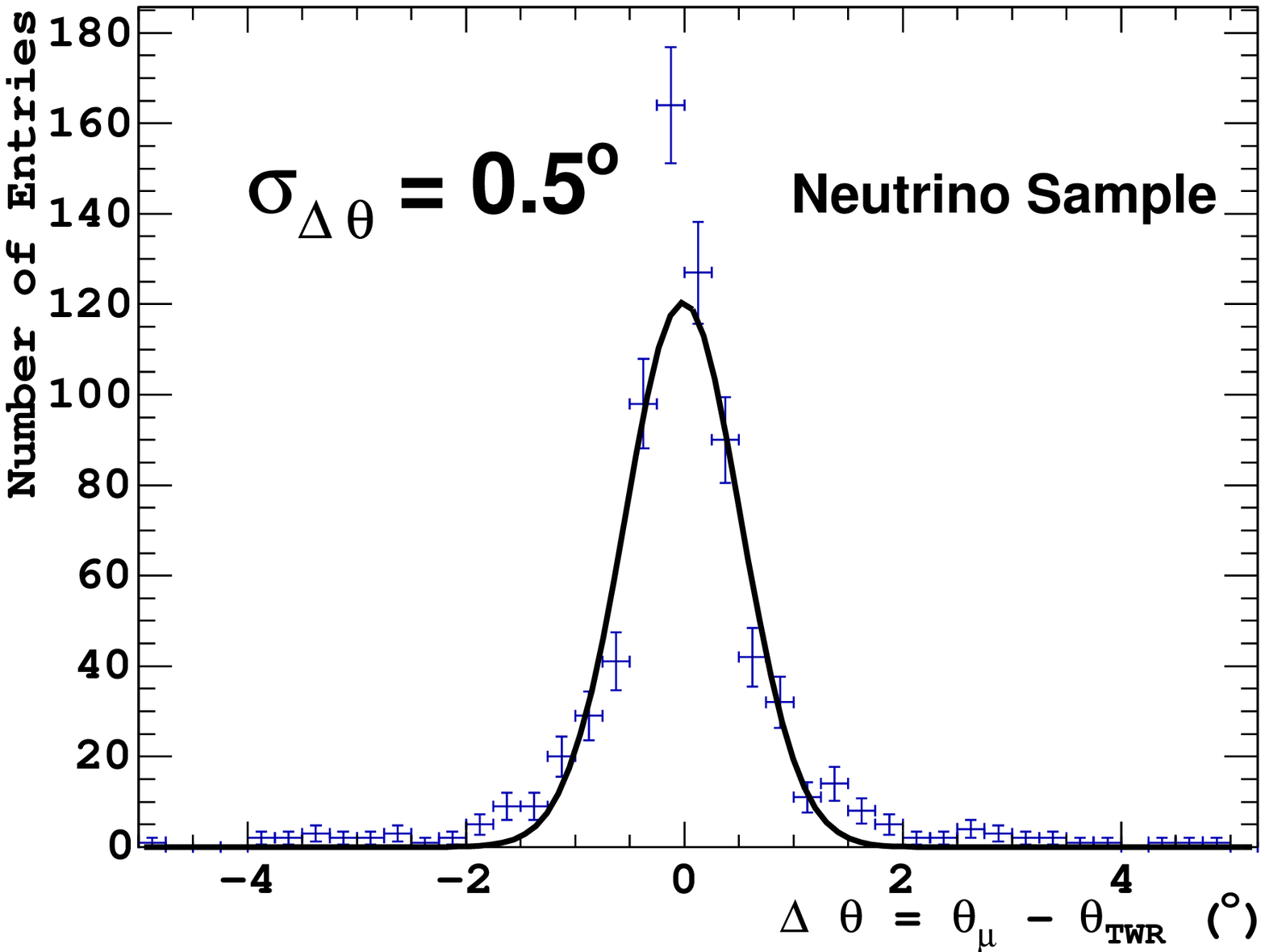}
\includegraphics*[width=0.49\textwidth,angle=0,clip]{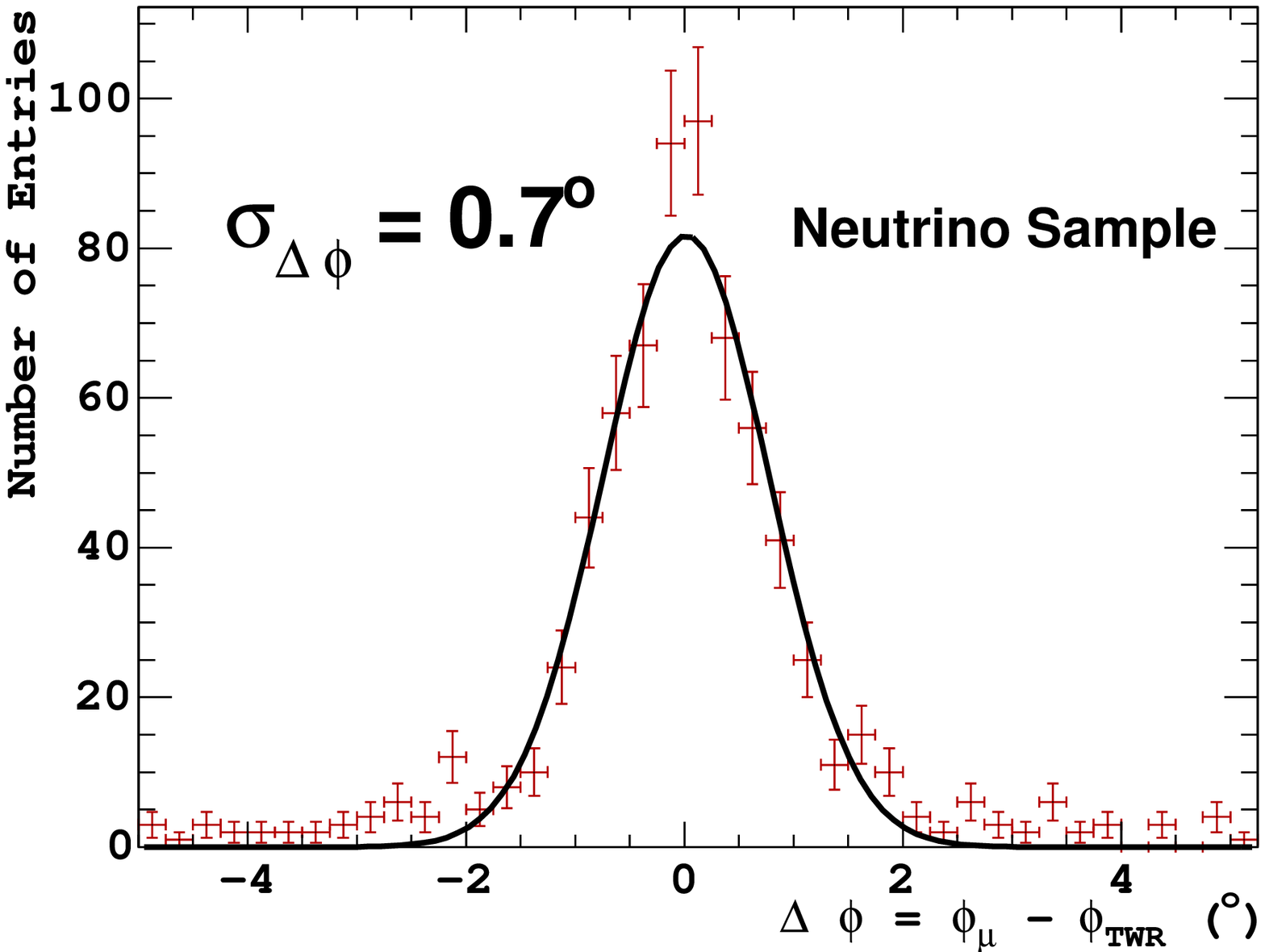}
\caption{\label{fig2e} (Left) The zenith $\Delta \theta = \theta_{\mu} -
  \theta_{TWR}$ difference distribution between the {$\mu$-{\tt DAQ}}
  and the {\tt TWR-DAQ} systems, (right) the azimuthal $\Delta \phi = \phi_{\mu}
  - \phi_{TWR}$ difference distribution.}
\end{center}
\end{figure}
Figure \ref{fig2e} shows that the angular mismatch for the final neutrino sample, 
$\sigma_{\Delta \theta}~{\rm is}~0.5^{\circ}$, where $\Delta \theta = \theta_{\mu}
- \theta_{TWR}$.
This value is expected from studies of the precision of the
global minimizer in the reconstruction program.
From the analysis based on {\tt TWR-DAQ} data,
1112 neutrinos are observed compared to 1026 neutrinos from the
{$\mu$-{\tt DAQ}} data analysis. The small differences in the
event rate are compatible with the small differences in analysis
procedures described in Section \ref{processing}.\par
\begin{figure}[ht]
\begin{center}
\includegraphics*[width=1.0\textwidth,angle=0,clip]{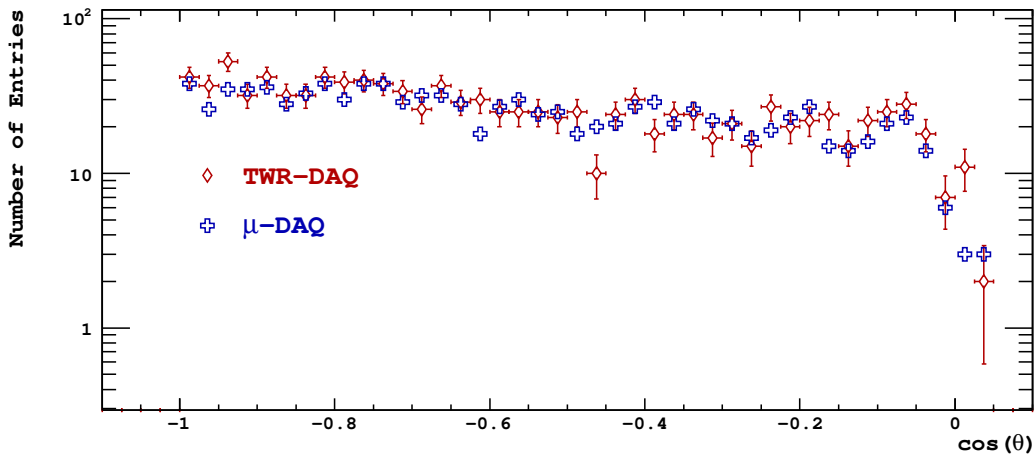}
\caption{\label {fig3b} cos$\theta$ distribution after final cut level
representing the atmospheric neutrino sample observed from the
{\tt TWR-DAQ} and the {$\mu$-{\tt DAQ}} analyses.}
\end{center}
\end{figure}

Figure \ref{fig3b} shows the cos$\theta$ distribution of the
atmospheric neutrino sample extracted from the {\tt TWR-DAQ} and
{$\mu$-{\tt DAQ}} data analysis.
Satisfactory agreement can be seen for the
cos$\theta$ distribution of the atmospheric neutrinos samples obtained
by the different analyses.

\section{Discussion and Conclusion}
The atmospheric neutrino analysis provides the first detailed
evaluation of the performance of the {\tt TWR-DAQ} system.
The preceding discussion demonstrated that the 
{\tt TWR-DAQ} produces similar event rates and angular distribution as the data 
collected by the standard {$\mu$-{\tt DAQ}} system. This encouraging result was 
obtained by first converting the waveforms into the more restricted 
information collected by the {$\mu$-{\tt DAQ}}. At that point, the
standard AMANDA analysis tools were applied to data from both of the
electronic systems, although some of the procedures were reformulated
to process waveform information.\par
We are developing new software tools to exploit the 
full information contained in the waveform. We expect that the additional
information will improve energy and angular resolution crucial for the
search of high energy phenomena.

\section{Acknowledgements}
The author acknowledges support from the U.S. National Science
Foundation Physics Division, the NSF-supported TeraGrid systems at
the San Diego Supercomputer Center (SDSC), and the National Center for
Supercomputing Applications (NCSA).

\newpage
\setcounter{section}{0}
\section*{\Large A software trigger for the AMANDA neutrino detector}

\vskip 0.05cm
{\large T. Messarius$^a$ for the IceCube Collaboration}
{\it (a) University of Dortmund, Otto-Hahn-Str. 4, 44221 Dortmund, Germany\\}

\vskip 0.05cm
{\large  Presenter: T. Messarius (messarius@physik.uni-dortmund.de), \
ger-messarius-T-abs1-og27-oral}

\title[A software trigger for the AMANDA neutrino detector]{A software trigger for the AMANDA neutrino detector}
\author[T. Messarius for the IceCube Collaboration]{T. Messarius$^a$ for the IceCube Collaboration
	\newauthor\\
	(a) University of Dortmund, Otto-Hahn-Str. 4, 44221 Dortmund, Germany\\
	}

\presenter{Presenter: T. Messarius (messarius@physik.uni-dortmund.de), \
ger-messarius-T-abs1-og27-oral}

\maketitle

\begin{abstract}
In the last few years a new Data Acquisition System (DAQ) for the
AMANDA-II detector was built and commissioned. The new system uses Flash ADCs
and works nearly dead time free (0.015\%), compared to 15\% dead time for the
old DAQ. Up to now, this new DAQ was triggered solely by the existing trigger system. 
Recently a software trigger was developed to take advantage of the new hardware. 
The first advantage is the ability to define more complex trigger-settings. 
A local coincidence trigger will improve the acceptance for low energy neutrinos 
The second advantage is that the new system can more readily integrate the 
existing 19 AMANDA strings into the new IceCube observatory.
\end{abstract}

\section{Introduction}
\begin{floatingfigure}[h]{3.9cm}
\vspace*{-0.5cm}
\includegraphics[height=7.8cm]{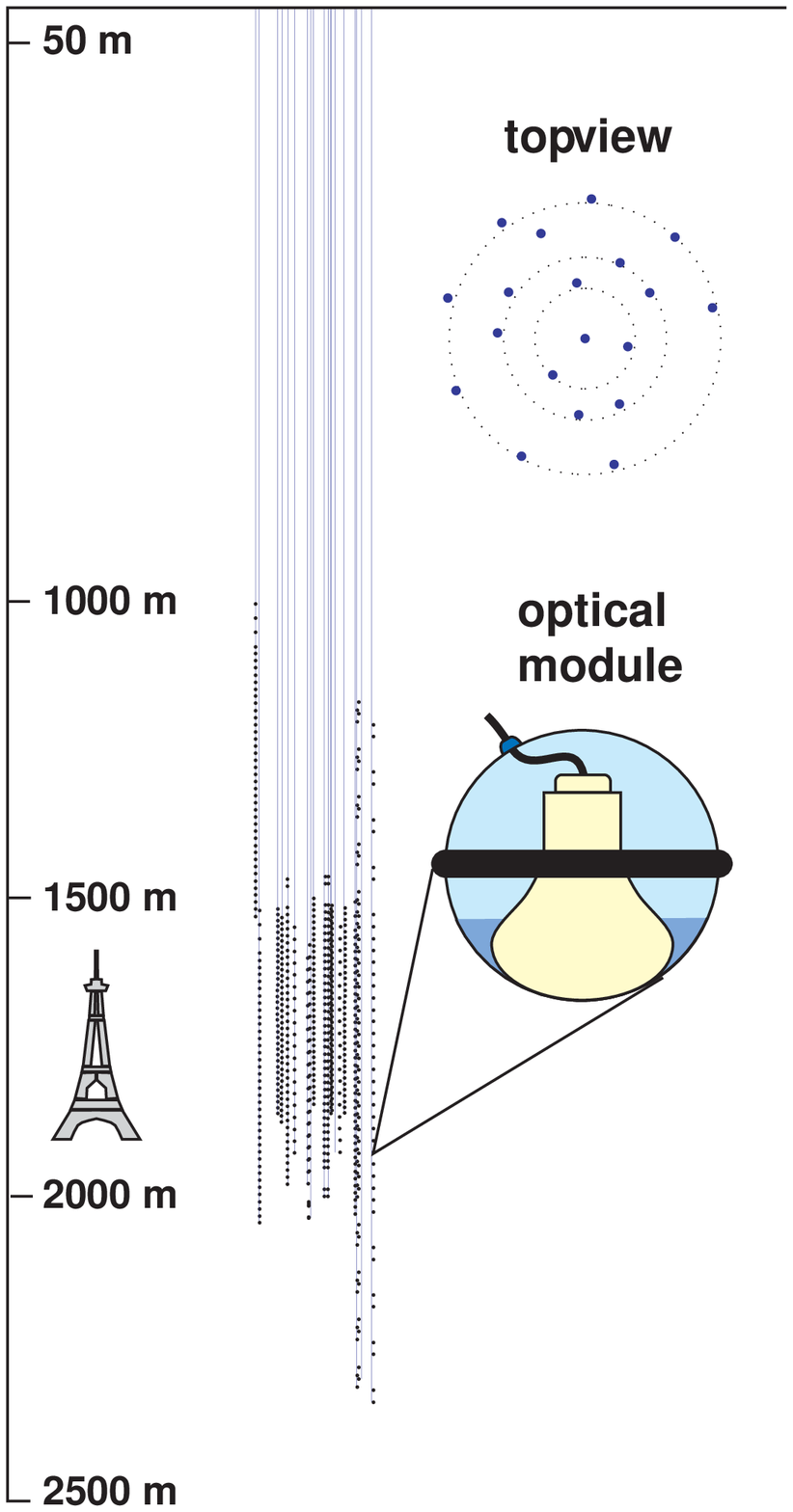} 
\vspace*{-0.8cm}
\protect\caption{
Amanda II. 
}
\label{AMANDA_detector}
\end{floatingfigure}
Since construction first began in 1996, the Antarctic Muon And Neutrino Detector Array (AMANDA)
has detected high energy muons and neutrinos. Neutrinos are observed indirectly by 
measuring the Cherenkov light from secondary leptons
generated in neutrino-nucleon interactions. AMANDA uses the ice sheet at the geographical
South Pole as its active volume and consists of 19 strings of Optical Modules
(OM) each containing a photomultiplier tube (PMT) and electronics in a glass
sphere. Most of the 677 OMs of the detector are located at a depth of
1.5 to 2 km below the ice surface. The detector is displayed schematically in
Fig.~\ref{AMANDA_detector}.

The analog PMT signals are transmitted via electrical cables or optical
fibers to the surface electronics. For the reconstruction of the energy and
direction of the particles, the arrival times of the pulses and the number of
photons contained in each PMT pulse are used. 

The original Data Acquisition system ($\mu$-DAQ) is triggered by a hardware based
trigger logic, the DMADD\footnote{Discriminator and Multiplicity ADDer}, 
which identifies events through a pre-set global multiplicity condition:, i.e., more than M OMs with a signal in a 2.5 $\mu$sec time window.
At low multiplicities, random coincidence of PMT noise pulses ("noise events")
dominate the muon signal. Lower multiplicity signals are triggered by local 
coincidence on a single string ("string trigger").

AMANDA has produced scientific results for many years and is a working neutrino
detector~\cite{andres00,results}. 
\section{The new TWR-DAQ system and science goals}
The 
$\mu$-DAQ system uses TDCs and peak sensing 
ADCs to extract information from an analog PMT pulse. 
It uses the DMADD trigger logic with a multiplicity M=24 which results in a trigger rate of $\sim100$ Hz.
The vast majority of these triggered events are due to atmospheric muons.
The settings are a compromise between the energy threshold and deadtime of the detector.
This and other limitations~\cite{Wolle03,wolledis} of the 
$\mu$-DAQ led to the 
development of a new DAQ system based on Transient Waveform Recorders~\cite{struck} (TWR), 
which are flash ADCs (FADC) capable of recording
the complete waveform of a PMT in a time window of 10.24 $\mu$sec with a
re\-solution of 12 bit and a sampling frequency of 100 MHz. The new system -
TWR-DAQ - has a significantly lower dead time (approx 0.015\%) compared to the
$\mu$-DAQ system (15\% deadtime), and a better readout performance leading to
a higher data bandwidth. Since Feb 2003, the TWR-DAQ used the DMADD as a
trigger source. In 2003, a M=24 level was used while in 2004 the trigger multiplicity
was lowered to M=18 with a corresponding trigger rate of $\sim$140Hz. The
$\mu$-DAQ still uses a M=24 multiplicity level.
 For each PMT pulse, the whole waveform is written out resulting in a larger amount of
data compared to the 
$\mu$-DAQ system. A first reduction is
performed by a Digital Signal Processor (DSP) collecting the data from the
TWRs in each crate and performing a so called Feature Extraction (FE) to extract
the rather short PMT pulses from the waveform. The new TWR-DAQ system is
running and taking data continuously for two years~\cite{Wolle03}. In
2003 and 2004, the system produced a data rate of $\sim$65~GB per day, nearly exceeding the
capability of network communication at the South Pole. For
a further extension of the capabilities of the detector with respect to the
intended integration of AMANDA into the IceCube detector~\cite{dima05,icecube}, the TWR-DAQ will be
further upgraded. The goals for the next upgrade are:
\begin{itemize}
\item to reduce the energy threshold of the detector in order to increase the
  sensitivity for low energy particles,
\item to introduce fast algorithms to search for patterns, i.e. local
  coincidences, in low multiplicity events to distinguish reconstructable
  events from noise events, and,
\item to send trigger signals to the IceCube trigger system in order to
  integrate AMANDA into the IceCube detector.
\end{itemize}  
\begin{floatingfigure}[lh]{8.8cm}
\vspace*{-0.5cm}
\includegraphics[height=5.0cm]{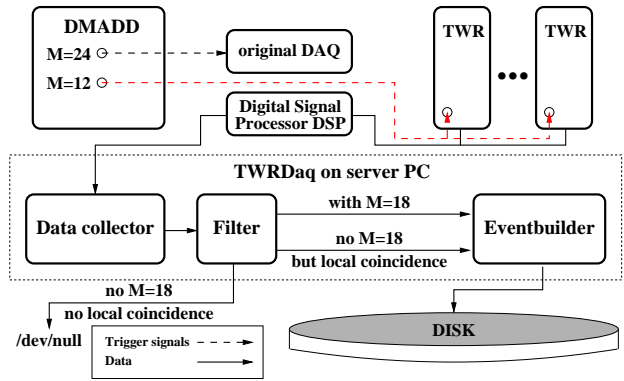} 
\protect\caption{Structure of the AMANDA DAQ system in 2005.}
\label{sftw_trig}
\end{floatingfigure}
These goals can be achieved by implementing software trigger algorithms into
the TWR-DAQ system. The layout of the trigger system is displayed in
Fig.~\ref{sftw_trig}. The DMADD system is still used as a pre-trigger with a
low multiplicity threshold. The global
multiplicity level can be lowered from M=18 to M=12. Events with a
multiplicity of M$>$18 are directly written to disk, while the remaining
events are tested for local coincidences in software. The algorithm is described in the next section.
A total trigger rate of $\sim$195 Hz is reached and noise events are mostly rejected.
During the last update of the TWR-DAQ in January 2005 an online monitoring was
implemented to the TWR-DAQ. Changes in the trigger rates and noise rates of each PMT 
can be monitored for each file.
\vspace{-0.5cm}
\section{Performance of the new TWR-DAQ system including the software trigger}
The TWR-DAQ including the software trigger is running stably since mid -
February 2005. Particles with high energy (100 GeV to TeV range) produce a sufficient amount of
light in the detector to be triggered reliably by the DMADD multiplicity
trigger while the detection efficiency for smaller energies is low. One
challenge for the software trigger is an improvement of the sensitivity of the array for low
energy particles by distinguishing between noise induced and particle induced
events. The PMT pulses of particle induced events are correlated in time and
space. A dedicated trigger algorithm searching for local coincidences
identifies these events in the detector. The so-called next neighborhood
algorithm (NEXT) measures the level of local coincidence in the event. For
each OM with a 
pulse, a sphere of 90 m radius around the OM is searched for
further OMs with a 
pulse. If there is at least one additional OM with a
pulse, the number of modules with at least one local coincidence pair
$N_{1hit}$ is increased by 1. Alternatively, the number of local coincidence
pairs is counted in $N_{pair}$. Double counts are excluded in these numbers.
In 2005, the software trigger uses $N_{pair} > 8$ or $N_{1hit} > 5$ to trigger event readout.\newline
While muons produce long tracks in the detector, electrons and taus will lose
their energy contained in a small volume inside the detector producing a
cascade like event. Figure~\ref{vol_trig} shows that the NEXT trigger is sensitive to both of the event types.

The pre-trigger rate from the DMADD is $\sim$250 Hz. First, 
multiplicity (M=18) events are written to disk and excluded from further processing. 
The software trigger keeps about 50\% of the remaining events.
%
The distribution of the number of PMT pulses in the triggered event,
$N_{hit}$~\footnote{Total number of PMT pulses in a time window of 10.24 $\mu$sec.}, are shown in Fig.~\ref{nhit_plot}. 
For this analysis a set of random data was taken using a frequency generator as trigger. Since
the frequency generator is not correlated to any physical events the data can
be used to investigate the trigger settings without any pre-trigger
thresholds. At small $N_{hit}$ ($N_{hit}<25$) the
distribution of the random data is dominated by Poissonian noise events.
At $N_{hit}>25$ the distribution is dominated by the slowly decreasing contribution of $\mu$-events. The NEXT trigger filters at small $N_{hit}$ down to $N_{hit}=12$ the reconstructable~\footnote{These events pass the standard track reconstruction of the AMANDA analysis chain.} events out of the event stream leading to an
increase of low-energy events.
\begin{floatingfigure}[lh]{8.8cm}
\vspace*{-0.5cm}
\includegraphics[height=7.0cm]{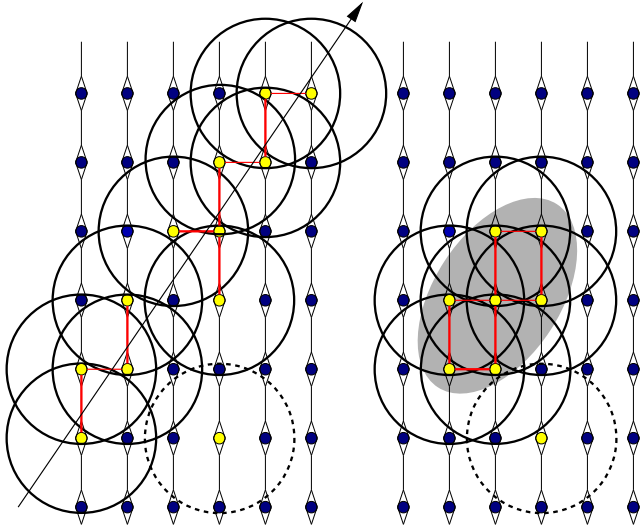} 
\protect\caption{2D schematic of the functionality of the next neighborhood trigger. The red/gray lines between the OMs show counted pairs of hit OMs. The left picture shows a track like event, the right picture a cascade like event. The black circles show the search volume.}
\label{vol_trig}
\end{floatingfigure}
The dotted line in Fig.~\ref{nhit_plot} shows the distribution of reconstructable events. 
At $N_{hit}>24$ the percentage of reconstructable events is 
about 90\% to 100\%. It decreases to $\sim$50\% for $N_{hit}=16$ and is down to
$\sim$15\% for $N_{hit}\le14$.\newline
The update of the TWR-DAQ would result in an increase of the data volume to $\sim$95
GB/day. To decrease the data volume an online compression, based on the Huffman algorithm, is used.
The performance of this algorithm depends
strongly on the distribution of values in the uncompressed data. In order to
improve the performance of the compression, a linear prediction of the
waveform values is used. Only the difference between the true value and the
prediction is stored.  
These steps reduce the data volume to 35 GB/day.  
Compared to 2004, the average number of bytes per
event decreased by 50\% while the trigger rate increased by 40\%.
%
\vspace{-0.4cm}
\section{Integration of AMANDA into IceCube}
\vspace{-0.2cm}
The existing AMANDA-II detector will be incorporated into the IceCube detector,
which is currently under construction. The first IceCube string was installed in the ice in Jan 2005 
and additional strings are planned for the
next deployment season (Nov. 2005 -Feb. 2006).
With the AMANDA TWR-DAQ system it is
more straigthforward to integrate information from the AMANDA strings into the IceCube trigger and data formats.
After the succsessful completion of the integration of the TWR-DAQ with the IceCube DAQ,
the original AMANDA DAQ will be decomissioned.
%
The AMANDA and IceCube strings will be synchronized via a common GPS clock. The TWR-DAQ will send its
trigger information to IceCube Global Trigger (GT), which identifies physical events in
the data stream with a software trigger algorithm.
The TWR-DAQ sends its trigger information via TCP/IP to GT which handles the
triggers from the IceCube sub-detectors. 
The IceCube DAQ is based on a custom IP-like communication with the DOMs
(Digital Optical Modules). There can be a latency of several 100 milliseconds before
the signals from the DOMs reach the surface. 
Once all sub-detectors have produced and send their own triggers to the GT, 
the GT orders them in time and then starts producing higher triggers which are 
sent to the IceCube event builder (EB). Upon receipt of the triggers from the 
GT, the EB requests and reads raw data including waveform information from the
string processors (SP) which can hold raw data for about 30 seconds.
Due to the different DAQ and
Trigger architecture, only the trigger information from the TWR-DAQ will 
be sent to IceCube GT but not vice versa.
Both data streams will be combined off\-line at the South Pole or in the northern hemisphere.
\vspace{-0.35cm}
\section{Outlook}
\vspace{-0.2cm}
\begin{floatingfigure}[rh]{9.5cm}
\vspace*{-0.7cm}
\includegraphics[height=6.5cm]{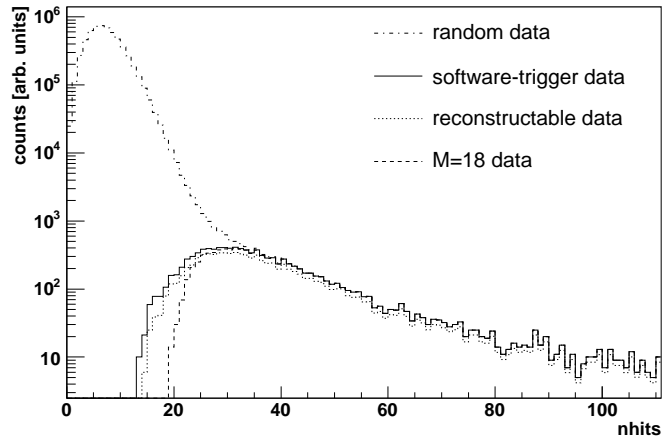} 
\vspace*{-0.8cm}
\protect\caption{The
distribution of $N_{hit}$. The solid line shows the triggered data with the
NEXT trigger and the global multiplicity. The dotted line shows the
reconstructable events and the dashed line shows the triggered events of the
year 2004 as comparison. The dash-dotted line represents $N_{hit}$
distribution obtained from random data.}
\label{nhit_plot}
\end{floatingfigure}
The new software trigger and a future implementation of the hardware string
trigger as a pre-trigger gives a great opportunity 
to lower the energy threshold for several analyses as well as increasing the sensitivity due to the 40\% gain in trigger rate.
It is intended to implement further filter algorithms to flag events
containing down-going $\mu$ tracks to accelerate an off\-line analysis.
A further idea is to implement a compensation for the cable delays of the detector to be able to reduce the time window and multiplicity level for a trigger decision.
The TWR-DAQ system will be equipped with a special Gamma
Ray Burst (GRB) trigger. A GRB trigger will be send from the GCN network
and will be received via satellite at the Pole. As soon as a GRB alert
arrived, the TWR-DAQ will store all data in the buffer and all the subsequent data for several hours.
%


%

\newpage
\setcounter{section}{0}
\section*{\Large Air showers with IceCube: First Engineering Data}

\vskip 0.05cm
{\large T.K. Gaisser$^{\it a}$, for the IceCube Collaboration$^{\it b}$}\\
{\it    (a) Bartol Research Institute, University of Delaware,
            Newark, DE USA \\ 
        (b) Members of the IceCube Collaboration are listed in the appendix to the Proceedings
        }

\vskip 0.05cm
{\large  Presenter: T.K. Gaisser (gaisser@bartol.udel.edu), \
usa-gaisser-TK-abs1-he15-oral}

\title[Air showers with IceCube ...]{
Air showers with IceCube: First Engineering Data}
\author[T.K. Gaisser for the IceCube Collaboration] {T.K. Gaisser$^a$, for the IceCube Collaboration$^b$
        \\
        (a) Bartol Research Institute, University of Delaware,
            Newark, DE USA \\ 
        (b) Members of the IceCube Collaboration are listed in the appendix to the Proceedings
        }
\presenter{Presenter: T.K. Gaisser (gaisser@bartol.udel.edu), \
usa-gaisser-TK-abs1-he15-oral}

\maketitle
\section{Introduction}

The IceTop km$^2$ air shower array of 80 pairs of
ice Cherenkov tanks is an integral part of the design of the
IceCube neutrino telescope at the South Pole.  The neutrino
telescope will consist of 80 strings, each instrumented
with 60 digital optical modules (DOMs) between 1450 and 2450 
m depth.  Thus IceCube will be a three-dimensional air shower detector 
as well as a neutrino telescope.  The power of a surface
array for calibration and background studies for the neutrino
telescope, as well as the physics potential for cosmic-ray
studies of the combined detector, are described in Ref.~\cite{ICRC2003}.
At ultra-high energy, where the Earth is opaque to neutrinos from
below, the surface array will give IceCube a significant
ability to discriminate between downward or horizontal neutrinos
and cosmic-ray induced background.  With the full detector we
anticipate 
study of the primary composition to EeV energies where the transition
from galactic to extra-galactic cosmic rays may occur.

\begin{figure}[htb]
\includegraphics[width=8.5cm,height=6.cm]{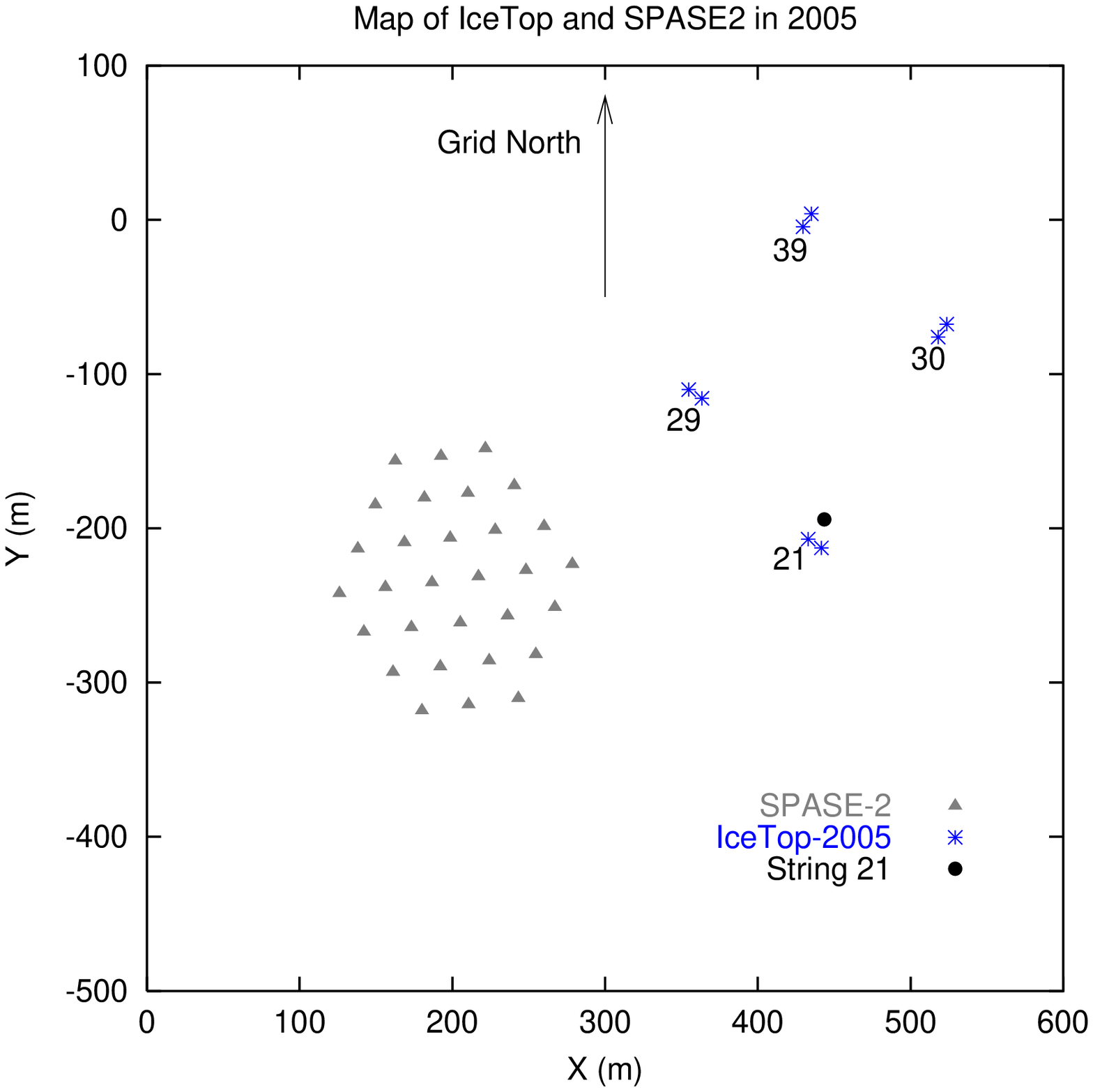}
\includegraphics[width=6cm,height=6.cm]{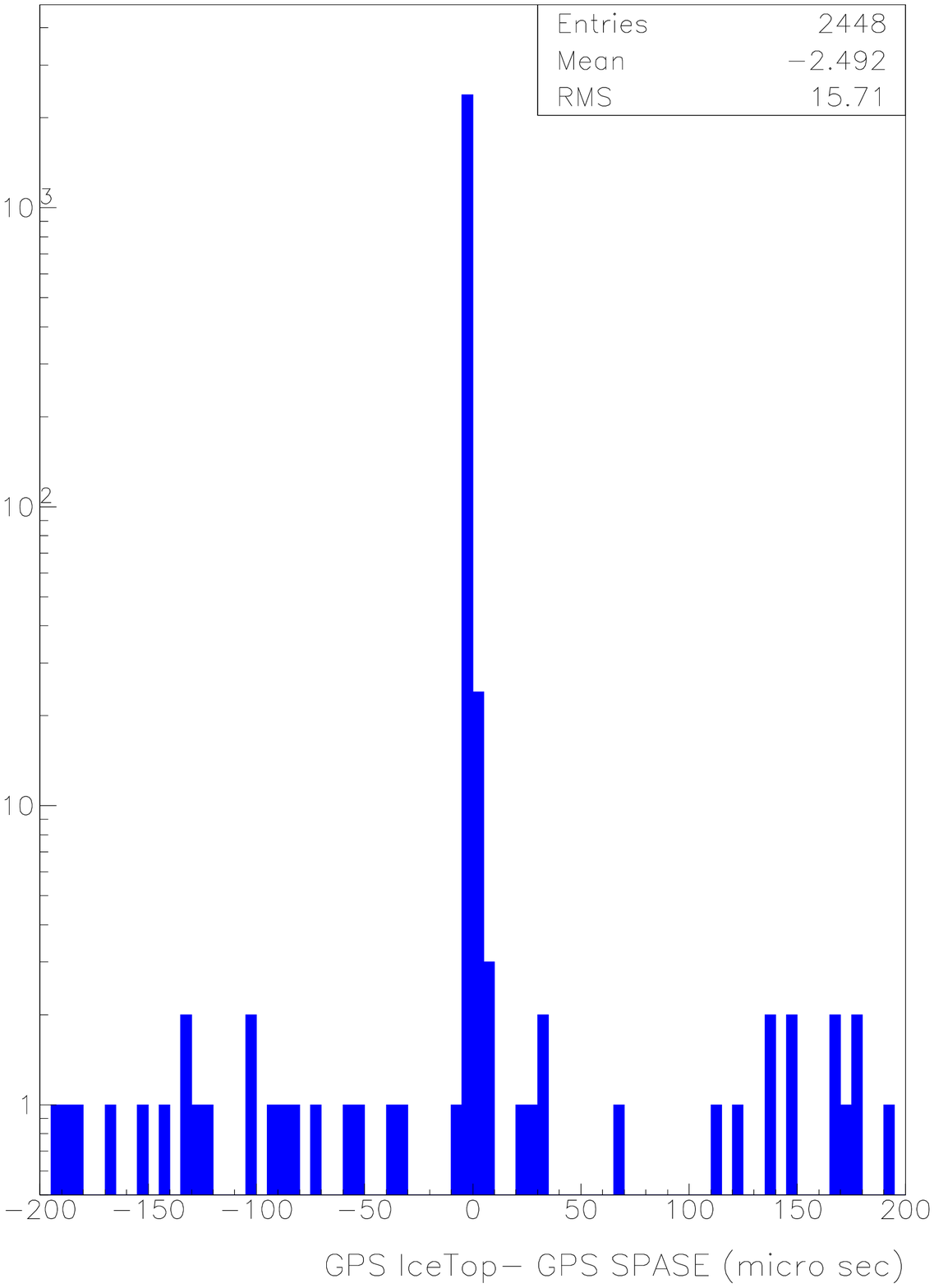}
\caption{Map of IceTop and SPASE in 2005.  The peak in the distribution
of time differences between IceTop and SPASE triggers contains events
seen by both arrays.}
\label{figure1}
\end{figure}

Deployment of IceCube
at the Amundsen-Scott South Pole Station
began in the 2004/2005 Austral summer season with the installation
of one string of 60 DOMs in the deep ice and four IceTop
stations on the surface.   Fig.~\ref{figure1} shows
the configuration of the first four stations relative
to the existing
South Pole Air Shower Experiment (SPASE)~\cite{SPASE},
which is still in operation.
In this paper we present engineering data from the first four surface
stations of IceCube, including coincidences
with SPASE.  Initial data from the deep string, including 
coincidences between surface and deep detectors, are presented
separately~\cite{Dima}.

\section{The IceTop Array}

\begin{figure}[htb]
\includegraphics[width=12cm]{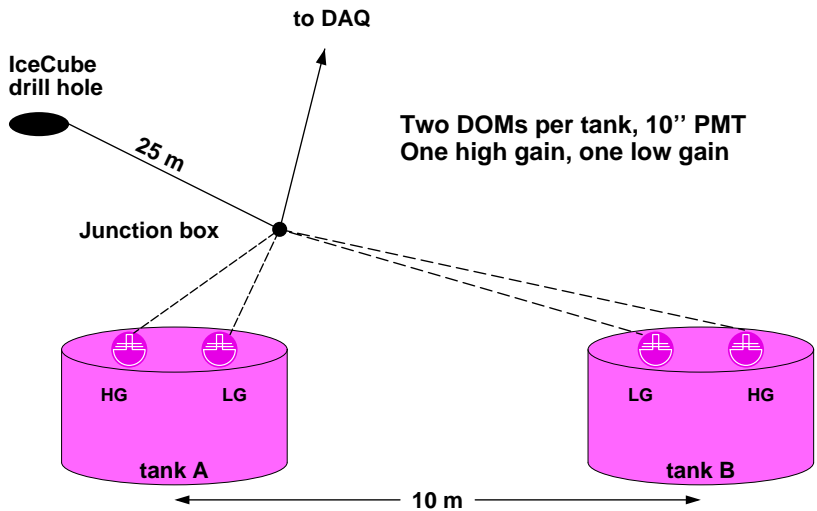}
\caption{Diagram of an IceTop station.}
\label{figure2}
\end{figure}

Each pair of IceTop tanks is associated with an IceCube string.
The result is a surface array in a triangular
pattern with a grid spacing of approximately
125 m.  Each IceTop station consists of two
ice Cherenkov tanks separated by 10 meters, as shown
schematically in Fig.~\ref{figure2}.  The two DOMs
in each tank face downward and respond primarily
to light reflected from the diffusely reflecting,
white liner of the tank.  The ice surface is in 
contact with granulated insulation, so that
the top inner surface is partially reflecting.
The tanks are cylindrical with surface area $2.7\,{\rm m}^2$
and ice thickness 90 cm.

Operation of the DOMs (which is described in more detail
in~\cite{Dima}) is controlled by
digital electronics, which can be set remotely.  The tanks
currently operate 
with one DOM set at low gain ($5 \times 10^5$)
and one at high gain ($5 \times 10^6$).  Three channels
of the analog transient waveform digitizer (ATWD)
on the DOM main board sample the waveforms
at 3.3 ns with different amplification, for a dynamic
range of more than 5000.

The rate of events detected per tank
depends on trigger and threshold settings.
The amplitude threshold is currently set
at a voltage corresponding to ten times the peak
voltage of a single photo-electron.
At this threshold and with no coincidence
requirement, the rate per DOM is about 3 kHz.
This rate includes approximately 1 kHz 
of muons.  Simulations~\cite{Clem} 
show that the balance is due primarily to
low energy electrons and to gamma-rays converting
in the tank.  The characteristic shape of the 
spectrum of single hits, which has a prominent
muon peak, is shown in the left plot of Fig.~\ref{figure3}.

For comparison with the air shower waveform shown below in Fig.~\ref{figure4},
we also show in Fig.~\ref{figure3} the average waveform of a throughgoing muon
in a high-gain DOM.  This is obtained by averaging over a sample of
waveforms corresponding to events in the muon peak of the distribution of
signals shown on the left panel of Fig.~\ref{figure3}.  With 90 cm of ice
thickness, typical energy-deposition by throughgoing muons
is 160 to 200 MeV, depending on zenith angle.  Comparison of
integrated single photo-electron waveforms with the integral of the average
muon waveform allows us to estimate that the typical through-going
muon produces enough light to generate approximately 160 photo-electrons 
in the photomultiplier.  The data shown in
Fig.~\ref{figure3} were taken during initial fast calibration
runs at the South Pole in January.  High statistics calibration
runs currently underway will provide the basis for detailed
calibration of amplitude and integrated charge.  Periodic sampling
of the single tank spectra will also serve to monitor performance
of each tank.  In addition, calibration runs with events tagged
by a muon telescope are planned for next season.

\begin{figure}[htb]
\includegraphics[width=7.7cm]{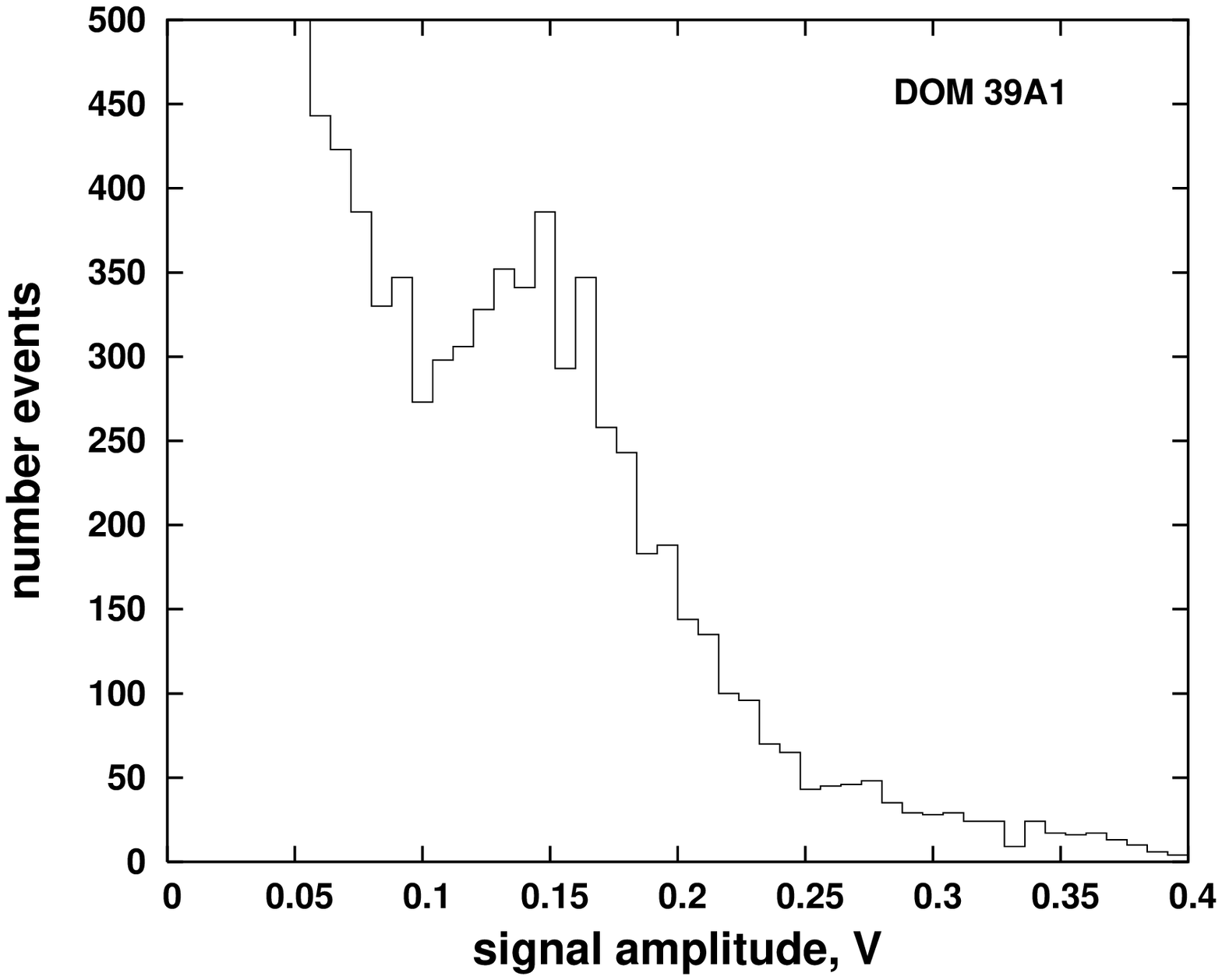}
\includegraphics[width=7.7cm]{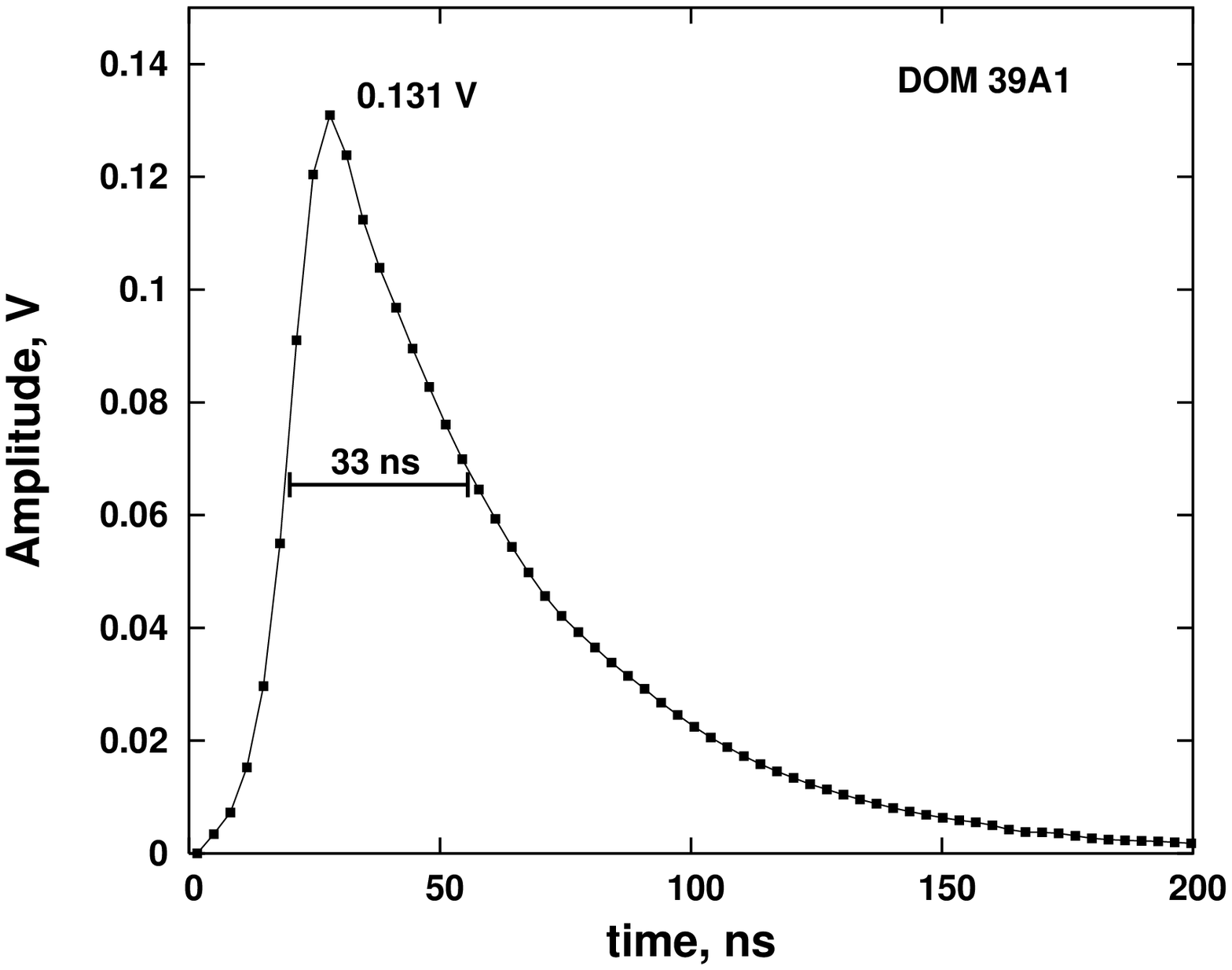}
\caption{Left: Peak amplitudes of single-tank signals
showing a muon peak (high-gain DOM);  right: average $\mu$ waveform.
}
\label{figure3}
\end{figure}

The spacing
between detectors at a station is chosen so that
a coincidence of two tanks at a station without hits
in adjacent stations corresponds to small showers
from cosmic-ray protons with primary energies
in the few to ten TeV range.  Such events have
a high probability of producing 
only one muon that reaches the
deep detector.  Tagging such events will be useful for
understanding the principal background of atmospheric
muons in the deep detector.

\section{Air Showers in IceTop}

For a series of runs taken February through May of 2005
the IceTop trigger was set to require 10 DOMs
above threshold within 2 $\mu sec$.  
During these initial runs there was a local 
coincidence requirement within a station
such that a DOM only reported
if at least one of the DOMs in the other tank
at the station was hit.  With this setup, only
air showers with at least three stations reporting
were recorded.  The rate of air showers with
this trigger is $0.7$~Hz,
divided approximately 60/40 between 3-station
events and 4-station events.  This rate is
consistent with the expected threshold of 300 TeV
and is due to events in the PeV and sub-PeV
energy range.

\begin{figure}[htb]
\includegraphics[width=15.5cm]{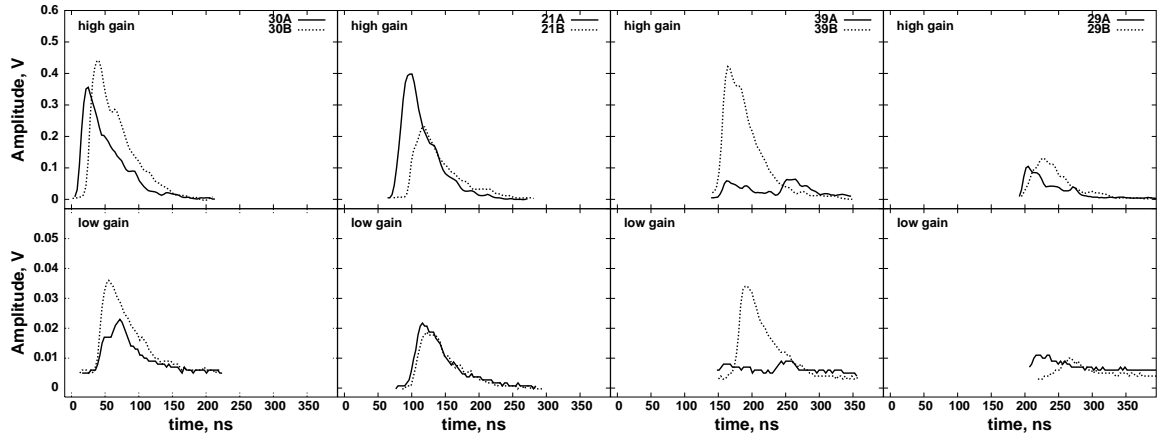}
\caption{Waveforms for a 16-fold event with a reconstructed zenith angle
of 23$^\circ$.}
\label{figure4}
\end{figure}

With only 4 stations, most triggers are from
air showers with cores
outside the perimeter of the array, and the
core locations cannot therefore be determined.
External events typically have one station with
a much bigger signal than the others.  Events
with cores inside the array will display a
more symmetric pattern. 
Fig.~\ref{figure4} shows a waveform display for
one such ``contained" event.  In this figure the
upper row shows the high-gain DOMs and the lower row
the low-gain DOMs.  
Solid lines show tank A and broken lines tank B
at each station.  Successive
panels (from left to right) show the event passing
across the array from station 30 to 29.
Low-gain waveforms generally track the high-gain
waveforms.  Shower-front fluctuations can cause significant
differences between tanks at the same station (e.g. 39),
and details of particle location and DOM response
within a tank also cause differences.  
When all high-gain channels saturate,
the low-gain channels may be used to extend the dynamic
range of the tank response to beyond that achievable
with the three ATWD channels.  We expect an overall dynamic range
for the tank response of approximately $10^5$.

Off-line we can find coincidences with the SPASE array
by matching up GPS times, as shown in Fig.~\ref{figure1}b.  
The rate of coincidences
between IceTop and SPASE triggers is about 1 per minute, which
corresponds to showers in the 10 PeV range.  

\vspace{-.2cm}

\section{Conclusion}
The four IceTop stations are recording air shower data.
During the remainder of this season, tank calibration
will be performed using the muon peak in the spectrum
of inclusive single tank hits.  In addition, fluctuations
in the shower front will be studied by comparing signals
in tanks at the same station, and detector fluctuations will
be studied by comparing DOMs set to the same gain in 
	one tank.  

The plan for the coming season at the South Pole is to
deploy 12 additional stations grid north of the present
array.  The result will be a 16 station air
shower array with an enclosed area of $0.12\;{\rm km}^2$.
For operation during 2006 we therefore expect to
cover a range of primary cosmic-ray energies from 300 TeV
to 100 PeV.  Approximately 10\% of IceTop triggers
will also give hits in the deep detectors, giving a
significant potential for calibration of the neutrino
telescope with the surface array, as well initial analysis
of composition in the knee region of the primary cosmic-ray
spectrum.

%

\newpage
\setcounter{section}{0}
\section*{\Large IceCube: Initial Performance}

\vskip 0.05cm
{\large D. Chirkin$^{\it a}$ for the IceCube Collaboration
        }\\
{\it 	(a) Lawrence Berkeley National Lab
	Berkeley, CA 94720-8158, U.S.A.
        }

\vskip 0.05cm
{\large Presenter: D. Chirkin (dchirkin@lbl.gov), \  
usa-chirkin-D-abs1-he15-oral}

\title[IceCube: initial performance]{IceCube: Initial Performance}
\author[D. Chirkin for the IceCube Collaboration] {D. Chirkin$^a$ for the 
IceCube Collaboration \\
	(a) Lawrence Berkeley National Lab
	Berkeley, CA 94720-8158, U.S.A.
        }
\presenter{Presenter: D. Chirkin (dchirkin@lbl.gov), \  
usa-chirkin-D-abs1-he15-oral}

\maketitle

\begin{abstract}
The first new optical sensors of the IceCube neutrino observatory - 60 on one string and
16 in four IceTop stations - were deployed during the austral summer of
2004-05. We present an analysis of the first few months of data collected by
this configuration. We demonstrate that hit times are
determined across the whole array to a precision of a few nanoseconds. We also look
at coincident IceTop and deep-ice events and verify the capability to
reconstruct muons with a single string. Muon events are compared to a simulation.
The performance of the sensors meets or exceeds the design requirements.

\end{abstract}
\vspace{-0.5cm}

\section{Introduction}
\vspace{-0.2cm}

\begin{wrapfigure}{l}{3.1cm}
\vspace{-0.6cm}
\epsfig{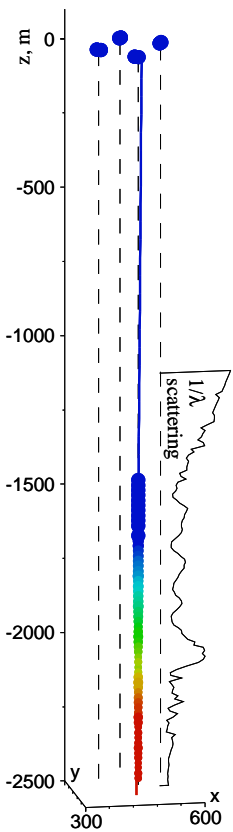}
\vspace{-0.7cm}
\caption[]{\label{ax} IceCube event }
\vspace{-0.8cm}
\end{wrapfigure}
The IceCube neutrino observatory at the South Pole will consist of 4800 optical sensors - digital optical modules (DOMs), installed on 80 strings between the depths of 1450 to 2450 meters in the antarctic ice, and 320 sensors deployed in 160 IceTop tanks on the ice surface directly above the strings. Each sensor consists of a 10 in.\ photomultiplier tube, connected to a waveform-recording data acquisition circuit capable of resolving pulses with sub-nanosecond precision and having a dynamic range of at least 250 photoelectrons per 10 ns. This year 76 such sensors were installed as a first part of the IceCube and IceTop \cite{icetop} arrays.


After a sensor acquires and digitizes an event trace, it transmits the data to the surface electronics. The events are time-stamped locally with an internal (to each sensor) clock, which has an estimated drift time of $\sim 1$ ns/s. All of the DOM clocks are time-calibrated with a special procedure, which involves sending an analog pulse from the surface to the DOM, where this pulse is received, digitized, and recorded. A similar analog pulse is sent from the DOM to the surface, where it is, in turn, digitized, and analyzed together with the pulse recorded by the DOM (which is transmitted to the surface digitally after the main ``round trip'' calibration procedure finishes).
In this report we demonstrate that events are time-stamped with a nanosecond-scale precision over the network of 76 deployed DOMs.

Fig.\ \ref{ax} shows an event involving all 76 DOMs. The circle size is proportional to the signal amplitude, while the color (from blue to red) indicates relative times of the hits recorded in the DOMs. All hits are consistent with an air shower on the surface coincident with a deep-ice muon, traveling down at a zenith angle of $3\pm2 ^{\circ}$.

From the ice scattering-length profile shown next to the detector string, one sees that most of the detector is located in very clear ice. In fact, the lower 25 DOMs are in ice that is up to 2 times clearer than that available to the AMANDA \cite{amanda1} sensors located at depths of 1500-2000 meters.

\vspace{-0.3cm}
\section{Time resolution and muon track reconstruction}

As a part of each sensor's time calibration, round trip times of the time calibration pulses are measured (Fig.\ \ref{a1}). The times are larger for progressively deeper sensors on the string (DOMs with numbers 1-60), and are essentially the same for the IceTop sensors (shown as DOMs with numbers 61-76). Calibrations are done automatically every few seconds. The round trip time varies slightly from one calibration to the next, and the size of the variation provides the basic measurement of the precision of the time calibration procedure (Fig.\ \ref{a2}).

\begin{figure}[!h]\begin{center}
\begin{tabular}{ccc}
\mbox{\epsfig{file=rt-mean.epsi,width=.45\textwidth}} & \ & \mbox{\epsfig{file=rt-rms.epsi,width=.44\textwidth}} \\
\parbox{.45\textwidth}{\caption[]{\label{a1} Round trip time of the time calibration pulse (IceTop DOMs are shown with numbers 61-76) }} & \ & \parbox{.44\textwidth}{\caption[]{\label{a2} The rms resolution of the round trip time of the time calibration pulse }} \\
\end{tabular}


\begin{tabular}{ccc}
\mbox{\epsfig{file=flasher-t.epsi,width=.45\textwidth}} & \ & \mbox{\epsfig{file=flasher-f.epsi,width=.44\textwidth}} \\
\parbox{.45\textwidth}{\caption[]{\label{ab} Hit time difference between 2 DOMs directly above the one flashing in clear ice }} & \ & \parbox{.44\textwidth}{\caption[]{\label{ac} Hit time resolution measured with flashers for 15 DOMs on the IceCube string }} \\
\end{tabular}
\end{center}\end{figure}

Each DOM contains an array of photodiode ``flashers'', which can be used for many types of calibrations. We used these to find the differences between the photon arrival times at a few DOMs directly above the ones flashing. Fig.\ \ref{ab} shows the distribution of such a difference for DOMs 59 and 58, when DOM 60 was flashing. The rms values for several such DOMs are shown in Fig.\ \ref{ac} and are best ($\sim 2$ ns) for the DOMs located in clearer ice.

\begin{wrapfigure}{r}{7.3cm}
\vspace{-0.4cm}
\epsfig{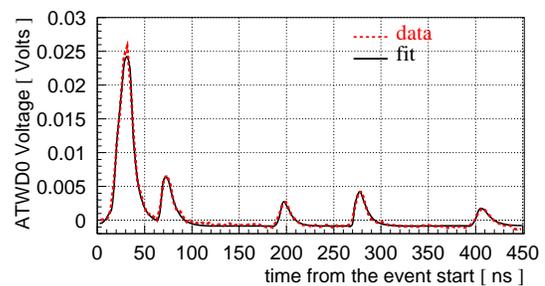}
\vspace{-0.2cm}
\caption[]{\label{ay} Captured hit event waveform}
\vspace{-0.8cm}
\end{wrapfigure}
A typical waveform captured by an IceCube sensor is shown in Fig.\ \ref{ay}. The waveforms are described very well by a waveform decomposition procedure, which yields single photon hit times.

A likelihood minimization algorithm for one-string track reconstruction in multi-layered ice was used to reconstruct the deep-ice data. The scattering and absorption values used were those measured with AMANDA and extrapolated to deeper ice using available ice core data and data collected by a dust measuring device used during the string deployment.

The track-fitting algorithm was tested on a simulated data sample of downgoing muons (Fig.\ \ref{a3}) and was found to reconstruct it rather well (Fig.\ \ref{a4}). The rms resolution of the muon track zenith angle reconstruction is 9.7$^{\circ}$ with an event hit multiplicity of 8 or more. The resolution improves rapidly as the multiplicity increases (3.0$^{\circ}$ at multiplicity 20, and 1.6$^{\circ}$ at multiplicity 40). This is similar to the one-string AMANDA analysis results \cite{klaus}.
\vskip -0.3cm
\begin{figure}[!h]\begin{center}
\begin{tabular}{ccc}
\mbox{\epsfig{file=mc-mult.epsi,width=.45\textwidth}} & \ & \mbox{\epsfig{file=mc-zdiff.epsi,width=.44\textwidth}} \\
\parbox{.45\textwidth}{\caption[]{\label{a3} Zenith angle distribution of simulated downgoing muons (red) vs.\ reconstructed tracks (green) }} & \ & \parbox{.44\textwidth}{\caption[]{\label{a4} Zenith angle difference distribution of reconstructed and simulated tracks }} \\
\end{tabular}


\begin{tabular}{ccc}
\mbox{\epsfig{file=ic-mc-mult.epsi,width=.45\textwidth}} & \ & \mbox{\epsfig{file=ic-mc-zenith.epsi,width=.45\textwidth}} \\
\parbox{.45\textwidth}{\caption[]{\label{a5} Muon hit multiplicity distribution of data and simulation }} & \ & \parbox{.45\textwidth}{\caption[]{\label{a6} Muon zenith angle distribution of data and simulation }} \\
\end{tabular}
\end{center}\end{figure}

\vspace{-0.289cm}

Fig.\ \ref{a5} compares the hit multiplicity distribution for 8 hours of data and a similar amount of simulated data. The zenith angle distribution of the reconstructed tracks in data is compared to the simulated data in Fig.\ \ref{a6}. The simulated data used in Fig.\ \ref{a3}-\ref{a6} was produced with standard AMANDA simulation, which was not tuned to the somewhat different trigger logic, ice conditions and different sensors of the deeper IceCube string. As the IceCube simulation matures, the apparent discrepancy observed in Fig.\ \ref{a6} is expected to become smaller.

Coincident deep-ice and IceTop events with a combined hit multiplicity of at least 14+14 hits collected during March, April, and May were reconstructed with both the IceTop shower reconstruction and the one-string muon track reconstruction discussed above. The resulting zenith angle distributions are compared in Fig.\ \ref{a7}. The directions obtained with the string reconstruction seem to be systematically closer to the vertical, which may indicate the need to improve the likelihood parameterization used in the track reconstruction. Alternatively it may be due to the shower front being curved and muons originating from a different part of the shower than that seen by IceTop. We measure a systematic offset of 2.1$^{\circ}$ with an rms deviation of 4.1$^{\circ}$ (Fig.\ \ref{a8}).

To measure systematic time offsets in the IceCube string we applied the one-string reconstruction to one day's worth of data 60 times. Each of the 60 DOMs was removed once during the reconstruction, and the time residuals of the hits in those DOMs to the expected direct (unscattered) hit times from the reconstructed tracks were evaluated. The residual time distributions are consistent with the expected distribution of hits coming from nearby muons (Fig.\ \ref{a9}). The maxima of such distributions indicate the time residuals of the most probable (in the current setup, direct) hits. In addition to systematic time calibration offsets these can be systematically removed from zero due to features of the DOM geometry still unaccounted for and scattering affecting photon propagation even at small distances. Most of these residuals are within 3 ns of each other, except for DOMs 35-43, which are located in dustier ice (Fig.\ \ref{aa}). This indicates that the DOM clock times for the whole array (currently 76 DOMs) are calibrated to within 3 ns of each other. An apparent large time offset of DOM 60 is currently under investigation.

\begin{figure}[!h]\begin{center}
\begin{tabular}{ccc}
\mbox{\epsfig{file=ic-it-zenith.epsi,width=.45\textwidth}} & \ & \mbox{\epsfig{file=ic-it-zdiff.epsi,width=.45\textwidth}} \\
\parbox{.45\textwidth}{\caption[]{\label{a7} Zenith angle distribution of string-reconstructed tracks (blue) and IceTop-reconstructed coincident showers (red) }} & \ & \parbox{.45\textwidth}{\caption[]{\label{a8} Zenith angle difference distribution between string-reconstructed tracks and IceTop-reconstructed coincident showers }} \\
\end{tabular}

\vspace{0.3cm}

\begin{tabular}{ccc}
\mbox{\epsfig{file=ic-mt-residual.epsi,width=.45\textwidth}} & \ & \mbox{\epsfig{file=ic-mt-adistr.epsi,width=.45\textwidth}} \\
\parbox{.45\textwidth}{\caption[]{\label{a9} Distribution of time residuals between the hits recorded by a DOM and time expectation for direct (unscattered) hits from nearby tracks reconstructed with the rest of the string }} & \ & \parbox{.45\textwidth}{\caption[]{\label{aa} Distribution of direct hit time residuals for all DOMs on the deployed IceCube string }} \\
\end{tabular}
\end{center}\end{figure}

\vspace{-0.7cm}

\section{Conclusions}

We have demonstrated that the newly deployed IceCube string is capable of detecting muons and muons coincident with IceTop air showers. The observed muon flux is compatible with the expectation from the simulation. The global detector time calibration uncertainty is 3 ns, which is better than the design requirement of 7 ns.



\newpage
\setcounter{section}{0}
\section*{\Large Calibration and characterization of photomultiplier tubes \\
of the IceCube neutrino detector}

\vskip 0.05cm
{\large H. Miyamoto$^{\it a}$ for the IceCube collaboration\\}
{\it        (a) Dept. of Physics, Chiba University, Chiba 263-8522 Japan
}\\

\vskip 0.05cm
{\large Presenter: H. Miyamoto (miya@hepburn.s.chiba-u.ac.jp), \  
jap-miyamoto-H-abs1-og25-poster}

%
\title[IceCube PMT calibration ...]
{Calibration and characterization of photomultiplier tubes \\
of the IceCube neutrino detector}

\author[H. Miyamoto for the IceCube Collaboration] {H. Miyamoto$^a$ for the IceCube collaboration\\
        (a) Dept. of Physics, Chiba University, Chiba 263-8522 Japan
}
\presenter{Presenter: H. Miyamoto (miya@hepburn.s.chiba-u.ac.jp), \  
jap-miyamoto-H-abs1-og25-poster}


\maketitle

\begin{abstract}

The IceCube neutrino observatory will consist of an InIce array of 4800 Digital 
Optical Modules (DOMs) located in the deep ice at the South Pole, and also an 
IceTop air shower array of 320 DOMs in 160 ice tanks located on the surface.
A 10 inch PMT is housed in each DOM for the detection of Cherenkov light. 
This paper describes the methods of calibration and characterization 
of the IceCube PMTs in the laboratory, which are germane to improving 
the detector resolution and reducing systematic uncertainties. 
Two dimensional scans on the entire photocathode to map out photon conversion 
efficiency have been carried out for 60 PMTs. 
The quantum/collection efficiency has been calibrated in an absolute manner 
using the Rayleigh scattered light from our newly built chamber filled 
with nitrogen gas. 
The charge response of the PMTs at temperatures below freezing has been extensively 
studied and found to be well represented by the analytical model. 
All these results have been combined and implemented in the detector Monte 
Carlo simulation.

\end{abstract}

\vspace{-0.5pc}
\section{\label{sec:intro}Introduction}
\vspace{-0.5pc}

IceCube\cite{IceCube} is a high energy cosmic neutrino detector which will
instrument a cubic kilometer of ice sheet covering the South Pole.
The InIce DOMs will be deployed on eighty strings of sixty DOMs each, and
these strings will be regularly spaced by 125 m over an area of approximately
one square kilometer, with the DOMs at depths ranging from 1.4 km to 2.4 km 
below the ice surface.
An IceCube event consists of Cherenkov photon profiles recorded in 
individual DOMs in the form of digitized waveform output from the PMT. 
An accurate understanding of the PMT response is, therefore, necessary in order
to determine both the geometry of these events as well as to reconstruct their
energies.
Moreover, having carefully calibrated PMTs would lead to a significant reduction
of systematic uncertainties because it is of great help in characterizing the optical
properties of glacial ice more accurately.
In this paper we describe our relative and absolute calibration of the IceCube PMTs. 
Their charge responses and photon detection efficiencies are measured and 
implemented in the detector Monte Carlo simulation.

\vspace{-0.5pc}
\section{\label{sec:DOM}The IceCube Optical Detector: DOM}
\vspace{-0.5pc}

Each DOM contains a 10 inch diameter PMT supported by coupling gel, 
a signal processing electronics board, an LED flasher board for 
calibration, and a high voltage base which powers the PMT, all of 
which are housed in a glass pressure sphere.
It contains its own processor, memory, flash file system, and real-time
operating system. 
Its new digital technology enables it to schedule communications in 
the background while acquiring data, to invoke all calibration functions 
under software control, and most notably, to digitize the PMT pulse and 
store the full waveform information. 

The IceCube PMT is the Hamamatsu R7081-02 with 10 dynodes.
This PMT exhibits excellent charge resolution and low noise.
The PMTs in InIce DOMs are operated nominally at a gain of $\sim 1 \times 10^7$.
They exhibit an excellent peak to valley ratio ($\geq 2.0$) and
remarkably low dark count rate ($\sim 500 Hz$) at sub-freezing temperatures,
both of which have a significant impact on the performance of the IceCube 
detector.

\vspace{-0.5pc}
\section{\label{sec:freezer}Fundamental PMT calibration in a freezer}
\vspace{-0.5pc}

The basic behavior of the PMTs at sub-freezing temperatures should be well 
understood. 
A PMT put in a freezer box is illuminated by diffuse light from a UV LED for 
measuring the PMT gain, single photoelectron (SPE) response, and dark count rate.
An SPE waveform has been found to be represented well by a gaussian with $\sigma\sim 2$ nsec.
Dark count rates above a trigger threshold of 0.3 PE were found to be distributed 
around  $600$ Hz for PMTs with a gain of $5\times 10^7$. 
This corresponds to $\sim 470$ Hz at the operating gain of $1\times 10^7$.

\begin{figure}
\begin{center}
\begin{tabular}{cc}
\includegraphics[width=0.4\textwidth,clip=true]{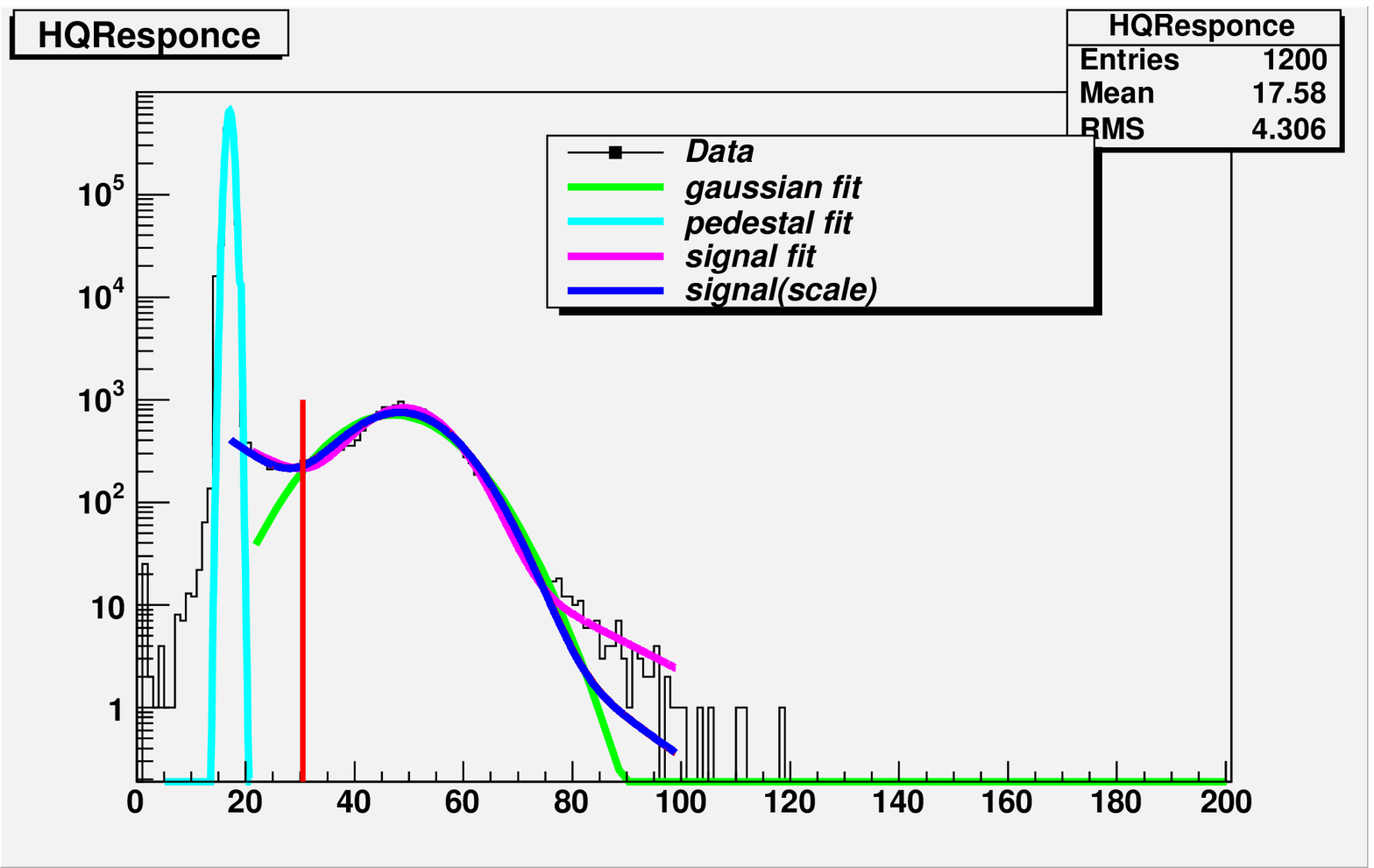} & \
\includegraphics[width=0.4\textwidth,clip=true]{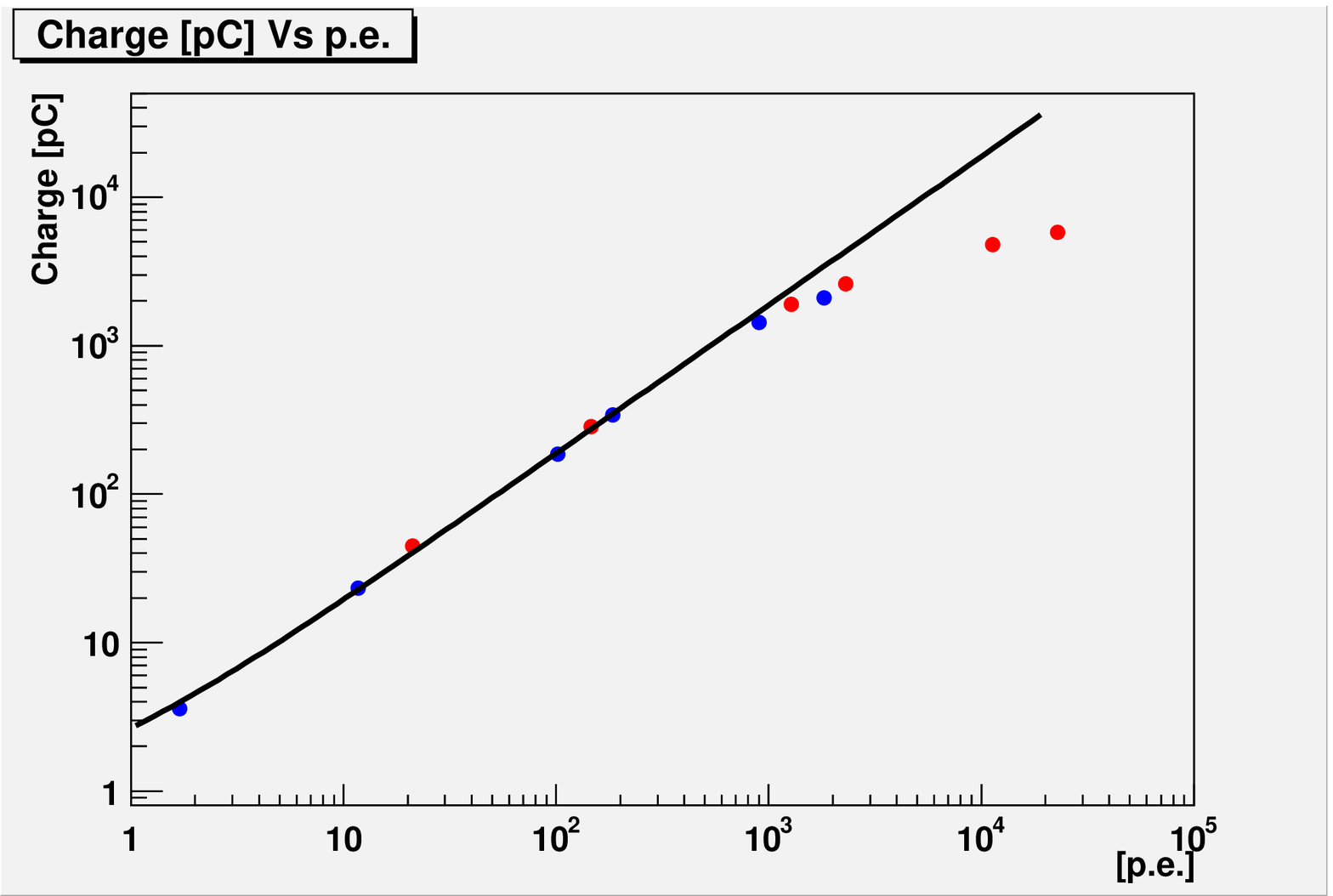} \\
\end{tabular}
\end{center}
 \vspace{-1.5pc}
\caption{Left panel:The charge histogram with the fitted SPE model function.
Right panel:PMT linearity data. 
The output charge is plotted as a function
of the number of photoelectrons corresponding to the LED luminosity.
The PMT gain is $1.2\times 10^7$.}
\label{fig:freezer_figs}
\end{figure}

The charge response for SPE detection has been extensively studied. 
In general it is assumed that the charge of an SPE pulse follows a gaussian distribution. 
However, a large PMT like those used in IceCube is occasionally observed to fail to stream 
secondary electrons out from the first to the second dynode, which results in a non-gaussian 
response.
Here the charge response of the IceCube PMTs is modeled by an exponential term 
in conjunction with the widely-used gaussian term.
The left panel of figure~\ref{fig:freezer_figs} illustrates how well this model 
function represents the charge histogram. 
One of the advantages in this model representation is that the scaling law to PMT 
gain appears valid in the model function.
The relevant model parameters are recorded into a database, which is
used by the detector Monte Carlo simulation.

It is also important to measure the linearity of the PMT response to 
multiple photoelectrons. 
A bright LED with a set of calibrated neutral density filters is used to 
provide various luminosities of UV photons for such a linearity measurement.
The result is shown in the right panel of Figure~\ref{fig:freezer_figs}.
The number of photoelectrons is estimated from the SPE peak and the attenuation
factor of each filter.
It is found that an IceCube PMT has a linear response up to approximately 
one thousand photoelectrons at the gain of $\sim 1 \times 10^7$. 
Since a representative FWHM of the LED light pulse is 30 nsec in this 
measurement, the corresponding current where the PMT deviates from the linear 
response is $\sim 60$ mA. 

\vspace{-0.5pc}
\section{\label{sec:2D} Two dimensional photocathode scan}
\vspace{-0.5pc}

The IceCube PMT collects photoelectrons generated in its relatively large
cathode area and its efficiency depends on the position on the cathode surface.
Moreover, the variance of the electric field inside the tube to collect photoelectrons 
at the dynode gives non-negligible differences in the charge collection efficiency 
from tube to tube. 
We have systematically analyzed the dependency of the overall PMT 
efficiency on the photocathode position, using a two dimensional scan 
system.
A UV LED moves along the curved PMT surface to scan the entire 
photocathode.
The light beam is controlled to be precisely perpendicular to the
cathode glass.
A collimator attached in front of the LED gives a spot size of $\sim$1 mm.
The light spot pointing accuracy is $\leq$ 4.8 mm which is sufficient to scan  
the cathode with its radius of $\sim$ 14 cm. 

\begin{figure}
\begin{center}
\begin{tabular}{ccc}
\includegraphics[width=0.3\textwidth,clip=true]{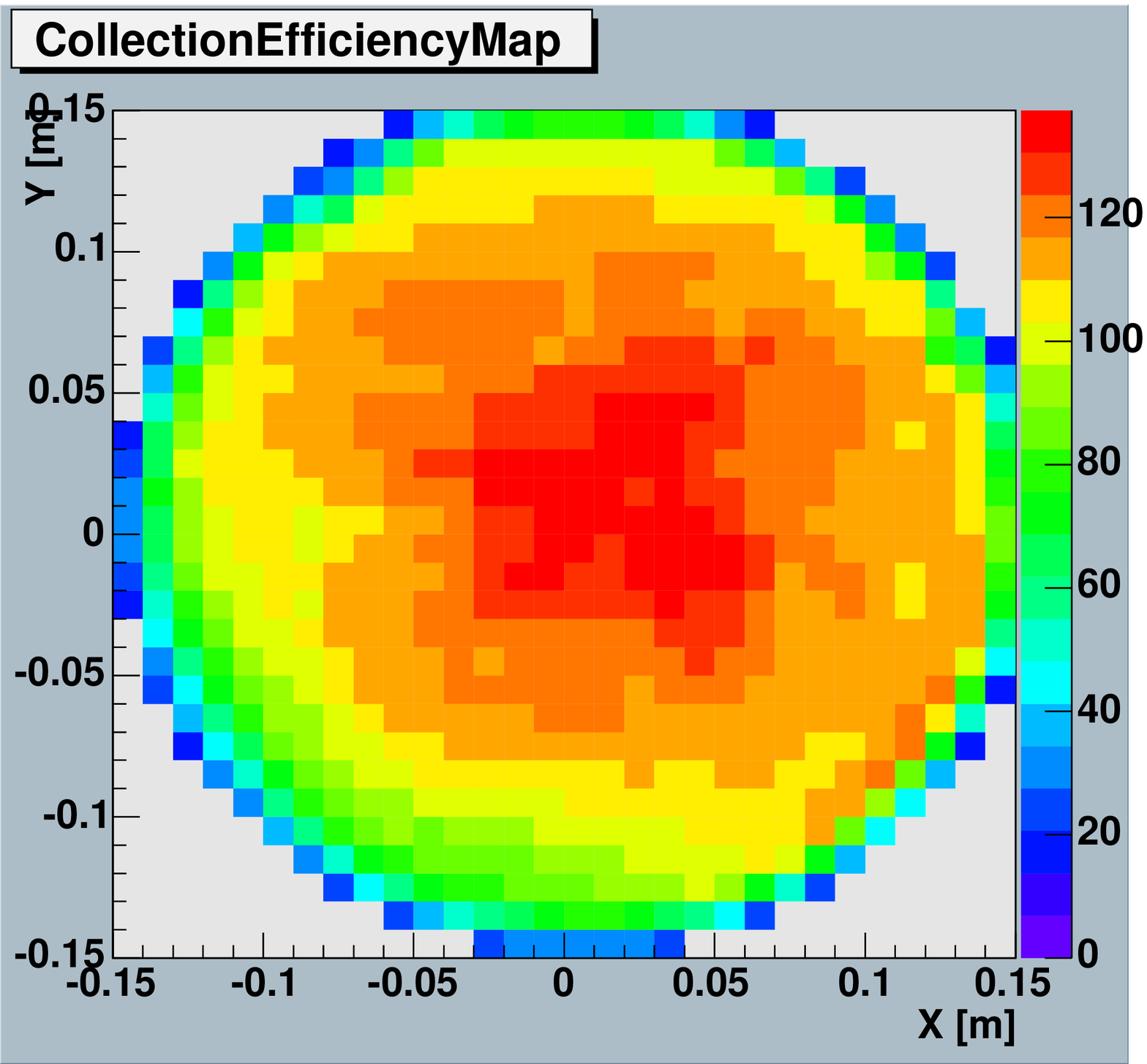} & \
\includegraphics[width=0.3\textwidth,clip=true]{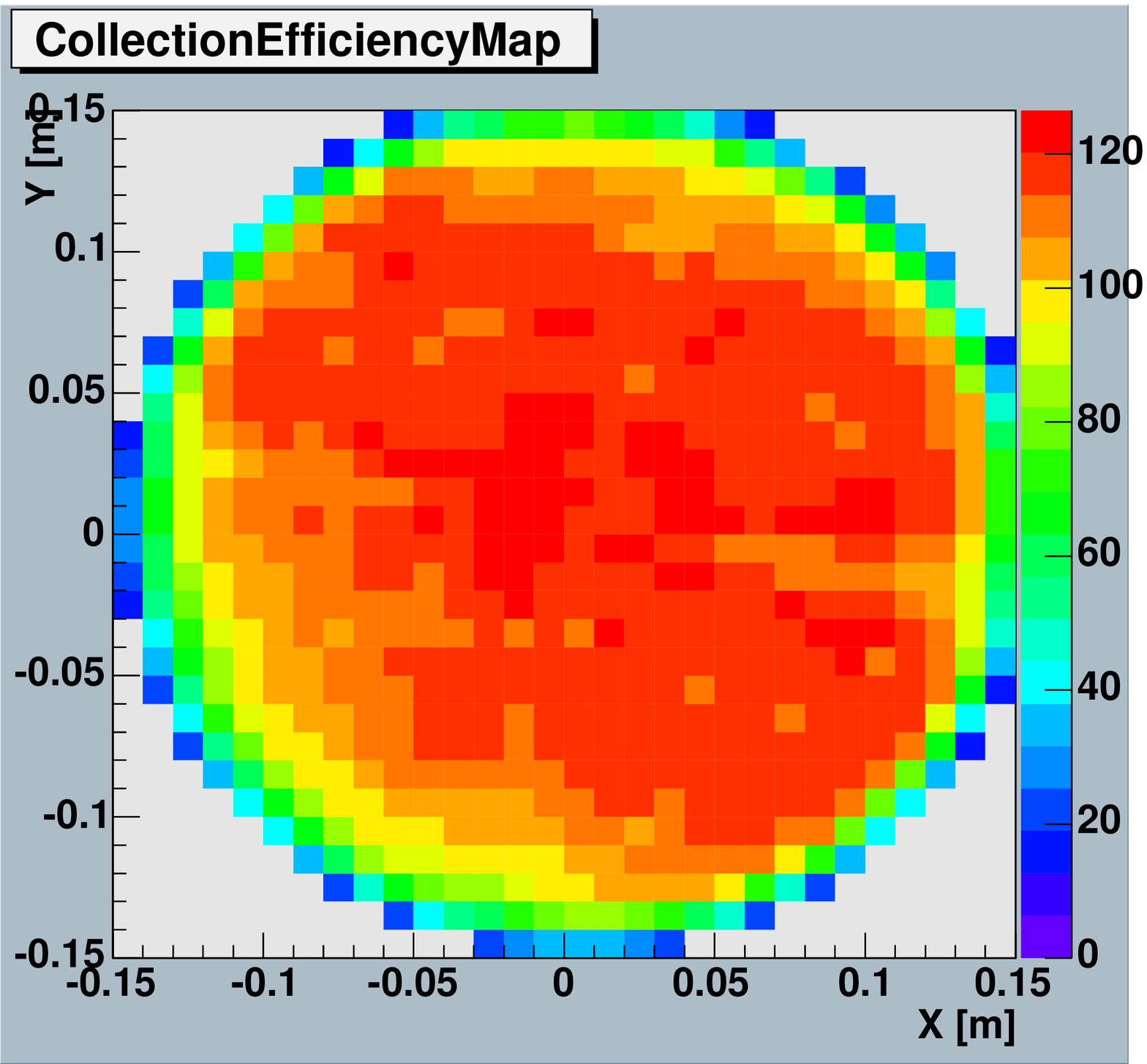} & \
\includegraphics[width=0.3\textwidth,clip=true]{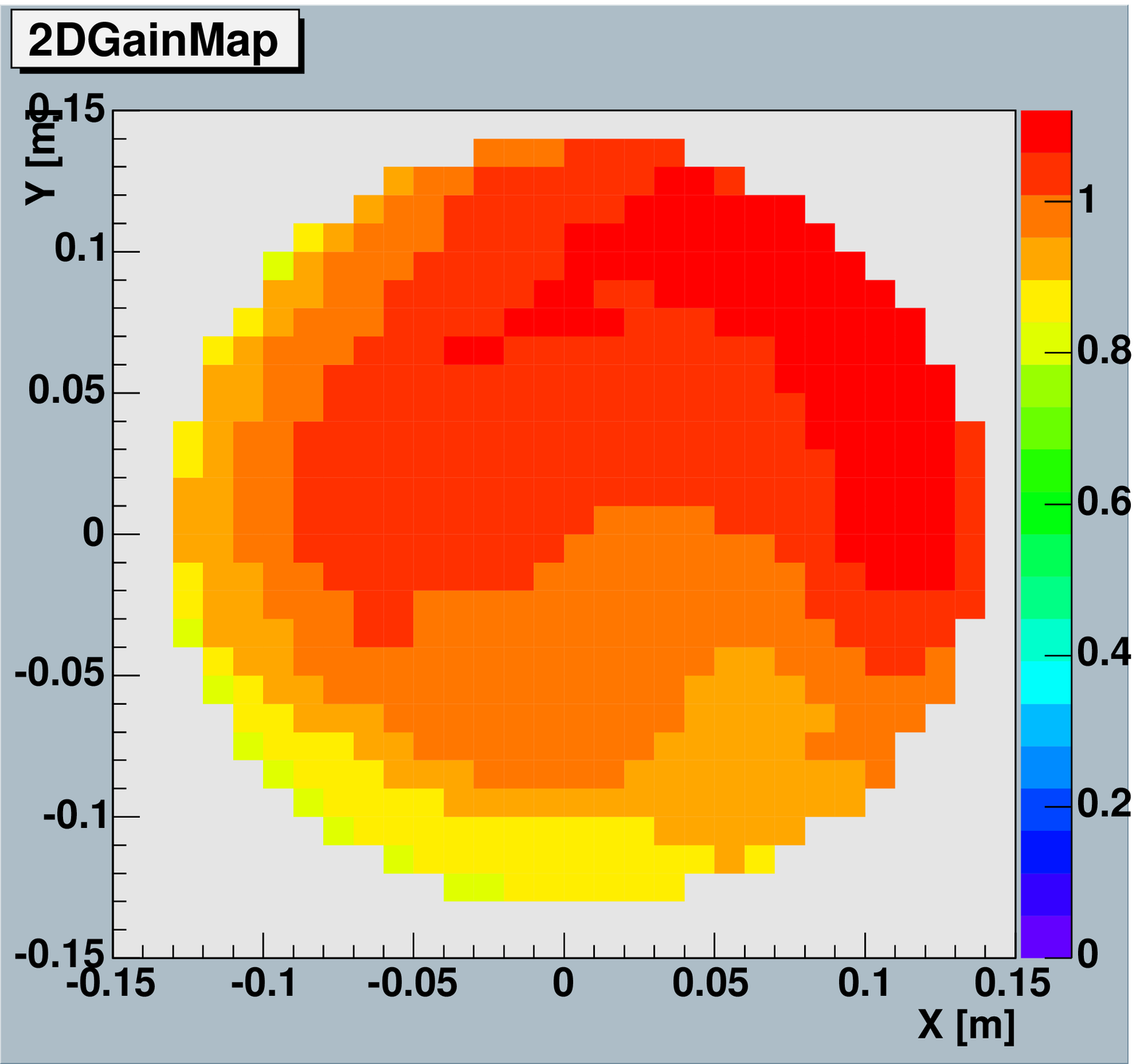} \\
\end{tabular}
\end{center}
 \vspace{-1.5pc}
\caption{Left and middle panels:
Examples of the position dependence of the collection efficiency.
The relative collection efficiency is mapped in the XY plane for two different
PMTs. Right panel: An example of the normalized position dependence of the PMT 
relative gain where the average gain over the cathode is 1.0. }
\label{fig:2d_map}
\end{figure}

Figure~\ref{fig:2d_map} shows some examples.
A relative collection efficiency is shown as a function of the projected X-Y location 
on the cathode surface.
A shield made of $\mu$-Metal is used to negate the geomagnetic field inside the PMT. 
From measuring the effects of rotating a PMT it was determined that the systematic
uncertainty due to the geomagnetic field was $\sim 5$ \%.
On average, an IceCube PMT exhibits non-uniformity at a level of $\sim$ 7 \%.
However, variations from PMT to PMT have been found to be $\sim 40$ \%. 
It is therefore necessary to evaluate how these variations would influence 
the detector resolution. 
The measured data has been integrated into the detector simulation for such a study.

In contrast to the collection efficiency, the gain should be rather uniform. 
Scanning the PMT using a dim LED provides an SPE-based gain map along the photocathode 
surface. 
The right panel of Figure ~\ref{fig:2d_map} shows an example.
The actual (absolute) gain for this particular PMT is approximately 
$4.4 \times 10^7$. 
As expected, it is quite uniform compared to the collection efficiency. 
Moreover, there is no significant variation from PMT to PMT. 
The gain uniformity behavior can be considered universal among the IceCube 
PMTs, which simplifies the detector response simulation.

\vspace{-0.5pc}
\section{\label{sec:absolute_calib} Absolute calibration}
\vspace{-0.5pc}

\begin{figure}
\begin{center}
\begin{tabular}{ccc}
\includegraphics[width=0.3\textwidth,clip=true]{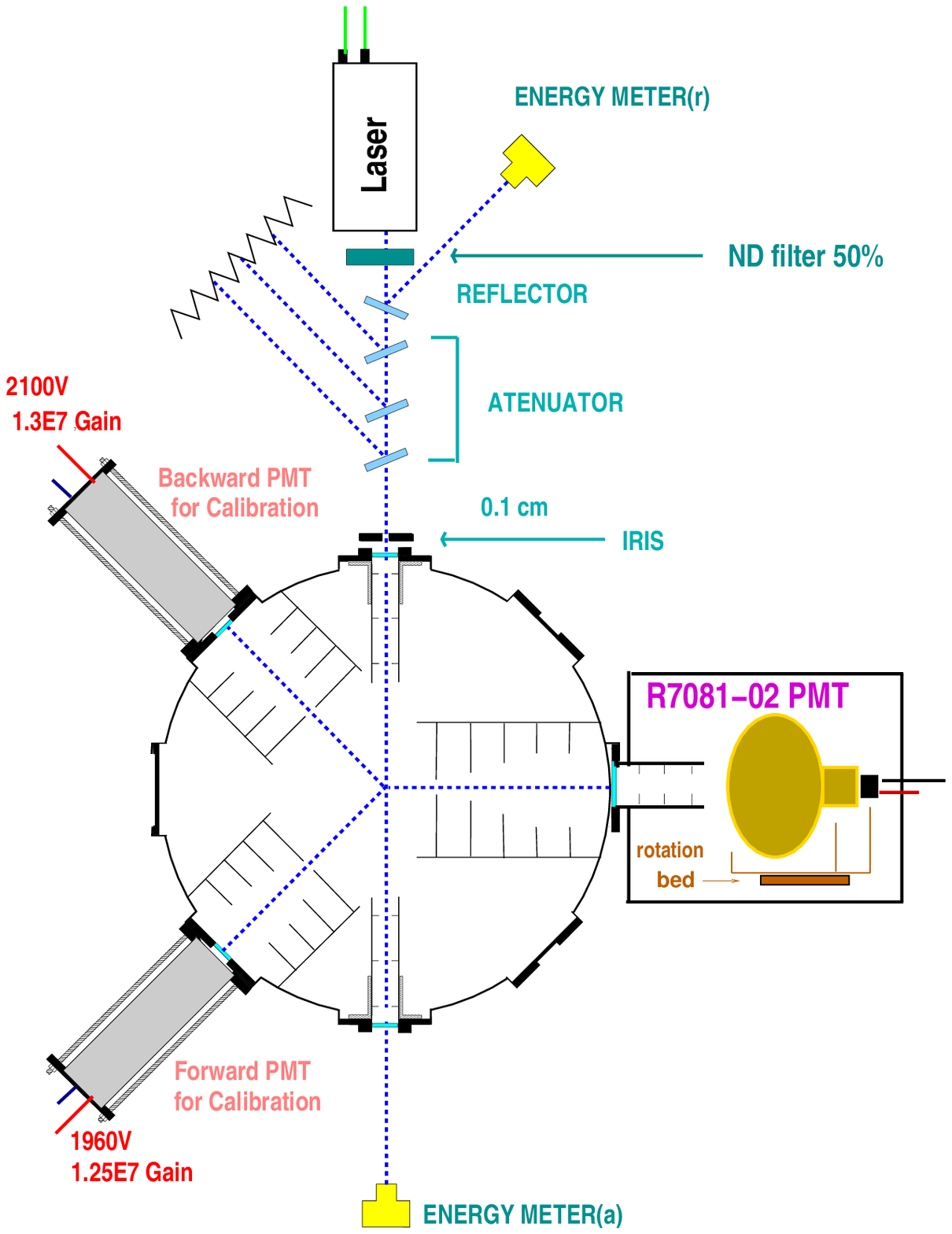} & \
\includegraphics[width=0.4\textwidth,clip=true]{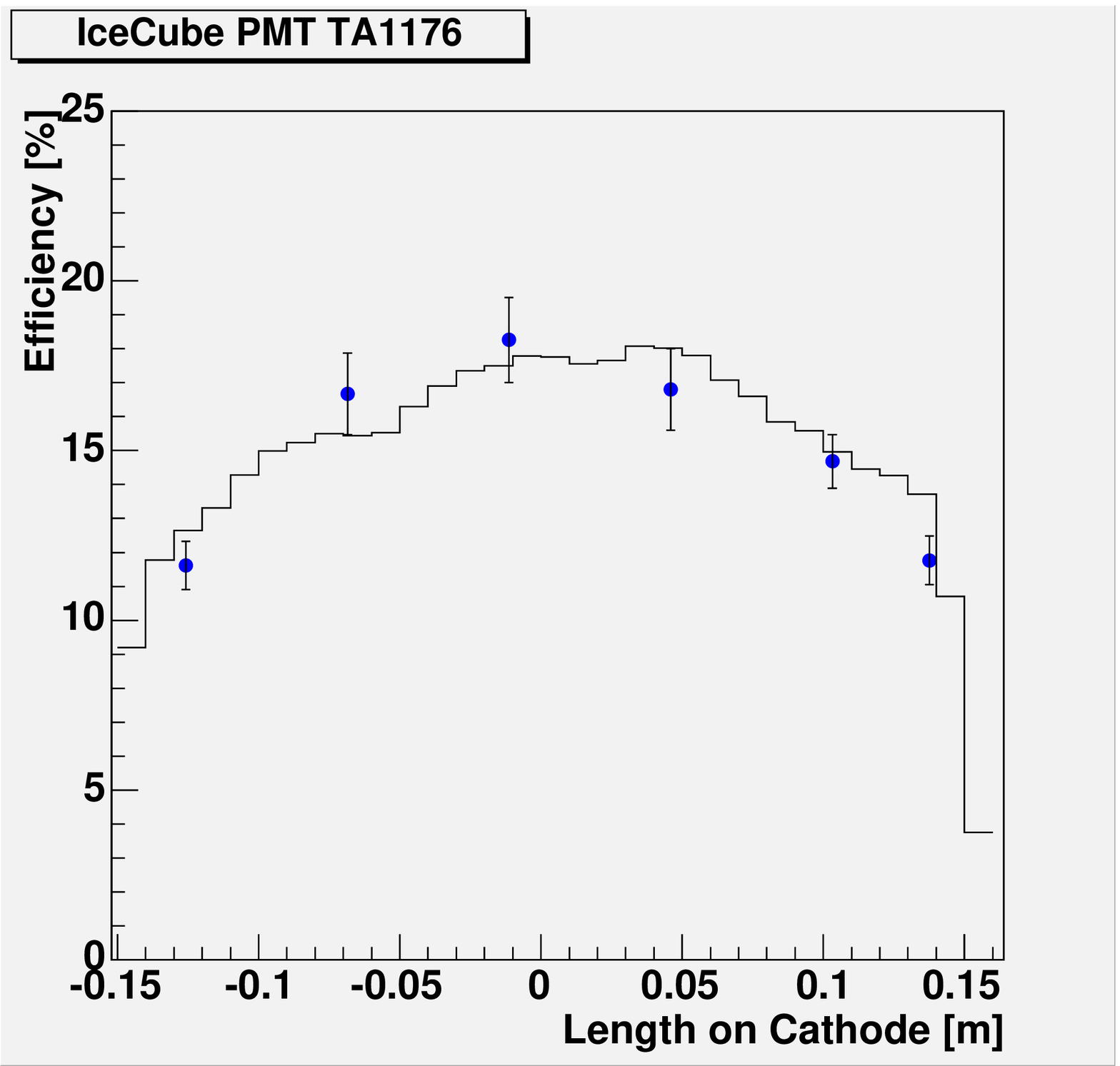} \\
\end{tabular}
\end{center}
 \vspace{-1.5pc}
\caption{Left panel: Schematic view of the absolute calibration system.
Right panel: The detection efficiency for photons with a wavelength of 337 nm 
for one of the IceCube PMTs.
The efficiency is plotted as a function of the distance from the cathode center.
The light spot size is approximately 1.0 cm. 
The histogram shows the two dimensional scan data described in
Section~\ref{sec:2D}, after applying the appropriate normalization.}
\label{fig:absolute_calib_qe}
\end{figure}

The left panel of Figure~\ref{fig:absolute_calib_qe} 
illustrates our absolute calibration system.
A tiny fraction of laser beam photons are scattered inside the pressurized chamber 
filled with nitrogen gas. 
A pressure gauge and a temperature sensor estimate the density of the gas. 
The laser beam energy is measured by the silicon energy meter. 
Because the wavelength of the laser photon is a monochromatic 337 nm, 
this energy can be converted to the number of photons injected by the laser. 
Only photons with scatter angles of $\sim 90^{\circ}$ will reach the IceCube 
PMT photocathode to provide an SPE signal, while scattered photons in the other 
directions will eventually be absorbed by a number of baffles inside the chamber. 
The spot size of scattered photons which illuminate the PMT cathode is approximately 1.5 cm. 
The PMT is rotated inside the dark box so that the detection efficiency can be 
measured from position to position on the photocathode surface.
The number of scattered photons which reach the PMT cathode, $N_{\gamma}$, 
is determined by the geometrical acceptance of the PMT, the well-understood
Rayleigh scattering cross section, the number density of the nitrogen gas, 
and the number of photons injected by the laser.

The average number of photoelectrons $N_{pe}$ detected by a PMT is obtained
from the ratio of the number of SPE and multiple photoelectron events $N_{SPE}$
to all events externally triggered by the synchronized gate pulse from the laser. 
Our definition of an SPE event here is those events which produce charge $q$ greater 
than $q_{th}=0.5$ photoelectrons, which can be clearly discriminated in the charge 
histogram taken by the CAMAC ADC. 
The values for $q_{th}=0.0$ photoelectrons have been extrapolated from the
measured values using above described charge response model. 
The PMT photon detection efficiency is then calculated from $N_{pe}$ and $N_{\gamma}$.
The relative statistical and systematic errors, $\Delta Q_{eff}/Q_{eff}$ of the 
present calibration are $\sim$ 5 \% and $\sim$ 7 \%, respectively.

\begin{table}
\caption{The evaluated absolute photon detection efficiency for 4 different PMTs.}
\label{table:qe_summary}
\begin{tabular}{@{}lllll}
\hline
PMT number & $Q_{eff}^{center}(q_{th}=0.5 p.e.)$ & \
$Q_{eff}^{whole}(q_{th}=0.5 p.e.)$ & $Q_{eff}^{whole}(q_{th}= 0.0 p.e.)$ & 
Error (stat) \\\hline
TA1052 & 19.4 \% & 16.4 \%  & 23.5 \% & $\pm 1.3$ \% \\
TA1062 & 19.2 \% & 16.7 \%  & 22.9 \% & $\pm 1.3$ \% \\
TA1176 & 17.5 \% & 13.9 \%  & 19.7 \% & $\pm 1.2$ \% \\
TA1167 & 18.2 \% & 16.4 \%  & 22.9 \% & $\pm 1.3$ \% \\
\hline
\end{tabular}\\[1pt]
\end{table}

The data points in the right panel of figure ~\ref{fig:absolute_calib_qe} show the 
determined detection efficiency $Q_{eff}$ as a function of the distance from the cathode 
center.
The detection efficiency in the central area of the photocathode is measured to be 
18 - 20 \%.
This is consistent with measurements by Hamamatsu which are based on DC light exposure.
The histogram in figure~\ref{fig:absolute_calib_qe} shows the two dimensionally scanned relative collection efficiency.
The locally measured $Q_{eff}$ data points agree very well with this histogram.
Consequently, this allows us to estimate the averaged $Q_{eff}$ over a certain area of 
the photocathode surface.
Table~\ref{table:qe_summary} shows $Q_{eff}$ in the central point
and those averaged over the entire photocathode.

As mentioned earlier, the charge response model for calibrating
PMTs in the dark freezer box described in Section~\ref{sec:freezer}
can determine $Q_{eff}$ with lower thresholds of photoelectron charge.
The non-biased efficiency $Q_{eff}(q_{th}= 0.0 p.e.)$ would be a useful
quantity in implementing the detector Monte Carlo simulation
and these values are also listed in the table.



\newpage
\setcounter{section}{0}
\section*{\Large Simulation of a Hybrid Optical/Radio/Acoustic Extension to IceCube for EeV Neutrino Detection}

\vskip 0.05cm
{\large D. Besson$^{\it a}$, S. B\"{o}ser$^{\it b}$, R. Nahnhauer$^{\it b}$, P.B. Price$^{\it c}$, and
	J. A. Vandenbroucke$^{\it c}$ (justin@amanda.berkeley.edu) for the IceCube Collaboration\\}
{\it        (a) Dept. of Physics and Astronomy, University of Kansas, Lawrence,
            KS 66045-2151, USA \\
        (b) DESY, D-15738 Zeuthen, Germany \\
        (c) Dept. of Physics, University of California, Berkeley, CA 94720, USA
        }

\vskip 0.05cm
{\large Presenter: R. Nahnhauer, \
ger-nahnhauer-R-abs1-og25-oral}

\title[Simulation of a Hybrid Extenstion to IceCube ...]{Simulation of a Hybrid Optical/Radio/Acoustic Extension to IceCube for EeV Neutrino Detection}

\author[D. Besson et al.] { D. Besson$^a$, S. B\"{o}ser$^b$, R. Nahnhauer$^b$, P.B. Price$^c$, and
	\newauthor
	J. A. Vandenbroucke$^c$ (justin@amanda.berkeley.edu) for the IceCube Collaboration\\
        (a) Dept. of Physics and Astronomy, University of Kansas, Lawrence,
            KS 66045-2151, USA \\
        (b) DESY, D-15738 Zeuthen, Germany \\
        (c) Dept. of Physics, University of California, Berkeley, CA 94720, USA
        }
\presenter{Presenter: R. Nahnhauer, \
ger-nahnhauer-R-abs1-og25-oral}

\maketitle

\begin{abstract}

Astrophysical neutrinos at $\sim$EeV energies promise to be an interesting source for 
astrophysics and particle physics. Detecting the predicted 
cosmogenic (``GZK'') neutrinos at 10$^{16}$ - 10$^{20}$ eV would test models of 
cosmic ray production at these energies and probe particle physics at $\sim$100~TeV
center-of-mass energy. While IceCube could detect $\sim$1 GZK event per year, it 
is necessary to detect 10 or more events per year in order to study temporal, 
angular, and spectral distributions. The IceCube observatory may be able to 
achieve such event rates with an extension including optical, radio, and 
acoustic receivers.  We present results from simulating such a hybrid detector. 

\end{abstract}

\section{Introduction}
Detecting and characterizing astrophysical neutrinos in the 10$^{16}$~eV to 10$^{20}$~eV range is a central 
goal of astro-particle
physics.  The more optimistic flux models in this range involve discovery physics including topological defects and
relic neutrinos.  Detecting the smaller flux of cosmogenic 
(or Greisen, Zatsepin, and Kusmin, ``GZK'')
neutrinos produced via ultra-high energy cosmic ray interaction with the cosmic microwave background 
 would test models
of cosmic ray production and propagation and of particle physics at extreme energies.
  With $\sim$100 detected events, their angular distribution
would give a measurement of the total neutrino-nucleon cross section at $\sim$100 TeV center of mass,
probing an energy scale well beyond the reach of the LHC.  Hence, as a baseline, a detector capable 
of detecting $\sim$10 GZK events per year
has promising basic physics potential.  If any of the more exotic theories predicting greater EeV neutrino
fluxes is correct,
the argument in favor of such a detector is even stronger. 

To detect $\sim$10 GZK events per year,
a detector with an effective volume of $\sim$100~km$^3$ at EeV energies 
is necessary.
In addition to the possibility of identifying neutrino-induced inclined air showers, there are three 
methods of ultra-high energy neutrino detection in solid media: optical, 
radio, and acoustic.  Optical Cherenkov detection is a well-established 
technique that has detected 
atmospheric neutrinos up to 10$^{14}$ eV and set limits up to
10$^{18}$~eV \cite{Chirkin}.  Radio efforts 
have produced steadily improving upper limits on neutrino fluxes from  10$^{16}$~eV to
10$^{25}$~eV \cite{radio}. Acoustic detection efforts are at an earlier stage, with one limit published thus far 
from 10$^{22}$ to 10$^{25}$~eV
 \cite{Vandenbroucke05}.

The currently planned 1~km$^3$ 
optical neutrino telescopes expect a GZK event rate of $\sim$1 per year.
It is possible to extend this by adding more optical strings for a modest
additional cost \cite{Halzen1}, but it's difficult to imagine achieving 10 or more events per year with
optical strings alone.  The radio and acoustic methods have potentially large effective volumes with relatively
few receivers, but the methods are unproven in that they have never detected a neutrino.  Indeed, if
radio experiments claim detection of a GZK signal, it may be difficult to confirm that it is really a
neutrino signal.
However, it may be possible to bootstrap the large effective volumes of radio and acoustic detection
with the optical method, by building a hybrid detector that can detect a large
rate of radio or acoustic events, a fraction of which are also detected by an optical detector.  
A signal seen
in coincidence between any two of the three methods
would be convincing. 
The information from multiple methods can be combined for hybrid reconstruction, yielding improved
angular and energy resolution.

We simulated the sensitivity of a detector that could be constructed by expanding the IceCube observatory 
currently under construction 
at the South Pole.  The ice at the South Pole is likely well-suited for all three methods:  Its optical clarity
has been established by the AMANDA experiment \cite{Chirkin}, and its radio clarity and suitability for radio detection in the 
GZK energy range has been established by the RICE experiment \cite{radio}.  
Acoustically, the signal in ice is ten times greater than that in water.
Theoretical estimates indicate 
low attenuation and noise
\cite{Price05}, and efforts are planned to measure both \cite{SPATS05} 
with sensitive transducers
developed for glacial ice \cite{Nahnhauer05}.
Here we estimate the sensitivity of such a detector by exposing all three components to a common
Monte Carlo event set and counting events
detected by each method alone and by each combination of multiple methods.

\section{Simulation}

IceCube will have 80 strings arrayed hexagonally with a horizontal spacing of 125~m.  In \cite{Halzen1},
the GZK sensitivity achieved by adding more optical strings at larger distances (``IceCube-Plus'')
was estimated, and the possibility of also adding radio and acoustic modules was mentioned. 
Here we consider an IceCube-Plus configuration consisting of 
a ``small'' optical array overlapped by a ``large'' acoustic/radio array
with a similar number of strings but larger horizontal spacing.  
The optimal string spacing for GZK detection was found to be $\sim$1~km
for both radio and acoustic strings.
This coincidence allows the two methods to share hole drilling and cable
costs, both of which are dominant costs of such arrays.

\begin{figure}
\begin{minipage}[t]{7cm}
\includegraphics[width=1.0\linewidth,height=5cm]{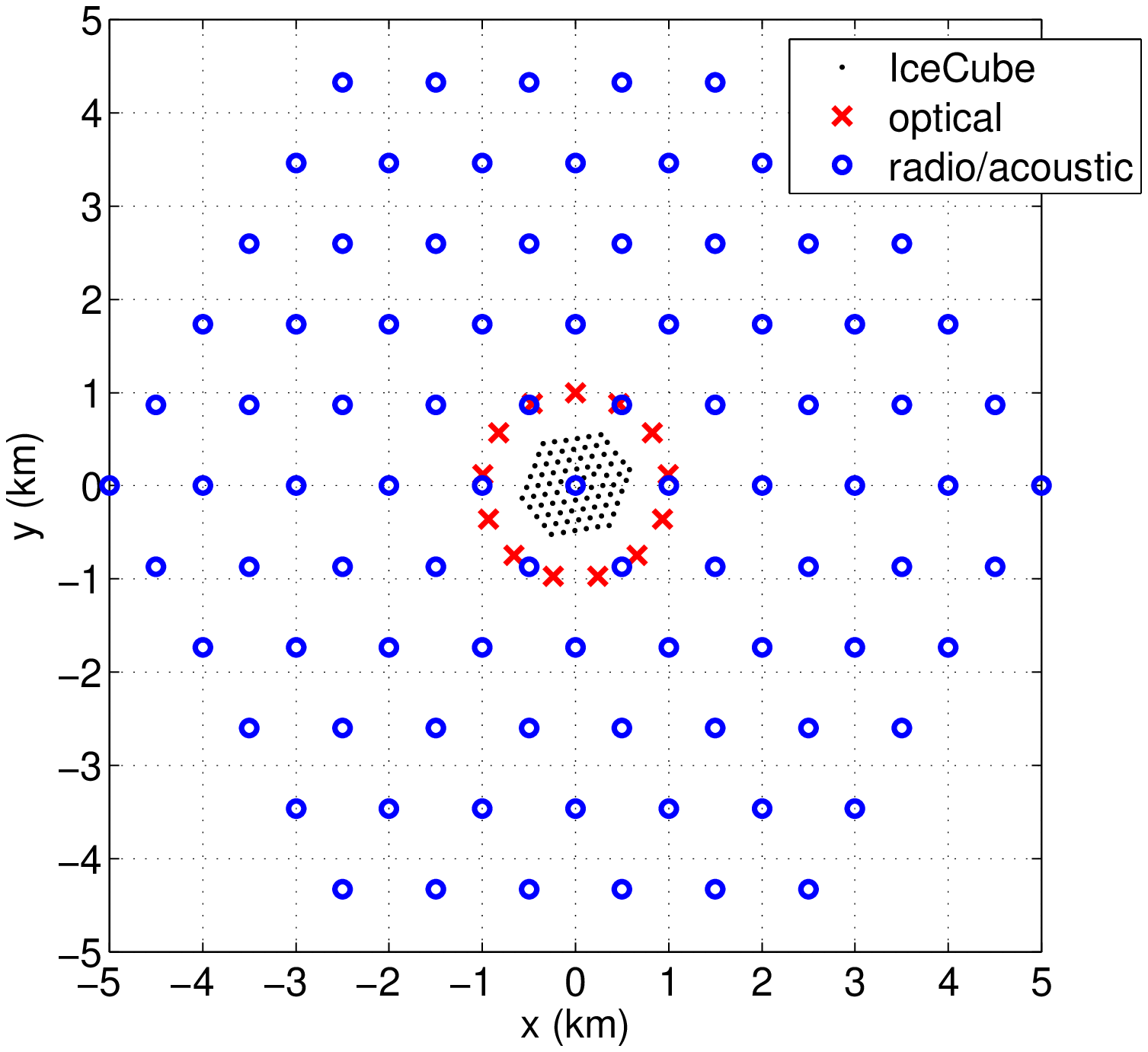}
\caption{\label {fig1e} Geometry of the simulated hybrid array.}
\end{minipage}
\hfill
\begin{minipage}[t]{8.5cm}
\includegraphics[width=1.0\linewidth,height=5.cm]{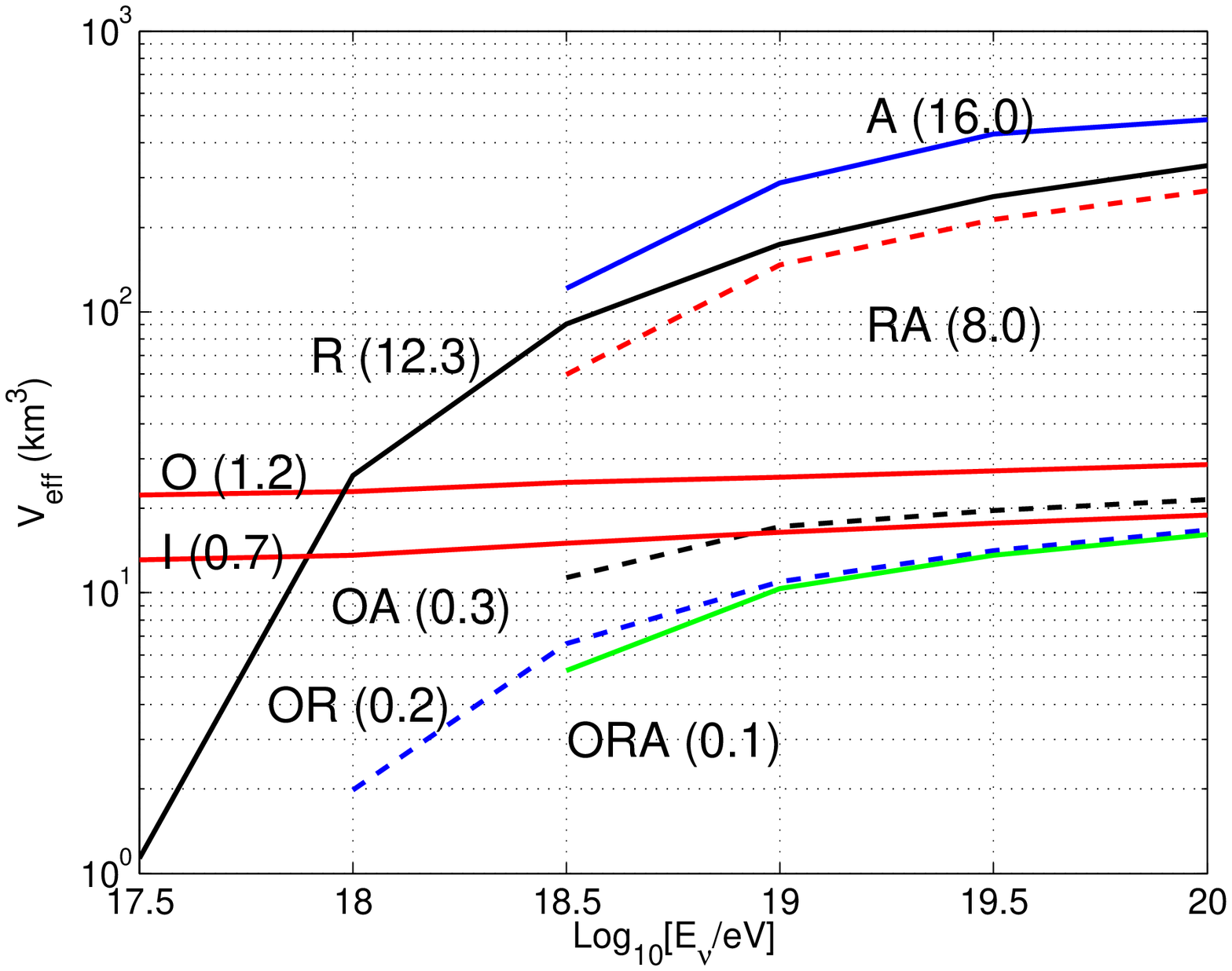}
\vskip -0.5cm
\caption{\label{fig2f} Effective volume for each of the seven
combinations of detector components, as well as for IceCube alone (``I'').
GZK event rates per year are given in parenthesis.  Note that different channels
were used for different combinations (see text).}
\end{minipage}
\hfill
\end{figure}

The geometry of the simulated array is shown in Fig. \ref{fig1e}.
We take the optical array to be IceCube as well as a ring
of 13 optical strings with a 1~km radius, surrounding IceCube.
All optical strings have standard IceCube geometry: 60 modules per
string, spaced every 17 m, from 1.4 to 2.4~km depth.
Encompassing this is a hexagonal array of 91 radio/acoustic strings
with 1~km spacing.  Each radio/acoustic hole has 5 radio receivers, spaced every 100~m from 
200~m to 600~m depth, and 300 acoustic receivers, spaced every 5~m from 5~m to 1500~m depth.  
At greater depths both methods suffer increased absorption due to the warmer ice.  The 
large acoustic density per string is necessary because the acoustic
radiation pattern is thin (only $\sim$10~m thick) in the direction along the shower.  
The array geometry was designed to seek
an event rate of $\sim$10 GZK events per year detectable with both radio and acoustic
independently.
To obtain rough event rate estimates, a very simple Monte Carlo generation scheme was chosen.  
Between 10$^{16}$ and 10$^{20}$~eV, the neutrino interaction length ranges between 6000 and 200~km
\cite{Gandhi},
so upgoing
neutrinos are efficiently absorbed by the Earth and only downgoing events are detectable.  A
full simulation would include the energy-dependent slow rolloff at the horizon.  Here we
assume all upgoing neutrinos are absorbed before reaching the fiducial volume, and no
downgoing neutrinos are; we generate incident neutrino directions isotropically in 2$\pi$~sr.
Vertices are also generated uniformly in a fiducial cylinder of radius 10 km, extending
from the surface to 3 km depth.  

The Bjorken parameter $y = E_{had}/E_{\nu}$ varies
somewhat with energy and from event to event, but we choose the mean value, $y = $ 0.2, for simplicity.
The optical method can detect both muons and showers, but here we only
consider the muon channel; simulation of the shower channel is in progress.  
The radio and acoustic methods cannot detect muon tracks but
can detect electromagnetic and hadronic showers. Under our assumptions of constant $y$
and no event-to-event fluctuations, 
all flavors interacting via both CC and NC produce the same hadronic shower.  Electron
neutrinos interacting via the charged current also produce  an 
electromagnetic cascade which produces radio and acoustic
signals superposed on the hadronic signals.  However, at the energies of interest here,
electromagnetic showers are lengthened to hundreds of meters by the Landau-Pomeranchuk-Migdal effect.  
This weakens their radio and acoustic signals significantly, and we assume
they are negligible.

For simulation of the optical response, the standard Monte Carlo chain 
used in current AMANDA-IceCube analyses \cite{Chirkin} was performed.
After the primary trigger requiring any 5 hits in a 2.5~$\mu$s 
window, a
local coincidence trigger was applied: Ten local coincidences were required, where
a local coincidence is at least two hits on neighboring or next-to-neighboring 
modules within 1~$\mu$s. Compared with \cite{Halzen1}, we used an updated ice
model with increased absorption, which may account for our factor of $\sim$2 lower effective
volume.

Each simulated radio ``receiver'' consists of two vertical
half-wave dipole antennas separated vertically by 5~m to allow local rejection
of down-going anthropogenic noise. We assume an effective height at the
peak frequency (280~MHz in ice) equal to 10~cm, with
$\pm$20\% bandwidth to the --3 dB points. As currently under development
for RICE-II, we assume optical fiber transport of the signal to the
DAQ, with losses of 1~dB/km (measured) through the fiber.
The electric field strength $E(\omega)$ is calculated
from the shower according to the ZHS prescription \cite{ZHS91,Alvarez-Muniz}.  Frequency-dependent 
ice attenuation effects are incorporated using measurements at
South Pole Station \cite{Barwick04}. The signal at the surface electronics 
is then transformed into the time domain, resulting in a waveform 10~ns long,
sampled at 0.5~ns intervals, at each antenna. Two
receivers with signals exceeding 3.5 times the
estimated rms noise temperature $\sigma_{kT}$ (thermal plus a system
temperature of 100~K) within a time window of 30~$\mu$s are required to trigger. 

The unattenuated acoustic pulse $P(t)$ produced 
at arbitrary position with
respect to a hadronic cascade is calculated by integrating over the cascade energy distribution.
The cascade is parametrized with the Nishimura-Kamata-Greisen parametrization, with 
$\lambda$ (longitudinal
tail length) parametrized from \cite{Alvarez-Muniz}.
The dominant mechanism of acoustic wave absorption in South Pole ice is theorized \cite{Price05} to be 
molecular reorientation, which increases with ice temperature.  Using a temperature profile 
measured at the South Pole
along with laboratory absorption measurements, an absorption vs. depth profile was estimated.
The predicted absorption length ranges from 8.6~km at the surface to 4.8~km at 1~km depth to 0.7~km at 
2~km depth.
The frequency-independent absorption is integrated from source to receiver and applied in the time domain.

South Pole ice is predicted to be much quieter than
ocean water at the relevant frequencies ($\sim$10-60 kHz), because there are no waves, currents, or animals.  Anthropogenic
surface noise will largely be waveguided back up to the surface due to the sound speed gradient
in the upper 200 m of uncompactified snow (``firn'').
For the current simulation we assume ambient noise
is negligible compared to transducer self-noise.  
Work is underway to produce transducers with self-noise at the 2-5~mPa level \cite{Nahnhauer05}. 
For comparison, ambient noise in the ocean is $\sim$100 mPa \cite{Vandenbroucke05}.
The acoustic trigger used in this simulation required that 3
receivers detect pressure pulses above a threshold of 9~mPa. 

\section{Results and Conclusion}

Ten-thousand events were generated at each half-decade in neutrino energy 
in a cylinder of volume 942~km$^3$.  For each
method and combination of methods, the number of detected events was used to
calculate effective volume as a function of neutrino energy (Fig. \ref{fig2f}).  This was folded with the
GZK flux model of \cite{ESS,Seckel} and the cross-section
parametrizations of \cite{Gandhi} to estimate detectable
event rates (Fig. \ref{fig2f}).  We use a flux model which assumes source evolution
according to $\Omega_{\Lambda}=$~0.7.  This model is a factor of $\sim$2
greater than that for $\Omega_{\Lambda}=$~0 evolution; it is unclear which model is
correct \cite{Seckel}. For radio and acoustic, and their combination,
all flavors and both interactions were included.  For those combinations including
the optical method, only the muon channel has been simulated thus far; including
showers will increase event rates for these combinations.  




It may be possible to build an extension like that considered here for a relatively small cost.
Holes for radio antennas and acoustic transducers can be narrow and shallow, and both
devices are simpler than photo-multiplier tubes.
The necessarily large acoustic channel multiplicity is partially offset by the fact that the 
acoustic signals are slower by five orders of magnitude, making data acquisition and processing easier.

The IceCube observatory will observe the neutrino universe from 10's of 
GeV to 100's of PeV.
Our simulations indicate that extending it with radio and acoustic strings could produce a neutrino detector
competitive with other projects optimized for high-statistics measurements of GZK neutrinos but with the unique 
advantage of cross-calibration via
coincident optical-radio, optical-acoustic, and radio-acoustic events.

\end{document}